\documentclass[11pt]{article}
\usepackage[utf8]{inputenc}
\usepackage{textcomp}
\usepackage{amsfonts}
\usepackage{amsmath}
\usepackage{amssymb}
\usepackage{mathrsfs}
\usepackage{bbold}
\usepackage[margin=2.2cm]{geometry}
\usepackage{longtable}
\usepackage{lscape,graphicx}
\usepackage{graphics}
\usepackage{color}
\usepackage{array}
\usepackage{makecell}
\usepackage{upgreek}
\usepackage{physics}
\usepackage{bm}
\usepackage{comment}
\usepackage{multicol}
\usepackage{blindtext}
\usepackage{float}
\usepackage{hyperref}
\usepackage{cleveref}
\usepackage{hhline}
\usepackage{multirow}

\usepackage[numbers, sort&compress]{natbib}

\usepackage{slashed}

\usepackage{sidecap}
\usepackage{caption}
\usepackage{subcaption}

\newcommand{\ba}{\begin{array}{c}}
\newcommand{\ea}{\end{array}}

\usepackage[dvipsnames]{xcolor}

\newcommand{\pN}{{\rm p}_N}
\newcommand{\cprod}{c_{\rm prod}}
\newcommand{\cdec}{c_{\rm dec}}
\newcommand{\BR}{{\rm B}^{Z}_\nu}

\renewcommand{\aa}{{\rm a}}

\newcommand{\ratio}[1]{\frac{U_{#1}^2}{U^2}}
\newcommand{\decay}{l}

\begin{document}
\begin{titlepage}
\vspace*{-1cm}
\phantom{hep-ph/***}
\flushright

\vskip 1.5cm
\begin{center}
{\LARGE\bf 
Low-scale seesaw with flavour and CP symmetries\\-- from colliders to leptogenesis
}
\vskip .3cm
\end{center}
\vskip 0.5  cm
\begin{center}
{\large M.~Drewes}$^{1}$,
{\large Y.~Georis}$^{1}$,
{\large C.~Hagedorn}$^{2}$,
{\large J.~Klarić}$^{3,4,5,1}$
\\
\vskip .7cm
{\footnotesize
$^{1}$ Centre for Cosmology, Particle Physics and Phenomenology, Université catholique de Louvain, Louvain-la-Neuve B-1348, Belgium\\[0.3cm]
$^{2}$ Instituto de F\'isica Corpuscular, Universidad de Valencia and CSIC,
Edificio Institutos Investigaci\'on, Catedr\'atico Jos\'e Beltr\'an 2, 46980 Paterna, Spain\\[0.3cm]
$^{3}$ Institute of Physics and Delta Institute for Theoretical Physics, University of Amsterdam, Science Park 904, 1098 XH Amsterdam, The Netherlands\\[0.3cm]
$^{4}$Theory Group, Nikhef, Science Park 105, 1098 XG, Amsterdam, The Netherlands\\[0.3cm]
$^{5}$Department of Physics, Faculty of Science, University of Zagreb, 10000 Zagreb, Croatia\\[0.3cm]
\vskip .5cm
\begin{minipage}[l]{.9\textwidth}
\begin{center}
\textit{E-mail:}
\tt{marco.drewes@uclouvain.be}, \tt{yannis.georis@uclouvain.be}, \tt{claudia.hagedorn@ific.uv.es}, \tt{juraj.klaric@nikhef.nl}
\end{center}
\end{minipage}
}
\\
\end{center}
\vskip 1cm
\begin{abstract}
We consider an extension of the Standard Model with three right-handed neutrinos, endowed with a flavour symmetry $G_f$, $G_f=\Delta (3 \, n^2)$ or $G_f=\Delta (6 \, n^2)$, $n \geq 2$, and CP. For large active-sterile mixing, we study the properties of the (nearly mass-degenerate) heavy neutrinos, such as their lifetimes and branching ratios. In doing so, we examine the four different cases, called Case 1) through Case 3 b.1), that lead to distinct lepton mixing patterns, all potentially compatible with current data. Furthermore, we comprehensively explore for each case the region of parameter space in which a sufficient amount of baryon asymmetry of the Universe can be generated via leptogenesis, while being testable at accelerator-based and potentially also precision flavour experiments.
\end{abstract}
\end{titlepage}
\setcounter{footnote}{0}

\setcounter{tocdepth}{2}
\tableofcontents

\newpage
\section{Introduction}
\label{intro}

The Standard Model (SM) of particle physics is very successful in describing gauge interactions, but it does not provide insights into the origin of neutrino masses, the organising principle that governs the flavour sector nor into the generation of the observed baryon asymmetry of the Universe (BAU).

The extension of the SM with at least two gauge singlet fermions, also called right-handed (RH) neutrinos, offers probably the most straightforward way to generate neutrino masses via the type-I seesaw mechanism~\cite{Minkowski:1977sc,Yanagida:1979as,Glashow:1979nm,Gell-Mann:1979vob,Mohapatra:1979ia}. These gauge singlet fermions have masses that preclude any (direct) test in laboratory experiments, if their couplings are of order one and no symmetry is implemented. If instead (a generalised) lepton number is nearly preserved~\cite{Shaposhnikov:2006nn,Kersten:2007vk,Moffat:2017feq}, their masses can be within the reach of accelerator-based experiments and tested with precision flavour facilities. A prototype of such an extension is the so-called Neutrino Minimal SM ($\nu$MSM)~\cite{Asaka:2005pn,Asaka:2005an} that, in addition to producing neutrino masses, also addresses the generation of the BAU and Dark Matter (DM) \cite{Canetti:2012kh,Ghiglieri:2020ulj}. In~\cite{Abada:2018oly,Drewes:2021nqr} the viable parameter space for leptogenesis has been explored in a model with three RH neutrinos. Recent studies of resonant leptogenesis with three RH neutrinos can be found in  e.g.~\cite{daSilva:2022mrx,Kang:2022psa,Zhao:2024asa}. In~\cite{Drewes:2024bla} a study of signals at accelerator-based experiments has been performed in a model with three RH neutrinos.

Regarding the flavour structure of the lepton sector, a promising approach is to impose a discrete non-abelian symmetry $G_f$ on the flavour space, often combined with CP, that is broken to different residual groups in the charged lepton and the neutrino sector; for reviews see~\cite{Ishimori:2010au,King:2013eh,Feruglio:2019ybq,Grimus:2011fk}. In particular, the combination of a flavour and a CP symmetry can lead to lepton mixing that only depends on one free angle~\cite{Feruglio:2012cw,Holthausen:2012dk,Chen:2014tpa} (for earlier works see~\cite{Grimus:1995zi,Ecker:1983hz,Ecker:1987qp,Neufeld:1987wa,Harrison:2002kp,Grimus:2003yn}). Prime examples are the series of groups $\Delta (3 \, n^2)$~\cite{Luhn:2007uq} and $\Delta (6 \, n^2)$~\cite{Escobar:2008vc}, $n \geq 2$ integer, combined with CP, that lead to four different types of lepton mixing patterns corresponding to Case 1), Case 2), Case 3 a) and Case 3 b.1), if the residual groups are $G_l=Z_3$ among charged leptons and $G_\nu=Z_2 \times CP$ for neutrinos~\cite{Feruglio:2012cw,Holthausen:2012dk,Chen:2014tpa}; see e.g.~\cite{Ding:2013hpa,Feruglio:2013hia,King:2014rwa,Ding:2014ora,Ding:2015rwa} for further literature on fermion mixing arising from these groups. As is known~\cite{Feruglio:2012cw,Holthausen:2012dk,Chen:2014tpa}, the CP symmetry is related to an automorphism of the flavour group. 

In the present work, we analyse a scenario with three RH neutrinos in which a member of the series $\Delta (3 \, n^2)$ or $\Delta (6 \, n^2)$ and a CP symmetry, both broken non-trivially, are responsible for the flavour structure of the Lagrangian. While the charged lepton mass matrix, shaped by $G_l=Z_3$, is diagonal in the chosen basis, $G_f$ and CP are preserved by the Majorana mass matrix of the RH neutrinos such that these are degenerate in mass. The symmetries are broken to $G_\nu$ by the neutrino Yukawa coupling matrix which carries the non-trivial flavour information among the neutral fermions. In this way, the scenario contains as  parameters three couplings, two angles and the scale of the RH neutrino masses, apart from the charged lepton masses. With these, the light neutrino masses, i.e.~either normally (NO) or inversely ordered (IO) and the lightest neutrino mass $m_0$, as well as the lepton mixing angles can be described correctly. Furthermore, the leptonic CP phases are predicted. We consider corrections to the Majorana mass matrix of the RH neutrinos that induce different splittings, $\kappa$ and $\lambda$, among their masses. For heavy neutrinos that are potentially testable in laboratory experiments, we scrutinise their lifetimes in different parts of the parameter space of the scenario and analyse their branching ratios to the different flavours $\alpha$ for Case 1) through Case 3 b.1). Successful leptogenesis can be achieved for vanishing splittings in some cases, though not in all of them. 
We note that certain aspects of this scenario have already been studied in~\cite{Drewes:2022kap}; see also~\cite{Curtin:2018mvb,Chauhan:2021xus}. In the current analysis, we extend these by comprehensively scanning over the parameters, characterising the flavour and CP symmetry and their residuals. In doing so, we put emphasis on the part of the parameter space that can be explored in different laboratory experiments, especially if it also leads to the generation of a sufficient amount of BAU.

The remainder of the paper is organised as follows: in section~\ref{sec2prelim} we recap the basics of the scenario, the choice of symmetries and the special values of the angle $\theta_R$ as well as several facts about heavy neutrino searches at accelerators and the phenomenological framework with one heavy neutrino. Section~\ref{lifetime} is dedicated to the lifetimes  of the heavy neutrinos and to the matching onto the phenomenological framework. We discuss in detail the branching ratios of the heavy neutrinos and the corresponding ternary plots, both for a generic lepton mixing matrix and if the flavour structure is determined by one of the cases, Case 1) through Case 3 b.1), in section~\ref{triangleplots}. In section~\ref{lepto}, we study  leptogenesis for each of the cases, underlining the possibility to test the viable parameter space with different accelerator-based and precision flavour experiments. We summarise in section~\ref{summ}. Two appendices, appendix~\ref{appA} and~\ref{appB}, contain the form of the matrices $\Omega ({\bf 3})$, $\Omega ({\bf 3^\prime})$ and $R_{ij} (\theta)$ and additional information related to section~\ref{lepto}, respectively.

\section{Preliminaries}
\label{sec2prelim}

In this section, we review the prerequisites needed for this work. We start by outlining the underlying framework. We then specify the employed flavour and CP symmetries as well as residual groups and comment on special values of the angle $\theta_R$. More details can be found in \cite{Drewes:2022kap}. Finally, we highlight the different types of heavy neutrino searches at accelerator-based experiments.

\subsection{Scenario}
\label{sec2}

\paragraph{Framework} We consider the type-I seesaw framework with three generations of RH neutrinos. The relevant renormalisable Lagrangian reads as follows
\begin{eqnarray}
	\label{L}
	\mathcal L \supset
	\mathrm{i} \, \overline{\nu_R} \, \slashed\partial \, \nu_{R}
	- \frac{1}{2}
	\overline{\nu^c_R}\, M_R\, \nu_{R}
	- \overline{l_{L}} \, Y_D \, \varepsilon H^* \, \nu_{R}
	+ {\rm h. c.} \, .
\end{eqnarray}
The matrix $M_R$ is the Majorana mass matrix of the RH neutrinos $\nu_R$ (which we label with the index $i=1,2,3$), $Y_D$ the neutrino Yukawa coupling matrix, $H$ the SM Higgs doublet, $\varepsilon$ the totally antisymmetric SU(2) tensor  and $l_{L \,}=\left(\begin{array}{c} \nu_{L\,} \\ e_{L\,} \end{array}\right)$ are the three left-handed (LH) lepton doublets (labelled with the lepton flavour index $\alpha=e,\, \mu,\, \tau$). 

While the matrices $M_R$ and $Y_D$ encode 18 new parameters, if no additional assumptions on their form are made, their structure as well as the one of the charged lepton mass matrix $m_l$ can be strongly constrained by flavour and CP symmetries and their residual groups, as we discuss below.

There are two sets of neutrino mass eigenstates, the light neutrinos $\upnu_i$ and the heavy neutrinos $N_i$, after electroweak symmetry breaking. These can be described by the Majorana spinors
\begin{align}
	\upnu_i &
	= \left[
		V_{\nu}^{\dagger}\nu_L-U_{\nu}^{\dagger}\uptheta \nu_R^c+V_{\nu}^T\nu_L^c-U_{\nu}^T\uptheta^{\ast} \nu_R
	\right]_i
	\ ,
	& N_i &
	= \left[
		V_N^\dagger\nu_R+\Uptheta^T \nu_L^c + V_N^T\nu_R^c+\Uptheta^{\dagger}\nu_L
	\right]_i
	\ . \label{HeavyMassEigenstates}
\end{align}
The matrix $\uptheta$, $\uptheta=\langle H \rangle \, Y_D \, M_R^{-1}$, encodes the mixing between LH and RH neutrinos, with 
$\langle H \rangle$ being the vacuum expectation value (VEV) of the SM Higgs doublet $H$, fixed to $\langle H \rangle \approx 174 \, \mathrm{GeV}$. The matrix $V_\nu$,
$V_\nu = (1 - \frac{1}{2}\uptheta\uptheta^\dagger ) U_\nu$, denotes the light neutrino mixing matrix, while the matrix $V_N$, $V_N = (1 - \frac{1}{2} \uptheta^T \uptheta^*) U_N$, is its equivalent in the heavy neutrino sector.
The unitary matrices $U_\nu$ and $U_N$ diagonalise the matrices
\begin{eqnarray}
\label{eq:blocks_mass_matrix}
m_\nu
= - \uptheta M_R \uptheta^T
&&\;\; \mbox{and} \;\;
M_N
= M_R + \Delta M_{\theta\theta}
\\
&&
\label{eq:definitionDMthetatheta}
\phantom{\;\; \mbox{and} \;\;}
\mbox{with} \;\;
\Delta M_{\theta\theta}=\frac{1}{2} (\uptheta^\dagger \uptheta M_R + M_R^T \uptheta^T \uptheta^{*})
\end{eqnarray}
as $ U_\nu^\dagger m_\nu U_\nu^*$ and $ U_N^T M_N U_N$, respectively.\footnote{The matrix $U_\nu$ coincides with the Pontecorvo-Maki-Nakagawa-Sakata (PMNS) mixing matrix in the charged lepton mass basis.
Differences between $V_\nu$ and $U_\nu$ can be ignored in this study.
}
The squared masses of the light neutrinos $\upnu_i$ and of the heavy neutrinos $N_i$  are given at tree level by the eigenvalues of the matrices $m_\nu m_\nu^\dagger$ and $M_N M_N^\dagger$, respectively.
The eigenvalues of the matrix $M_N$, labelled $M_{N_i}$, are close to those of $M_R$, $M_{R_i}$, but in the regime of quasi-degenerate masses 
corrections of order $\uptheta^2$ to the splitting between them can impact both  leptogenesis~\cite{Shaposhnikov:2008pf} and lepton number violating (LNV) event rates at colliders~\cite{Drewes:2019byd}.
Furthermore, we can introduce
a measure for the difference between the unitary matrices $U_R$ and $U_N$ by defining
\begin{equation}
\label{eq:URN}
U_{RN}= U_R^\dagger \, U_N \; ,
\end{equation}
where $U_R$ is the matrix that diagonalises $M_R$ as
\begin{equation}
\hat{M}_R = U_R^T M_R U_R\,,
\end{equation}
 with the corresponding eigenstates given by
\begin{equation}
    \hat{\nu}_{R} = U_R^\dagger \, \nu_{R}\,.
\end{equation}
All quantities are denoted by a hat in the RH neutrino mass basis.

The suppression of the  weak interactions of the heavy neutrinos, relative to those of the light neutrinos, is given by the elements of the active-sterile neutrino mixing matrix
\begin{equation}
\label{eq:asnumix}
\Uptheta = \uptheta\,  U_N^*.
\end{equation}
It is convenient to define the quantities
\begin{eqnarray}
\label{U2defs}
U_{\alpha i}^2 = |\Uptheta_{\alpha i}|^2 \ , \ U_i^2 = \sum_\alpha U_{\alpha i}^2 \ , \ U_\alpha^2 = \sum_i U_{\alpha i}^2 \;\; \mbox{and} \;\; U^2 = \sum_i U_i^2 = \sum_\alpha U_\alpha^2 \; .
\end{eqnarray}
Indeed, the number of events at accelerator-based experiments is mostly sensitive to these combinations in the limit where production and decay can be seen as independent processes, see  section~\ref{subsec:effLagrangian}. The expected magnitude of the active-sterile mixing angles can be estimated from Eq.~\eqref{eq:blocks_mass_matrix} as 
\begin{equation}
	\label{eq:naiveseesawformula}
	U_i^2 \sim \frac{\sum_j m_{j}}{M_{N_i}} \sim 10^{-10} \frac{\mbox{GeV}}{M_{N_i}} \, ,
\end{equation}
without taking into account the matrix structure in flavour space. 
Note that $m_j$ stand for the light neutrino masses.
The estimate in Eq.~\eqref{eq:naiveseesawformula} is called the \textit{naive seesaw limit} and it would strongly constrain
the possibility of any direct detection of the heavy neutrinos in the near future. However, in low-scale seesaw models in which the smallness of the light neutrino masses is due to an approximate generalised $B-L$ symmetry~\cite{Shaposhnikov:2006nn,Kersten:2007vk,Moffat:2017feq}, large active-sterile mixing angles, corresponding to order-one Yukawa couplings, can be achieved for heavy neutrinos with sub-TeV masses.

\paragraph{Symmetries}  In order to explain the observed lepton mixing pattern, we endow the type-I seesaw framework with a flavour and a CP symmetry
and assume the residual groups $G_l$ and $G_\nu$ to be present in the charged lepton and neutrino sector, respectively.
The flavour symmetry $G_f$ is chosen to be a group of the series $\Delta (6 \, n^2)$, $n$ even,\footnote{This requirement is necessary to ensure that the group has an irreducible real three-dimensional representation.} $n \geq 2$ and $3 \nmid n$ ($n$ is not divisible by three)~\cite{Escobar:2008vc}. The CP symmetry
corresponds to an automorphism of $G_f$ and its action is represented by the CP transformation $X$~\cite{Feruglio:2012cw,Holthausen:2012dk,Chen:2014tpa,Grimus:1995zi,Ecker:1983hz,Ecker:1987qp,Neufeld:1987wa,Harrison:2002kp,Grimus:2003yn}. 
In addition to these symmetries, we assume the existence of an auxiliary $Z_3$ symmetry, denoted by $Z_3^{(\mathrm{aux})}$, 
which allows us to distinguish between the three generations of RH charged leptons and hence to accommodate three different charged lepton masses.

The three LH lepton doublets $l_{L \, \alpha}$, $\alpha=e, \mu, \tau$, transform as irreducible, faithful, complex
three-dimensional representation ${\bf 3}$ of the flavour symmetry, while the three RH neutrinos $\nu_{R \, i}$, $i=1,2,3$, are assigned to an irreducible, unfaithful (if $n > 2$),
real three-dimensional representation ${\bf 3^\prime}$. Lastly, RH charged leptons are in the trivial singlet ${\bf 1}$ of $G_f$, but are assigned to $1$, $\omega$ and $\omega^2$
under $Z_3^{(\mathrm{aux})}$, respectively, where $\omega=e^{2\pi i/3}$. In contrast, both LH lepton doublets and RH neutrinos do not transform under this auxiliary symmetry.  

The residual symmetry $G_l$ is taken to be a $Z_3$ group which corresponds to the diagonal subgroup $Z_3^{(\mathrm{D})}$ of a $Z_3$ symmetry in $G_f$ and the auxiliary symmetry $Z_3^{(\mathrm{aux})}$, 
while $G_\nu$ is the direct product of a $Z_2$ symmetry, contained in $G_f$, and CP.  In the following, we call the generator of this $Z_2$ group $Z$.

The form of the resulting charged lepton mass matrix $m_l$, the neutrino Yukawa coupling matrix $Y_D$ and the Majorana mass matrix $M_R$ of the RH neutrinos turns out
to be strongly constrained. Indeed, the matrix $m_l$ is diagonal with three distinct entries (and we assume here that the (11)-element is the electron, the (22)-element the muon
and the (33)-element the tau lepton mass).
In the limit of unbroken $G_f$ and CP, the matrix $M_R$ is non-zero and takes the form
\begin{equation} 
\label{eq:MR0}
M_R=M_R^0= M \, \left( \begin{array}{ccc}
1 & 0 & 0\\
0 & 0 & 1\\
0 & 1 & 0
\end{array}
\right)
\end{equation}
with $M>0$. The free parameter $M$ sets the mass scale of the RH neutrinos, i.e.~$M_{R_i}=M$ for $i=1,2,3$. The matrix $U_R$ is of the form 
\begin{equation}
U_R = \frac{1}{\sqrt{2}} \, \left(
\begin{array}{ccc}
\sqrt{2} & 0 & 0\\
0 & 1 & i\\
0 & 1 & -i
\end{array}
\right) \; .
\end{equation}
 The matrix $Y_D$ instead only preserves the residual symmetry $G_\nu$, i.e.~it fulfils 
\begin{equation}
Z({\bf 3})^\dagger \, Y_D \, Z({\bf 3^\prime}) = Y_D \;\; \mbox{and} \;\; X({\bf 3})^* \, Y_D \, X({\bf 3^\prime}) = Y_D^* \; ,  
\end{equation}
where $Z (\mathrm{\mathbf{r}})$ and $X (\mathrm{\mathbf{r}})$ refer to the generator $Z$ of the $Z_2$ symmetry (in $G_\nu$) and the CP transformation $X$ in the representation $\mathrm{\mathbf{r}}$.
For explicit forms, see~\cite{Drewes:2022kap}. As has been discussed in~\cite{Hagedorn:2016lva,Drewes:2022kap}, the matrix $Y_D$ can be written as 
\begin{equation}
\label{eq:formYD}
Y_D = \Omega ({\bf 3}) \, R_{ij} (\theta_L) \, \mbox{diag} \, (y_1, y_2, y_3) \, P^{ij}_{kl} \, R_{kl} (-\theta_R) \, \Omega ({\bf 3^\prime})^\dagger
\end{equation}
and thus contains five free real parameters -- the couplings $y_f$, $f=1,2,3$, the angle $\theta_L$, $0 \leq \theta_L \leq \pi$, and the angle $\theta_R$, $0 \leq \theta_R \leq 2 \, \pi$. These correspond to the masses
of the three light neutrinos and the free parameter contained in the PMNS mixing matrix. Furthermore, the angle $\theta_R$ is not fixed and associated with the three RH neutrinos.
 The matrices $\Omega ({\bf 3})$ and $\Omega ({\bf 3^\prime})$ are determined by the CP transformation $X$ in the representation ${\bf 3}$ and ${\bf 3^\prime}$, respectively.
 The planes, the $(ij)$- and $(kl)$-plane, in which the rotation matrices $R_{ij} (\theta_L)$, $i<j$, and $R_{kl} (\theta_R)$, $k<l$, act, respectively, are, like the matrices $\Omega ({\bf 3})$ and 
 $\Omega ({\bf 3^\prime})$, fixed for a given choice of residual group $G_\nu$. The permutation matrix $P^{ij}_{kl}$ is only necessary in case these two planes do not coincide. This only happens
  for certain combinations of parameters in Case 3 a) and Case 3 b.1). All this information is given, for completeness, for each case, Case 1) through Case 3 b.1), in appendix~\ref{appA}. The matrix $Y_D$
 in the RH neutrino mass basis reads 
 \begin{equation}\label{YhatDef}
 \hat{Y}_D = Y_D \, U_R \, .
 \end{equation}
 Given the form of $Y_D$, the Dirac neutrino mass matrix $m_D$ reads
 \begin{equation} 
m_D = Y_D \, \langle H \rangle \; .
\end{equation}
The light neutrino mass matrix $m_\nu$ arises from the type-I seesaw mechanism,
\begin{equation}
m_\nu = -m_D \, M_R^{-1} \, m_D^T \, .
\end{equation}
Since the charged lepton sector does not contribute to lepton mixing, the matrix $m_\nu$ solely determines the PMNS mixing matrix, i.e.~
\begin{equation}
U_{\mbox{\scriptsize{PMNS}}}^\dagger \, m_\nu \, U_{\mbox{\scriptsize{PMNS}}}^* = \mbox{diag} \, (m_1, m_2, m_3)  \; .
\end{equation}
If the combination 
\begin{equation}
\label{eq:combdiag}
R_{kl} (-\theta_R) \, \Omega ({\bf 3^\prime})^\dagger \, M_R^{-1} \, \Omega ({\bf 3^\prime})^* \, R_{kl} (\theta_R)
\end{equation}
is diagonal, the PMNS mixing matrix is of the form
\begin{equation}
\label{eq:UPMNScombdiag}
U_{\mbox{\scriptsize{PMNS}}} = \Omega ({\bf 3}) \, R_{ij} (\theta_L) \, K_\nu \, .
\end{equation}
While the matrix $\Omega ({\bf 3})$ and the $(ij)$-plane are fixed by the residual symmetry $G_\nu$,
 the matrix $K_\nu$ is diagonal with non-zero entries $\pm 1$ and $\pm i$ and encodes the CP parities of the light neutrinos. 
 Its exact form is relevant for the sign of the sines of the Majorana phases $\alpha$ and $\beta$.
If instead the combination in Eq.~(\ref{eq:combdiag}) is not diagonal, the angle $\theta_L$ is replaced by an effective angle $\widetilde{\theta}_L$ which is the sum of $\theta_L$ and an additional angle, depending on the couplings $y_f$ and the angle $\theta_R$.
For more details see below and~\cite{Drewes:2022kap}.

In explicit models corrections to this symmetry breaking pattern are expected, see e.g.~reviews~\cite{Ishimori:2010au,King:2013eh,Feruglio:2019ybq,Grimus:2011fk}. 
 Here, we focus on the ones contributing to the Majorana mass matrix $M_R$ of the RH neutrinos.
Like in~\cite{Drewes:2022kap}, we mainly consider corrections to $M_R$ that are invariant under the residual symmetry $G_l$, present in the charged lepton sector. These are of the form
\begin{equation}
\label{eq:deltaMR}
\delta M_R = \kappa \, M \, \left(
\begin{array}{ccc}
2 & 0 & 0\\
0 & 0 & -1\\
0 & -1 & 0
\end{array}
\right)
\end{equation}
with $\kappa$ being a (small) dimensionless parameter. This correction leads to (small) shifts in the RH neutrino masses 
\begin{equation}
\label{eq:Mkappa}
M_{R_1}= M \, (1+2 \, \kappa) \;\; \mbox{and} \;\; M_{R_2}=M_{R_3}= M \, (1-\kappa) \; .
\end{equation}
Interestingly enough, this correction still leaves the second and third RH neutrino mass degenerate. An example for a correction which splits all three RH neutrino masses is
\begin{equation}
\Delta M_R = \lambda \, M \, \left(
\begin{array}{ccc}
0 & 0 & 0\\
0 & 1 & 0\\
0 & 0 & 1
\end{array}
\right) \; ,
\end{equation}
with $\lambda$ being a (small) dimensionless parameter, 
such that the eigenvalues of the Majorana mass matrix $M_R$, $M_R=M_R^0+\delta M_R+ \Delta M_R$, read
\begin{equation}
\label{eq:Mlambda}
M_{R_1}= M \, (1+2 \, \kappa) \; , \;\; M_{R_2}= M \, (1-\kappa+\lambda) \;\; \mbox{and} \;\; M_{R_3}= M \, (1-\kappa-\lambda) \; .
\end{equation}
We note that for the specific choices $\lambda = \pm \, 3\, \kappa$ the RH neutrino masses become partially degenerate, see also~\cite{Drewes:2022kap}. 

\mathversion{bold}
\subsection{Choice of symmetries and special values of angle  \texorpdfstring{$\theta_R$}{Lg}}
\mathversion{normal}
\label{sec3}

In the first part of this section, we list constraints on the choices of symmetries as well as the undetermined angle in the PMNS mixing matrix
that are imposed by the requirement to accommodate the lepton mixing angles well in the different cases, Case 1) through Case 3 b.1). We also give some numerical examples.
In its second part, we comment on special values of the angle $\theta_R$ which lead to one large coupling $y_f$. Such special values are encountered for certain combinations of the parameters in the different
cases. 
For details about the choice of symmetries and numerical examples see the original work~\cite{Hagedorn:2014wha} as well as~\cite{Drewes:2022kap}. 
More information about special values of $\theta_R$ can also be found in~\cite{Drewes:2022kap}.

\subsubsection{Symmetries and examples}
\label{sec31}

\paragraph{Case 1)} The $Z_2$ symmetry, contained in $G_\nu$, is generated by $c^{n/2}$. The explicit form of this generator in the representations ${\bf 3}$ and ${\bf 3^\prime}$ can be found in~\cite{Drewes:2022kap}.
Apart from $n$ even and not divisible by four no further constraints can be inferred. The CP symmetry depends on a single parameter $s$ which ranges between
$0$ and $n-1$. It is convenient to introduce the parameter $\phi_s$ as 
\begin{equation}
\phi_s=\frac{\pi \, s}{n} \; . 
\end{equation}
The matrices $\Omega (s)({\bf 3})$ and $\Omega (s)({\bf 3^\prime})$, derived from the CP transformation $X (s)$ in the representation ${\bf 3}$ and ${\bf 3^\prime}$, respectively, are given for
completeness in appendix~\ref{appA}. In this appendix the rotation matrices $R_{ij} (\theta_L)$ and $R_{kl} (\theta_R)$, see Eq.~(\ref{eq:formYD}), are also specified.

The undetermined angle $\theta_L$ in the PMNS mixing matrix,\footnote{As explained in~\cite{Drewes:2022kap}, in general this undetermined angle is an effective angle $\widetilde{\theta}_L$ which is the sum of $\theta_L$
and an additional angle which depends on the couplings $y_1$ and $y_3$ and the angle $\theta_R$ for Case 1). This clearly does not affect the results of the fit to the global fit data~\cite{Esteban:2020cvm} and is taken into account, 
where necessary, in the presented numerical analysis.
}
\begin{equation}
U_{\mbox{\scriptsize{PMNS}}} = \Omega (s)({\bf 3}) \, R_{13} (\theta_L) \, K_\nu \; ,
\end{equation}
is restricted to take values either close to $0$ or $\pi$. In particular, for small values of $\theta_L$ the best agreement with the global fit data on the lepton mixing angles~\cite{Esteban:2020cvm} is achieved,\footnote{Throughout this study we use the global fit results v5.1 of NuFIT. These are (very) similar to the updated results v6.0~\cite{Esteban:2024eli}.} for light 
neutrino masses with NO (IO), for
\begin{equation}
\label{eq:Case1thetaLfit}
\theta_L \approx 0.183 \, (0.184) \; .
\end{equation}
The parameter $s$, reflecting the choice of the CP symmetry, fixes the value of the Majorana phase $\alpha$, i.e.~the magnitude of $\sin \alpha$ equals the magnitude of $\sin 6 \, \phi_s$,
while the other two CP phases, the Dirac phase $\delta$ and the second Majorana phase $\beta$, only take trivial values, i.e.~$\sin \delta=0$ and $\sin\beta=0$, see e.g.~\cite{Drewes:2022kap} for the convention of the CP phases. 
This prediction for the CP phase $\delta$, $\delta=0$ or $\delta=\pi$, is compatible at the $3 \, \sigma$ level and at the $1 \, \sigma$ level with the global fit data~\cite{Esteban:2020cvm}, respectively,
for light neutrino masses with NO, while for IO only $\delta=0$ is in agreement with the data at the $3 \, \sigma$ level. 

For a numerical example the index $n$ of the flavour group can be chosen as $n=10$ so that $0 \leq s \leq n-1=9$, see~\cite{Drewes:2022kap}.

\paragraph{Case 2)} The same residual $Z_2$ symmetry is present in the neutrino sector as in Case 1). Thus, the constraints on the index $n$ are the same, i.e.~$n$ should be even and not be divisible by four.  
The CP symmetry is described by two parameters, $s$ and $t$, which both vary between $0$ and $n-1$. As discussed in~\cite{Hagedorn:2014wha}, it is more convenient to use the parameters $u$ and $v$
in the phenomenological analysis
\begin{equation}
\label{eq:def_u_v}
u= 2 \, s-t \;\; \mbox{and} \;\; v= 3 \, t \; .
\end{equation}
These range between $-(n-1) \leq u \leq 2 \, (n-1)$ and $0 \leq v \leq 3 \, (n-1)$, respectively.
Furthermore, we define 
\begin{equation}
\phi_u=\frac{\pi \, u}{n} \;\; \mbox{and} \;\;  \phi_v=\frac{\pi \, v}{n} \, .
\end{equation}
Viable choices of the matrices $\Omega (s,t)({\bf 3})=\Omega (u,v)({\bf 3})$ and $\Omega (s,t)({\bf 3^\prime})$ as well as the rotation matrices $R_{ij} (\theta_L)$ and $R_{kl} (\theta_R)$ are given in appendix~\ref{appA}.

The form of the PMNS mixing matrix is given by
\begin{equation}
U_{\mbox{\scriptsize{PMNS}}} = \Omega (u,v)({\bf 3}) \, R_{13} (\theta_L) \, K_\nu \; .
\end{equation}
For $t$ odd and three massive light neutrinos, the angle $\theta_L$ is replaced by an effective angle $\widetilde{\theta}_L$ which is the sum of $\theta_L$ and an additional angle, depending on the couplings $y_1$ and $y_3$ as well as the angle 
$\theta_R$, like for Case 1). The lepton mixing angles depend on this undetermined angle and $u/n$ ($\phi_u$). The smallness of the reactor mixing angle $\theta_{13}$~\cite{Esteban:2020cvm}
requires $\theta_L$ close to $0$ or $\pi$ and $u/n$ ($\phi_u$) small, 
\begin{equation}
\label{eq:Case2unrange}
-0.10 \lesssim u/n \lesssim 0.12 \;\; \mbox{corresponding to} \;\; -0.314 \lesssim \phi_u \lesssim 0.377 \; ,
\end{equation}
up to symmetry transformations~\cite{Hagedorn:2014wha};  see~\cite{Drewes:2022kap} for these updated values. The Dirac phase $\delta$ and the Majorana phase $\beta$ depend on $\phi_u$ (and $\theta_L$),
while the magnitude of the sine of the Majorana phase $\alpha$ is determined by the magnitude of $\sin \phi_v$. 
The constraints on the CP phase $\delta$ only have a moderate impact on the choice of $\phi_u$ and $\theta_L$ for light neutrino masses with NO, but 
they substantially reduce the parameter space favoured by the global fit data~\cite{Esteban:2020cvm} for IO.
For details, see figure 1 in~\cite{Drewes:2022kap}. 

One viable choice is $n=14$, $u=0, \pm 1$ 
and the corresponding values of $v$, see~\cite{Drewes:2022kap}. For $u=\pm 1$, we have $t$ odd and thus potentially large values of the active-sterile neutrino mixing angles $U^2_{i \alpha}$, see also section~\ref{sec32}. 
Numerical results for the values of $\theta_L$ can be found in table 1 in~\cite{Drewes:2022kap}.

\paragraph{Case 3 a) and Case 3 b.1)} For these cases, the residual $Z_2$ symmetry in the neutrino sector is generated by $b \, c^m \, d^m$ with $0 \leq m \leq n-1$, see~\cite{Drewes:2022kap} for the form of this generator
in the representations ${\bf 3}$ and ${\bf 3^\prime}$. The value of $m$ is strongly restricted by the requirement to accommodate the experimental data on lepton mixing angles well, i.e.~for Case 3 a) we have
$m\approx 0$ and $m\approx n$, while for Case 3 b.1) $m$ is close to $n/2$, see Eqs.~(\ref{eq:mnCase3a}) and (\ref{eq:mnCase3b1}). The CP symmetry depends on a single parameter $s$, ranging between $0$ and $n-1$. For convenience, we introduce the parameters 
$\phi_m$ and $\phi_s$
\begin{equation}
\phi_m=\frac{\pi \, m}{n} \;\; \mbox{and} \;\;  \phi_s=\frac{\pi \, s}{n} \, .
\end{equation}
The matrices $\Omega (s,m)({\bf 3})$ and $\Omega (s)({\bf 3^\prime})$, deduced from the form of the CP transformation in ${\bf 3}$ and ${\bf 3^\prime}$, respectively, the rotation matrices 
$R_{ij} (\theta_L)$ and $R_{kl} (\theta_R)$ and the permutation matrix $P^{ij}_{kl}$, where needed, can be found in appendix~\ref{appA}.

For Case 3 a), the PMNS mixing matrix is of the form
\begin{equation}
U_{\mbox{\scriptsize{PMNS}}} = \Omega (s,m)({\bf 3}) \, R_{12} (\theta_L) \, K_\nu 
\end{equation}
with the angle $\theta_L$ being replaced by an effective angle $\widetilde{\theta}_L$ that is the sum of $\theta_L$ and an additional angle, depending on the couplings $y_1$ and $y_2$ as well as the angle $\theta_R$, for $m$ even and $s$ odd
or vice versa and non-zero lightest neutrino mass $m_0$. Both the reactor mixing angle $\theta_{13}$ and the atmospheric mixing angle $\theta_{23}$ are determined by $m/n$ ($\phi_m$). In particular,
the smallness of $\theta_{13}$ requires that
\begin{equation}
\label{eq:mnCase3a}
 0.056 \lesssim m/n \lesssim 0.061 \;\; \mbox{corresponding to} \;\;  0.175 \lesssim \phi_m \lesssim  0.193 
\end{equation}
with the same range for $1-m/n$ and $\pi-\phi_m$, respectively, 
see~\cite{Hagedorn:2014wha}. Because of this, the index $n$ of the flavour group has to be larger than ten. In~\cite{Hagedorn:2014wha} and~\cite{Drewes:2022kap} examples are given which fulfil
this requirement, for $n$ even, as necessary in this scenario, e.g.~$n=16$ and $m=1$ ($m=15$) and $n=34$ and $m=2$. All choices of $s$ lead to a viable result for the solar mixing angle $\theta_{12}$
for at least one value of the angle $\theta_L$; explicit values can be found in tables 2, 7 and 8 of~\cite{Drewes:2022kap}. 

As regards the CP phases they all depend in general on the parameters $n$, $m$, $s$
and the angle $\theta_L$.  Adding the constraints from the CP phase $\delta$ hardly changes the results for preferred values of the parameters $s/n$ ($\phi_s$) and $\theta_L$,
if light neutrino masses have NO, whereas the parameter space, compatible with the global fit data~\cite{Esteban:2020cvm} at the $3 \, \sigma$ level or better, is roughly halved for light neutrino masses with IO,
compare figure 2 in~\cite{Drewes:2022kap}.
About the Majorana phases, we only note that the sine of the Majorana phase $\alpha$ approximately equals in magnitude $\sin 6 \, \phi_s$ and refer for more discussion to~\cite{Hagedorn:2014wha}.    

In Case 3 b.1) the form of the PMNS mixing matrix is given by
\begin{equation}
\label{eq:PMNSCase3b1}
U_{\mbox{\scriptsize{PMNS}}} = \Omega (s,m)({\bf 3}) \, R_{12} (\theta_L) \, P \, K_\nu \;\; \mbox{with} \;\; P=\left(
\begin{array}{ccc}
0 & 1 & 0\\
0 & 0 & 1\\
1 & 0 & 0
\end{array}
\right) \; .
\end{equation}
Like for Case 3 a), $\theta_L$ has to be read as effective angle $\widetilde{\theta}_L$ that is the sum of $\theta_L$ and an additional angle, determined by $y_1$, $y_2$ and $\theta_R$, if $m$ is even and $s$ odd or vice versa and $m_0$ non-vanishing. 
The ratio $m/n$ ($\phi_m$) is strongly restricted by the measured value of the solar mixing angle $\theta_{12}$, $\sin^2 \theta_{12}\approx 1/3$, such that
\begin{equation}
\label{eq:mnCase3b1}
0.44 \lesssim m/n \lesssim 0.56 \;\; \mbox{corresponding to} \;\;  1.38 \lesssim \phi_m \lesssim  1.76 \; ,
\end{equation}
see also~\cite{Hagedorn:2014wha}. The value of the angle $\theta_L$ is constrained as well, i.e.~it has to be close to $\pi/2$ in order to explain the smallness of $\theta_{13}$. The atmospheric mixing angle $\theta_{23}$ 
 also depends on the choice of the CP symmetry, meaning on the parameter $s$. 
 
 The CP phases are determined by a combination of the parameters $n$, $m$, $s$ and $\theta_L$. 
 The effect of imposing the constraints on the CP phase $\delta$ from the global fit~\cite{Esteban:2020cvm} is for Case 3 b.1) similar to Case 3 a), i.e.~the impact
is only very mild for light neutrino masses with NO, while the allowed parameter space is reduced by at least a factor of two for IO, as can be clearly seen in figure 3 in~\cite{Drewes:2022kap}.
  Furthermore, for $m=\frac{n}{2}$,
 the magnitude of the sines of both Majorana phases $\alpha$ and $\beta$ equals the magnitude of $\sin 6 \, \phi_s$. 
 
 Numerical examples can be found in~\cite{Hagedorn:2014wha} and~\cite{Drewes:2022kap}. In the
 latter, updated results are given for the choice $n=8$ and $m=4$ and the allowed values of $s$ (only the choices $s=1$, $s=4$ and $s=7$ lead to an agreement with the measured values of the three
  lepton mixing angles at the $3 \, \sigma$ level or better) as well as for $n=20$ and $m=9$, $m=10$ and $m=11$ and several values of $s$. For details see tables 3, 9 and 10 in~\cite{Drewes:2022kap}.  

\vspace{0.2in}
\noindent Since the splitting $\kappa$ (and $\lambda$) is always expected to be small, we neglect its effect on the presented results for lepton mixing as well as on the light neutrino masses, see also the comment at the end of section~\ref{sec32}.

\mathversion{bold}
\subsubsection{Special values of angle $\theta_R$}
\mathversion{normal}
\label{sec32}

We mention the formulae for the light neutrino masses, if they depend on the angle $\theta_R$, and comment on the phenomenological relevance of special values of this angle.

\paragraph{Case 1)}  The light neutrino masses $m_1$ and $m_3$ depend in general on the couplings $y_1$ and $y_3$ as well as on the angle $\theta_R$, i.e.~the light neutrino masses read for all values of the parameter $s$
\begin{equation}
m_2 = \frac{y_2^2 \, \langle H \rangle^2}{M} \;\; \mbox{and} \;\; m_{1,3} = \frac{\langle H \rangle^2}{2 \, M} \, \Big| (y_1^2-y_3^2) \, \cos 2 \, \theta_R \pm \sqrt{4 \, y_1^2 \, y_3^2 + (y_1^2-y_3^2)^2 \, \cos^2 2 \, \theta_R} \Big|
\end{equation}
and the tangent of twice the additional angle (expressed in terms of $\theta_L$ and $\widetilde{\theta}_L$) is given by
\begin{equation}
\tan 2 \, (\widetilde{\theta}_L - \theta_L ) = - \frac{2 \, y_1 \, y_3}{y_1^2+y_3^2} \, \tan 2 \, \theta_R \, .
\end{equation}
If strong NO, i.e.~light neutrino masses follow NO and the lightest neutrino mass vanishes, $m_0=0$, is imposed by taking $y_1$ to be zero, the dependence on the
angle $\theta_R$ simplifies and the neutrino mass $m_3$ becomes proportional to $y_3$ and the magnitude of $\cos 2 \, \theta_R$ such that values of $\theta_R$ close to odd multiples of 
$\pi/4$ entail a large enhancement of the coupling $y_3$. The same is true for strong IO, i.e.~light neutrino masses follow IO and $m_0=0$, with the roles of the masses $m_1$ and $m_3$
as well as the couplings $y_1$ and $y_3$ being exchanged, see~\cite{Drewes:2022kap} for a detailed discussion,
\begin{equation}
\label{eq:Case1strongNOIO}
m_3 = \frac{y_3^2 \, \langle H \rangle^2}{M} \, |\cos 2 \, \theta_R| \;\, \mbox{(strong NO) and} \;\; m_1 = \frac{y_1^2 \, \langle H \rangle^2}{M} \, |\cos 2 \, \theta_R| \;\, \mbox{(strong IO).}
\end{equation} 

\paragraph{Case 2)} As discussed in~\cite{Drewes:2022kap}, for $t$ even the light neutrino masses $m_i$ are independent of the value of the angle $\theta_R$. For $t$ odd
instead the situation is very similar to Case 1), i.e.~the masses $m_1$ and $m_3$ depend in general on the couplings $y_1$ and $y_3$ as well as $\theta_R$. We have for all values of the parameter $s$
\begin{equation}
m_2 = \frac{y_2^2 \, \langle H \rangle^2}{M} \;\; \mbox{and} \;\; m_{1,3} = \frac{\langle H \rangle^2}{2 \, M} \, \Big| (y_1^2-y_3^2) \, \sin 2 \, \theta_R \pm \sqrt{4 \, y_1^2 \, y_3^2 + (y_1^2-y_3^2)^2 \, \sin^2 2 \, \theta_R} \Big|
\end{equation}
and the tangent of twice the additional angle is given by
\begin{equation}
\tan 2 \, (\widetilde{\theta}_L - \theta_L ) = \frac{2 \, y_1 \, y_3}{y_1^2+y_3^2} \, \cot 2 \, \theta_R \, .
\end{equation}
In particular, we find
\begin{equation}
m_3 = \frac{y_3^2 \, \langle H \rangle^2}{M} \, |\sin 2 \, \theta_R| \;\, \mbox{(strong NO) and} \;\; m_1 = \frac{y_1^2 \, \langle H \rangle^2}{M} \, |\sin 2 \, \theta_R| \;\, \mbox{(strong IO).}
\end{equation} 
Thus, for $\theta_R$ close to multiples of $\pi/2$ one coupling becomes large for fixed values of the light neutrino masses.

\paragraph{Case 3 a) and Case 3 b.1)} If $m$ and $s$ are both even or both odd, none of the light neutrino masses $m_i$ depends on the value of the angle $\theta_R$, while for $m$ even
and $s$ odd or $m$ odd and $s$ even two of these reveal a dependence on $\theta_R$ as well as on the couplings $y_1$ and $y_2$. For Case 3 a) these two masses are $m_1$ and $m_2$, i.e.~we find for the light neutrino mass spectrum
\begin{equation}
m_3 = \frac{y_3^2 \, \langle H \rangle^2}{M} \;\; \mbox{and} \;\; m_{1,2} = \frac{\langle H \rangle^2}{2 \, M} \, \Big| (y_1^2-y_2^2) \, \cos 2 \, \theta_R \pm \sqrt{4 \, y_1^2 \, y_2^2 + (y_1^2-y_2^2)^2 \, \cos^2 2 \, \theta_R} \Big| \; ,
\label{eq:numassesCase3a}
\end{equation}
whereas for Case 3 b.1) they are $m_2$ and $m_3$, i.e.~the light neutrino mass spectrum reads
\begin{equation}
m_1= \frac{y_3^2 \, \langle H \rangle^2}{M} \;\; \mbox{and} \;\; m_{2,3} = \frac{\langle H \rangle^2}{2 \, M} \, \Big| (y_1^2-y_2^2) \, \cos 2 \, \theta_R \pm \sqrt{4 \, y_1^2 \, y_2^2 + (y_1^2-y_2^2)^2 \, \cos^2 2 \, \theta_R} \Big| \; .
\label{eq:numassesCase3b1}
\end{equation}
The tangent of twice the additional angle is always given by
\begin{equation}
\tan 2 \, (\widetilde{\theta}_L - \theta_L ) = (-1)^{(s \, \mbox{\scriptsize mod \normalsize} 2)} \, \frac{2 \, y_1 \, y_2}{y_1^2+y_2^2} \, \tan 2 \, \theta_R \, .
\end{equation}
For Case 3 b.1) and $m$ even and $s$ odd these results should be compared with equations (94) and (95) in~\cite{Drewes:2022kap}.
 Especially, for Case 3 a) we have for strong NO
\begin{equation}
m_2 = \frac{y_2^2 \, \langle H \rangle^2}{M} \, |\cos 2 \, \theta_R| \;\, \mbox{and} \;\; m_2 = \frac{y_1^2 \, \langle H \rangle^2}{M} \, |\cos 2 \, \theta_R|
\end{equation}
in the case of strong IO for Case 3 b.1).
Thus, like for Case 1), values of $\theta_R$ close to odd multiples of 
$\pi/4$ lead to a large enhancement of one coupling, as long as the light neutrino masses are kept fixed.

\vspace{0.1in}
\noindent We note that in several occasions these special values are related to an enhancement of the residual symmetry of the combination $Y_D^\dagger \, Y_D$ of the neutrino Yukawa coupling matrix $Y_D$.
Thus, the angle $\theta_R$ being close to one such value corresponds to a small breaking of the enhanced symmetry. For details see section 7 in~\cite{Drewes:2022kap}.

\vspace{0.1in}
\noindent Furthermore, we would like to emphasise that one of the couplings $y_f$, $f=1,2,3$, being large -- for $\theta_R$ approaching the mentioned special values -- is equivalent to large active-sterile mixing, since in this case some entries in the matrix $\uptheta$ become large and, thus, in $\Uptheta$, compare below Eq.~(\ref{HeavyMassEigenstates}) and Eq.~(\ref{eq:asnumix}). Consequently, large values of $U_{\alpha i}^2$ can be achieved, see Eq.~(\ref{U2defs}).

\vspace{0.2in}
\noindent Eventually, we comment on the effect of the splittings $\kappa$ and $\lambda$, in case the angle $\theta_R$ is close to a special value. E.g.~in Case 1) for $\cos 2 \, \theta_R \approx 0$
the leading correction to the light neutrino mass $m_3$ reads for strong NO at order $\kappa$ and $\lambda$ 
\begin{equation}
\label{eq:m3kappalambda}
m_3 \approx \frac{y_3^2 \, \langle H \rangle^2}{M} \, \Big|\kappa+\frac 23 \, \lambda +\cos 2 \, \theta_R\Big| \, .
\end{equation}
Requiring these corrections to be small imposes as condition
\begin{equation}
\label{eq:kappalambdathetaR}
|\kappa| \; ,\; |\lambda| \ll |\cos 2 \, \theta_R| \, .
\end{equation} 
We can also formulate a condition on the size of $U^2$, see Eq.~(\ref{U2defs}), depending on the size of the splittings $\kappa$ and $\lambda$
\begin{equation}
\label{eq:U2kappa}
U^2 \ll \frac{1}{M}\left(m_2+\frac{m_3}{|\kappa|, |\lambda|}\right) \; .
\end{equation}
The conditions in Eqs.~(\ref{eq:kappalambdathetaR}) and (\ref{eq:U2kappa}), and alike for the other cases, Case 2) through Case 3 b.1), are taken into account in the following numerical analysis
and their impact is highlighted in the figures, see e.g.~Fig.~\ref{fig:kappavsU2}. Eqs.~(\ref{eq:m3kappalambda}) and (\ref{eq:kappalambdathetaR}) generalise the results given in~\cite{Drewes:2022kap}, see equations
(96) and (97).

\subsection{Heavy neutrino searches at accelerators}
\label{ColliderSearches}

From a phenomenological viewpoint the low-scale seesaw framework is particularly interesting, because the heavy neutrinos $N_i$ can be light enough to be discovered in particle physics experiments~\cite{Atre:2009rg,Boyarsky:2009ix,Drewes:2013gca,Deppisch:2015qwa,Antusch:2016ejd,Cai:2017mow,Agrawal:2021dbo,Abdullahi:2022jlv}.

\subsubsection{Effective Lagrangian at collider-accessible energies}
\label{subsec:effLagrangian}

Starting from the Lagrangian in Eq.~(\ref{L}),\footnote{Assuming that this Lagrangian represents a valid effective field theory~\cite{delAguila:2008ir} at all relevant scales is a self-consistent framework~\cite{Bezrukov:2012sa,Shaposhnikov:2007nj}, see~e.g.~section 5.1 in~\cite{Agrawal:2021dbo} for a discussion. } 
the production and decay of the heavy neutrinos $N_i$ are both controlled by their suppressed weak interactions, proportional to $\Uptheta_{\alpha i}$, see Eq.~(\ref{HeavyMassEigenstates}).
The interactions of the heavy neutrinos with the weak gauge bosons $W$ and $Z$ and the Higgs boson $h$ are  effectively described by the Lagrangian~\cite{Atre:2009rg}
\begin{equation}
\label{eq:weak WW}
 \mathcal L
\supset
- \frac{m_W}{ \langle H\rangle} \overline{N_i} \,\Uptheta^*_{\alpha i} \,\gamma^\mu e_{L\,\alpha} W^+_\mu
- \frac{m_Z}{\sqrt{2}\, \langle H\rangle} \overline{N_i} \,\Uptheta^*_{\alpha i} \,\gamma^\mu \nu_{L\,\alpha} \,Z_\mu
- \frac{M_{N_i}}{\sqrt{2}\,\langle H\rangle}  \Uptheta_{\alpha i} \,h \, \overline{\nu_{L\,\alpha}} \,N_i
+ \text{h.c.},
\end{equation}
with $m_W$ and $m_Z$ being the masses of the $W$ and the $Z$ boson, respectively.
The Lagrangian in Eq.~(\ref{eq:weak WW}) is based on the assumption that heavy neutrinos  
are produced and decay exclusively via SM weak interactions, mediated through the mixing $\Uptheta$.\footnote{If the RH neutrinos have additional gauge interactions, relevant at the TeV scale, this could considerably enhance the production cross section at the LHC.}

Searches for heavy neutrinos whose interactions with SM particles are described by Eq.~(\ref{eq:weak WW}) have a long record, with recent examples including results by ATLAS~\cite{ATLAS:2022atq}, CMS~\cite{CMS:2023jqi,CMS:2024ita,CMS:2024xdq,CMS:2024ake,CMS:2024hik}, T2K~\cite{T2K:2019jwa}, Belle~\cite{Belle:2024wyk} and NA62~\cite{NA62:2020mcv}.
The sensitivity of collider bounds will considerably improve during the High Luminosity LHC (HL-LHC) runs~\cite{Pascoli:2018heg,Drewes:2019fou}.  Further improvement becomes possible, in case one or several of the proposed dedicated detectors are added~\cite{Beacham:2019nyx,Agrawal:2021dbo,Antel:2023hkf}. 
Out of these, we use the expected sensitivities from MATHUSLA~\cite{Curtin:2018mvb,MATHUSLA:2020uve} and FASER~\cite{Ariga:2018uku,Feng:2024zfe}. 
For heavy neutrino masses below the various meson masses, one can e.g.~operate the existing NA62 experiment in dump mode~\cite{Drewes:2018gkc}; an improvement of one to two orders of magnitude can be expected from the DUNE near detector~\cite{Krasnov:2019kdc,Ballett:2019bgd} or a similar facility at T2HyperK and the recently supported SHiP experiment~\cite{Alekhin:2015byh,SHiP:2018xqw,Gorbunov:2020rjx}. 

As already pointed out in~\cite{Kersten:2007vk}, observing at least a few events usually requires that the quantities $U_i^2$, see Eq.~(\ref{U2defs}), exceed the naive seesaw limit, given in Eq.~(\ref{eq:naiveseesawformula}), like we illustrate with the simple estimates in Eqs.~(\ref{eq:sigmaN}) and~(\ref{Nobsa}). As mentioned, this is possible in a technically natural way in the presence of an approximate generalised $B-L$ symmetry~\cite{Shaposhnikov:2006nn,Kersten:2007vk,Moffat:2017feq}.

In this case the heavy neutrinos tend to be long-lived enough to decay with a measurable displacement in collider experiments in the parameter region where leptogenesis successfully produces the BAU~\cite{Drewes:2022kap}.
If their masses are sufficiently different from each other that one can kinematically distinguish them, one would generically expect that rapid oscillations between different heavy neutrino states remove any quantum correlation between their production and decay, cf.~\cite{Drewes:2019byd,Antusch:2023nqd} and references therein. Consequently, their production and decay can be treated independently and one can express all event rates in terms of the quantities $U_{\alpha i}^2$ defined in Eq.~(\ref{U2defs}).  
This is not always the case in the scenario under consideration, as the expressions for the mass splittings and lifetimes in sections~\ref{sec2} and~\ref{lifetime} show, implying that certain collider observables can be sensitive to the phases in  $\Uptheta_{\alpha i}$, see e.g.~\cite{Abada:2022wvh}. In the present work, we restrict ourselves to total event numbers to individual lepton flavours and the displacement of the decays. In contrast to more intricate observables (such as angular or energy distributions, LNV or CP-violating signatures, etc.), these quantities can still be written in terms of $U_{\alpha i}^2$ alone, even if there is no loss of coherence between production and decay. 
This justifies the use of analytic formulae for the event numbers presented in section~\ref{Sec:OneFlavourModel}.

It is convenient to describe these event rates in terms of the ratios $\frac{U_\alpha^2}{U^2}$, since the sensitivity of a given experiment in the mass-mixing plane can only be determined for fixed values of these (and can strongly depend on them~\cite{SHiP:2018xqw,Drewes:2018gkc,Tastet:2021vwp}). Moreover, it is well-known that  realisations of the type-I seesaw mechanism make testable predictions for these ratios (cf.~e.g.~\cite{Hernandez:2016kel,Drewes:2016jae,Chrzaszcz:2019inj,Krasnov:2023jlt,Drewes:2024bla}), and in section~\ref{triangleplots} we find that such predictions tighten in the analysed scenario.

\subsubsection{Estimates of event numbers for one heavy neutrino}
\label{Sec:OneFlavourModel}

Phenomenological studies and interpretations of experimental data are often performed in a simplified framework with only one heavy neutrino species, taking $i=1$ in Eq.~\eqref{eq:weak WW}. We adapt this assumption in the remainder of this subsection. 
While it is clear that this cannot correctly reproduce the observed neutrino masses, 
it can capture relevant phenomenological aspects of the studied scenario, as we discuss in section~\ref{Sec:PhenoRealsiticModels}.

 In the simplified framework with only one heavy neutrino species we can treat its production and decay independently.
The production cross section of the heavy neutrino $N$ is parametrically given by 
\begin{eqnarray}
\label{eq:sigmaN}
\sigma_N = \cprod \frac{U^2}{3} \, \sigma_{\nu} \, \Pi \; ,
\end{eqnarray}
with $\sigma_{\nu}$ being the production cross section for SM neutrinos in a given experiment, $\Pi$ a potentially complicated kinematical  factor that takes into account the mass $M$ of the heavy neutrino and in general depends on the masses of all final state particles, and $\cprod$ a coefficient that can be used to effectively mimic phenomenological aspects of models with several heavy neutrino species.  
It is implicitly assumed in Eq.~\eqref{eq:sigmaN} that both the neutrino production cross section and the kinematic factor $\Pi$ are independent of the final neutrino flavour. This is true, if the heavy neutrinos are produced from $Z$ boson decays as is the case at FCC-ee/CEPC. 
 For later convenience, we define the quantity
\begin{eqnarray}
\label{GammaNgeneric}
\Gamma_{N} = \frac{\cdec}{96\,\pi^3} \, U^2 \, M^5 G_F^2 \, \aa \; ,
\end{eqnarray}
to parameterise the decay length of the heavy neutrino; in the framework of a single heavy neutrino considered in this subsection Eq.~\eqref{GammaNgeneric} 
with $\aa\simeq 12$ and $\cdec=1 \ (1/2)$ for Majorana (Dirac) heavy neutrinos corresponds to the inclusive heavy neutrino decay width in vacuum, when all final state masses are neglected (which is justified for $m_B \ll M \ll m_W$)~\cite{Gorbunov:2007ak,Atre:2009rg}.\footnote{Since the same interaction controls production and decay of the heavy neutrinos, 
the total mixing $U^2$ can be extracted independently from the total event numbers as well as from the distribution of the displacement $\decay$ in individual decays, i.e.~from the normalisation of Eq.~\eqref{DifferentialDecay} 
as well as from the slope of $\mathrm{d}\log N_\text{obs}/\mathrm{d}\log\decay$. 
Comparing the two allows to determine $\cdec$ and has been proposed as a way to distinguish Dirac from Majorana heavy neutrinos~\cite{Blondel:2022qqo}.
However, this interpretation can be hampered by effects arising from the presence of several heavy neutrino species in realistic models~\cite{Drewes:2022rsk}, see also section~\ref{lifetime}.}

In the following, we primarily consider two observables, the heavy neutrino lifetime (see section~\ref{lifetime}) and the branching ratios of heavy neutrino decays into different flavours $\alpha$ (see section~\ref{triangleplots}). Both of these can have the potential to distinguish between the employed flavour and CP symmetries and their residuals in the considered scenario, providing an important step towards testing the fundamental theory behind the mechanism of neutrino mass generation and the origin of the BAU. The accuracy with  which they can be measured mainly depends on the total number of heavy neutrino decays that are observed. For the mass $M$ around a few tens of GeV, one can expect the largest event numbers at future lepton colliders~\cite{Antusch:2016ejd}, such as FCC-ee~\cite{FCC:2018evy} or CEPC~\cite{CEPCStudyGroup:2018ghi}. 
For these the total event numbers  produced in $Z$ boson decays~\cite{Blondel:2014bra} and the sensitivity region in the mass-mixing plane can be computed analytically in good approximation~\cite{Drewes:2022rsk}.\footnote{
Analytic estimates also exist for the event numbers at beam dump experiments~\cite{Bondarenko:2019yob} and at the LHC~\cite{Drewes:2019vjy}, but are more complicated and less accurate.} 
The number $N_\mathrm{HNL}$ of heavy neutrinos produced from the decay of on-shell $Z$ bosons is given by
\begin{equation}
\label{Nobsa} 
N_\mathrm{HNL}
\simeq 
2 \,
\frac{U^2}{3} \,
\cprod \,
N_Z \,
\BR \,
\Pi \; , \;
\Pi=\left(\frac{2\,\pN}{m_Z}\right)^2\,
\left(
1+\frac{(M/m_Z)^2}{2}
\right) \; , \; \pN = \frac{m_Z}{2}\left(1 - (M/m_Z)^2\right) 
\end{equation}
  with the branching ratio $\BR=\operatorname{BR}(Z\mathpunct\to\nu \overline\nu) \simeq \frac{1}{5}$.
  The boost factor is simply given by  
 $\upbeta\upgamma= \pN/M$ for the $1\to 2$ decay, and the total number of $Z$ bosons $N_Z$ is related to Eq.~\eqref{eq:sigmaN} via $N_Z\BR = 
\sigma_{\nu} \mathfrak{L}$, with $\mathfrak{L}$ being the integrated luminosity (assuming that the heavy neutrino production is dominated by on-shell $Z$ boson decays).

Given $N_\mathrm{HNL}$, the number of decays with a displacement between $\decay$ and $\decay + \mathrm{d}\decay$ reads
\begin{eqnarray}
\label{DifferentialDecay}
    \frac{\mathrm{d} N_\text{obs}}{\mathrm{d} \decay} 
\simeq 
\frac{N_\mathrm{HNL}}{\lambda_{N}} \,
\exp(-\decay/\lambda_{N}) \; ,
\end{eqnarray}
with the decay length in the laboratory given by
\begin{eqnarray}
\lambda_{N} = \frac{\upbeta \upgamma}{\Gamma_{N}}
\simeq \frac{1.6}{ U^2 \, \cdec} \, \left(\frac{M}{\rm GeV}\right)^{-6} \, \left(1-(M/m_Z)^2\right) \, {\rm cm} \; .
\end{eqnarray}
Hence, the total number of decays within a cylindrical detector of length $l_{\rm cyl}$ and diameter $d_{\rm cyl}$
can in good approximation be estimated as
\begin{equation} 
\label{eq:observed events}
N_\text{obs} 
\simeq 
N_\mathrm{HNL} \, 
\left[\exp(-l_0/\lambda_{N}) - \exp(-l_1/\lambda_{N}) \right] \; ,
\end{equation}
with $l_0$ being the minimal displacement needed to remove backgrounds and $l_1 = \frac{1}{2} \, (3/2)^{1/3} \, 
 d_{\rm cyl}^{2/3} \, l_{\rm cyl}^{1/3}$. 
The expected number of observed semi-leptonic decays into the flavour $\alpha$, $N \rightarrow \ell_\alpha jj$, reads
 \begin{align}
 \label{eq:observed events_alpha}
	 N_\text{obs}^\alpha = \operatorname{BR}(N \rightarrow \ell jj) \cdot \ratio{\alpha} \, N_\text{obs}\; ,
 \end{align}
 where $\ell_\alpha$ is a charged lepton of flavour $\alpha$ and $\operatorname{BR}(N \rightarrow \ell jj) \approx 0.5$ is the fraction of semi-leptonic heavy neutrino decays~\cite{Antusch:2017pkq}. In practice, these events are particularly useful as they allow for a straightforward reconstruction of the flavour ratio $\frac{U_\alpha^2}{U^2}$ by tagging the flavour of the charged lepton $\ell_\alpha$.
 Here, we have assumed idealised detectors with $100\%$ reconstruction efficiencies for all final states, using the analytic formulae from~\cite{Drewes:2022rsk}. A more realistic study~\cite{Ajmal:2024kwi} appears to be roughly consistent with~\cite{Drewes:2022rsk}. With the above formulae one finds that a number of heavy neutrinos as large as $10^5$ can be produced at FCC-ee or CEPC,  if $U^2$ lies just below the current upper limits~\cite{Drewes:2022rsk}, compare Fig.~\ref{fig:flavratiovsU2Case1}.

In realistic models with several heavy neutrino species, phenomenologically interesting heavy neutrino couplings that exceed the naive seesaw limit, see Eq.~\eqref{eq:naiveseesawformula}, require an approximate generalised $B-L$ symmetry to protect the light neutrino masses~\cite{Shaposhnikov:2006nn,Kersten:2007vk,Moffat:2017feq}, cf.~section~\ref{sec2}. In the minimal extension of the SM with only two heavy neutrino generations that can fit neutrino oscillation data (and also in the $\nu$MSM) this enforces $U_{\alpha 1}\simeq U_{\alpha 2}$ and hence $\Gamma_{N_1}\simeq \Gamma_{N_2}$, implying that the distribution of events in space still follows a simple exponential decay law, see Eq.~\eqref{DifferentialDecay}, though $\cprod$ and $\cdec$ then become effective parameters that can take values other than $1/2$ and $1$~\cite{Drewes:2022rsk}. For three mass-degenerate heavy neutrinos the protection of the light neutrino masses from radiative corrections does not require $U_{\alpha i}\simeq U_{\alpha j}$~\cite{Drewes:2021nqr}. Hence,  deviations from a simple exponential distribution of events, as discussed in section~\ref{lifetime}, are possible without fine-tuning.
 
\section{Lifetimes  of heavy neutrinos}
\label{lifetime}

In the following, we first present results for the heavy neutrino masses, then for their lifetimes in different limits and lastly we discuss how these can be related to the simplified framework with one heavy neutrino species, used in many searches.

The heavy neutrino production and decay rates in laboratory experiments are governed 
by the complex active-sterile neutrino mixing matrix  $\Theta_{\alpha i}$, encoding the mixing angles between the SM neutrino $\nu_{L \, \alpha}$ 
and the physical heavy neutrino $N_i$.
To determine this mixing matrix we have to identify the latter. The heavy neutrino mixing $U_i^2$ can then be used to estimate their lifetimes by employing  Eq.~\eqref{GammaNgeneric} with $\cdec=\cprod=1$.
We focus on the regime in which $U^2$ is much larger than in the naive seesaw limit, compare Eq.~(\ref{eq:naiveseesawformula}),
and neglect corrections of order $m_\nu/M$, unless explicitly stated.
As this regime corresponds to only one large coupling $y_f$,  $y_f^2 \gg m_\nu M/\langle H \rangle^2$, see section~\ref{sec32}, the neutrino Yukawa coupling matrix $Y_D$ and the corresponding mixing matrix $\Theta$ are approximately of rank one. This allows us to factorise the mixing as~\cite{Drewes:2024bla}
\begin{align}
	\label{eq:factorization}
	U^2_{\alpha i} \approx \frac{U^2_{\alpha} \cdot U^2_{i}}{U^2}\; .
\end{align}
Effectively, all three heavy neutrinos have the same branching ratios into leptons (neglecting the kinematic corrections from the different heavy neutrino masses)
and the main observables of interest in laboratory experiments are the quantities $U_i^2$ and $U_\alpha^2$.
This fact can be used to estimate the 
individual decay rates $\Gamma_{N_i}$  of the heavy neutrinos in relation to the quantity $\Gamma_N$,  defined in Eq.~\eqref{GammaNgeneric}, as
\begin{equation}
\label{eq:GammaNi}
	\Gamma_{N_i} \approx \Gamma_N \,\frac{U^2_i}{U^2}\; .
\end{equation}

\subsection{Physical heavy neutrinos and their mixing}
\label{Sec:MassStatesAndLifetimes}

Heavy neutrinos owe their masses both to the Lagrangian mass term $M_R$ as well as to the Higgs mechanism --- which gives them the additional mass term $\Delta M_{\theta \theta}$, compare Eq.~(\ref{eq:definitionDMthetatheta}).
This contribution is usually subdominant to the Lagrangian mass term and can be neglected kinematically. 
However, as the mass splitting between the heavy neutrinos is typically small in low-scale seesaw scenarios,
the contribution from the Higgs mechanism can lead to significant perturbations to $M_R$.
The physical heavy neutrinos are therefore eigenstates of the mass matrix
\begin{align}
	\hat{M}_N = \hat{M}_R + \Delta \hat{M}_{\theta \theta}\; ,
	\label{def:physMass}
\end{align}
written in the RH neutrino mass basis. 
The matrix $\hat{M}_N$ is diagonalised by $U_{RN}$, defined in Eq.~\eqref{eq:URN}, with its  eigenvalues $M_{N_i}$ corresponding to the states $N_i$.
Depending on the interplay between the symmetry breaking terms, $\delta M_R$ and $\Delta M_R$, parametrised by the splittings $\kappa$ and $\lambda$, respectively,
and the contribution  induced by the Higgs mechanism, $\Delta \hat{M}_{\theta \theta}$,
one can find very different heavy neutrino phenomenology in laboratory  experiments as well as in the early Universe.

At leading order in the splittings $\kappa$ and $\lambda$,
the structure of $\Delta \hat{M}_{\theta \theta}$ is completely determined by the choice of the matrix $\Omega(\mathbf{3^\prime})$,
the couplings $y_f$ and the angle $\theta_R$
\begin{align}
\nonumber
	\Delta \hat{M}_{\theta \theta} &\approx
	\frac{\langle H \rangle^2}{M} \mathrm{Re} \Big( \hat{Y}_D^\dagger \, \hat{Y}_D \Big)
	\\
	& = \frac{\langle H \rangle^2}{M} \, \mathrm{Re} \Big( U_R^\dagger \, \Omega({\bf 3^\prime}) \, R_{kl}(\theta_R) \, (P^{ij}_{kl})^T \, \mathrm{diag}(y_1^2, y_2^2, y_3^2) \, P^{ij}_{kl} \, R_{kl}(-\theta_R) \, \Omega({\bf 3^\prime})^\dagger \, U_R \Big)\; ,
\end{align}
where all the dependence on the matrix $\Omega(\mathbf{3})$ cancels.
Parametrically, one may expect it to be negligible --- of the order of the light neutrino masses.
Hence, for any choice of $\kappa$ and $\lambda$ large enough, $|\kappa|,\,|\lambda| > U^2$, the contribution $\Delta \hat{M}_{\theta \theta}$ is  irrelevant.  
However, in situations that are interesting for collider searches, one of the couplings $y_f$ exceeds this naive expectation,
which causes a sizeable contribution to the heavy neutrino mass splittings.
Specifically we find that $\Delta \hat{M}_{\theta \theta}/M \sim U^2$, which can be orders of magnitude larger than the splittings $\kappa$ and $\lambda$ required for successful leptogenesis, see section~\ref{lepto}. 

\vspace{0.05in}
In the limit of unbroken flavour and CP symmetries, the masses induced by the Lagrangian mass term $M_R$ are exactly degenerate, and the only perturbation to the heavy neutrino mass spectrum comes from the term $\Delta \hat{M}_{\theta \theta}$.
In this case, we always find two states that have a large mixing angle, with the third heavy neutrino approximately decoupled from the light neutrinos. 

\vspace{0.05in}
If the flavour and CP symmetries are broken, i.e.~at least one of the splittings $\kappa$ and $\lambda$ is non-zero, the heavy neutrino mixing depends on the perturbations to the degeneracy limit and, in principle, it can take very different values.

We first consider Case 1) and Case 2), $t$ odd.
The matrix $\hat{M}_N$ coincides for these two cases. It is independent of the choice of the parameter $s$. It also does neither depend on the light neutrino mass spectrum nor on the chosen special value of $\theta_R$. Its form reads
\begin{align}
	\hat{M}_N \approx
	\begin{pmatrix}
		M & 0 & 0\\
		0 & M & 0\\
		0 & 0 & M
	\end{pmatrix} +
	M \begin{pmatrix}
		\frac{U^2}{3} + 2 \kappa & - \frac{U^2}{3 \sqrt{2}} & 0 \\
		-\frac{U^2}{3 \sqrt{2}} & \frac{U^2}{6} - \kappa + \lambda & 0 \\
		0 & 0 & \frac{U^2}{2} - \kappa - \lambda
	\end{pmatrix}\,.
	\label{eq:MphysCase1even}
\end{align}
After diagonalising the mass matrix by $U_{RN}$, we can determine the mixing angles $\Theta_{\alpha i}$. This allows us to find the ratios of the heavy neutrino mixing
\begin{align}
	U_1^2: U_2^2 : U_3^2 = \cos^2 \alpha_N : \sin^2 \alpha_N : 1\; ,
	\label{eq:ratiosCase1even}
\end{align}
where the angle $\alpha_N$ is given by
\begin{align}
	\tan 2 \, \alpha_N = -\frac{4 \, \sqrt{2} \, (3 \, \kappa -\lambda)}{3 \, U^2 +2 \, (3 \, \kappa -\lambda)} \; .
\end{align}
The limiting values are 
\begin{align}
	U_1^2: U_2^2 : U_3^2 =
	\begin{cases}
		1 : 0 : 1 \text{ for } U^2 \gg |3 \, \kappa - \lambda|\; ,\\
		2 : 1 : 3 \text{ for } U^2 \ll |3 \, \kappa - \lambda|\; .
	\end{cases}
	\label{eq:limitRatiosCase1even}
\end{align}
Two comments are in order: first, 
the ordering of $U_1^2$, $U_2^2$ and $U_3^2$ is conventional,
as it is not possible to label the individual heavy neutrinos without specifying $\kappa$ and $\lambda$. Second, the heavy neutrino  corresponding to $0$ is not exactly decoupled from the rest, but instead has a significantly longer lifetime.
A deviation from these limiting values could be used to indirectly measure $|3 \, \kappa-\lambda|$.
 For Case 3 a) and Case 3 b.1) for $m$ even and $s$ odd, we get similar results as for Case 1) and Case 2). Concretely, we have
\begin{equation}
	\hat{M}_N \approx \left( \begin{array}{ccc} 
 M & 0 & 0 \\
 0 & M & 0 \\
 0 & 0 & M
 \end{array}
 \right) + M \, 
 \left(
	\begin{array}{ccc}
		\frac{U^2}{2} + 2 \, \kappa & 0 & 0 \\
		0 & \frac{U^2}{8} - \kappa + \lambda & -\frac{\sqrt{3} \, U^2}{8} \\
		0 & -\frac{\sqrt{3} \, U^2}{8} & \frac{3 \, U^2}{8} - \kappa - \lambda
	\end{array} 
 \right) \; .
  \label{eq:Case3a3b1mevensoddMNhat}
\end{equation}
This result holds for both light neutrino mass orderings as well as all special values of $\theta_R$.
The general form of the ratios is
\begin{equation}
U_1^2 \, : \, U_2^2 \, : \, U_3^2 = 1 \, : \, \sin^2 \alpha_N \, : \, \cos^2 \alpha_N\; ,
\end{equation}
with
\begin{equation}
\tan 2 \, \alpha_N = -\frac{2 \sqrt{3} \, \lambda}{U^2 - 2 \, \lambda} \; .
\end{equation}
As two distinct limits, we identify  
$U^2 \gg |\lambda|$ and $U^2 \ll |\lambda|$, independent of the value of $\kappa$.
For $U^2 \gg |\lambda|$ we have 
\begin{equation}
U_1^2 \, : \, U_2^2 \, : \, U_3^2 = 1 \, : \, 0 \, : \, 1 \; ,
\end{equation}
and for $U^2 \ll |\lambda|$ the result is 
\begin{equation}
U_1^2 \, : \, U_2^2 \, : \, U_3^2 = 4 \, : \, 1 \, : \, 3 \; .
\end{equation}
Eventually, for Case 3 a) and Case 3 b.1) for $m$ odd and $s$ even we find
\begin{equation}
	\hat{M}_N \approx \left( \begin{array}{ccc} 
 M & 0 & 0 \\
 0 & M & 0 \\
 0 & 0 & M
 \end{array}
 \right) + M \, 
 \left(
	\begin{array}{ccc}
		\frac{U^2}{6} + 2 \, \kappa & -\frac{U^2}{6 \, \sqrt{2}} & \frac{U^2}{2 \, \sqrt{6}} \\[0.15cm]
		-\frac{U^2}{6 \, \sqrt{2}} & \frac{11 \, U^2}{24} - \kappa + \lambda & \frac{U^2}{8 \, \sqrt{3}} \\[0.15cm]
		\frac{U^2}{2 \, \sqrt{6}} & \frac{U^2}{8 \, \sqrt{3}} & \frac{3 \, U^2}{8} - \kappa - \lambda
	\end{array} 
 \right) \; .
 \label{eq:Case3a3b1moddsevenMNhat}
\end{equation}
The different possible limits can be found in Tab.~\ref{tab:U2i_ratios}, together with the results for the other cases.

\begin{table}[!t]
\begin{center}
\begin{tabular}{|l|c|c|c|}
\hline
Case & $|\lambda| \gg U^2$ & $|\kappa| \gg U^2 \gg |\lambda|$ & $U^2 \gg |\kappa|, |\lambda|$\\
\hline
Case 1) & $2:1:3$ & $2:1:3$ & $1:0:1$\\
\hline
Case 2), $t$ odd & $2:1:3$ & $2:1:3$ & $1:0:1$\\
\hline
Case 3 a) and Case 3 b.1), & \multirow{2}{*}{$4:1:3$} & \multirow{2}{*}{$1:0:1$} & \multirow{2}{*}{$1:0:1$} \\
$m$ even and $s$ odd &  & & \\
\hline
Case 3 a) and Case 3 b.1), & \multirow{2}{*}{$4:11:9$} & \multirow{2}{*}{$1:3:2$} & \multirow{2}{*}{$0:1:1$}\\
$m$ odd and $s$ even & & & \\
\hline
\end{tabular}
\end{center}
\caption{{\bf Ratios of heavy neutrino mixing,  \mathversion{bold}$U_1^2: U_2^2: U_3^2$\mathversion{normal}}, for the different cases, Case 1) through Case 3 b.1), depending on the relative size of the splittings $\kappa$ and $\lambda$ and the total mixing $U^2$. With these, the heavy neutrino lifetimes can be estimated according to Eq.~(\ref{eq:GammaNi}). These results are valid for both light neutrino masses with NO and with IO. Note that the heavy neutrino corresponding to $0$ does not exactly decouple, but has a negligible coupling compared to the other two ones.
}
\label{tab:U2i_ratios}
\end{table}

We remark that special care needs to be taken when $\Gamma_N \sim |M_{N_i} - M_{N_j}|$, where $M_{N_{i,j}}$ are the heavy neutrino masses and $\Gamma_N$ is the lifetime from Eq.~\eqref{GammaNgeneric}. 
In this occasion, the oscillations between the heavy neutrinos are neither slow enough to be neglected nor fast enough to be averaged out,
but instead a thorough analysis of heavy neutrino oscillations has to be carried out to obtain accurate predictions~\cite{Antusch:2017ebe,Cvetic:2018elt,Tastet:2019nqj}.

\subsection{Limiting cases of heavy neutrino lifetimes}

In general, we find three phenomenologically interesting regimes: $|\lambda| \gg U^2$, $|\kappa| \gg U^2 \gg |\lambda|$ and $U^2 \gg |\kappa|,|\lambda|$, compare Tab.~\ref{tab:U2i_ratios}, which we discuss in more detail.  

\paragraph{Limit of large mass splittings, $|\lambda| \gg U^2$}
This is the situation  most commonly studied in the literature, in  which the eigenstates of the Majorana mass term $M_R$ approximately coincide with the physical heavy neutrino states. This is also consistent with both splittings being large, $|\kappa|\,,|\lambda| \gg U^2$.\footnote{We,  however, avoid the accidental partial mass degeneracy, occurring for $|3\, \kappa-\lambda| \ll U^2$, while $|\kappa|\,,|\lambda| \gg U^2$, see Eq.~\eqref{eq:Mlambda} and text below. The predicted heavy neutrino lifetimes would be modified in that instance.}
In this case, the effect of the matrix $U_{RN}$ can be neglected and we have $\Theta \approx \uptheta \, U_R^*$, compare Eq.~(\ref{eq:asnumix}). 

This can lead to a wide range of heavy neutrino lifetimes, crucially depending on the choice of the angle $\theta_R$~\cite{Curtin:2018mvb,Chauhan:2021xus}. Certain choices of this parameter can lead to very long-lived states, $U^2_i \ll m_\nu/M$~\cite{Curtin:2018mvb,Chauhan:2021xus},
which may give rise to interesting collider signatures, if the heavy neutrinos can be produced through additional effective interactions (see e.g.~\cite{Cottin:2021lzz}).
However, in the scenario considered here, their mixing is too small to ensure sufficient production in any planned laboratory experiment.

\begin{figure}[!t]
  \centering
  \includegraphics[angle=0,width=0.45\textwidth]{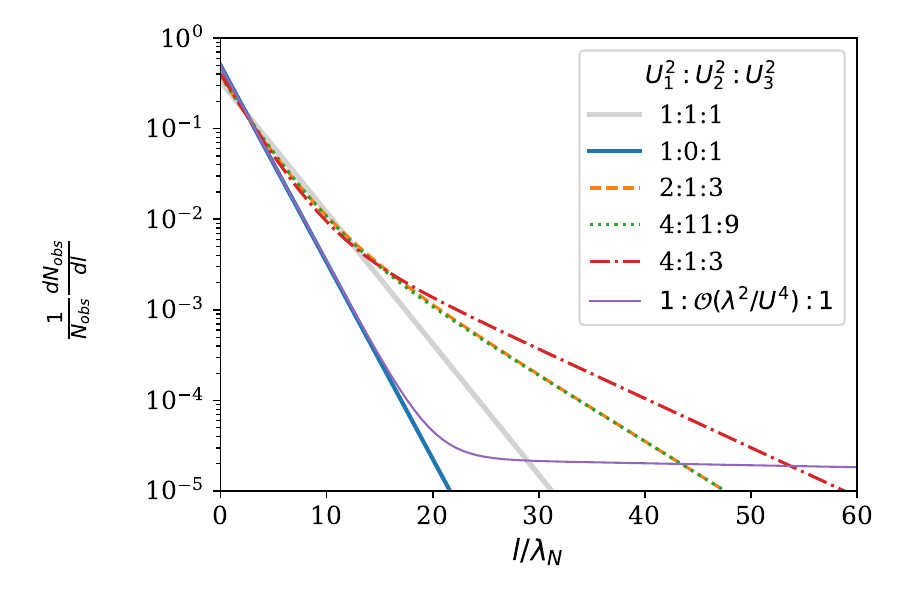}
  \includegraphics[angle=0,width=0.45\textwidth]{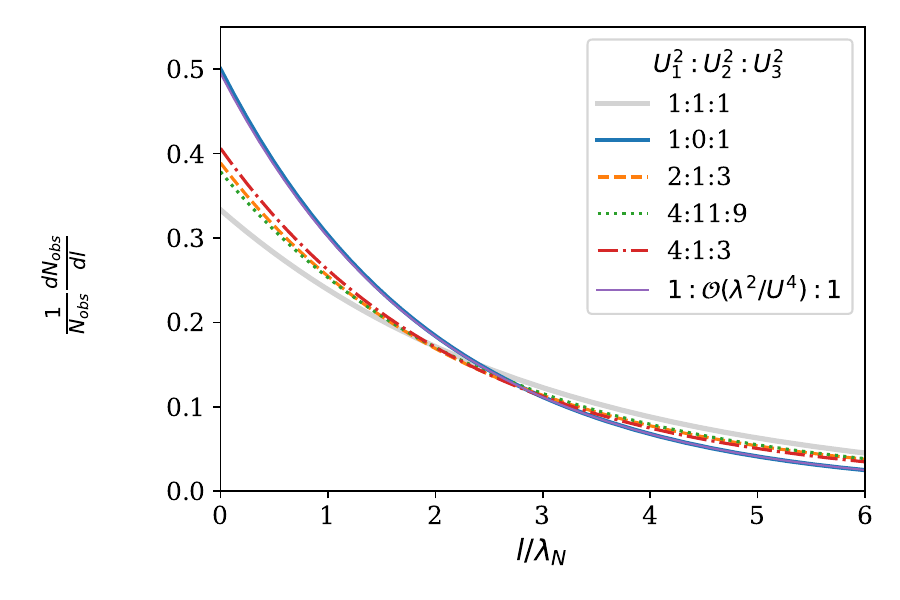}
	\caption{{\bf Expected differential distribution of heavy neutrino decays as a function of \mathversion{bold}$l/\lambda_N$\mathversion{normal}} assuming the different ratios from Tab.~\ref{tab:U2i_ratios}.
		We show the distribution both on a log scale (left) and on a linear scale (right). The following ratios of heavy neutrino mixing are considered: $2:1:3$ (orange, dashed), $4:11:9$ (green, dotted), $4:1:3$ (red, dot-dashed) as well as $1:0:1$ (blue, solid), all found in Tab.~\ref{tab:U2i_ratios}, together with $1:1:1$ (gray, solid). 
   For comparison, we also include a case with a decoupled state, corresponding to Eq.~\eqref{eq:decoupledLambda} with $\kappa=0$ and $\lambda \approx 0.2\cdot U^2$ (violet, solid).}
  \label{Fig:decayDist}
\end{figure}

While in general the heavy neutrino mixing $U_{i}^2$ can differ by orders of magnitude, in the studied scenario the form of the matrices $\Omega(\mathbf{3^\prime})$
involving the TB mixing matrix, see appendix~\ref{appA}, leads to 
 all three heavy neutrinos having  large mixing and to comparable lifetimes in the collider-testable limit, $U^2_i \gg m_\nu/M$.
Based on the results of the preceding section, we find that in Case 1) and Case 2), $t$ odd, the heavy neutrino mixing has simple ratios, $U_1^2: U_2^2 : U_3^2 = 2:1:3$.
In Case 3 a) and Case 3 b.1) these ratios are different and depend on whether $m$ is even and $s$ odd or $m$ odd and $s$ even, see Tab.~\ref{tab:U2i_ratios}.
 Measuring the ratios of the heavy neutrino lifetimes can therefore offer insight into which one of the cases is realised, even if the individual heavy neutrino masses cannot be kinematically distinguished.
The fact that the lifetimes can have non-trivial ratios means that the distribution of the observed lifetimes differs from a simple exponential distribution,
allowing for a possibility of reconstructing the individual heavy neutrino lifetimes by fitting the observed distribution of the heavy neutrino displacements.
An idealised example of such a distribution is shown in Fig.~\ref{Fig:decayDist}.
We display in the left plot the distribution for different ratios of the heavy neutrino mixing $U_1^2 : U_2^2 : U_3^2$ on a log scale, which emphasises the differences between their tails. In the right plot the same results are given on a linear scale, which highlights the differences in the shape of the distributions at a moderate displacement $l$. We see that 
 while the branching ratios $2:1:3$ (orange, dashed) and $4:11:9$ (green, dotted) lead to  practically indistinguishable results, there is a clear difference to the distribution for the ratios $4:1:3$ (red, dot-dashed) which correspond to Case 3 a) and Case 3 b.1) for $m$ even and $s$ odd. Additionally, we find that typical ratios such as $1:0:1$ (blue, solid) and $1:1:1$ (gray, solid) also lead to distinguishable distributions.
  The feasibility of such a measurement depends on the expected event numbers and  experimental details, going beyond the scope of the present work.

  Lastly, we note that the discussed limit, $|\lambda| \gg U^2$, is not always consistent with the condition found in Eq.~(\ref{eq:U2kappa}) and alike, and that for $U^2 \gtrsim \sqrt{m_\nu/M}$, higher order corrections  should be included.

\paragraph{Limit of one large mass splitting, $|\kappa| \gg U^2 \gg |\lambda|$}
In this limit, the state $\hat{\nu}_{R_1}$ is split by $\kappa$ from the remaining states, compare Eq.~(\ref{eq:Mkappa}). Depending on how the contribution $\Delta M_{\theta \theta}$ is aligned with $\delta M_R$, see Eq.~(\ref{eq:deltaMR}), the other two states may or may not remain degenerate.
To good approximation, the matrix $U_{RN}$ therefore only mixes the states $\hat{\nu}_{R_2}$ and $\hat{\nu}_{R_3}$, while $\hat{\nu}_{R_1}$ coincides with an eigenstate of the mass matrix $\hat{M}_N$ in Eq.~(\ref{def:physMass}). 

For $U_i^2 \gg m_\nu/M$, we find for Case 1) and Case 2), $t$ odd, that the term $\Delta \hat{M}_{\theta \theta}$ only affects $\hat{\nu}_{R_1}$ and $\hat{\nu}_{R_2}$, see Eq.~(\ref{eq:MphysCase1even}), effectively splitting $\hat{\nu}_{R_3}$ from these two. As the contribution proportional to the splitting $\kappa$ leads to $M_{R_1} \neq M_{R_2} = M_{R_3}$, the degeneracy between all three states is broken. The eigenstates of $\hat{M}_R$ and $\hat{M}_N$ approximately coincide, and the ratios of the heavy neutrino mixing are the same as for $|\lambda| \gg U^2$.

For Case 3 a) and Case 3 b.1) we find a different behaviour, depending on whether $m$ is even and $s$ odd or vice versa.

If $m$ odd and $s$ even is assumed, we see that the contribution $\Delta M_{\theta \theta}$ affects the entire heavy neutrino mass matrix, compare Eq.~(\ref{eq:Case3a3b1moddsevenMNhat}).
As the state $\hat{\nu}_{R_1}$ becomes separated by the splitting $\kappa$, we only have to identify the eigenstates of the $(23)$-block of the mass matrix $\hat{M}_N$. The heavy neutrino mixing of the state $\hat{\nu}_{R_1}$ therefore remains unchanged, while the one of the other two is modified, leading to the ratios
$U_1^2: U_2^2 : U_3^2 = 1:3:2$, which are the same as for Case 1) and Case 2), $t$ odd.

For $m$ even and $s$ odd, the contribution $\Delta M_{\theta \theta}$ only affects the states $\hat{\nu}_{R_2}$ and $\hat{\nu}_{R_3}$, see Eq.~(\ref{eq:Case3a3b1mevensoddMNhat}).
The heavy neutrino mixing of $\hat{\nu}_{R_1}$ remains the same, while one of the two other states effectively decouples, compare Tab.~\ref{tab:U2i_ratios}.
So, we expect that only two heavy neutrinos can be produced at colliders, with approximately equal mixing, $U_1^2: U_2^2 : U_3^2 = 1:0:1$, see Fig.~\ref{Fig:decayDist} for the resulting distribution.
This decoupling is, however, not exact, but this state can still mix with $U_2^2 \sim \lambda^2/U^2$ 
or in the most pessimistic situation with $U_2^2 \sim m_\nu/M$,
which is beyond the reach of most currently planned collider experiments. 

\paragraph{Limit of small mass splittings, $U^2 \gg |\kappa|\, , |\lambda|$}
In this case, the mass eigenstates are primarily determined by the contribution $\Delta M_{\theta \theta}$.
In the limit of vanishing $\kappa$ and $\lambda$,
one generically finds a pair of states characterised by a large coupling of the order of $U^2/2$,
and a third state that is nearly decoupled.

The mixing of the nearly decoupled state does not necessarily vanish, but is highly suppressed compared to the other two.
If we consider as example Case 1), we find that this state still has a heavy neutrino mixing
\begin{align}
	\label{eq:decoupledLambda}
	U^2_2 \approx \frac{4}{9} \,\frac{(3 \,\kappa - \lambda)^2}{U^2}\;.
\end{align}
This mixing cannot be made arbitrarily small, since the remaining Yukawa couplings also contribute to it. In Fig.~\ref{Fig:decayDist} we display an example of a distribution for the ratios of the  heavy neutrino mixing being $U_1^2:U_2^2:U_3^2 = 1:\mathcal{O} (\lambda^2/U^4):1$ for $\kappa=0$ and $\lambda \approx 0.2 \cdot U^2$ (violet, solid). Its characteristic shape could be used to measure the size of the splitting  $\lambda$.

In order to obtain a lower bound on the mixing of the nearly decoupled heavy neutrino, we include the effect of the smaller couplings.
In Case 1) and Case 2), $t$ odd, this corresponds to
\begin{align}
	U^2_2 \geq \frac{y_2^2 \, \langle H \rangle^2}{M^2} = \frac{m_2}{M}\;,
\end{align}
giving rise to a state that is long-lived compared to the remaining two heavy neutrinos. Nevertheless, it is not stable on cosmological time scales, and such a state typically decays before Big Bang nucleosynthesis (BBN)~\cite{Boyarsky:2020dzc,Sabti:2020yrt}.
For Case 3 a) and Case 3 b.1) we find instead a nearly  decoupled state with
\begin{align}
	U^2_j \geq \frac{y_3^2 \, \langle H \rangle^2}{M^2}\; 
\end{align}
with $j=2$ for $m$ even and $s$ odd, while $j=1$ for $m$ odd and $s$ even, compare Tab.~\ref{tab:U2i_ratios}.
The coupling $y_3$ corresponds to the coupling responsible for the lightest neutrino mass $m_0$ in Case 3 a), if light neutrino masses follow IO, and in Case 3 b.1) in case of NO, compare Eqs.~(\ref{eq:numassesCase3a}-\ref{eq:numassesCase3b1}).
In these cases, the minimal mixing of this heavy neutrino is therefore closely related to $m_0$, i.e.~
\begin{align}
	U^2_j \gtrsim \frac{m_0}{M}\;.
\end{align}
The cosmological constraints on the heavy neutrino lifetime become much more important in this situation, as this state could be sufficiently long-lived to interfere with standard BBN.
 For light neutrino masses with strong IO and Case 3 a) as well as strong NO and Case 3 b.1), we see that the considered scenario can effectively be reduced to the framework with only two heavy neutrinos, since one of the heavy neutrinos can completely decouple.

It is worth noting that although we find nearly decoupled states, this scenario is distinct from the one considered in~\cite{Curtin:2018mvb,Chauhan:2021xus},
as it results from a different choice of the angle $\theta_R$, for which the other two heavy neutrinos can have couplings sizeable enough to be produced at colliders. The latter might not be exactly degenerate in mass, since they can still receive a mass splitting of the order of the light neutrino masses. This can cause heavy neutrino oscillations at colliders~\cite{Drewes:2019byd}, that could be sufficiently fast to lead to LNV signals~\cite{Anamiati:2016uxp,Anamiati:2017rxw,Das:2017hmg,Dib:2017iva,Abada:2019bac,Tastet:2019nqj,Blondel:2021mss,Drewes:2022rsk,Antusch:2023jsa,Antusch:2024otj}. For example, for Case 1) and Case 2), $t$ odd, we expect that this mass splitting is of the order of $|m_3 - m_1|$, which is typically larger than the splitting found in the minimal seesaw model, possibly extending the region of parameter space where LNV is observable at colliders.

We note that the large number of events expected at FCC-ee/CEPC could allow for a measurement of the ratios of the heavy neutrino mixing, even if these cannot be  distinguished kinematically.
While ratios of their lifetimes of order one would require a detailed statistical analysis, it is instructive to remark that the long-lifetime tail in Fig.~\ref{Fig:decayDist} could be distinguished from a simple exponential decay law, if of the order of $10^5$ heavy neutrino decays are detected, since it could lead to an observable number of events in the bin $l/\lambda_N > 30$.
This could in turn make an indirect measurement of a combination of the splittings $\kappa$ and $\lambda$, as indicated in Eq.~(\ref{eq:decoupledLambda}), possible.

\subsection{Mapping to framework with one heavy neutrino}
\label{Sec:PhenoRealsiticModels}

To date, almost all phenomenological studies and experimental searches
for heavy neutrinos that exclusively couple through their mixing are performed assuming only one heavy neutrino species, see section~\ref{ColliderSearches}, while realistic neutrino mass models must contain at least two heavy neutrinos to explain the observed mass squared differences among light neutrinos (and at least three, if $m_0$ is non-zero).

In this subsection we focus on how the fact that we deal with three heavy neutrinos affects the interpretation of experimental results that have been  obtained under the assumption of a single heavy neutrino species, in particular the extraction of the lifetime. 

We consider two observable quantities, the total event number and the distribution of heavy neutrino decays in space, as characterised by their displacement $\decay$.
 In the case of a single heavy neutrino, both quantities are determined by the simple exponential, described by Eq.~(\ref{DifferentialDecay}).
This is in general not the case in the scenario considered here, 
because there can be up to three distinct decay lengths $\Gamma_{N_i}$, whose relation to the quantity $\Gamma_N$, defined in Eq.~(\ref{GammaNgeneric}), has to be clarified.

One can broadly distinguish three instances.
$i)$ If the differences between the different heavy neutrino masses, given by the eigenvalues of $M_N M_N^\dagger$, see Eq.~(\ref{eq:blocks_mass_matrix}) and text below, are larger than the experimental (kinematic) mass resolution, then the phenomenology of any given heavy neutrino species $N_i$ is to good approximation independent of the properties of the other heavy neutrinos. 
In this case we can simply apply the framework with one heavy neutrino, introduced in section~\ref{Sec:OneFlavourModel}, to each heavy neutrino species individually. 
 If the splitting of the heavy neutrino masses is smaller than the experimental resolution, then there are two possibilities. Firstly,
$ii)$ for the splitting being larger than the widths of the heavy neutrino resonances, these fundamentally behave like distinct particle species. Secondly, 
$iii)$ if the splitting is smaller, then there are interferences between them.\footnote{One of the most known examples is the potential suppression of LNV signatures~\cite{Kersten:2007vk}, which can be observed directly in fully reconstructed final states or indirectly through angular distributions and spectra~\cite{Arbelaez:2017zqq,Balantekin:2018ukw}.
However, whether and how these effects manifest themselves cannot predicted from the knowledge of $U_{\alpha i}^2$ alone~\cite{Anamiati:2016uxp,Drewes:2019byd,Antusch:2023nqd}, and their simulation requires non-standard tools~\cite{Antusch:2022ceb}.
 Hence, while they, in principle, provide an interesting target for collider searches, see e.g.~\cite{Anamiati:2017rxw,Das:2017hmg,Dib:2017iva,Cvetic:2018elt,Hernandez:2018cgc,Abada:2019bac,Abada:2022wvh,Antusch:2023jsa}, a detailed analysis goes beyond the scope of this work.} 

In the cases $ii)$ and $iii)$, in which the heavy neutrinos are  kinematically indistinguishable, the total number of produced heavy neutrinos is fixed by the value of $U^2$, while the lifetime is determined by the individual decay rates, $\Gamma_{N_i}$. This can lead to an apparent mismatch between the event numbers and their mean lifetime, when data is interpreted within the framework with a single heavy neutrino, found in section~\ref{Sec:OneFlavourModel}.
 There are several ways to map such scenarios onto the phenomenological framework.

Within the framework with one heavy neutrino, such lifetimes can be interpreted as specific values of 
$\cdec$ and $\cprod$, see~\cite{Drewes:2022rsk}. For pseudo-Dirac heavy neutrinos, this effect can also be interpreted as $c_\mathrm{dec} = 1/2$, if $U^2$ is extracted from the total event numbers.  In fact, in the case with three heavy neutrinos that all decay within the detector volume,
the mean lifetime is $1/3$ of the quantity $\Gamma_N$ expected from Eq.~(\ref{GammaNgeneric}). This could be interpreted as a modified value of $c_\mathrm{dec}$,  $c_\mathrm{dec}=1/3$, if the deviations from a simple exponential decay law are too subtle to be detected, cf.~Fig.~\ref{Fig:decayDist}.

In addition, many studies presume that the heavy neutrinos couple only to one flavour. While this assumption is suitable to define a set of commonly agreed benchmarks for a reasonable comparison between experiments~\cite{Beacham:2019nyx}, the observed neutrino oscillations require mixing with different flavours (cf.~\cite{Drewes:2022akb} for an updated discussion). This matters for collider searches insofar as a careful interpretation of experimental results reveals a strong dependence on the relative sizes $\frac{U_{\alpha}^2}{U^2}$~\cite{Tastet:2021vwp}, which determine the branching ratios of heavy neutrinos into the different flavours.
In the following and in section~\ref{triangleplots} 
 we show that these branching ratios can be used to distinguish between the different cases, Case 1) through Case 3 b.1), in the scenario under consideration. 

\begin{figure}[!t]
    \centering
    \includegraphics[width=.49\textwidth]{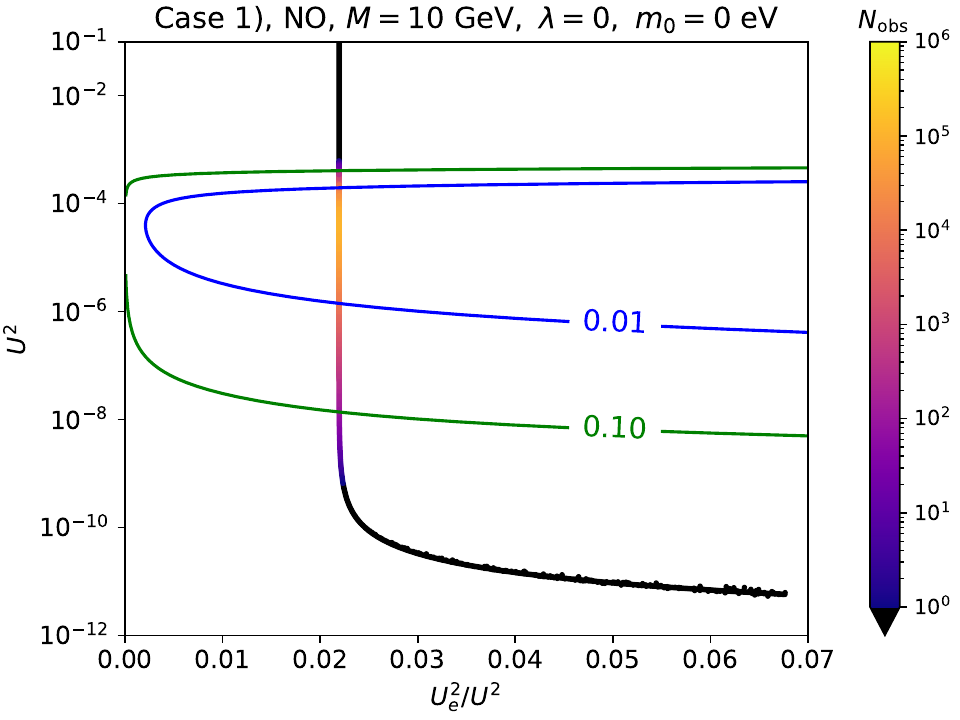}
    \includegraphics[width=.49\textwidth]{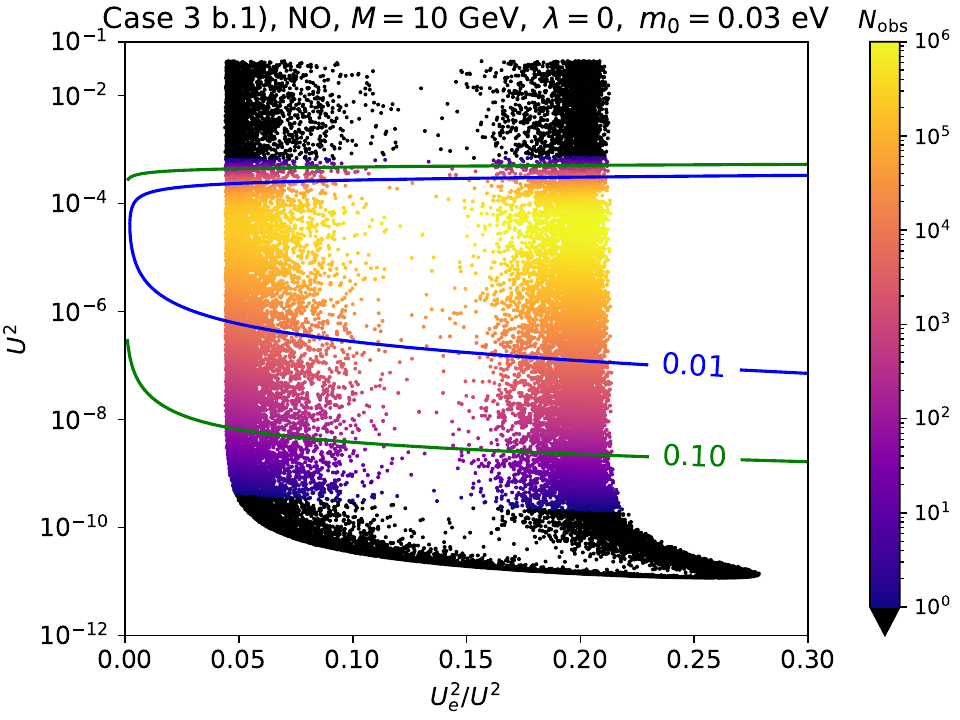}
    \caption{{\bf Event numbers displayed in the  \mathversion{bold} $\frac{U_e^2}{U^2}-U^2$\mathversion{normal}-plane} for semi-leptonic decays expected at FCC-ee/CEPC for $10^{12}$ produced $Z$ bosons. The Majorana mass $M$ is set to $M = 10$ GeV, the splitting $\lambda$ to $\lambda=0$ and we assume light neutrino masses follow NO, while marginalising over $\kappa$. The left plot refers to Case 1) with the lightest neutrino mass being zero, $m_0=0$, while in the right plot results for Case 3 b.1) and $m_0=0.03$ eV are displayed. For both cases we marginalise over $\frac{s}{n}$, and for Case 3 b.1) also over $\frac{m}{n}$. The colour bar indicates the obtained event numbers; in particular, black points refer to numbers smaller than one. The green (blue) contour highlights the region in which $\frac{U_e^2}{U^2}$ can be measured with $10 \%$ ($1\%$) relative accuracy. For further details see text.}
    \label{fig:flavratiovsU2Case1}
\end{figure}

In Fig.~\ref{fig:flavratiovsU2Case1} (left plot), we show for a fixed value of the Majorana mass $M=10$ GeV the expected event numbers assuming semi-leptonic decays for Case 1) during the $Z$-pole run of FCC-ee/CEPC as function of the ratio $\frac{U_e^2}{U^2}$, presuming $10^{12}$ produced $Z$ bosons. We only consider this channel and do not make any assumption regarding the ratios $\frac{U_\mu^2}{U^2}$ and $\frac{U_\tau^2}{U^2}$. The results are based on the simplified formula given in Eq.~(\ref{eq:observed events_alpha}). We marginalise over $\frac{s}{n}$.
 Furthermore, light neutrino masses follow strong NO and we set $\lambda=0$, while we marginalise over the splitting $\kappa$. However, both $\kappa$ and $\lambda$ do not impact the value of the ratios $\frac{U_\alpha^2}{U^2}$ as long as the condition in Eq.~(\ref{eq:kappalambdathetaR}) is fulfilled.

Black points in Fig.~\ref{fig:flavratiovsU2Case1} correspond to combinations of $\frac{U_e^2}{U^2}$ and $U^2$, for which the observed event number is smaller than one, while the colour bar indicates that event numbers as large as $10^5$ could be reached. From these, it is possible to estimate the statistical uncertainty on the measurement of the ratios
$\frac{U_\alpha^2}{U^2}$, neglecting any systematic uncertainty, as
\begin{equation}
\delta\left(\frac{U_\alpha^2}{U^2}\right)/\left(\frac{U_\alpha^2}{U^2}\right) \sim \sqrt{1/N_{\mathrm{obs}}^\alpha-1/N_{\mathrm{obs}}} \; ,
\end{equation}
 compare appendix B in~\cite{Antusch:2017pkq}. The green and blue contours in Fig.~\ref{fig:flavratiovsU2Case1} enclose  the region in which the ratio $\frac{U_e^2}{U^2}$ can be measured with a relative accuracy better than $10 \%$ and $1\%$, respectively. 
 
 For comparison, we display in the right plot in Fig.~\ref{fig:flavratiovsU2Case1} an example for Case 3 b.1), still assuming that $M=10 \, \mathrm{GeV}$, light neutrinos follow NO as well as $\lambda=0$, while marginalising over both $\frac{s}{n}$ (in its admitted range) and the splitting $\kappa$. Here, we marginalise as well over $\frac{m}{n}$ in its allowed range. We remind that the combinations of $m$ and $s$ that can lead to large active-sterile mixing $U^2$ are $m$ even and $s$ odd or vice versa. 
 In contrast, we take the lightest neutrino mass to be $m_0=0.03 \, \mathrm{eV}$. The colour-coding is the same as in the left plot. As we observe, a larger variety and also larger size of the values of $\frac{U_e^2}{U^2}$ are  anticipated.

These examples show that the ratios $\frac{U_\alpha^2}{U^2}$ could, indeed, be quite precisely measured, potentially permitting to distinguish between the different cases, Case 1) through Case 3 b.1), as well as among the light neutrino mass orderings and making it possible to extract the lightest neutrino mass $m_0$, as we comment towards the end of the following section.

\section{Branching ratios of heavy neutrinos and ternary plots}
\label{triangleplots}

In this section, we study the flavour ratios $\frac{U_\alpha^2}{U^2}$, determining the decay of the heavy neutrinos into the different flavours $\alpha$ in the limit
of mass-degenerate states and large active-sterile mixing $U^2$. 
As can be seen from Fig.~\ref{fig:flavratiovsU2Case1}, the latter assumption holds to (very) good accuracy in the region of parameter space, for which heavy neutrinos can potentially be discovered. 

We present results for the ternary plots, analytically and numerically, and for the different cases, Case 1) through Case 3 b.1).
Since we focus on the situation of large active-sterile mixing $U^2$, the angle $\theta_R$ is
always assumed to be close to a special value, see section~\ref{sec32} and, consequently, one of the couplings $y_f$, $f=1,2,3$, is large, while we can neglect
the other two couplings (in the analytic study). Before discussing the different cases in turn, it is instructive to analyse the instance in which the lepton mixing matrix is not assumed to be constrained
by any symmetry, but rather by experimental data, as summarised in the global fit~\cite{Esteban:2020cvm}. 

\subsection{Generic lepton mixing matrix}
\label{sec:gencase}

\begin{figure}[!t]
	\centering
	\includegraphics[width=0.49\textwidth]{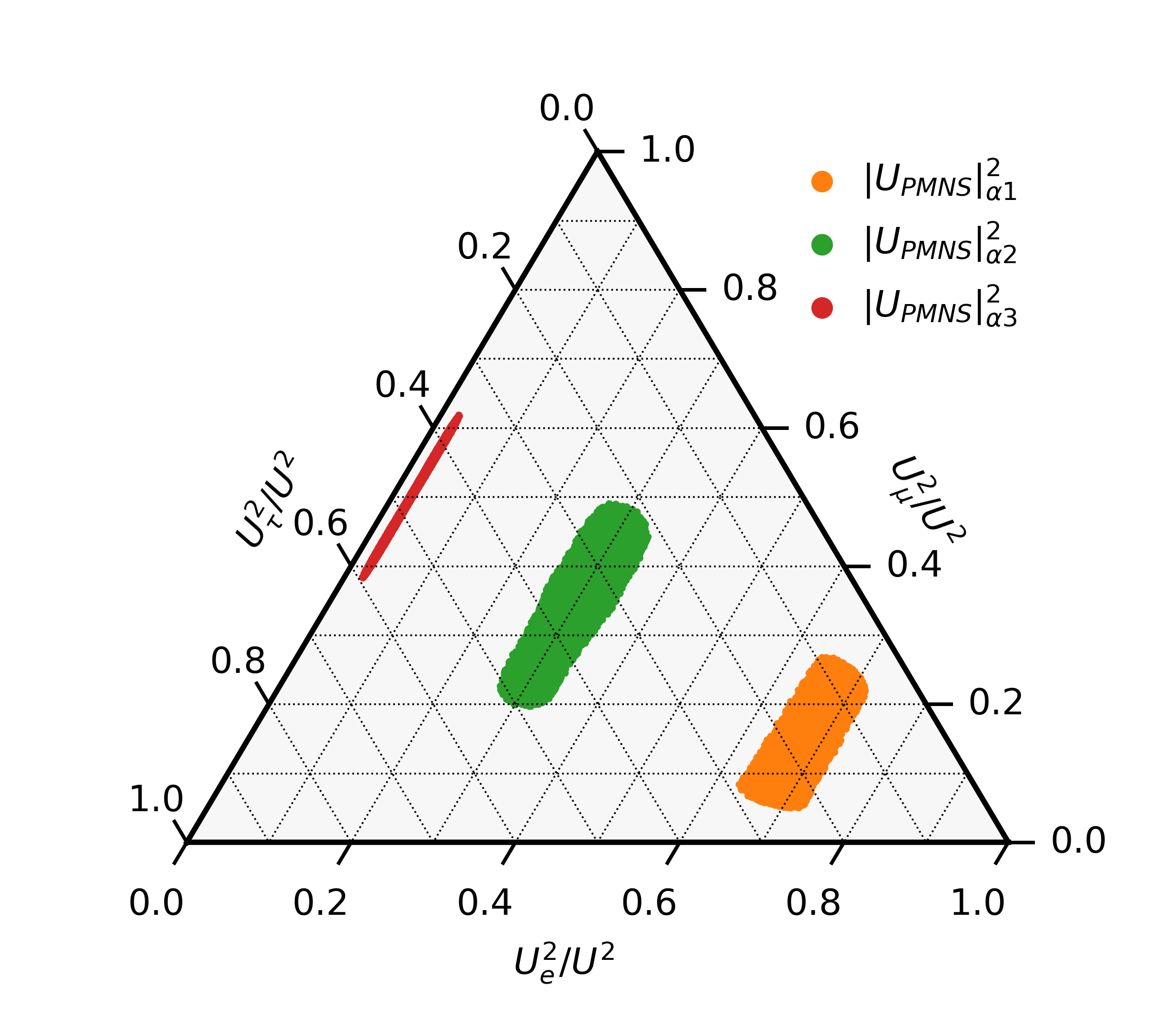}
	\caption{
        {\bf Generic lepton mixing matrix} 
Results for the ratios $\frac{U_\alpha^2}{U^2}$ in case the lepton mixing matrix is not constrained by a symmetry, but by experimental data only~\cite{Esteban:2020cvm}. We marginalise over the three lepton mixing angles $\theta_{12}$, $\theta_{13}$, $\theta_{23}$, as well as the CP phase $\delta$. We use the tabulated $\chi^2$ values for light neutrino masses with NO.
}
\label{fig:ternaryPMNS}
\end{figure}

We still assume that the structure of the matrices $M_R$ and $Y_D$ is as given in Eqs.~(\ref{eq:MR0},\ref{eq:formYD}).
With the definition of $U_{\alpha i}^2$, see Eq.~(\ref{U2defs}), we can write $U_\alpha^2$ and $U_i^2$ as
\begin{equation} 
U_\alpha^2 = \left( \Theta \, \Theta^\dagger \right)_{\alpha \alpha} \;\;\; \mbox{and} \;\;\; U_i^2 = \left( \Theta^\dagger \, \Theta \right)_{i i}^\star \; . 
\end{equation}
Using the form of $M_R$ in Eq.~(\ref{eq:MR0}), i.e.~we neglect possible mass splittings, we have
\begin{equation}
\Theta \, \Theta^\dagger = \frac{\langle H \rangle^2}{M^2} \, Y_D \, Y_D^\dagger \; ,
\end{equation}
and, furthermore, with the form of $Y_D$, as in Eq.~(\ref{eq:formYD}), we find 
\begin{equation}
\Theta \, \Theta^\dagger  = \frac{\langle H \rangle^2}{M^2} \,  \Omega ({\bf 3}) \, R_{ij} (\theta_L) \, \mbox{diag} \, (y_1^2, y_2^2, y_3^2) \, \left( \Omega ({\bf 3}) \, R_{ij} (\theta_L) \right)^\dagger \; .
\end{equation}
We can distinguish two cases, namely the matrix in Eq.~(\ref{eq:combdiag}) is diagonal and thus the lepton mixing matrix is given by Eq.~(\ref{eq:UPMNScombdiag}) and, hence, $\Theta \, \Theta^\dagger$ is (for Case 1) through Case 3 a), we comment on Case 3 b.1) subsequently)
\begin{equation}
\label{eq:noaddrotation}
\Theta \, \Theta^\dagger =  \frac{\langle H \rangle^2}{M^2} \, U_{\mbox{\scriptsize{PMNS}}} \, \mbox{diag} \, (y_1^2, y_2^2, y_3^2) \, U_{\mbox{\scriptsize{PMNS}}}^\dagger \; ,
\end{equation}
where the matrix $K_\nu$ drops out. If instead the matrix in Eq.~(\ref{eq:combdiag}) is not diagonal, we have 
\begin{equation}
\label{eq:addrotation}
\Theta \, \Theta^\dagger =  \frac{\langle H \rangle^2}{M^2} \, U_{\mbox{\scriptsize{PMNS}}} \, K_\nu^\dagger \, R_{ij} (-(\widetilde{\theta}_L - \theta_L)) \, \mbox{diag} \, (y_1^2, y_2^2, y_3^2) \, R_{ij} (\widetilde{\theta}_L - \theta_L) \, K_\nu \, U_{\mbox{\scriptsize{PMNS}}}^\dagger \; .
\end{equation}
If the form of $\Theta \, \Theta^\dagger$ is as in Eq.~(\ref{eq:noaddrotation}), we get, e.g.~for only $y_1$ non-vanishing (which corresponds to the limit in which this coupling is large)
\begin{equation}
\left( \Theta \, \Theta^\dagger \right)_{\alpha \beta} = \frac{\langle H \rangle^2}{M^2} \, y_1^2 \, \left( U_{\mbox{\scriptsize{PMNS}}}\right)_{\alpha 1} \, \left( U_{\mbox{\scriptsize{PMNS}}}\right)_{\beta 1}^\star \; 
\end{equation} 
as well as (taking the lepton mixing matrix to be unitary)
\begin{equation}
U^2 = \sum_\alpha U_\alpha^2 = \frac{\langle H \rangle^2}{M^2} \, y_1^2 \, \sum_\alpha \left| \left( U_{\mbox{\scriptsize{PMNS}}}\right)_{\alpha 1} \right|^2 = \frac{\langle H \rangle^2}{M^2} \, y_1^2 
\end{equation}
and, thus, we end up with
\begin{equation}
\label{eq:Ua2U21stcolumn}
\frac{U_\alpha^2}{U^2} =  \left| \left( U_{\mbox{\scriptsize{PMNS}}}\right)_{\alpha 1} \right|^2 \; ,
\end{equation}
meaning that the ratios $\frac{U_\alpha^2}{U^2}$ are solely determined by the values of the elements of the first column of the lepton mixing matrix. Numerically we get 
\begin{equation}
0.64 \lesssim \frac{U_e^2}{U^2} \lesssim 0.71 \; , \;\;  0.05 \lesssim \frac{U_\mu^2}{U^2} \lesssim 0.26 \; , \;\; 0.07 \lesssim \frac{U_\tau^2}{U^2} \lesssim 0.28 \; ,
\end{equation}
using the results of the NuFIT collaboration~\cite{Esteban:2020cvm}. This is shown as orange area in the ternary plot in Fig.~\ref{fig:ternaryPMNS}.
Similarly, we find for $y_2$ being the only non-negligible coupling that
\begin{equation}
\label{eq:Ua2U22ndcolumn}
\frac{U_\alpha^2}{U^2} =  \left| \left( U_{\mbox{\scriptsize{PMNS}}}\right)_{\alpha 2} \right|^2 \; ,
\end{equation}
as well as for $y_3$ being the only relevant coupling
\begin{equation}
\label{eq:Ua2U23rdcolumn}
\frac{U_\alpha^2}{U^2} =  \left| \left( U_{\mbox{\scriptsize{PMNS}}}\right)_{\alpha 3} \right|^2 \; .
\end{equation}
Plugging in the results from~\cite{Esteban:2020cvm}, we have
\begin{equation}
0.26 \lesssim \frac{U_e^2}{U^2} \lesssim 0.34 \; , \;\;  0.21 \lesssim \frac{U_\mu^2}{U^2} \lesssim 0.48 \; , \;\; 0.22 \lesssim \frac{U_\tau^2}{U^2} \lesssim 0.49 
\end{equation}
and
\begin{equation}
\frac{U_e^2}{U^2} \approx 0.02 \; , \;\;  0.40 \lesssim \frac{U_\mu^2}{U^2} \lesssim 0.61 \; , \;\; 0.37 \lesssim \frac{U_\tau^2}{U^2} \lesssim 0.58 \; ,
\end{equation}
respectively. These correspond to the green and red areas in the ternary plot in Fig.~\ref{fig:ternaryPMNS}, respectively. As one can see, the three areas do not overlap in the ternary plot
and most strikingly the ratio $\frac{U_e^2}{U^2}$ is very small and entirely fixed by the (well-measured) reactor mixing angle in the last case (i.e.~only $y_3$ is large).
We emphasise that any dependence on the Majorana phases drops out in the formulae in Eqs.~(\ref{eq:Ua2U21stcolumn},\ref{eq:Ua2U22ndcolumn},\ref{eq:Ua2U23rdcolumn}). 

In case $\widetilde{\theta}_L$ and $\theta_L$ do not coincide, i.e.~the matrix in Eq.~(\ref{eq:combdiag}) is not diagonal, the form of $\Theta \, \Theta^\dagger$
simplifies to Eq.~(\ref{eq:noaddrotation}) if the indices $i$ and $j$ of the plane of rotation are the same as the indices of the two couplings $y_f$, $f=1,2,3$, that can be neglected, e.g.~for $(ij)=(12)$ only the coupling $y_3$
should be sizeable in order to obtain the same result as above, see Eq.~(\ref{eq:Ua2U23rdcolumn}).

For Case 3 b.1),
we have to take into account the permutation $P$, compare Eq.~(\ref{eq:PMNSCase3b1}), and then obtain for only $y_1$ large that
\begin{equation}
\frac{U_\alpha^2}{U^2} =  \left| \left( U_{\mbox{\scriptsize{PMNS}}}\right)_{\alpha 2} \right|^2 
\end{equation}
holds instead of the relation in Eq.~(\ref{eq:Ua2U21stcolumn}). Similarly, we get
 \begin{equation}
 \frac{U_\alpha^2}{U^2} =  \left| \left( U_{\mbox{\scriptsize{PMNS}}}\right)_{\alpha 3} \right|^2 \;\;\; \mbox{and} \;\;\; \frac{U_\alpha^2}{U^2} =  \left| \left( U_{\mbox{\scriptsize{PMNS}}}\right)_{\alpha 1} \right|^2 
 \end{equation}
 instead of Eqs.~(\ref{eq:Ua2U22ndcolumn}) and (\ref{eq:Ua2U23rdcolumn}), respectively.

\subsection{Symmetry-determined lepton mixing matrix}
\label{sec:symmcase}

We now discuss the different cases, Case 1) through Case 3 b.1), both analytically and numerically.
In doing so, the lightest neutrino mass $m_0$ can vary between $0$ and the maximal value ($m_0=0.03$ eV for NO and $m_0=0.015$ eV for IO) compatible with cosmological observations~\cite{Planck:2018vyg}. For $m_0$ being non-zero and for Case 3 a) and Case 3 b.1) in certain instances also for vanishing $m_0$, the angle $\theta_L$ appearing in the formulae for $\frac{U^2_\alpha}{U^2}$ does not coincide with the free angle $\widetilde{\theta}_L$,
present in the lepton mixing matrix, see section~\ref{sec2}. We take this effect into account by computing/estimating the size of the additional rotation. 
The value of the angle $\widetilde{\theta}_L$ is obtained by fitting the different lepton mixing patterns to the experimentally
preferred best-fit values~\cite{Esteban:2020cvm}. In the following, we use the results from~\cite{Drewes:2022kap}.

\begin{figure}[!t]
	\includegraphics[width=0.49\textwidth]{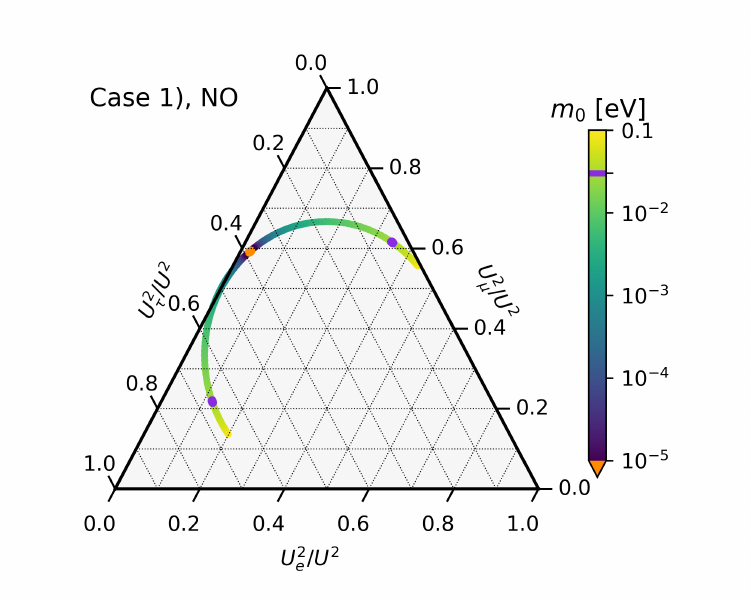}
	\includegraphics[width=0.49\textwidth]{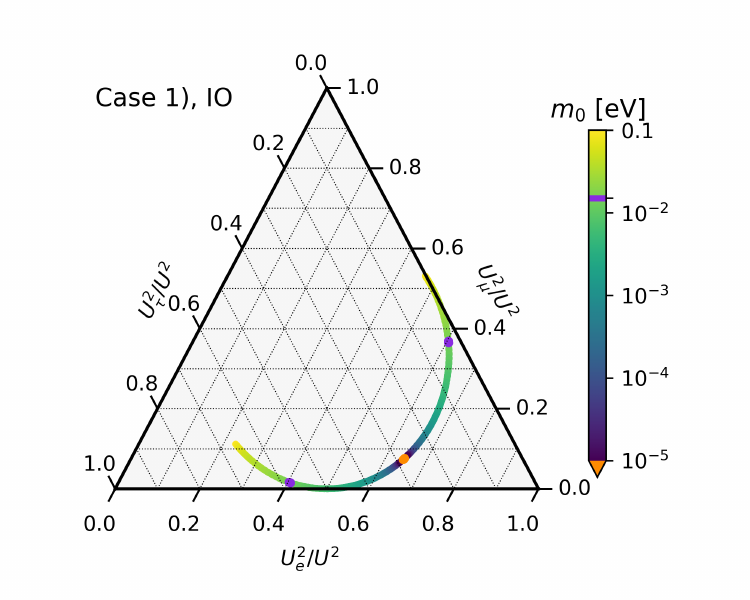}
	\caption{
{\bf Case 1)} Results for ratios $\frac{U_\alpha^2}{U^2}$ assuming light neutrino masses with NO (left plot) and IO (right plot), respectively. The different colours indicate different values of the
lightest neutrino mass $m_0$. The point shown in orange corresponds to the choice $m_0=0$, while the points in violet correspond to the current upper bound on $m_0$ from cosmology~\cite{Planck:2018vyg}.}
\label{fig:Case1ternary}
\end{figure}

\paragraph{Case 1)} In the large $U^2$ limit, we find for the ratios
\begin{equation}
\label{eq:Ua2U2Case1}
    \frac{U_\alpha^2}{U^2} \approx 
    \begin{cases}
    \frac23 \sin^2 \, \theta_{L, \alpha} \; & \text{for NO} \; , \\
    \frac23 \cos^2 \, \theta_{L, \alpha} & \text{for IO} \; ,
    \end{cases}
\end{equation}
with $\theta_{L,\alpha}$ being defined as in~\cite{Drewes:2022kap}, i.e.~$\theta_{L,\alpha} = \theta_L + \rho_\alpha \, \frac{4\pi}{3}$, where $\rho_e = 0$, $\rho_\mu = +1$ and $\rho_\tau = -1$. We note that these results are independent of the 
choice of the parameter $s$ ($s$ even or $s$ odd). 

For $m_0=0$, $\theta_L$ also appears in the lepton mixing matrix and we can immediately derive numerical values for $\frac{U_\alpha^2}{U^2}$. For light neutrino masses with strong NO, we have for $\alpha=e$ that the ratio
is given by the reactor mixing angle,
\begin{equation}
\label{eq:Case1Ue2U2strongNO}
\frac{U_e^2}{U^2} = \sin^2 \theta_{13} \approx 0.022 \; ,
\end{equation}
compare the analytic formulae given in~\cite{Hagedorn:2014wha}. Furthermore, we know that $\theta_L$ should be close to $0.18$, in order to accommodate the rest of the lepton mixing angles well, see~\cite{Drewes:2022kap} 
and Eq.~(\ref{eq:Case1thetaLfit}), such that
\begin{equation}
\frac{U_\mu^2}{U^2} \approx 0.59 \;\;\; \mbox{and} \;\;\; \frac{U_\tau^2}{U^2} \approx 0.39 \; .
\end{equation}
These ratios agree very well with the point marked in orange in the ternary plot in the left of Fig.~\ref{fig:Case1ternary}. Analogously, we find for strong IO that for $\alpha=e$
\begin{equation}
\label{eq:Case1Ue2U2strongIO}
\frac{U_e^2}{U^2} = \frac 23 - \sin^2 \theta_{13} \approx 0.64 
\end{equation} 
as well as, using $\theta_L \approx 0.18$ as needed for an acceptable fit to the lepton mixing angles, 
\begin{equation}
\frac{U_\mu^2}{U^2} \approx 0.08 \;\;\; \mbox{and} \;\;\; \frac{U_\tau^2}{U^2} \approx 0.28 \; .
\end{equation}
These values also agree well with the orange point highlighted in the ternary plot in the right of Fig.~\ref{fig:Case1ternary}.
Including the effect of non-zero $m_0$ requires taking into account that the angles $\widetilde{\theta}_L$ and $\theta_L$ do not coincide. We first note that the special values of $\theta_R$ which allow for large active-sterile mixing $U^2$
fulfil $\cos 2 \,\theta_R \approx 0$. Thus, for large $m_0$ the relation between the two angles is
\begin{equation}
\label{eq:relthetaLtildethetaL}
\theta_L \approx \widetilde{\theta}_L - k \, \frac{\pi}{4} \;\;\; \mbox{with} \;\;\; k \;\; \mbox{being odd.}
\end{equation}
The angle $\widetilde{\theta}_L$, appearing in the lepton mixing matrix, must be close to $0.18$. Considering the different values of $k$ we have two possible results for the ratios $\frac{U_\alpha^2}{U^2}$, either
\begin{equation}
\frac{U_e^2}{U^2} \approx 0.21 \; , \;\; \frac{U_\mu^2}{U^2} \approx 0.12 \;\;\; \mbox{and} \;\;\; \frac{U_\tau^2}{U^2} \approx 0.66 \; ,
\end{equation}
or
\begin{equation}
\frac{U_e^2}{U^2} \approx 0.45 \; , \;\; \frac{U_\mu^2}{U^2} \approx 0.54 \;\;\; \mbox{and} \;\;\; \frac{U_\tau^2}{U^2} \approx 0.004 \; .
\end{equation}
These values are very similar to those obtained for $m_0=0.1$ eV which is the largest value shown in yellow in Fig.~\ref{fig:Case1ternary}. Performing the same analysis for light neutrino masses with IO and large $m_0$, 
we obtain very similar values, compare the yellow region in the ternary plot in the right of Fig.~\ref{fig:Case1ternary}. 
For intermediate values of $m_0$ the difference between the angles $\theta_L$ and $\widetilde{\theta}_L$ is e.g.~smaller than $\frac{\pi}{4}$, and thus the different points on the arcs in Fig.~\ref{fig:Case1ternary}, for NO and IO, can be accessed.  
As we can see, assuming Case 1) leads to a very reduced parameter space in the ternary plots with no overlap between the areas obtained for light neutrino masses with NO or IO, if the constraints on the lightest neutrino mass $m_0$ arising from cosmological observations~\cite{Planck:2018vyg} are taken into account.

\paragraph{Case 2)} We proceed with the analysis of this case in a similar way as for Case 1). We first mention the approximate results for the ratios $\frac{U_\alpha^2}{U^2}$
\begin{equation}
\label{eq:Ua2U2Case2}
    \frac{U_\alpha^2}{U^2} \approx 
    \begin{cases}
    \frac13 \, \left( 1 -\cos 2 \, \theta_L \, \cos \phi_{u, \alpha} \right) \; & \text{for NO} \; , \\
    \frac13 \, \left( 1 + \cos 2 \, \theta_L \, \cos \phi_{u, \alpha} \right) & \text{for IO} \; . 
    \end{cases}
\end{equation}
We remind that only for $t$ odd large active-sterile mixing $U^2$ can be achieved, as only in this case the light neutrino masses also depend on the choice of the angle $\theta_R$. Furthermore, the formulae in Eq.~(\ref{eq:Ua2U2Case2})
are valid for all choices of $s$. The parameter $\phi_{u,\alpha}$ is defined as $\phi_{u,\alpha} = \phi_u + \rho_\alpha \frac{4\pi}{3}$, where $\rho_e=0$, $\rho_\mu=-1$ and $\rho_\tau = +1$, see~\cite{Drewes:2022kap}.  
Note that these results do not depend on the parameter $v$ that is proportional to $t$, see Eq.~(\ref{eq:def_u_v}).

\begin{figure}[!t]
	\includegraphics[width=0.49\textwidth]{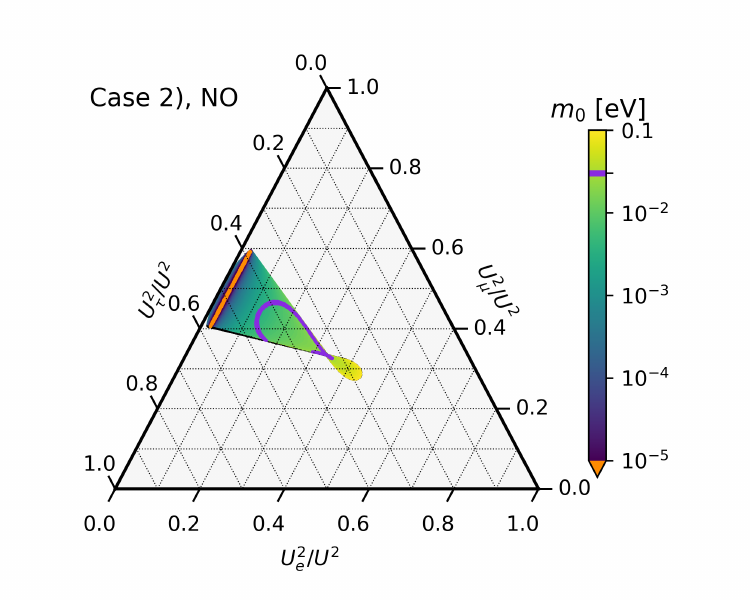}
	\includegraphics[width=0.49\textwidth]{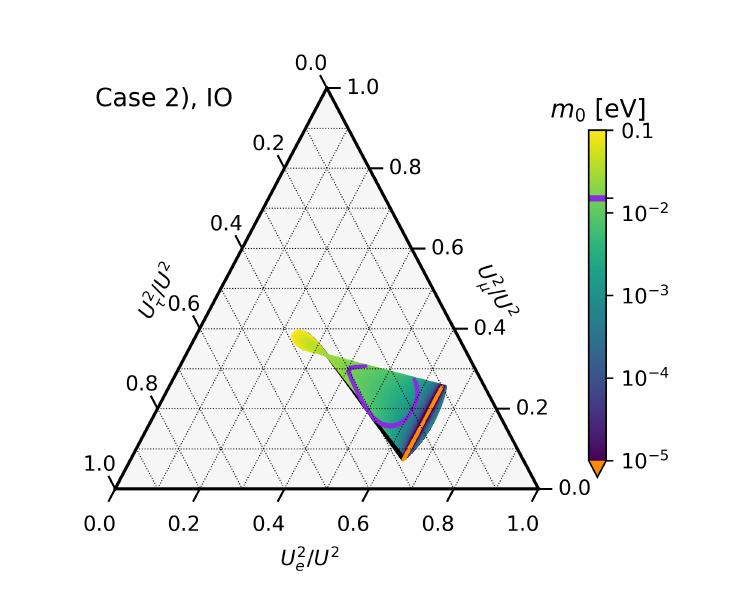}
	\caption{{\bf Case 2)} Results for ratios $\frac{U_\alpha^2}{U^2}$ assuming light neutrino masses with NO (left plot) and IO (right plot), respectively. The colour-coding is the same as in Fig.~\ref{fig:Case1ternary}. Note that the colours indicate the minimal value of $m_0$ that allows for a particular $\frac{U_\alpha^2}{U^2}$. The black regions correspond to regions of parameter space in which only the lepton mixing angles can
be fitted at the $3 \, \sigma$ level or better, but not the CP phase $\delta$~\cite{Esteban:2020cvm}.  
}
\label{fig:Case2ternary}
\end{figure}

Again, we begin with the study of light neutrino masses with strong NO. For $\alpha=e$, we can directly express the ratio in terms of the reactor mixing angle, see Eq.~(\ref{eq:Case1Ue2U2strongNO}); compare~\cite{Hagedorn:2014wha}
for the relation of $\sin^2 \theta_{13}$ and the parameters $\phi_u$ and $\theta_L$. For the other
two flavours we obtain 
\begin{equation}
\frac{U_\mu^2}{U^2} \approx 0.49 + 0.27 \, \tan \phi_u \;\;\; \mbox{and} \;\;\; \frac{U_\tau^2}{U^2} \approx 0.49 - 0.27 \, \tan \phi_u \; ,
\end{equation}
using the relation between the reactor mixing angle and the parameters $\phi_u$ and $\theta_L$ and the experimentally preferred best-fit value of $\sin^2 \theta_{13}$. If we take into account that $\phi_u$ must take values between $-0.314$ and $0.377$ 
to ensure compatibility with the data on lepton mixing angles, see Eq.~(\ref{eq:Case2unrange}), we get very similar admitted intervals for these two ratios,
\begin{equation}
0.40 \lesssim \frac{U_\mu^2}{U^2} \lesssim 0.60 \;\;\; \mbox{and} \;\;\; 0.38 \lesssim \frac{U_\tau^2}{U^2} \lesssim 0.58 \; .
\end{equation}
These ranges should be confronted with the orange stripes shown in the ternary plot in the left of Fig.~\ref{fig:Case2ternary}. Similarly, we find for light neutrino masses with strong IO that the ratio for the electron flavour, $\alpha=e$, is directly determined by the measured
value of the reactor mixing angle, see Eq.~(\ref{eq:Case1Ue2U2strongIO}), while the ratios for the muon and tau flavours also depend on the value of the parameter $\phi_u$, i.e.~
\begin{equation}
\frac{U_\mu^2}{U^2} \approx 0.18 - 0.27 \, \tan \phi_u \;\;\; \mbox{and} \;\;\; \frac{U_\tau^2}{U^2} \approx 0.18 + 0.27 \, \tan \phi_u 
\end{equation}
such that
\begin{equation}
0.07 \lesssim \frac{U_\mu^2}{U^2} \lesssim 0.27 \;\;\; \mbox{and} \;\;\; 0.09 \lesssim \frac{U_\tau^2}{U^2} \lesssim 0.29 \; .
\end{equation}
These ranges correspond to the orange line shown in the ternary plot in the right of Fig.~\ref{fig:Case2ternary}. In order to also study the limit of large $m_0$ analytically, we have to take into account the effect of the 
difference between the angles $\widetilde{\theta}_L$ and $\theta_L$. We first remind that for large active-sterile mixing $U^2$, the angle $\theta_R$ is constrained by $\sin 2 \, \theta_R \approx 0$. This translates into the same difference between 
$\widetilde{\theta}_L$ and $\theta_L$ as mentioned in Eq.~(\ref{eq:relthetaLtildethetaL}). Rewriting $\widetilde{\theta}_L$ in terms of the reactor mixing angle and the parameter $\phi_u$, see~\cite{Hagedorn:2014wha}, leaves only the latter as independent parameter, which is varied in the interval shown in Eq.~(\ref{eq:Case2unrange}). In this way, we get for both light neutrino mass  orderings as approximate ranges of the flavour ratios
\begin{equation}
0.21 \lesssim \frac{U_e^2}{U^2} \lesssim 0.45 \;\;\; \mbox{and} \;\;\; 0.26 \lesssim \frac{U_\mu^2}{U^2} \, , \, \frac{U_\tau^2}{U^2} \lesssim 0.40 \; .
\end{equation}
We see that, in particular, the attainable values of $\frac{U_e^2}{U^2}$ considerably change with the value of $m_0$.
The ranges match the yellow regions (for $m_0=0.1$ eV) in the ternary plots in Fig.~\ref{fig:Case2ternary}. The possibility to distinguish between light neutrino masses with NO and IO crucially depends on the size of $m_0$ and for larger
values the results for the ratios $\frac{U_\alpha^2}{U^2}$ become more similar and closer to $\frac{U_\alpha^2}{U^2} \approx \frac 13$. In Fig.~\ref{fig:Case2ternary}, 
the black areas indicate the region of parameter space in which the lepton mixing angles can
be fitted at the $3 \, \sigma$ level or better, whereas this is not possible for the CP phase $\delta$. As we can see, for Case 2) this only very mildly affects the allowed parameter space.

\begin{figure}[!t]
    \centering
    \includegraphics[width=.49\textwidth]{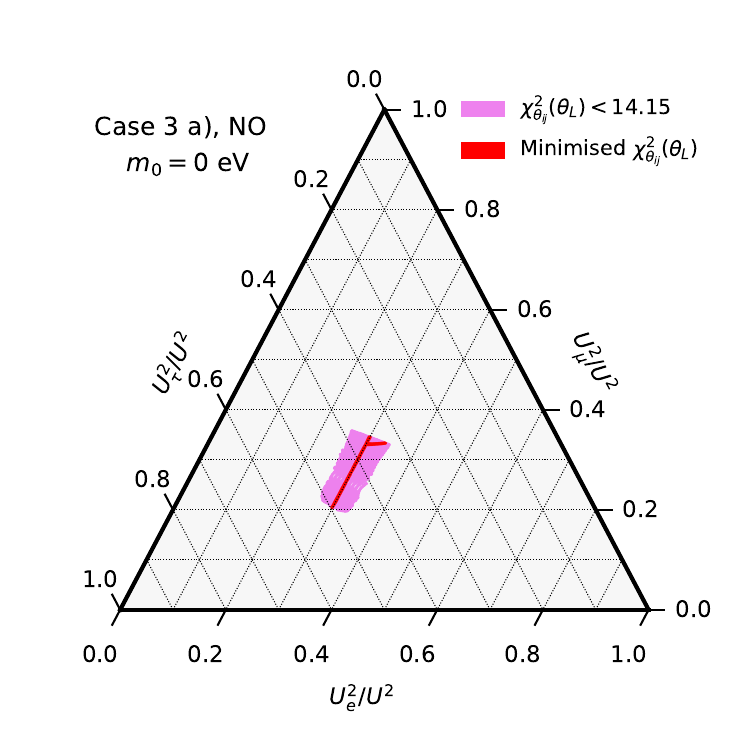}
    \caption{
   {\bf Case 3 a)} Results for ratios $\frac{U_\alpha^2}{U^2}$, assuming that light neutrino masses follow strong NO. In case we only consider results with the minimum value of $\chi^2$ for the lepton mixing angles, we obtain the two red lines, while the pink area represents the results, if we demand that the lepton mixing angles are reproduced within the $3 \, \sigma$ level.}
    \label{fig:flavratiobestfitvsallthetaL}
\end{figure}

\paragraph{Case 3 a)} The ratios $\frac{U_\alpha^2}{U^2}$ are given, for $m$ even and $s$ odd or $m$ odd and $s$ even, by 
\begin{equation}
\label{eq:Ua2U2Case3a}
    \frac{U_\alpha^2}{U^2} \approx \frac 13 \, \left( 1 + \sin^2 \theta_L \, \cos 2 \, \phi_{m, \alpha} + \sqrt{2}  \, \sin 2 \, \theta_L \, \cos \phi_{m,\alpha} \, \cos 3 \, \phi_s \right) \, ,
\end{equation}
independent of the ordering of the light neutrino mass spectrum. We note that for $m$ and $s$ both even or both odd, large active-sterile mixing $U^2$ cannot be reached, see~\cite{Drewes:2022kap}. 
We have defined $\phi_{m,\alpha} = \phi_m + \rho_\alpha \frac{4\pi}{3}$ with $\rho_e = 0, \rho_\mu = +1$ and $\rho_\tau = -1$, according to~\cite{Drewes:2022kap}.

\begin{figure}[!t]
	\includegraphics[width=0.49\textwidth]{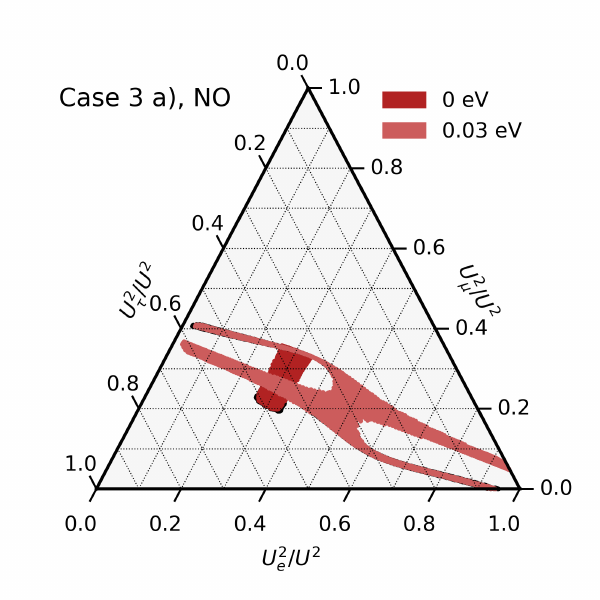}
	\includegraphics[width=0.49\textwidth]{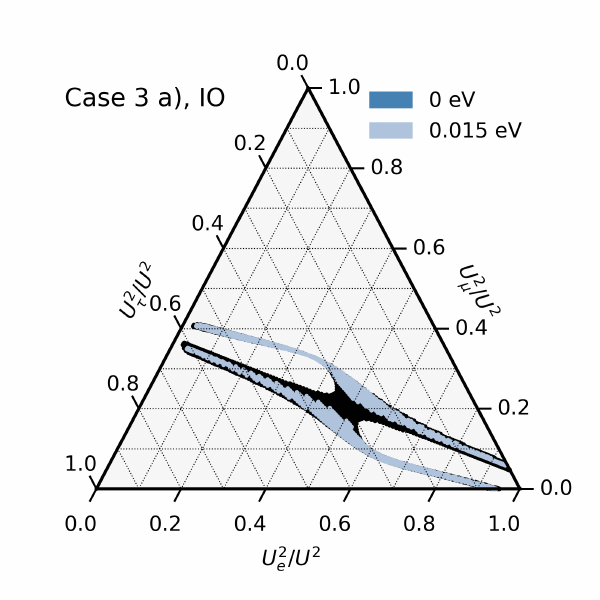}
	\caption{
 {\bf Case 3 a)} Results for ratios $\frac{U_\alpha^2}{U^2}$ assuming light neutrino masses with NO (left plot) and IO (right plot), respectively, for two different values of the lightest neutrino mass $m_0$: $m_0=0$ (darker red area) and $m_0=0.03$ eV (lighter red) for NO and $m_0=0$ (darker blue area) and $m_0=0.015$ eV (lighter blue) for IO. We marginalise over $\phi_m$ and $\phi_s$ in their allowed intervals. The black regions correspond to regions of parameter space in which only the lepton mixing angles can
be fitted at the $3 \, \sigma$ level or better, but not the CP phase $\delta$~\cite{Esteban:2020cvm}.}
 \label{fig:Case3aternary}
\end{figure}

We begin with the simplest situation (where $\theta_L$ also appears in the lepton mixing matrix) and, thus, compute the ranges of $\frac{U_\alpha^2}{U^2}$ in the case of light neutrino masses with strong NO. Using the dependence of the solar mixing
angle on the parameters $\phi_m$, $\phi_s$ and $\theta_L$, see~\cite{Hagedorn:2014wha}, we find for $\alpha=e$ the following
\begin{equation}
\label{eq:Ue2U2strongNOCase3a}
\frac{U_e^2}{U^2} = \frac 13 \, \left( 2 + \cos 2 \, \phi_m \right) \, \sin^2 \theta_{12} \; .
\end{equation}
Taking into account that for a fixed value of $\phi_m$ the solar mixing angle can be fitted to its experimental best-fit value for most of the allowed $\phi_s$ for a certain value of $\theta_L$, compare the fits presented in~\cite{Drewes:2022kap}, we assume
$\sin^2 \theta_{12}=0.304$~\cite{Esteban:2020cvm} and vary $\phi_m$ in the entire interval, see Eq.~(\ref{eq:mnCase3a}), to obtain
\begin{equation}
0.296 \lesssim \frac{U_e^2}{U^2} \lesssim 0.298
\end{equation}
that is very close to $0.3$ and nicely confirmed by the (main) red line in Fig.~\ref{fig:flavratiobestfitvsallthetaL}. If the best-fit value of the solar mixing angle cannot be achieved for any choice of $\theta_L$ for some $\phi_s$, it has been shown that
the obtained value must be larger than this best-fit value, see~\cite{Hagedorn:2014wha}. Consequently, the value of $\frac{U_e^2}{U^2}$ becomes larger, i.e.~
\begin{equation}
0.296 \lesssim \frac{U_e^2}{U^2} \lesssim 0.336 \; .
\end{equation}
This possibility corresponds to the second red branch, displayed in the ternary plot in Fig.~\ref{fig:flavratiobestfitvsallthetaL}. For the experimentally preferred $3 \, \sigma$ range of the solar mixing angle, $0.269 \leq \sin^2 \theta_{12} \leq 0.343$, see~\cite{Esteban:2020cvm},
we find as range 
\begin{equation}
0.26 \lesssim \frac{U_e^2}{U^2} \lesssim 0.34 
\end{equation}
that coincides well with the extension of the pink area in the $\frac{U_e^2}{U^2}$-direction in Fig.~\ref{fig:flavratiobestfitvsallthetaL}. 
We can perform a similar analysis for $\frac{U_\mu^2}{U^2}$ and $\frac{U_\tau^2}{U^2}$. In the limit of $\frac{m}{n}$ being very small and taking into 
account that the proximity of the solar mixing angle to $1/3$ constrains the angle $\theta_L$, i.e~$\theta_L \approx 0, \, \pi$ or $\tan \theta_L \approx - 2 \, \sqrt{2} \, \cos 3 \, \phi_s$,\footnote{If one considered the limit of $\frac{m}{n}$ being very close to one, then the non-trivial solution for the angle $\theta_L$ would read $\tan \theta_L \approx 2 \, \sqrt{2} \, \cos 3 \, \phi_s$.} see~\cite{Hagedorn:2014wha}, we find that both $\frac{U_\mu^2}{U^2}$ and $\frac{U_\tau^2}{U^2}$ should be close to $1/3$. 
For $\phi_m$ lying in the range shown in Eq.~(\ref{eq:mnCase3a}) and using the best-fit value of the solar mixing angle, we can derive the numerical expressions
\begin{equation}
\label{eq:Case3a_Umu2U2Utau2U2num}
\frac{U_\mu^2}{U^2} \approx \frac 13 \, \left( 0.96 - 0.44 \, \sin^2 \theta_L \right) \;\;\; \mbox{and} \;\;\; \frac{U_\tau^2}{U^2} \approx \frac 13 \, \left( 1.04 + 0.49 \, \sin^2 \theta_L \right) \; .
\end{equation}
Considering also the experimentally allowed $3 \, \sigma$ range of $\sin^2 \theta_{12}$, these expressions lead, for the angle $\theta_L$ varying in the entire admitted range, to
\begin{equation}
0.17 \lesssim \frac{U_\mu^2}{U^2} \lesssim 0.35 \;\;\; \mbox{and} \;\;\; 0.32 \lesssim \frac{U_\tau^2}{U^2} \lesssim 0.51 \; .
\end{equation}
These results agree with the observed allowed intervals of these ratios shown in the ternary plot in Fig.~\ref{fig:flavratiobestfitvsallthetaL}. This area is also shown in darker red in the ternary plot in the left of Fig.~\ref{fig:Case3aternary}.

In order to discuss the results for light neutrino mass with NO and $m_0$ non-zero (and large), we have to take into account the difference between the angles $\theta_L$ and $\widetilde{\theta}_L$. One can check that this is given  
by Eq.~(\ref{eq:relthetaLtildethetaL}). Thus, we can write Eq.~(\ref{eq:Ua2U2Case3a}) approximately as
\begin{equation}
    \frac{U_\alpha^2}{U^2} \approx \frac 13 \, \left( 1 + \frac 12 \, (1 \mp \sin 2 \, \widetilde{\theta}_L) \, \cos 2 \, \phi_{m, \alpha} \mp \sqrt{2}  \, \cos 2 \, \widetilde{\theta}_L \, \cos \phi_{m,\alpha} \, \cos 3 \, \phi_s \right) \, ,
\end{equation}
with the different signs originating from the different possible values of $k$ in Eq.~(\ref{eq:relthetaLtildethetaL}). Using that $\widetilde{\theta}_L$ is constrained to be either $\widetilde{\theta}_L \approx 0, \, \pi$ or 
$\tan \widetilde{\theta}_L \approx - 2 \, \sqrt{2} \, \cos 3 \, \phi_s$, the smallness of $\phi_m$ as well as that $\cos 3 \, \phi_s$ varies between $-1$ and $1$, we find that the ratio $\frac{U_e^2}{U^2}$ can take values in almost its entire
allowed range between $0$ and $1$, as can be deduced from the area in lighter red in the ternary plot in the left of Fig.~\ref{fig:Case3aternary}. This result is very different from the constrained range obtained for vanishing $m_0$.
Furthermore, we can see that the ratios $\frac{U_\mu^2}{U^2}$ and $\frac{U_\tau^2}{U^2}$ are (more) restricted, namely 
\begin{equation}
0 \lesssim \frac{U_\mu^2}{U^2} \lesssim 0.41 \;\;\; \mbox{and} \;\;\; 0 \lesssim \frac{U_\tau^2}{U^2} \lesssim 0.62 \; .
\end{equation}
These ranges agree with those of the area in light red in the ternary plot in the left of Fig.~\ref{fig:Case3aternary}, where $m_0=0.03$ eV is used, since it is the largest value of $m_0$ compatible with cosmological observations~\cite{Planck:2018vyg}
for light neutrino masses with NO.
In order to obtain the results for light neutrino masses with IO, both with $m_0$ being zero and $m_0$ large, we have to take into account the difference between the angles $\theta_L$ and $\widetilde{\theta}_L$
which is again approximated by Eq.~(\ref{eq:relthetaLtildethetaL}). We, thus, get the same results as for light neutrino masses with NO and $m_0$ non-zero. This is confirmed by the ternary plot in the right of Fig.~\ref{fig:Case3aternary},
where the area in darker blue represents the result for $m_0=0$ (lying below the area in lighter blue) and the one in lighter blue for $m_0=0.015$ eV. We note the considerable reduction of the allowed parameter space that can be achieved by taking into account the 
experimental constraints on the value of the CP phase $\delta$ in the case of light neutrino masses wit IO, as indicated by the black area.
Eventually, we remark that for light neutrino masses with strong IO the result shown in Fig.~\ref{fig:Case3aternary} (right plot) can be compared with the available parameter space found for two RH neutrinos in~\cite{Drewes:2022akb}.

\begin{figure}[!t]
 	\includegraphics[width=0.49\textwidth]{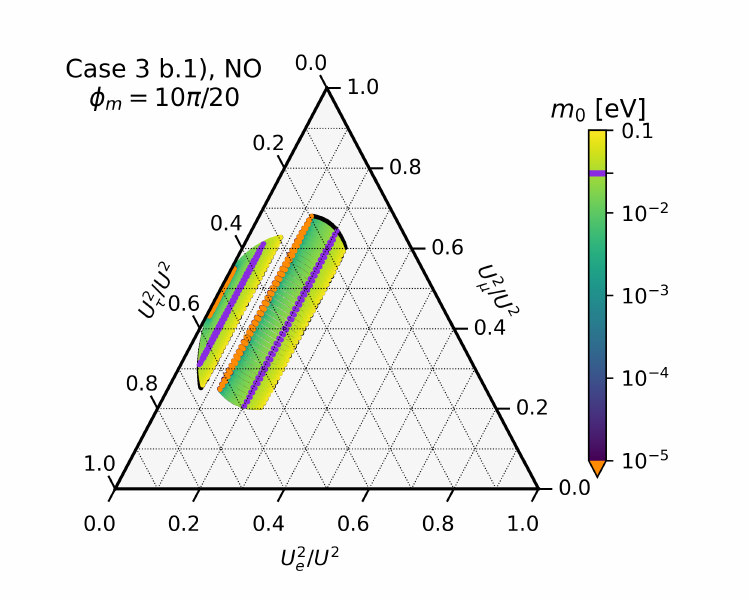}
	\includegraphics[width=0.49\textwidth]{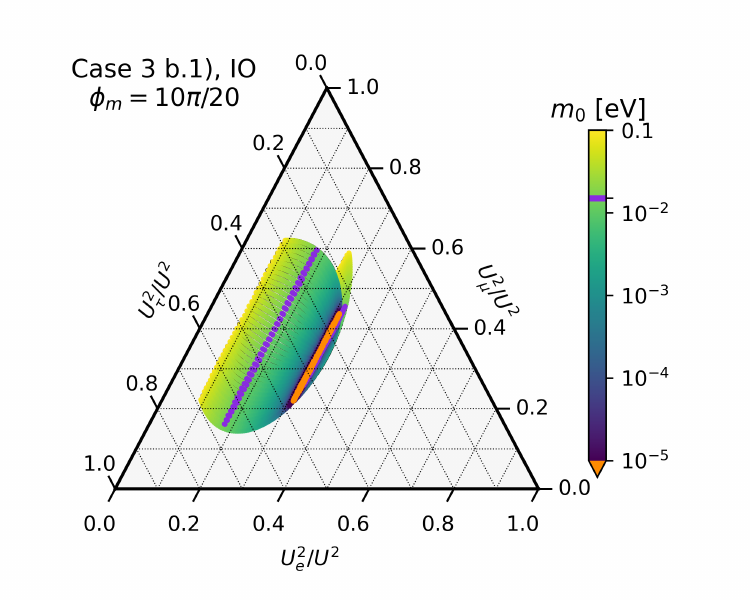}\\
	\includegraphics[width=0.49\textwidth]{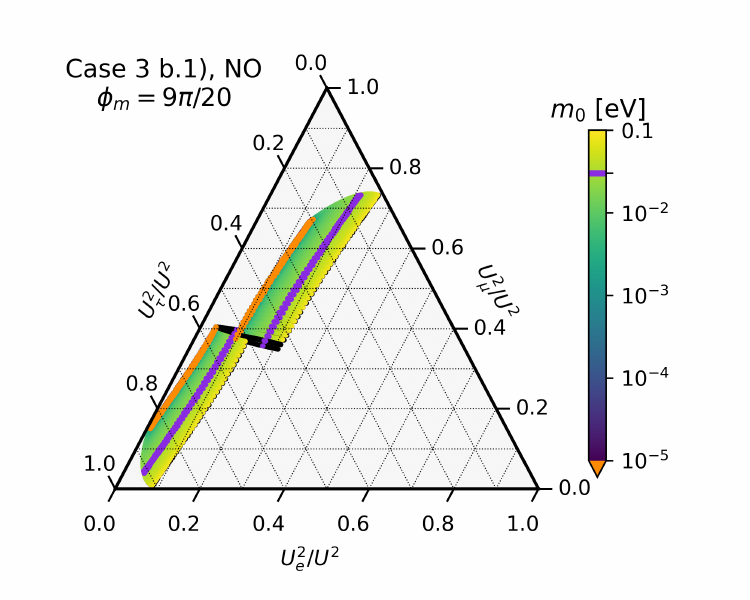}
	\includegraphics[width=0.49\textwidth]{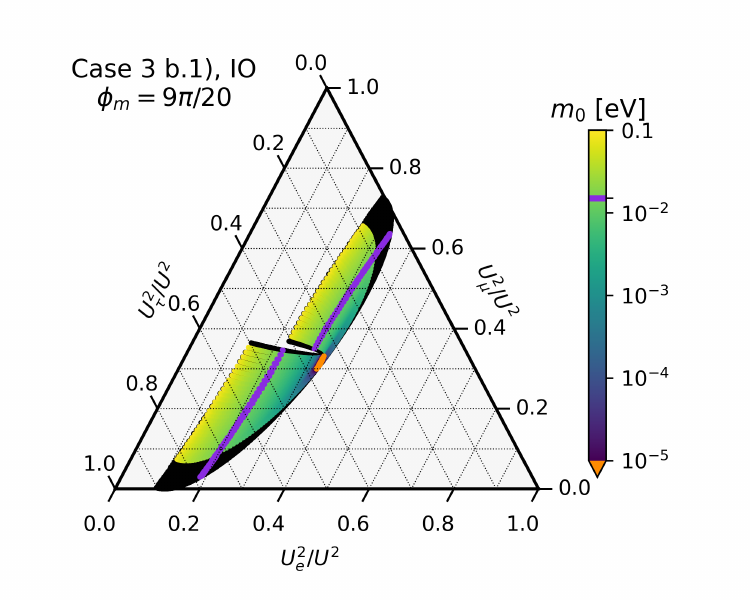}
	\caption{
 {\bf Case 3 b.1)} Results for ratios $\frac{U_\alpha^2}{U^2}$ assuming light neutrino masses with NO (left plots) and IO (right plots), respectively, for a fixed value of $\phi_m$, $\phi_m=\frac{10 \, \pi}{20}$ (upper plots) and $\phi_m=\frac{9 \, \pi}{20}$ (lower plots), while it is marginalised over the allowed values of $\phi_s$. The colour-coding is the same as in Fig.~\ref{fig:Case2ternary}.}
   \label{fig:Case3b1ternaryphim}
\end{figure}

\begin{figure}[!t]
 	\includegraphics[width=0.49\textwidth]{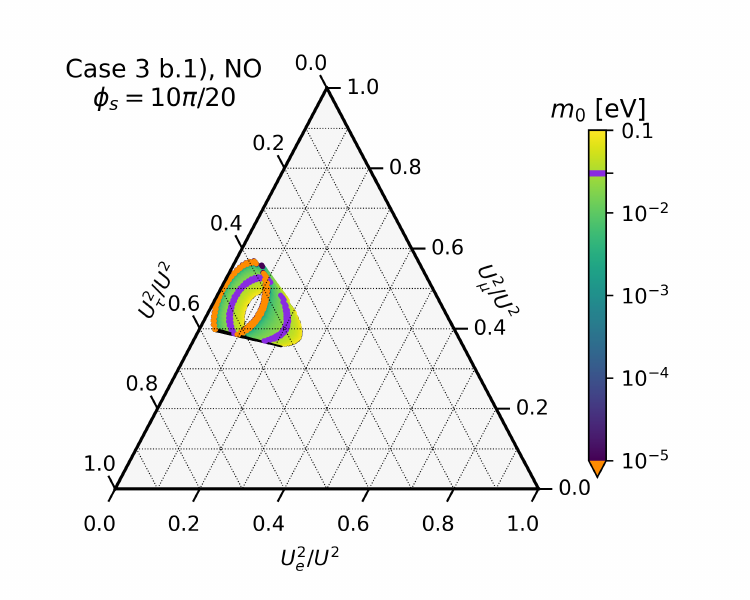}
	\includegraphics[width=0.49\textwidth]{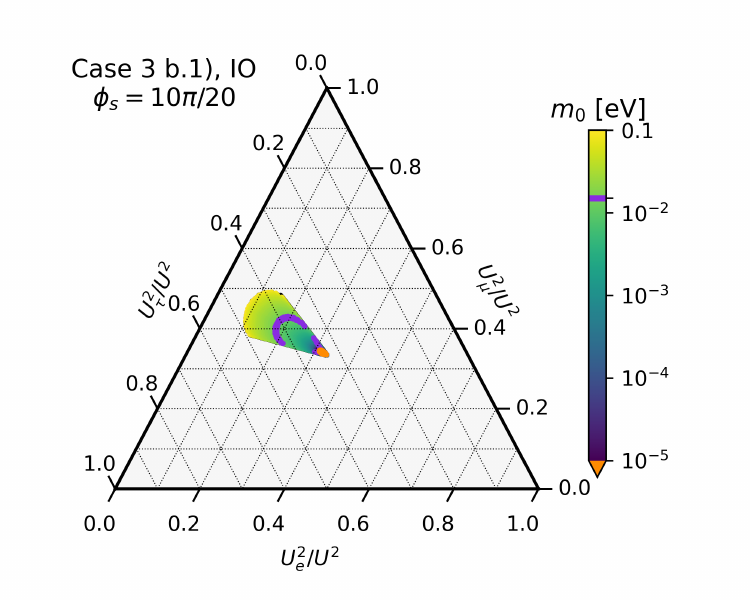}\\
	\includegraphics[width=0.49\textwidth]{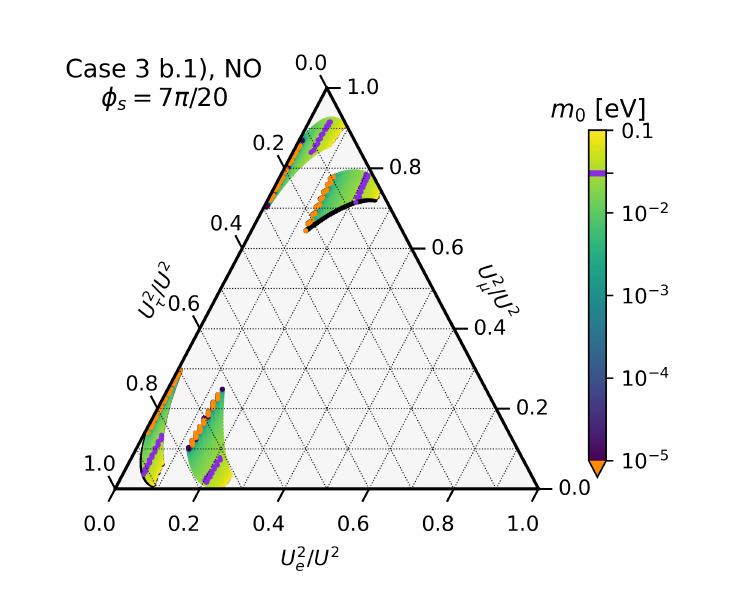}
	\includegraphics[width=0.49\textwidth]{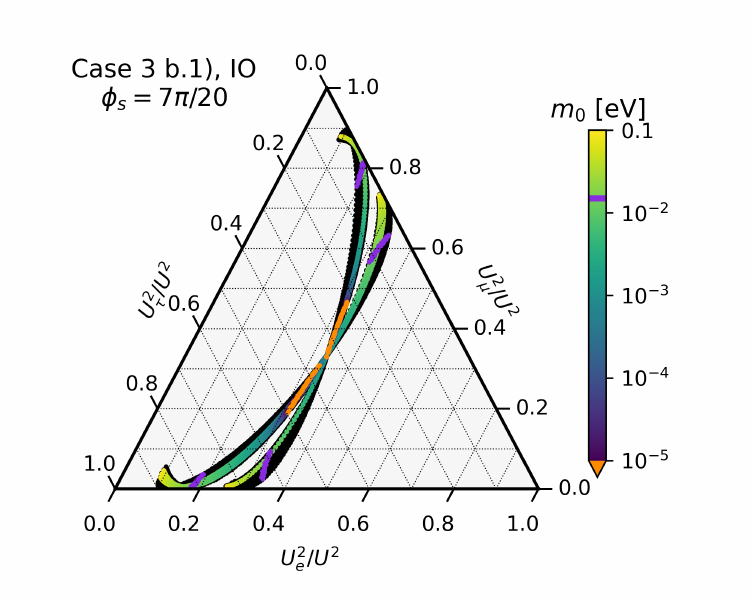}
	\caption{
 {\bf Case 3 b.1)} Results for ratios $\frac{U_\alpha^2}{U^2}$ assuming light neutrino masses with NO (left plots) and IO (right plots), respectively, for a fixed value of $\phi_s$, $\phi_s=\frac{10 \, \pi}{20}$ (upper plots) and $\phi_s=\frac{7 \, \pi}{20}$ (lower plots), while it is marginalised over the allowed values of $\phi_m$. The colour-coding is the same as in Fig.~\ref{fig:Case2ternary}.
 }
  \label{fig:Case3b1ternaryphis}
\end{figure}

\begin{figure}[!t]
	\includegraphics[width=0.49\textwidth]{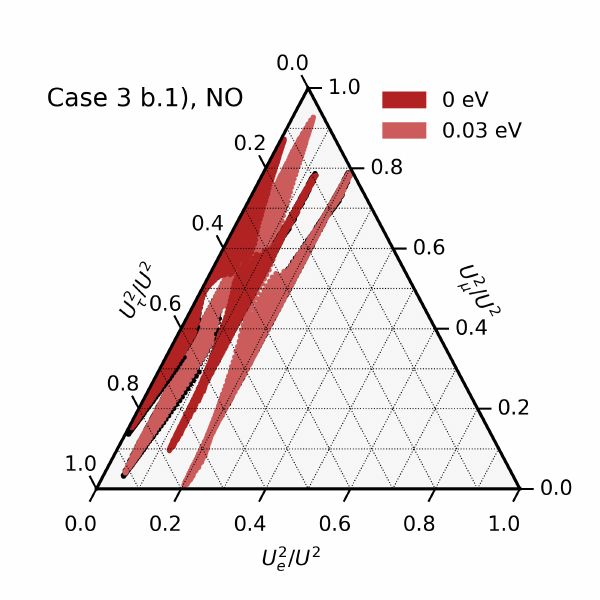}
	\includegraphics[width=0.49\textwidth]{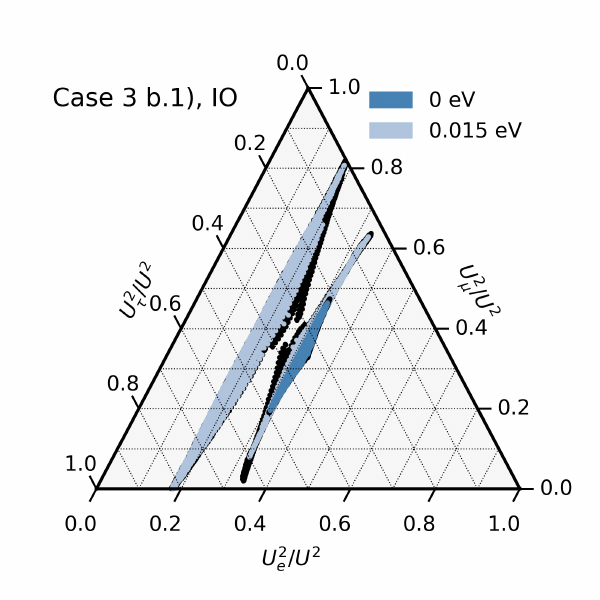}
	\caption{
 {\bf Case 3 b.1)} Results for ratios $\frac{U_\alpha^2}{U^2}$ assuming light neutrino masses with NO (left plot) and IO (right plot), respectively, for two different values of the lightest neutrino mass $m_0$: $m_0=0$ and $m_0=0.03$ eV for NO as well as $m_0=0$ and $m_0=0.015$ eV for IO. We marginalise over $\phi_m$ and $\phi_s$ in their allowed intervals. The colour-coding is the same as in Fig.~\ref{fig:Case3aternary}.}
   \label{fig:Case3b1ternary}
\end{figure}

\paragraph{Case 3 b.1)} In this case, the flavour ratios are
\begin{equation}
\label{eq:Ua2U2Case3b1}
    \frac{U_\alpha^2}{U^2} \approx 
    \begin{cases}
    \frac 13 \, \left( 1 + \sin^2 \theta_L \, \cos 2 \, \phi_{m, \alpha} + \sqrt{2}  \, \sin 2 \, \theta_L \, \cos \phi_{m,\alpha} \, \cos 3 \, \phi_s \right) \, & \text{for NO} \; ,\\
     \frac 13 \, \left( 1 + \cos^2 \theta_L \, \cos 2 \, \phi_{m, \alpha} - \sqrt{2}  \, \sin 2 \, \theta_L \, \cos \phi_{m,\alpha} \, \cos 3 \, \phi_s \right) \, & \text{for IO} \; .
    \end{cases}
\end{equation}
As for Case 3 a), these results are only valid for $m$ even, $s$ odd or vice versa, since these combinations can lead to large active-sterile mixing $U^2$ for special values of the angle $\theta_R$. 
 The parameter $\phi_{m,\alpha}$ is defined in the same way as for Case 3 a). Since the dependence of these ratios as well as of the lepton mixing parameters on $\phi_m$, $\phi_s$ and $\theta_L$ 
 (as well as $\widetilde{\theta}_L$) is rather involved, we begin the discussion of this case by fixing either $\phi_m$ or $\phi_s$ to a specific value, as displayed in the ternary plots in Figs.~\ref{fig:Case3b1ternaryphim} and~\ref{fig:Case3b1ternaryphis}.
 Similar to Case 3 a), we first focus on the situation in which $\theta_L$ also appears in the lepton mixing matrix. This happens for light neutrino masses with strong IO, see~\cite{Drewes:2022kap}. Then, we obtain for
 the electron flavour, $\alpha=e$,

 \begin{equation}
 \label{eq:Case3b1Ue2U2strongIO}
 \frac{U_e^2}{U^2} \approx 0.64 + 0.33 \, \cos 2 \, \phi_m \; ,
 \end{equation}
 where we have used the relation among the reactor mixing angle and the parameters $\phi_m$, $\phi_s$ and $\theta_L$, see~\cite{Hagedorn:2014wha}, as well as its best-fit value~\cite{Esteban:2020cvm}. 
  For $\frac mn= \frac 12$ and thus $\phi_m=\frac{\pi}{2}$ this leads to
  \begin{equation}
   \label{eq:Case3b1Ue2U2strongIOnum}
  \frac{U_e^2}{U^2} \approx 0.31 \; , \;\; \mbox{while} \;\; 0.31 \lesssim   \frac{U_e^2}{U^2} \lesssim 0.34
  \end{equation}
  is the interval corresponding to the allowed range for $\phi_m$, as mentioned in Eq.~(\ref{eq:mnCase3b1}). These results coincide with the extension in the $\frac{U_e^2}{U^2}$-direction of the orange areas/lines in the
  different ternary plots, shown in the right of Figs.~\ref{fig:Case3b1ternaryphim} and~\ref{fig:Case3b1ternaryphis}. It also agrees well with the extension in the $\frac{U_e^2}{U^2}$-direction of the area in darker blue in the ternary plot in the 
  right of Fig.~\ref{fig:Case3b1ternary}.
  In order to determine as well the admitted intervals of the ratios $\frac{U_\mu^2}{U^2}$ and $\frac{U_\tau^2}{U^2}$, we first fix $\frac mn=\frac 12$. Furthermore, we remind that the values of $\theta_L$
  which lead to good agreement with the experimental data on lepton mixing angles are $\theta_L \approx 1.31$ and $\theta_L \approx 1.83$~\cite{Hagedorn:2014wha,Drewes:2022kap}. Additionally, the constraints on $\phi_s$ arising from the experimentally preferred values 
  of $\sin^2 \theta_{23}$ lead to $-0.6 \lesssim \cos 3 \, \phi_s \lesssim 0.7$, see~\cite{Hagedorn:2014wha} and also~\cite{Drewes:2022kap}. Using this, we find 
 \begin{align}
0.2 \lesssim \frac{U_\mu^2}{U^2} \; , \; \frac{U_\tau^2}{U^2} \lesssim 0.49 \; .
\end{align}
As one can check, these ranges agree well with the ones which we can read off from the orange line in the ternary plot in the upper right of Fig.~\ref{fig:Case3b1ternaryphim}. Next, we fix $\frac sn = \frac 12$, i.e.~$\phi_s = \frac{\pi}{2}$. From 
Eq.~(\ref{eq:Ua2U2Case3b1}) we see that the last term becomes zero. Additionally, we know that $\cos^2 \theta_L \lesssim 0.07$ from the reactor mixing angle and the admitted interval of $\phi_m$, see
Eq.~(\ref{eq:mnCase3b1}). With this information and taking $0.14 \lesssim \cos 2 \, \phi_{m,\mu} \, , \, \cos 2 \, \phi_{m,\tau} \lesssim 0.79$, we obtain
\begin{equation}
0.33 \lesssim \frac{U_\mu^2}{U^2} \; , \;\frac{U_\tau^2}{U^2} \lesssim 0.35 \; ,
\end{equation}
together with the interval for $\frac{U_e^2}{U^2}$ as given in Eq.~(\ref{eq:Case3b1Ue2U2strongIOnum}). These estimates agree well with the small orange area shown in the ternary plot in the upper right of 
Fig.~\ref{fig:Case3b1ternaryphis}. The next special value of $\phi_m$ that we consider is given for $m=9$, $n=20$ such that $\frac mn= 0.45$ (which is close to the lower bound of the interval of admitted values of the ratio $\frac mn$, compare Eq.~(\ref{eq:mnCase3b1})). In order to evaluate the
ratios $\frac{U_\mu^2}{U^2}$ and $\frac{U_\tau^2}{U^2}$, we use analytic results from~\cite{Hagedorn:2014wha}. The parameter $\bar{\kappa} \approx -0.16$ quantifies the deviation of $\frac mn$ from $\frac 12$ ($\frac mn = \frac 12 + \frac{\bar{\kappa}}{\pi}$)
and $\epsilon \approx -\sqrt{2} \, \bar{\kappa} \, \cos3 \, \phi_s \approx 0.22 \, \cos 3 \, \phi_s$ is the deviation of $\theta_L$ from $1.31$ or $1.83$, i.e.~$\theta_L= \theta_0 + \epsilon$ with $\theta_0 \approx 1.31, \, 1.83$.
Since $\cos 3 \, \phi_s$ can take any value between $-1$ and $1$, we have to distinguish two possibilities: for $\cos 3 \, \phi_s \geq 0$ it is $\theta_L \approx 1.31 + 0.22 \, \cos 3 \, \phi_s$ and for $\cos 3 \, \phi_s \leq 0$
it is $\theta_L \approx 1.83 + 0.22 \, \cos 3 \, \phi_s$. Using this information, we arrive at
\begin{equation}
0.28 \lesssim \frac{U_\mu^2}{U^2} \lesssim 0.34 \;\;\; \mbox{and} \;\;\; 0.35 \lesssim \frac{U_\tau^2}{U^2} \lesssim 0.40 \; .
\end{equation}

These ranges together with the restriction of $\frac{U_e^2}{U^2}$ to the interval in Eq.~(\ref{eq:Case3b1Ue2U2strongIOnum}) describe well the small orange area displayed in the ternary plot in the lower right of Fig.~\ref{fig:Case3b1ternaryphim}.
Finally, to properly estimate the ratios $\frac{U_\alpha^2}{U^2}$ for $\frac sn = \frac{7}{20}$, leading to $\cos 3 \, \phi_s \approx -1$, we first have to calculate the values of $\frac mn$ ($\phi_m$) that allow for an acceptable fit to the lepton mixing angles. These can
be determined from the experimental constraints on the atmospheric mixing angle, $0.41 \lesssim \sin^2 \theta_{23} \lesssim 0.623$, together with the reactor mixing angle being $\sin^2 \theta_{13} \approx 0.022$, see~\cite{Esteban:2020cvm}. The relevant
equation is 
\begin{equation}
-0.21 \lesssim \sin \theta_L \, (\cos \phi_m \, \sin \theta_L + \sqrt{2} \, \cos \theta_L) \lesssim 0.15 \; .
\end{equation}
Furthermore, we also use the analytic relation between the deviation of $\frac mn$ from $\frac 12$ and the angle $\theta_L$ from $1.31$ or $1.83$, i.e.~either for $\frac mn \leq \frac 12$ it is 
$\theta_L \approx 1.83 + \sqrt{2} \, \pi \, (\frac mn - \frac 12)$ or for $\frac mn \geq \frac 12$ it is $\theta_L \approx 1.31 + \sqrt{2} \, \pi \, (\frac mn - \frac 12)$. The result for the compatible intervals of $\frac mn$ is then
\begin{equation}
0.445 \lesssim \frac mn \lesssim 0.484 \;\;\; \mbox{and} \;\;\; 0.523 \lesssim \frac mn \lesssim 0.56 \; .
\end{equation}
With this information, we can estimate well the ranges of the ratios $\frac{U_\alpha^2}{U^2}$. For the electron flavour, $\alpha=e$, we have
\begin{equation}
0.31 \lesssim \frac{U_e^2}{U^2} \lesssim 0.34 \; ,
\end{equation}
while for $\frac{U_\mu^2}{U^2}$ we obtain
\begin{equation}
0.193 \lesssim \frac{U_\mu^2}{U^2} \lesssim 0.323 \;\;\; \mbox{and} \;\;\; 0.328 \lesssim \frac{U_\mu^2}{U^2} \lesssim 0.471
\end{equation}
as well as for $\frac{U_\tau^2}{U^2}$ 
\begin{equation}
0.347 \lesssim \frac{U_\tau^2}{U^2} \lesssim 0.494 \;\;\; \mbox{and} \;\;\; 0.215 \lesssim \frac{U_\tau^2}{U^2} \lesssim 0.337 \; .
\end{equation}
As one can see, these intervals are disjunct and these small gaps can also be observed in the ternary plot in the lower right of Fig.~\ref{fig:Case3b1ternaryphis}, when looking at the orange area. 

The situation of $m_0$ non-zero can be analysed in a similar way,
taking into account the difference between the angles $\theta_L$ and $\widetilde{\theta}_L$. For $m_0=0.015$ eV and light neutrino masses with IO this difference can be estimated to be
\begin{equation}
\theta_L \approx \widetilde{\theta}_L - 0.48 \;\;\; \mbox{or} \;\;\; \theta_L \approx \widetilde{\theta}_L - \pi - 0.48 \; .
\end{equation}
We exemplify the analytic results for the fixed value $\frac mn= \frac 12$. In this case, the ratio $\frac{U_\alpha^2}{U^2}$ only depends on the angle $\theta_L$ for the electron flavour, $\alpha=e$,
\begin{equation}
\frac{U_e^2}{U^2} = \frac 13 \, (1-\cos^2 \theta_L) \; .
\end{equation}
This leads to two possible values of this ratio for $\widetilde{\theta}_L \approx 1.31$ or $\widetilde{\theta}_L \approx 1.83$, i.e.~
\begin{equation}
\frac{U_e^2}{U^2} \approx 0.18 \;\;\; \mbox{and} \;\;\; \frac{U_e^2}{U^2} \approx 0.32 \; .
\end{equation}
To these values correspond two different intervals for the ratios $\frac{U_\mu^2}{U^2}$ and $\frac{U_\tau^2}{U^2}$, since they display a further dependence on $\theta_L$ and also on $\phi_s$.
We find for $\frac{U_\mu^2}{U^2}$
\begin{equation}
0.12 \lesssim \frac{U_\mu^2}{U^2} \lesssim 0.65 \;\;\; \mbox{and} \;\;\; 0.22 \lesssim \frac{U_\mu^2}{U^2} \lesssim 0.44
\end{equation}
and for $\frac{U_\tau^2}{U^2}$
\begin{equation}
0.16 \lesssim \frac{U_\tau^2}{U^2} \lesssim 0.70 \;\;\; \mbox{and} \;\;\; 0.23 \lesssim \frac{U_\tau^2}{U^2} \lesssim 0.47 \; .
\end{equation}

Inspecting the two violet stripes in the ternary plot in the upper right of Fig.~\ref{fig:Case3b1ternaryphim} we can see reasonable agreement with these analytic estimates. Note that the second violet stripe nearly coincides with the orange stripe in this plot.

We can also study the instance of light neutrino masses with NO and $m_0=0$ or $m_0=0.03$ eV. In both cases, the difference between $\theta_L$ and $\widetilde{\theta}_L$ needs to be considered.
For strong NO, this difference is approximately
\begin{equation}
\theta_L \approx \widetilde{\theta}_L - 0.38 \;\;\; \mbox{or} \;\;\; \theta_L \approx \widetilde{\theta}_L - \pi - 0.38 \; .
\end{equation}  
Plugging this into the formula for the ratio $\frac{U_e^2}{U^2}$ that reads
\begin{equation}
\frac{U_e^2}{U^2} = \frac 13 \, (1- \sin^2 \theta_L)
\end{equation}
and using $\widetilde{\theta}_L \approx 1.31$ or $\widetilde{\theta}_L \approx 1.83$ for $\frac mn=\frac 12$, we get
\begin{equation}
\frac{U_e^2}{U^2} \approx 0.12 \;\;\; \mbox{and} \;\;\; \frac{U_e^2}{U^2} \approx 0.005 \; , 
\end{equation}
while for the other two ratios we obtain certain intervals
\begin{equation}
0.21 \lesssim \frac{U_\mu^2}{U^2} \lesssim 0.72 \;\;\; \mbox{and} \;\;\; 0.44 \lesssim \frac{U_\mu^2}{U^2} \lesssim 0.57
\end{equation}
and 
\begin{equation}
0.16 \lesssim \frac{U_\tau^2}{U^2} \lesssim 0.68 \;\;\; \mbox{and} \;\;\; 0.43 \lesssim \frac{U_\tau^2}{U^2} \lesssim 0.56 \; ,
\end{equation}
respectively.
These ranges agree with those indicated by the orange stripes found in the ternary plot in the upper left of Fig.~\ref{fig:Case3b1ternaryphim}. 

Eventually, we proceed in the same way in order to estimate the ranges for the ratios, if light 
neutrino masses follow NO and $m_0=0.03$ eV. First, we estimate the difference between the angles $\theta_L$ and $\widetilde{\theta}_L$ which is 
\begin{equation}
\theta_L \approx \widetilde{\theta}_L - 0.60 \;\;\; \mbox{or} \;\;\; \theta_L \approx \widetilde{\theta}_L - \pi - 0.60 \; .
\end{equation}
With this information and $\widetilde{\theta}_L \approx 1.31$ or $\widetilde{\theta}_L \approx 1.83$ for $\frac mn= \frac 12$, we end up with two possible values for the ratio $\frac{U_e^2}{U^2}$
\begin{equation}
\frac{U_e^2}{U^2} \approx 0.19 \;\;\; \mbox{and} \;\;\; \frac{U_e^2}{U^2} \approx 0.037 \; , 
\end{equation}
together with 
\begin{equation}
0.16 \lesssim \frac{U_\mu^2}{U^2} \lesssim 0.69 \;\;\; \mbox{and} \;\;\; 0.33 \lesssim \frac{U_\mu^2}{U^2} \lesssim 0.66
\end{equation}
and 
\begin{equation}
0.12 \lesssim \frac{U_\tau^2}{U^2} \lesssim 0.65 \;\;\; \mbox{and} \;\;\; 0.30 \lesssim \frac{U_\tau^2}{U^2} \lesssim 0.64 \; ,
\end{equation}
respectively.
Also in this case we find reasonable agreement with the ranges indicated by the two violet stripes in the ternary plot in the upper left of Fig.~\ref{fig:Case3b1ternaryphim}. 

For a fixed value of $\phi_m$ or $\phi_s$ the
allowed parameter space can be very different depending on the light neutrino mass ordering and the value of $m_0$, as we see in Figs.~\ref{fig:Case3b1ternaryphim} and~\ref{fig:Case3b1ternaryphis}. We observe, for example, that the accessible regions for $\phi_s=\frac{7 \, \pi}{20}$ and for $\phi_s=\frac{10 \, \pi}{20}$ hardly overlap, once the light neutrino mass ordering is fixed, see Fig.~\ref{fig:Case3b1ternaryphis}. Furthermore, we find that, for $\phi_s=\frac{7 \, \pi}{20}$ and IO
light neutrino masses, nearly half of the allowed parameter space can be excluded by taking into account the  available information on the CP phase $\delta$, compare the black regions in the ternary
plot in the lower right of Fig.~\ref{fig:Case3b1ternaryphis}.

Looking at the ternary plots in Fig.~\ref{fig:Case3b1ternary}, where $\phi_m$ and $\phi_s$ are both marginalised in their allowed intervals,
we notice that the attainable values of $\frac{U_e^2}{U^2}$ are the most striking difference between the parameter space allowed for light neutrinos with NO and those with IO, since
for the former $\frac{U_e^2}{U^2}$ is smaller than approximately $0.2$, while for IO we find the interval $0.15 \lesssim \frac{U_e^2}{U^2} \lesssim 0.35$. Case 3 b.1) could therefore be
falsified, if a ratio $\frac{U_e^2}{U^2}$ larger than $0.35$ is detected. The ratios $\frac{U_\mu^2}{U^2}$
and $\frac{U_\tau^2}{U^2}$ are instead hardly constrained. Furthermore, we remark that the available parameter space for $m_0=0$ (darker blue region) and light neutrino masses with IO is contained
in the one for $m_0=0.015$ eV (lighter blue region) such that it might not be possible to distinguish between these, see ternary plot in the right of Fig.~\ref{fig:Case3b1ternary}. Taking into account experimental information on the CP phase $\delta$ further constrains the
parameter space accessible in the case of IO, while it hardly affects the results obtained for NO, compare the black regions in Fig.~\ref{fig:Case3b1ternary}. Finally, we note that for light neutrino masses with strong NO the result shown in Fig.~\ref{fig:Case3b1ternary} (left plot) should be comprised within the available parameter space found for two RH neutrinos in~\cite{Drewes:2022akb}.

\paragraph{Comment on experimental testability}  We see from Fig.~\ref{fig:flavratiovsU2Case1} that $U^2$ should be about two orders of magnitude larger than the naive seesaw limit, see Eq.~(\ref{eq:naiveseesawformula}), in order to obtain at least $10\%$ relative accuracy (for the flavour ratio $\frac{U_e^2}{U^2}$). In this case, we can expect that it is possible to distinguish between certain cases, but also between different choices of the relevant group theory parameters. For example, by determining whether $\frac{U_e^2}{U^2}$ is larger than $0.35$ or $\frac{U_\mu^2}{U^2}$ larger than $0.45$ and $\frac{U_\tau^2}{U^2}$ larger than $0.65$, Case 3 b.1) or Case 3 a) could already be disfavoured. Additionally, it might be possible, depending on the value of the lightest neutrino mass $m_0$, to distinguish between light neutrino masses with NO or IO for Case 1) and Case 2) and also for Case 3 a) and Case 3 b.1), for certain combinations of the parameters $m$ and $s$. If Case 1) is realised, it should as well be possible to pinpoint $m_0$ with good accuracy.

\vspace{0.1in}
Results for the branching ratios of the heavy neutrinos have also been presented in~\cite{Chauhan:2021xus}. However, these have been obtained for a different part of the
parameter space of this scenario. 

\section{Leptogenesis}
\label{lepto}

In this section, we scrutinise the parameter space that allows for successful low-scale leptogenesis for the different cases, Case 1) through Case 3 b.1).
These results expand and complement those found in~\cite{Drewes:2022kap}. We first state the prerequisites for this analysis and then discuss the results for
each case. We also study the relevance of experiments searching for charged lepton flavour violation in $\mu-e$ transitions in constraining the viable parameter space for leptogenesis as well as comment on the impact of the splitting $\lambda$.

\subsection{Prerequisites}
\label{prereqlepto}

In the following, we establish the prerequisites of the numerical calculations, 
present analytical formulae for three CP-violating combinations and the 
flavoured washout parameter, characterise the parameter space considered in the 
numerical analysis, discuss how the different cases, Case 1) through Case 3 b.1), are explored, and lay out the conventions used in the plots, shown in 
this section. 

\paragraph{Quantum kinetic equations}  For heavy neutrinos with masses  
potentially testable in laboratory experiments, the evolution of positive 
(negative) helicity heavy neutrino densities $\rho_N$ ($\bar{\rho}_N$) 
and of the comoving lepton asymmetry number densities $n_{\Delta_\alpha}$, $\alpha=e,\mu,\tau$, 
is best described by the 
following set of quantum kinetic equations, see 
e.g.~\cite{Garbrecht:2018mrp} for a review on their derivation,\footnote{For simplicity, this set of equations does not take into account the expansion of the Universe. It can nonetheless be included following~\cite{Beneke:2010wd}.}
\begin{subequations}\label{QKE}
	\begin{align}
		i \frac{\mathrm{d} n_{\Delta_\alpha}}{\mathrm{d} t}
		&= -2 i \frac{\mu_{\alpha}}{T} \int \frac{\mathrm{d}^{3} k}{(2 \pi)^{3}} \operatorname{Tr}\left[\Gamma_{\alpha}\right] f_{N}\left(1-f_{N}\right) 
		+i \int \frac{\mathrm{d}^{3} k}{(2 \pi)^{3}} \operatorname{Tr}\left[\tilde{\Gamma}_{\alpha}\left(\bar{\rho}_{N}-\rho_{N}\right)\right],
		\label{kin_eq_a}
		\\
		i \, \frac{\mathrm{d}\rho_{N}}{\mathrm{d}t}
		&= \left[H_{N}, \rho_{N}\right]-\frac{i}{2}\left\{\Gamma, \rho_{N}-\rho_{N}^{eq} \right\} 
		-\frac{i}{2} \sum_{\alpha} \tilde{\Gamma}_{\alpha}\left[2 \frac{\mu_{\alpha}}{T} f_{N}\left(1-f_{N}\right)\right] ,
		\label{kin_eq_b}
		\\
		i \, \frac{\mathrm{d} \bar{\rho}_{N}}{\mathrm{d} t}
		&= -\left[H_{N}, \bar{\rho}_{N}\right]-\frac{i}{2}\left\{\Gamma, \bar{\rho}_{N}-\rho_{N}^{eq} \right\}
		+\frac{i}{2} \sum_{\alpha} \tilde{\Gamma}_{\alpha}\left[2 \frac{\mu_{\alpha}}{T} f_{N}\left(1-f_{N}\right)\right] \; .
		\label{kin_eq_c}
\end{align}
\label{eq:kin_eq}
\end{subequations}
In these equations, $H_N$ represents the effective Hamiltonian 
of the system, while $\Gamma$, $\Gamma_\alpha$ and 
$\tilde{\Gamma}_\alpha$ are different thermal interaction rates. 
 The quantity $f_N$ denotes the Fermi-Dirac distribution associated with 
 the heavy neutrinos and $\mu_{\alpha}$ are flavoured lepton chemical 
 potentials, related to $n_{\Delta_\alpha}$ by a susceptibility 
 matrix~\cite{Buchmuller:2005eh,Garbrecht:2019zaa}.

 In this study, we extrapolate the results of~\cite{Ghiglieri:2017gjz} for the rates to 
 the non-relativistic regime as has been performed 
 in~\cite{Klaric:2021cpi}; for further literature 
 see~\cite{Biondini:2017rpb,Laine:2022pgk}.

We remark that the angularity of some of the plots shown in this section,
see e.g.~Fig.~\ref{fig:CaseIMU2}, is due to the discretisation of the grid
in the RH neutrino mass at which the different rates are computed. 

\paragraph{CP-violating combinations and flavoured washout parameter} 
 As already discussed in~\cite{Drewes:2022kap}, we can solve the quantum kinetic 
equations in Eqs.~(\ref{eq:kin_eq}) perturbatively in both $H_N$ and $\Gamma$ in order to determine
the combinations of mass and Yukawa matrices that are relevant for CP violation. It 
turns out that there are three independent CP-violating combinations which read as
follows
\begin{eqnarray}
C_{\mathrm{LFV},\alpha} &=& i \, \mathrm{Tr} \Big( \left[ \hat{M}_R^2, \hat{Y}_D^\dagger \,  \hat{Y}_D\right] \,  \hat{Y}_D^\dagger \, P_\alpha \,  \hat{Y}_D \Big) \; ,
\label{eq:defLFVcombination}
\\
C_{\mathrm{LNV},\alpha} &=&  i \, \mathrm{Tr} \Big( \left[ \hat{M}_R^2, \hat{Y}_D^\dagger \,  \hat{Y}_D\right] \,  \hat{Y}_D^T \, P_\alpha \,  \hat{Y}_D^* \Big) \; ,
\label{eq:defLNVcombination}
\\
C_{\mathrm{DEG},\alpha} &=& i \, \mathrm{Tr} \Big( \left[ \hat{Y}_D^T \,  \hat{Y}_D^* , \hat{Y}_D^\dagger \,  \hat{Y}_D\right] \,  \hat{Y}_D^T \, P_\alpha \,  \hat{Y}_D^* \Big) \; ,
\label{eq:defCdegalpha}
\end{eqnarray}
where $P_\alpha$ stands for the projector on the lepton flavour $\alpha$,
\begin{equation}
P_e = \left(
\begin{array}{ccc}
1 & 0 & 0\\
0 & 0 & 0\\
0 & 0 & 0
\end{array}
\right) \;\; , \;\;
P_\mu = \left(
\begin{array}{ccc}
0 & 0 & 0\\
0 & 1 & 0\\
0 & 0 & 0
\end{array}
\right) \;\; , \;\;
P_\tau = \left(
\begin{array}{ccc}
0 & 0 & 0\\
0 & 0 & 0\\
0 & 0 & 1
\end{array}
\right) \; .
\end{equation}
Note that these quantities are defined in the mass basis of the RH neutrinos,
as indicated by the hat. While the CP-violating combinations $C_{\mathrm{LFV},\alpha}$
and $C_{\mathrm{DEG},\alpha}$ only give rise to a lepton flavour asymmetry, 
$C_{\mathrm{LNV},\alpha}$ directly violates lepton number. These combinations 
typically dominate in different regimes of the RH neutrino mass, 
see~\cite{Drewes:2022kap} for details. See also e.g.~\cite{Hernandez:2015wna,Hernandez:2022ivz} for
an analysis in the fully relativistic regime using CP invariants.

Furthermore, we consider the flavoured washout parameter
\begin{align}
	\label{eq:flvw}
	f_\alpha = \frac{(\hat{Y}_D \hat{Y}_D^\dagger)_{\alpha \alpha}}{ \Tr(\hat{Y}_D \hat{Y}_D^\dagger)}
\end{align}
that quantifies the rates at which the three different lepton flavour asymmetries 
are washed out. This is particularly important in the case in which the generation of
a lepton number asymmetry, encoded in the CP-violating combination  
$C_{\mathrm{LNV},\alpha}$, is suppressed.

The results for the different cases shown in~\cite{Drewes:2022kap} are complemented by an additional formula for Case 2), found in appendix~\ref{appB1}.

\paragraph{Parameter space explored} In this work, we consider two possible initial conditions, vanishing and thermal initial heavy neutrino abundances. 
While vanishing initial conditions (VIC) implicitly assume that heavy neutrinos can only be produced through their Yukawa couplings, thermal initial conditions (TIC) are, in particular, justified in the context of extensions of the SM in which RH neutrinos possess gauge interactions that can bring them in thermal equilibrium at early times. 
In practice, we however assume that these additional interactions are only efficient at high temperature such that one can safely neglect their impact on the leptogenesis dynamics (apart from the different initial conditions).
In general, VIC and TIC can lead to different allowed parameter space for successful leptogenesis, see e.g.~Fig.~\ref{fig:CaseIMU2}. 

We vary the heavy neutrino masses between $50 \, \mathrm{MeV}$
and $70 \, \mathrm{TeV}$, see Tab.~\ref{range of values parameters}. This covers the  
entire experimentally accessible mass range and, at the same time, takes into account 
constraints from cosmological considerations~\cite{Hernandez:2014fha,Vincent:2014rja,Sabti:2020yrt,Boyarsky:2020dzc,Domcke:2020ety,Mastrototaro:2021wzl}, supernovae~\cite{Mastrototaro:2019vug} and direct 
searches~\cite{Arguelles:2021dqn,Kelly:2021xbv}, combined with neutrino oscillation data~\cite{Drewes:2016jae,Bondarenko:2021cpc,Chrzaszcz:2019inj}, and the range of validity of 
the used quantum kinetic equations. In certain instances, we fix the Majorana mass $M$, see Eq.~(\ref{eq:MR0}), to one of the benchmark values, $M=10 \, \mathrm{GeV}$, 
$M=100 \, \mathrm{GeV}$ and $M=1 \, \mathrm{TeV}$. These correspond to the relativistic, 
intermediate and non-relativistic regime of leptogenesis. 

The two splittings $\kappa$ and $\lambda$ fix the size of the mass splittings of the RH  
neutrinos and are expected to lie in the interval between $0$ and $10^{-1}$
in order not to break the flavour and CP symmetry strongly. We remind that for non-vanishing $\kappa$ still two of the three RH neutrinos are degenerate
in mass, while a non-zero splitting $\lambda$ generically splits all three masses, compare Eqs.~(\ref{eq:Mkappa}) and (\ref{eq:Mlambda}). As mentioned in section~\ref{sec2}, the form of 
the splitting $\kappa$ is determined by the residual symmetry of the charged lepton
sector, whereas the one of $\lambda$ is generic. For these reasons, 
 we mainly focus on the effect of $\kappa$, whose absolute value is 
 chosen in the numerical scans to lie in the 
 interval $10^{-20} \lesssim |\kappa| \lesssim 10^{-1}$, while keeping $\lambda=0$. In 
 order to also address the impact of $\lambda$, we either use two benchmark values, 
 $\lambda=10^{-10}$ and $\lambda=10^{-4}$, or vary its absolute value between
 $10^{-20}$ and $10^{-1}$, while setting $\kappa=0$, compare 
 Figs.~\ref{fig:Case1lambdanonzero} and~\ref{fig:LambdavsU21TeV}. Only for certain cases, 
 we fix both these splittings, $\kappa$ and $\lambda$, to zero, see 
 e.g.~Fig.~\ref{CaseIIMU2m0heavy}, and can still produce a non-vanishing BAU.
 
 As commented at the end of section~\ref{sec3}, the size of the splittings $\kappa$ and 
 $\lambda$ should be small enough in order not to disturb the light neutrino mass 
 spectrum. This is encoded in equations like the ones in Eqs.~(\ref{eq:m3kappalambda}-
 \ref{eq:U2kappa}) and taken into account in the numerical analysis. 
 
 For the light neutrino masses we focus on two benchmark values, either $m_0=0$, 
 i.e.~strong NO or strong IO, or $m_0= 0.03 \, (0.015) \, \mathrm{eV}$ that is the maximal 
 value compatible with constraints on the sum of light neutrino masses from 
 cosmology~\cite{Planck:2018vyg} for a mass spectrum with NO (IO). The measured mass squared differences, $\Delta m_{21}^2=m_2^2 -m_1^2$ and
 $\Delta m_{31(2)}^2=m_3^2-m_{1(2)}^2$, are always fixed to their best-fit values~\cite{Esteban:2020cvm}.  Where sufficient, we 
 only discuss the results for the case of NO. 

In general, the light neutrino masses depend on the (real) couplings $y_f$, $f=1,2,3$, whose values are chosen in order to accommodate one of the mentioned forms of the light neutrino mass spectrum. In doing so, the couplings $y_f$ can take either sign, $y_f \geq 0$ and $y_f<0$.
 
The angle $\theta_R$ is not constrained by the light neutrino mass spectrum nor 
by lepton mixing and, thus, is taken between $0$ and $2 \, \pi$ with a log-prior, excluding special values
of $\theta_R$ at which the total mixing $U^2$ would diverge, e.g.~$\theta_R=\pi/4$ 
for Case 1), see Eq.~(\ref{eq:Case1strongNOIO}). 

Lastly, the angle $\theta_L$ is fixed by the requirement to accommodate the lepton mixing angles well, see e.g.~\cite{Esteban:2020cvm}. For concreteness, we always use the best-fit value of $\theta_L$ in the numerical scans. In case there are two such values for $\theta_L$ we use both of them.
\begin{table}[!t]
\begin{center}
\begin{tabular}{|c|c|c|}
\hline
Parameter & Range of values & Prior \\
\hline
$M$ & $\big[50 \mbox{ MeV},~70 \mbox{ TeV}\big]$ &  Log  
\\
\hline
$|\kappa|$ &  $\big[10^{-20},10^{-1}\big]$ & Log\\
\hline
$\theta_R$ &  $\big[0,2\pi\big]$ & $\mbox{Log on } |\theta_R-k\frac{\pi}{4}|$ \\
\hline
$\ba \underline{\mbox{Case 1)}} \\ \frac{s}{n}  \\ \underline{\mbox{Case 2)}} \\ \frac{u}{n} \\ \frac{v}{n} \\ \underline{\mbox{Case 3 a)}}\\ \frac{m}{n} \\ \frac{s}{n} \\ \underline{\mbox{Case 3 b.1)}} \\ \frac{m}{n} \\ \frac{s}{n} \ea$\vspace{0.01in} &  $\ba \\ \big[0,1\big] \\ \\ \big[-0.10,0.12\big] \\ \big[0,3\big] \\ \\ \big[0.056,0.061\big] \\ \big[0,1\big] \\ \\ \big[0.44,0.56\big] \\ \big[0,1\big] \ea$ & Linear\\
\hline
\end{tabular}
\end{center}
\caption{{\bf Range of values
for the
free parameters used in this analysis.} 
The allowed range of  $\frac{m}{n}$, relevant for Case 3 a) and Case 3 b.1), slightly depends on the value of $\frac{s}{n}$ and vice versa. In this table, we only mention 
the maximal ranges allowed. ``Prior'' refers here to the measure in parameter 
space taken for the randomisation and is not meant to indicate a Bayesian 
parameter fit. Note $k$ is an integer and its actual value depends on the case, 
Case 1) through Case 3 b.1). 
The remaining parameters are constrained as follows: $\theta_L$ is fitted in 
order to reproduce the measured lepton mixing angles; the splitting $\lambda$
is set to zero, if not otherwise stated; the lightest neutrino mass $m_0$ is either fixed to $m_0=0$ or $m_0=0.03 \, (0.015)$ eV for NO (IO). 
}
\label{range of values parameters}
\end{table}
\paragraph{Exploration of Case 1) through Case 3 b.1)} The parameters specifying the residual symmetries, in particular the CP symmetry, are chosen as follows for the different cases 
 \begin{itemize}
\item for Case 1), we vary $s$ in its entire allowed range
 corresponding to $0 \leq s \leq n-1$, since this parameter does not impact the lepton mixing angles, but determines the value of the Majorana phase $\alpha$, see section~\ref{sec3}. Indeed, we treat the ratio $\frac sn$, and similar ones for the other cases, as continuous and consider the interval $0 \leq \frac sn \leq 1$;
  \item for Case 2), the ratio $\frac un$ is constrained by the data on lepton mixing angles, see Eq.~(\ref{eq:Case2unrange}), while the parameter $v$ (also describing the choice of the CP symmetry) can take any admitted value and, thus, $0 \leq \frac vn \leq 3$, as it mainly fixes the Majorana phase $\alpha$;
 \item for Case 3 a), the choice of the residual $Z_2$ symmetry, characterised by the parameter $m$, is strongly restricted by the measured value of the reactor mixing angle, see Eq.~(\ref{eq:mnCase3a}), while the parameter $s$ corresponding to the choice of the CP symmetry remains unconstrained and, hence, we vary $\frac{s}{n}$ freely between $0$ and $1$. A slight dependence of the range of $\frac{s}{n}$ on the value of $\frac{m}{n}$ (and vice versa) is taken into account in the numerical scan;
 \item similarly for Case 3 b.1), the ratio $\frac mn$ is mainly fixed by the measurement of the solar mixing angle, compare Eq.~(\ref{eq:mnCase3b1}), whereas $\frac{s}{n}$ is taken to range 
 between $0$ and $1$. Again, a slight dependence of the allowed ranges of $\frac mn$ and $\frac sn$ on each other is observed and accounted for in the numerical scan.
 \end{itemize} 
 We note that all these parameters are varied linearly in the mentioned intervals. Details on the derivation of the latter can be found in~\cite{Hagedorn:2014wha}, see also section~\ref{sec3}. 

\paragraph{Conventions for plots} In the following, we list 
the conventions for the series of plots that we detail in 
section~\ref{numericslepto} and appendix~\ref{appB2}. Areas compatible with leptogenesis 
for VIC are indicated with green(-bordered) areas in the plots, while
those for TIC are shown as blue(-bordered) areas, see e.g.~
Figs.~\ref{fig:CaseIMU2},~\ref{fig:CaseIMU2alpha} and~\ref{fig:kappavsU2}, if not otherwise stated. 

The plots in the $M-U_\alpha^2$-plane contain various experimental constraints: in the upper left part the grey-shaded area summarises the existing limits from electroweak precision data and previous experiments that have been extracted from~\cite{Chrzaszcz:2019inj,Antel:2023hkf}, while the light grey-shaded
area (on the left) indicates the constraints from BBN~\cite{Boyarsky:2020dzc} and the lower left part is excluded, because light neutrino masses cannot be reproduced correctly in this parameter space,\footnote{Note that the upper boundary of this area (slightly) depends on the light neutrino mass ordering and the lepton flavour $\alpha$ of the active-sterile mixing, i.e.~$\frac{U_\alpha^2}{U^2}$.} compare e.g.~Figs.~\ref{fig:CaseIMU2},~\ref{fig:CaseIMU2alpha} and~\ref{CaseIMUeUmu}. 
Furthermore, limits, expected from current and future facilities,  
are displayed with differently coloured dashed lines in several plots, see e.g.~Figs.~\ref{fig:CaseIMU2alpha},~\ref{CaseIIMU2m0light} and~\ref{CaseIMUeUmu}. Especially, we highlight the reach of DUNE~\cite{Ballett:2019bgd}, FASER2~\cite{Ariga:2018uku}, FCC-ee/CEPC~\cite{Blondel:2022qqo}, displaced vertex~\cite{Izaguirre:2015pga,Drewes:2019fou} and prompt~\cite{Izaguirre:2015pga,Pascoli:2018heg} searches at LHC and its upgrade HL-LHC,\footnote{For further studies see e.g.~\cite{Das:2017gke}.} MATHUSLA~\cite{Curtin:2018mvb,MATHUSLA:2020uve}, NA62~\cite{Drewes:2018gkc}
and SHiP~\cite{SHiP:2018xqw,Gorbunov:2020rjx}; see \cite{Antel:2023hkf} for a recent review of these constraints. Constraints from direct searches are usually provided for a fixed flavour pattern only. Hence, when showing plots in the $M-U^2$-plane, we conservatively use the constraints from direct searches for $U_e^2:U_\mu^2:U_\tau^2 = 0:1:0$. In addition, we study the potential impact of searches for charged lepton flavour violation in $\mu-e$ transitions, as performed by MEG II~\cite{MEGII:2021fah}, Mu3e~\cite{Mu3e:2020gyw} (in particular, Mu3e in its Phase II), Mu2e~\cite{Mu2e:2014fns} and COMET~\cite{COMET:2018auw} and (a version of) PRISM/PRIME~\cite{Barlow:2011zza,KUNO:17072023}; see~\cite{Calibbi:2017uvl,Davidson:2022jai} for recent reviews on this topic. We also comment that signals of charged lepton flavour violating processes involving the tau lepton are usually too suppressed to lead to any relevant constraints on the parameter space.

We mostly show plots for the active-sterile mixing angle $U_\mu^2$, since the highest experimental sensitivity and most experimental reaches are given for the muon flavour. Furthermore, $U_\mu^2$ and $U^2$ turn out to be of similar order in the studied cases.  

The grey-shaded areas in the plots showing $U^2 \cdot M$ versus (the absolute value of) a splitting, $\kappa$ or $\lambda$, and the mass splitting $\Delta M$, respectively, see e.g.~Figs.~\ref{fig:kappavsU2} and~\ref{fig:LambdavsU21TeV}, indicate the part of 
the parameter space in which $\kappa$ and/or $\lambda$ are not small enough in order to fulfil the constraints given in equations such as Eqs.~(\ref{eq:m3kappalambda}-\ref{eq:U2kappa}). We emphasise that this  is an assumption made in the presented analysis, while this does not preclude that this region 
 of parameter space might still allow for viable leptogenesis.

 In the mentioned type of plots, see e.g.~Fig.~\ref{fig:kappavsU2}, we also show as guidance for the eye a red line that indicates the resonance condition  
 \begin{equation}
 \label{eq:naiveresoncance}
 \Gamma_\mathrm{naive} =\Delta M \; .
 \end{equation}
 The quantity $\Gamma_\mathrm{naive}$ represents the decay width of the heavy neutrinos in the symmetric phase and non-relativistic regime, i.e.~$\Gamma_\mathrm{naive} = \frac{M \, (\hat{Y}_D^\dagger \hat{Y}_D)_{ii}}{8\, \pi}$. Furthermore, we define $\Delta M$ as the largest RH neutrino mass splitting. Consequently, this resonance condition reads $\Gamma_\mathrm{naive} = \Delta M = 3\,\kappa \, M$, compare Eq.~(\ref{eq:Mkappa}) and see Figs.~\ref{fig:kappavsU2} and~\ref{CaseIIIb1MU2m0lightkappazero} (right plot), or $\Gamma_\mathrm{naive} = \Delta M =2\, \lambda M$, compare Eq.~(\ref{eq:Mlambda}) see Fig.~\ref{fig:LambdavsU21TeV}. We emphasise that the condition in Eq.~(\ref{eq:naiveresoncance}) does not include finite temperature and Higgs corrections to both the mass splitting and interaction rate and, as such, is not expected to reproduce exactly the region of parameter space which maximises the BAU.
 Such corrections are, however, taken into account in the numerical calculations. Apart from functioning as guideline, the red line indicates the boundary between the parameter space in which one typically anticipates the decay of heavy neutrinos at colliders to violate lepton number (to the left of the red line), $\Gamma_\mathrm{naive} \ll \Delta M$, and the region where one does not expect to observe any LNV (to the right of the red line), $\Gamma_\mathrm{naive} \gg \Delta M$, see e.g.~\cite{Drewes:2019byd,Deppisch:2015qwa,Anamiati:2016uxp}. For more precise estimates, the consistent procedure to extract both the heavy neutrino decay widths and their mass splittings is provided in section~\ref{lifetime}, which in general depends on the size of the splittings $\kappa$ and $\lambda$. 

\subsection{Numerical results}
\label{numericslepto}

We begin with Case 1), since it contains the least number of free parameters and, at the same time, allows us to exemplify different features.
For the rest of the cases, we mention distinctive properties. In particular, we display also results for Case 3 a) which has not been studied in~\cite{Drewes:2022kap}. Furthermore, we consider the possibility to constrain the parameter space with the help of experiments searching for charged lepton flavour violation in $\mu-e$ transitions as well as summarise in Tab.~\ref{tab:maxU2forallplots} the maximum value of the total mixing $U^2$ multiplied by the RH neutrino mass $M$ for all situations that we have studied numerically. Lastly, we comment on the effect of the splitting $\lambda$ which is set to zero otherwise.

\begin{figure}[!t]
    \centering
    \includegraphics[width=.49\textwidth]{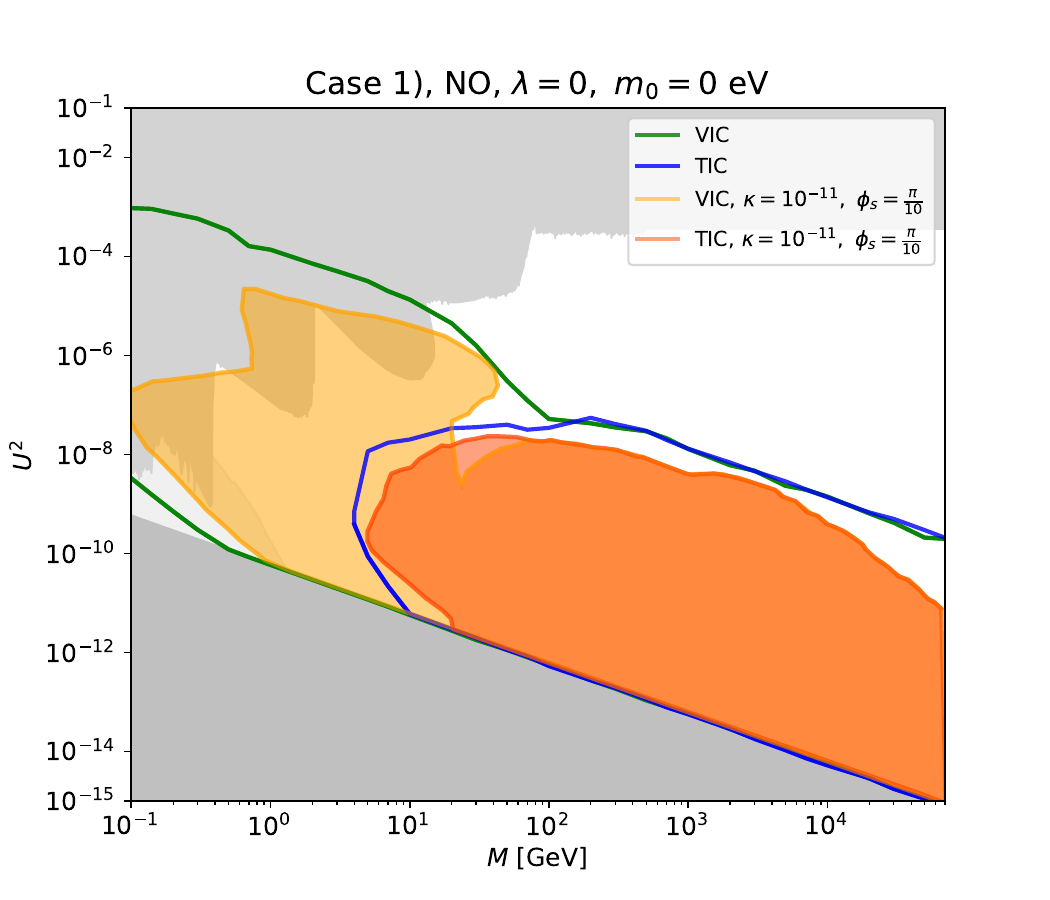}
     \includegraphics[width=.49\textwidth]{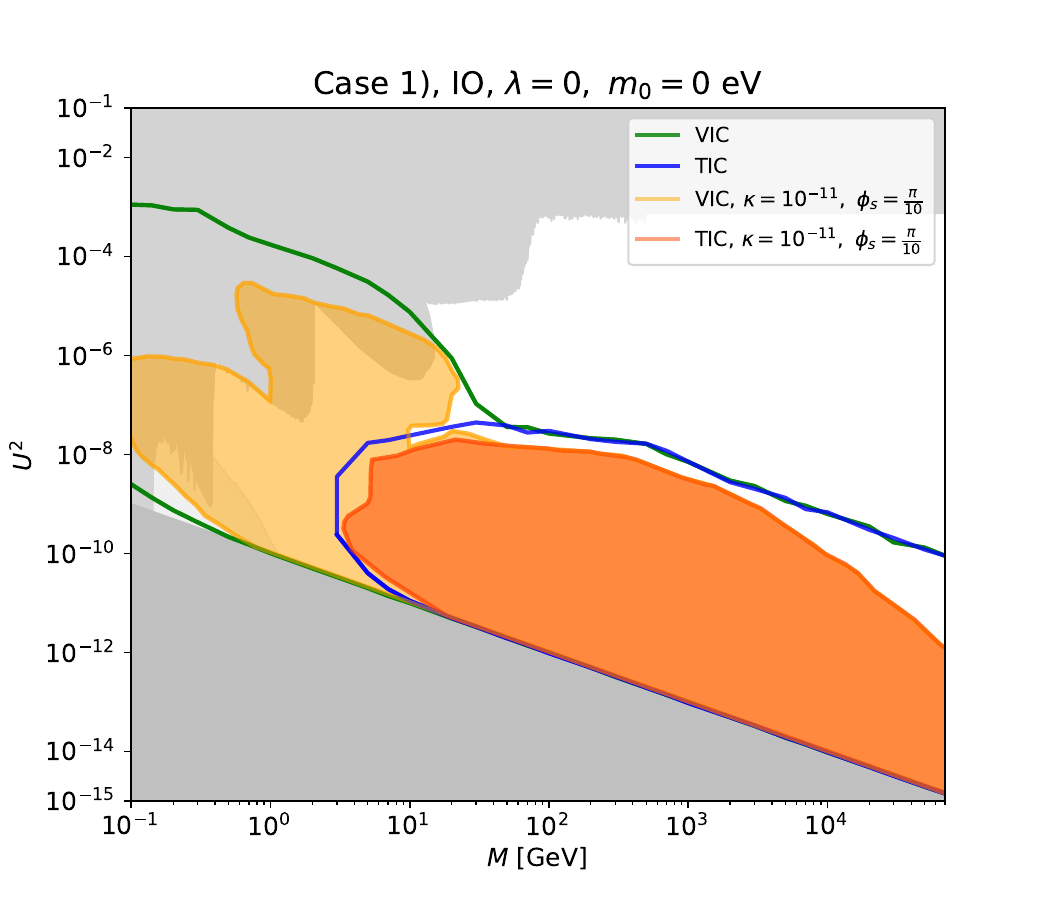}
    \caption{{\bf Case 1)} Comparison of parameter space consistent with leptogenesis in the $M-U^2$-plane in case of fully marginalising over the splitting $\kappa$ and the ratio $\frac sn$ (green- and blue-bordered regions) and for fixed $\kappa$ and $s$ ($\kappa=10^{-11}$ and $\frac sn=\frac{1}{10}$, see~\cite{Drewes:2022kap}, yellow- and orange-shaded areas) for both types of initial conditions, VIC and TIC, and both light neutrino mass orderings, strong NO (left plot) and strong IO (right plot).}
    \label{fig:CaseIMU2}
\end{figure}

\paragraph{Case 1)} 
We first discuss the results of the numerical scan shown in the $M-U^2$-plane, marginalising over the parameter $s$ and the splitting $\kappa$. These can be found in Fig.~\ref{fig:CaseIMU2} for light neutrino masses with strong NO (left plot) and with strong IO (right plot), respectively. The parameter space allowing for successful generation of the BAU is very similar for both orderings. This is also confirmed by the maximal attainable value of the combination $U^2 \cdot M$, given for both orderings as well as for both types of initial conditions, VIC and TIC, see Tab.~\ref{tab:maxU2forallplots}. Indeed, these values only differ by up to a factor of two at maximum. 

As expected the different initial conditions permit partly
different accessible parameter space, in particular VIC are required for heavy neutrino masses smaller than $M \lesssim 5 \, \mathrm{GeV}$. 

\begin{figure}[!t]
    \centering
    \includegraphics[width=.49\textwidth]{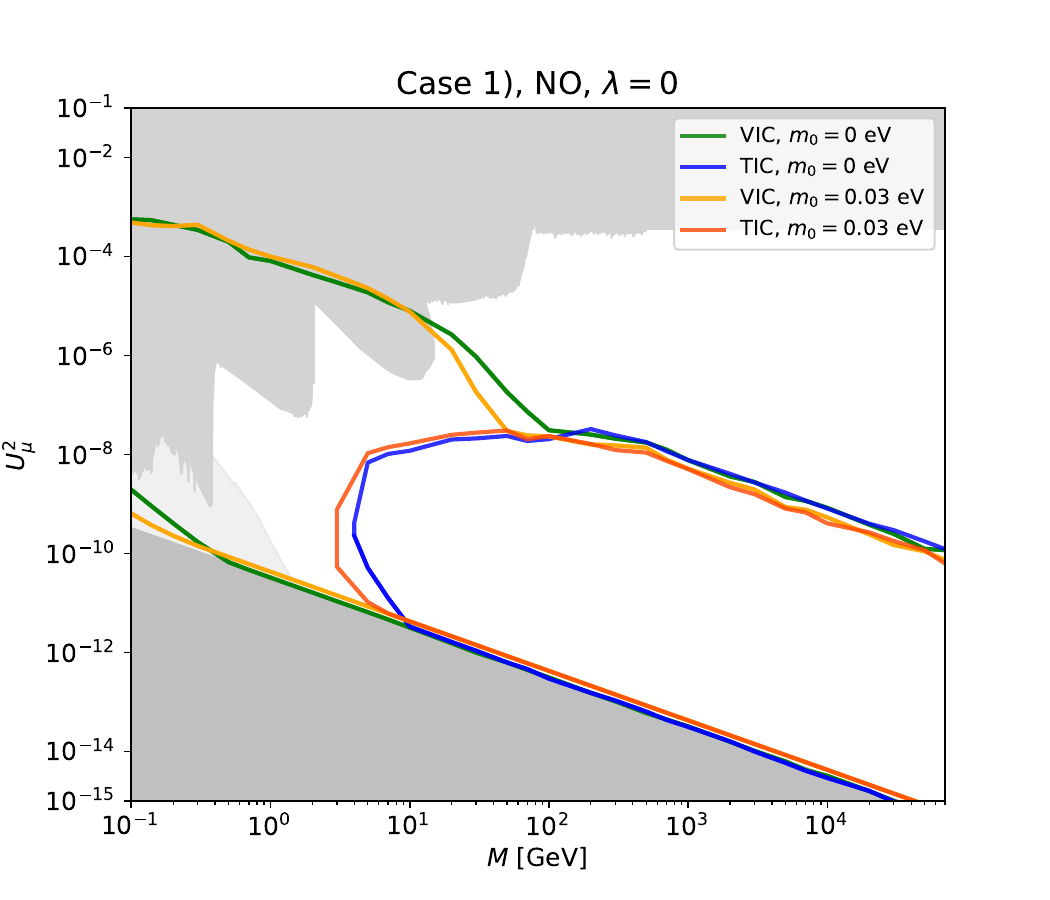}
    \includegraphics[width=.49\textwidth]{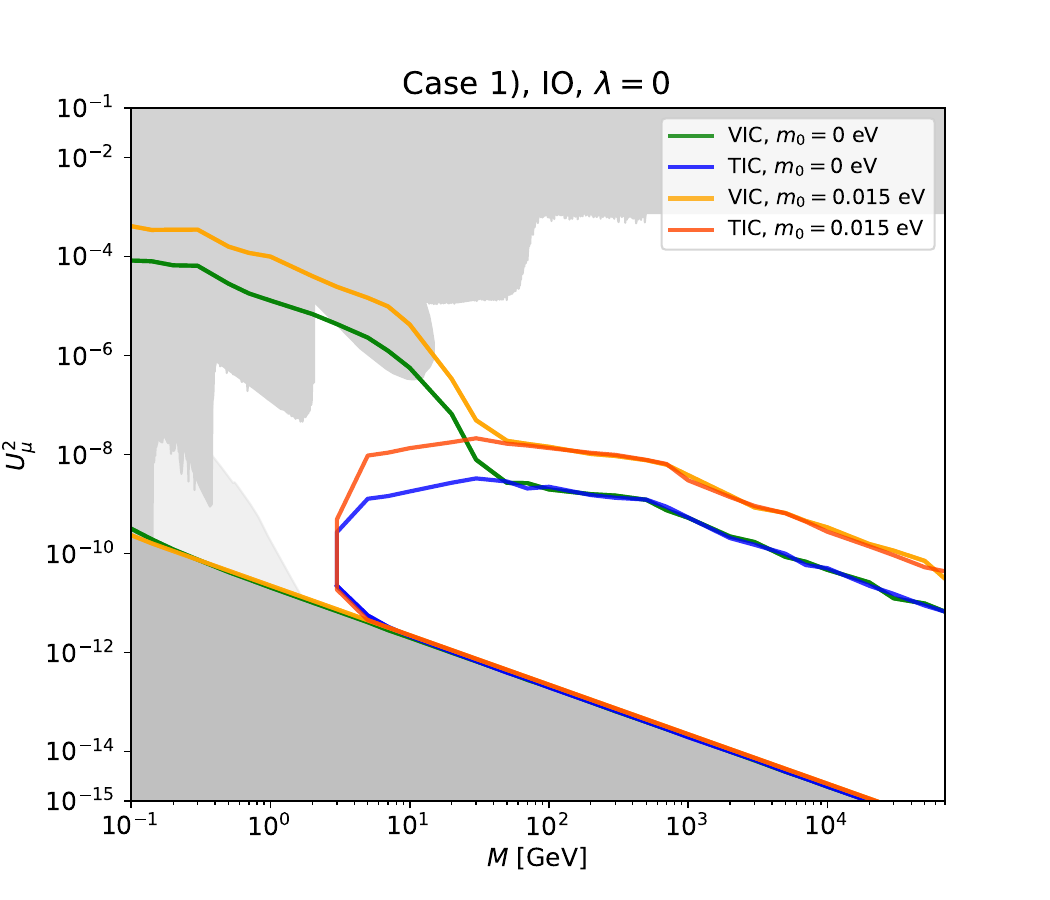}
    \caption{{\bf Case 1)} Comparison of parameter space viable for leptogenesis in the $M-U_\mu^2$-plane
    for vanishing lightest neutrino mass, $m_0=0$, and  its value, maximally allowed by cosmology, $m_0=0.03 \, (0.015)$ eV for NO (IO). Green and blue lines refer to $m_0=0$, while yellow and orange lines to non-zero $m_0$. Both, VIC and TIC, are considered. The left (right) plot assumes light neutrino masses with NO (IO).}
    \label{fig:bothm0Case1}
\end{figure}

\begin{figure}[!t]
    \centering
    \includegraphics[width=0.49\textwidth]{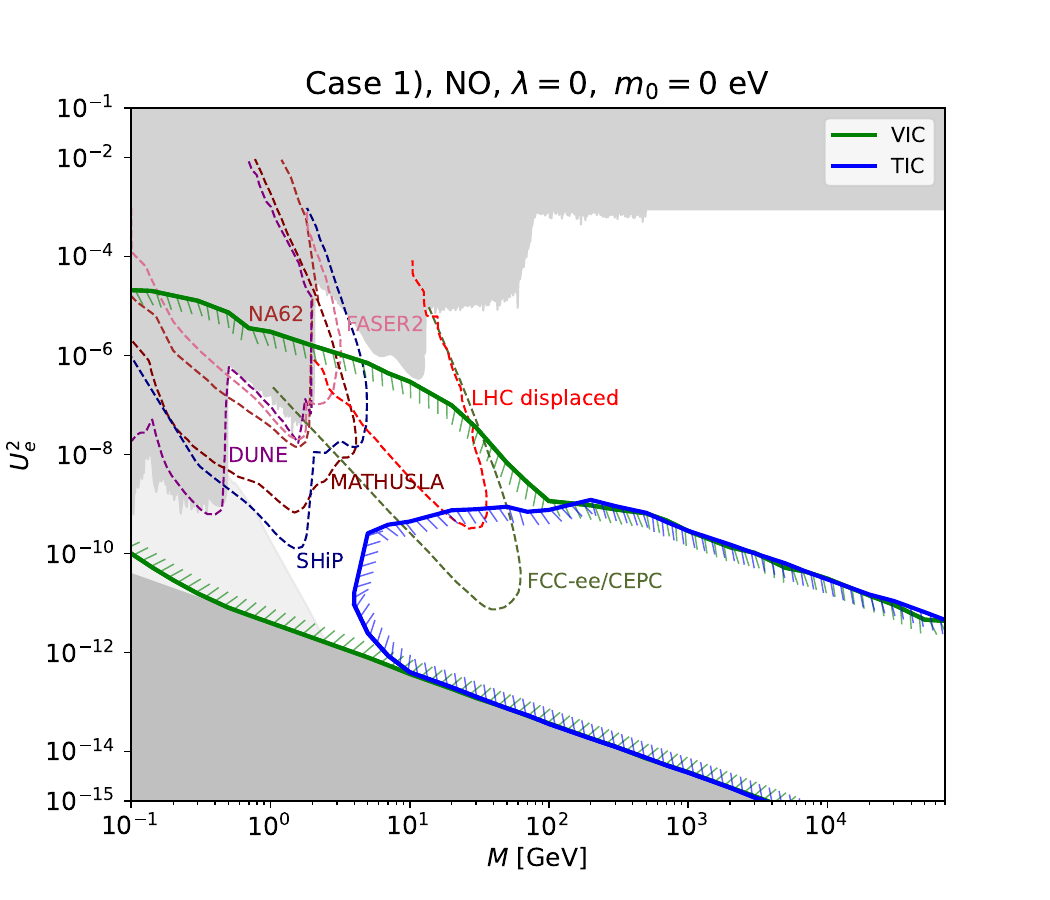}
    \includegraphics[width=0.49\textwidth]{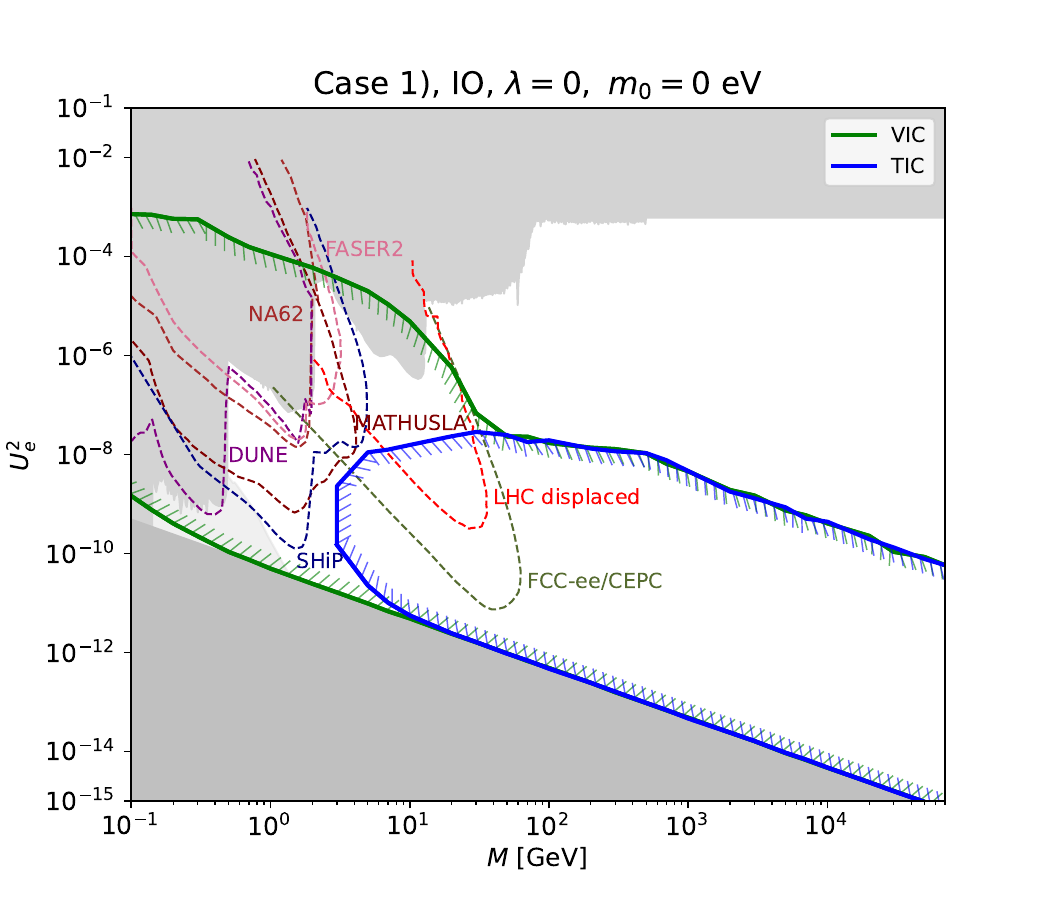}
    \includegraphics[width=0.49\textwidth]{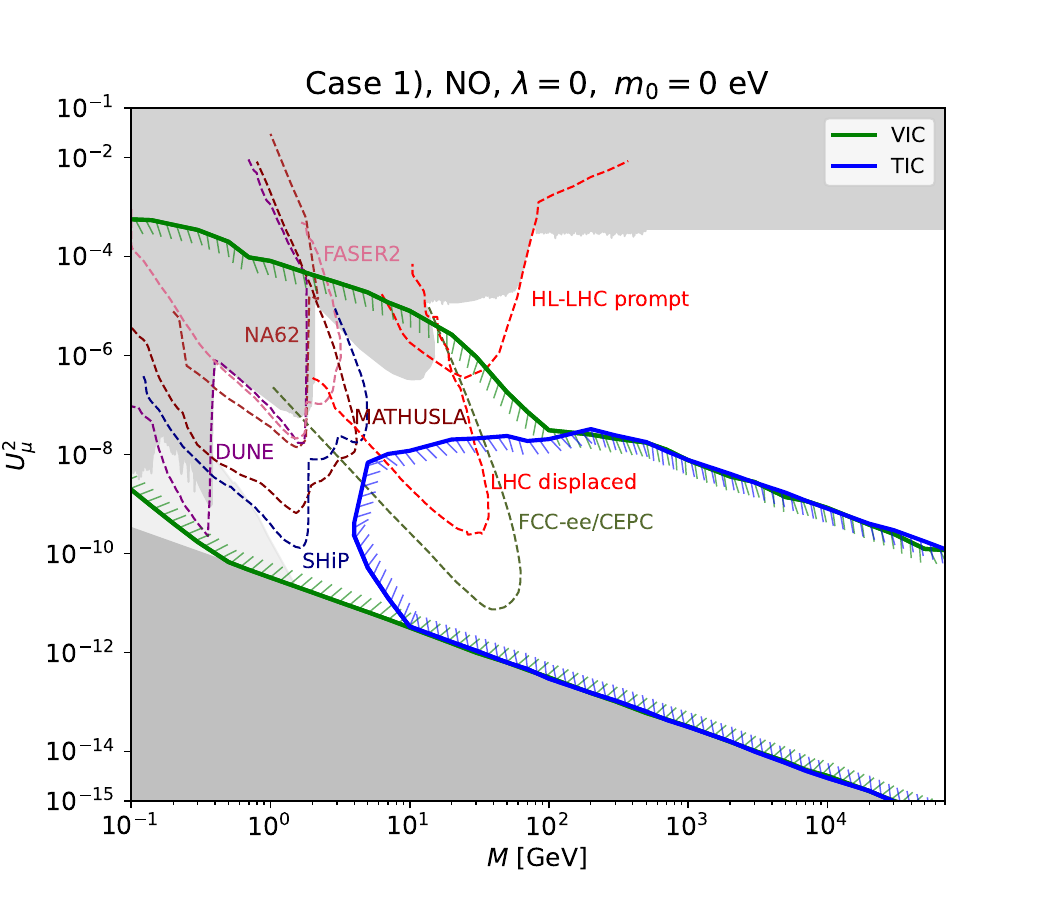}
     \includegraphics[width=0.49\textwidth]{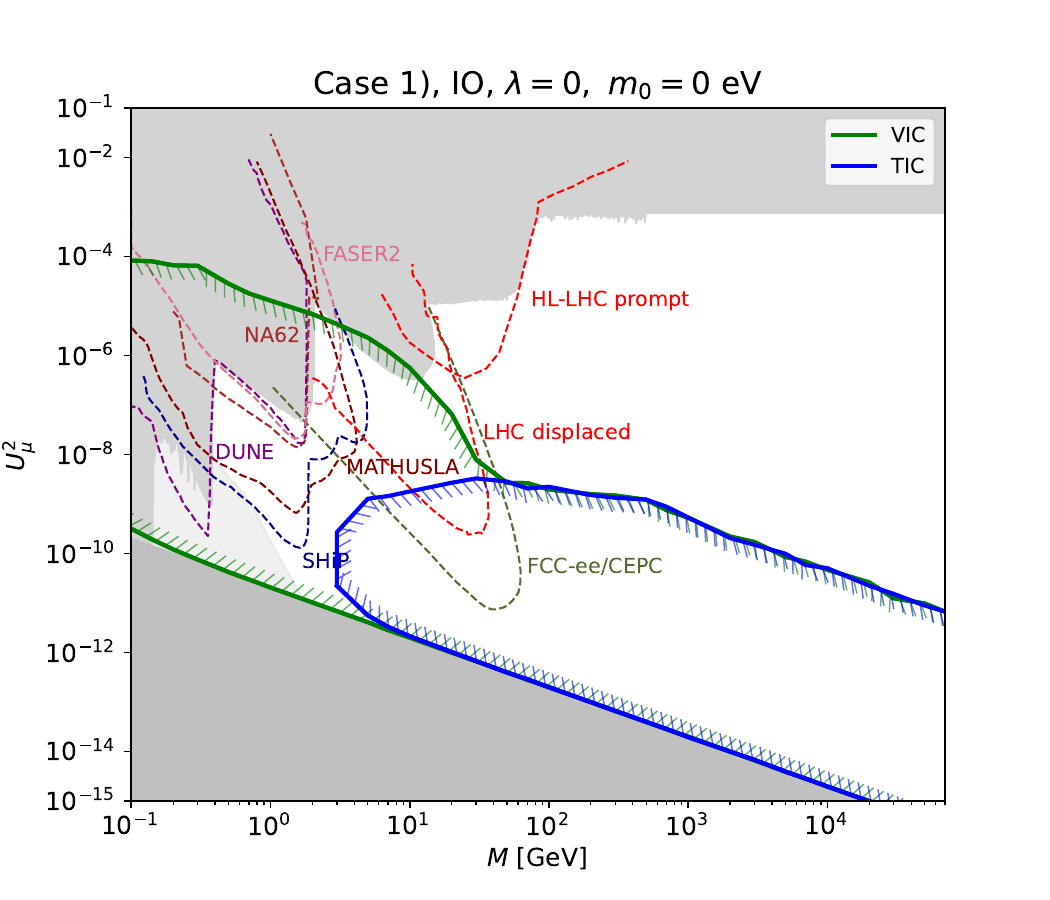}
    \includegraphics[width=0.49\textwidth]{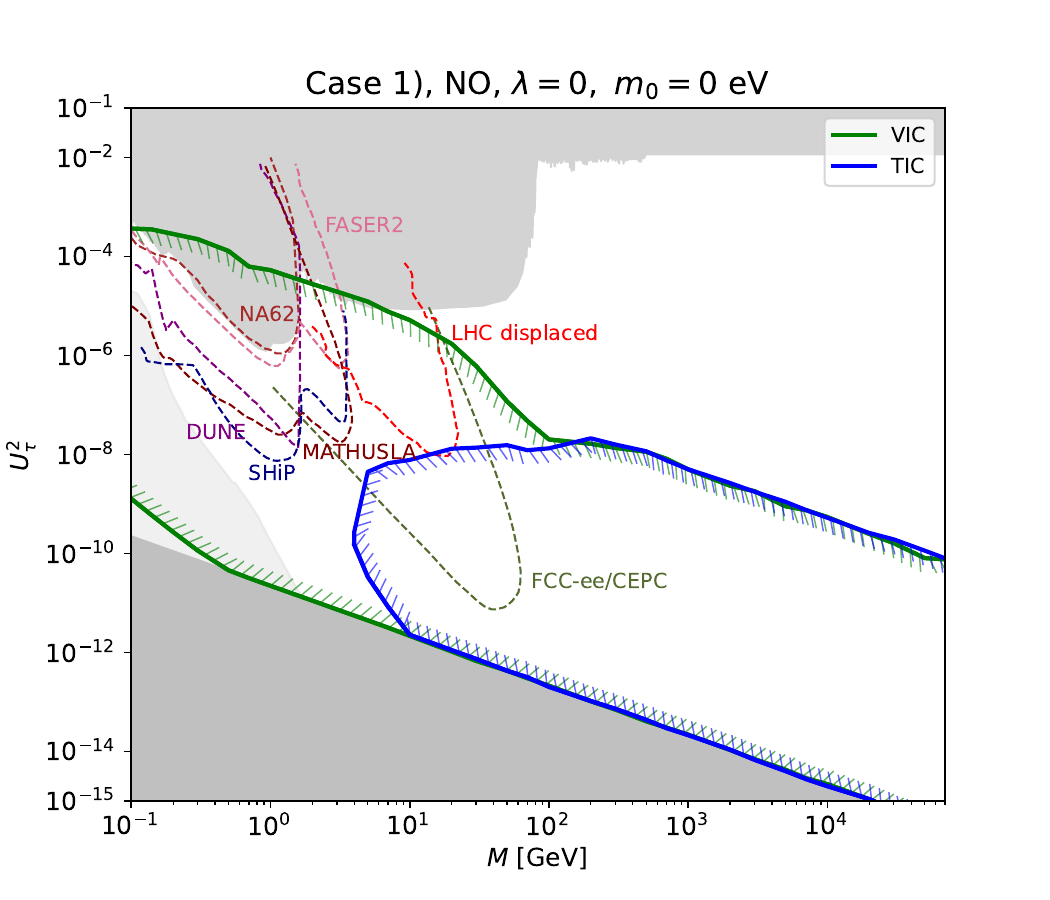}
    \includegraphics[width=0.49\textwidth]{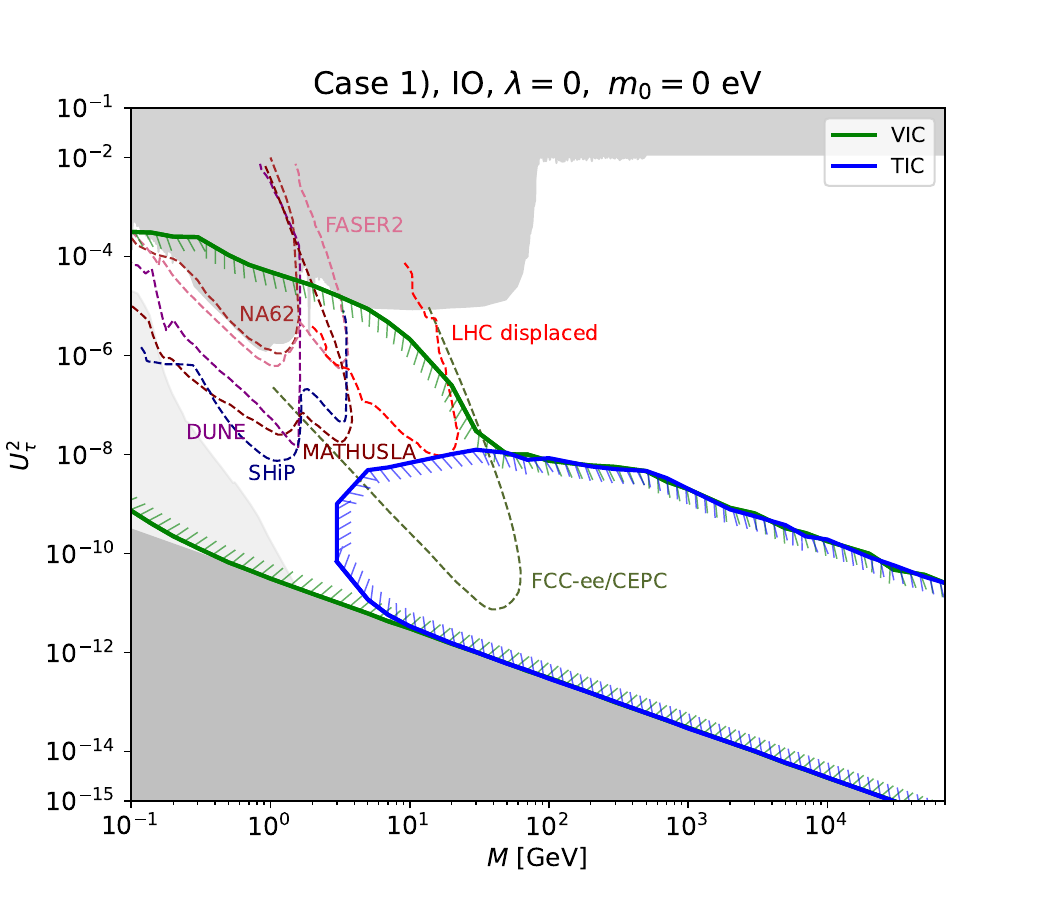}
    \caption{\label{fig:CaseIMU2alpha}
    {\bf Case 1)} Viable parameter space for leptogenesis displayed in the different planes $M-U_\alpha^2$, $\alpha=e,\mu,\tau$ (top, middle, bottom plots) for light neutrino masses with strong NO (left plots) and strong IO (right plots). Results for VIC and
    TIC are shown. 
    Expected sensitivities of various experiments, extracted from~\cite{Drewes:2019fou,Blondel:2022qqo,Antel:2023hkf}, are represented by different coloured dashed lines.}
\end{figure}

In the same figure, we compare the results of this fully marginalised scan with the viable parameter space obtained for a fixed value of the splitting $\kappa$, $\kappa = 10^{-11}$, and a certain value of $\frac sn$, $\frac sn=\frac{1}{10}$ (which is equivalent to $\phi_s = \frac{\pi}{10}$). We show the parameter space as a yellow-shaded area for VIC and as an orange-shaded one for TIC.
 These results have already been displayed in~\cite{Drewes:2022kap}, see figures 9 and 29. We observe an increase of the parameter space, especially for heavy neutrino masses below $1$ GeV and above approximately $2$ TeV.

With Fig.~\ref{fig:bothm0Case1} we explore the relevance of the value of the lightest neutrino mass $m_0$ for the size of the parameter space consistent with leptogenesis. As one can see, the differences between the allowed regions for $m_0$ being zero and for $m_0$ non-zero appear to be marginal for both, VIC and TIC, in the case of light neutrino masses with NO. For light neutrino masses with IO instead, we observe an increase of the size of the viable parameter space for non-zero $m_0$ compared to $m_0=0$. This increase is due to the potentially stronger hierarchy among the couplings of the heavy neutrinos to the different flavours $\alpha$, $\alpha=e,\mu,\tau$, see Fig.~\ref{fig:Case1ternary} (right plot), which allows a larger lepton asymmetry to be preserved from being washed out. In particular, we can read off from  Fig.~\ref{fig:Case1ternary} (right plot) that for $m_0$ being zero the ratios $\frac{U_\alpha^2}{U^2}$ take the values $0.644$,  $0.074$, $0.281$ for $\alpha=e, \mu, \tau$, respectively, whereas for the maximal value of $m_0$ considered, $m_0=0.015$ eV, these can be $0.404$, $0.016$, $0.580$. The maximally achieved value of the combination $U^2 \cdot M$ only slightly depends on the exact value of $m_0$; in the case of light neutrino masses with IO the enhancement of $U^2 \cdot M$ is less than a factor of two for non-zero $m_0$, compare Tab.~\ref{tab:maxU2forallplots}.

\begin{figure}[!t]
    \centering\includegraphics[width=.49\textwidth]{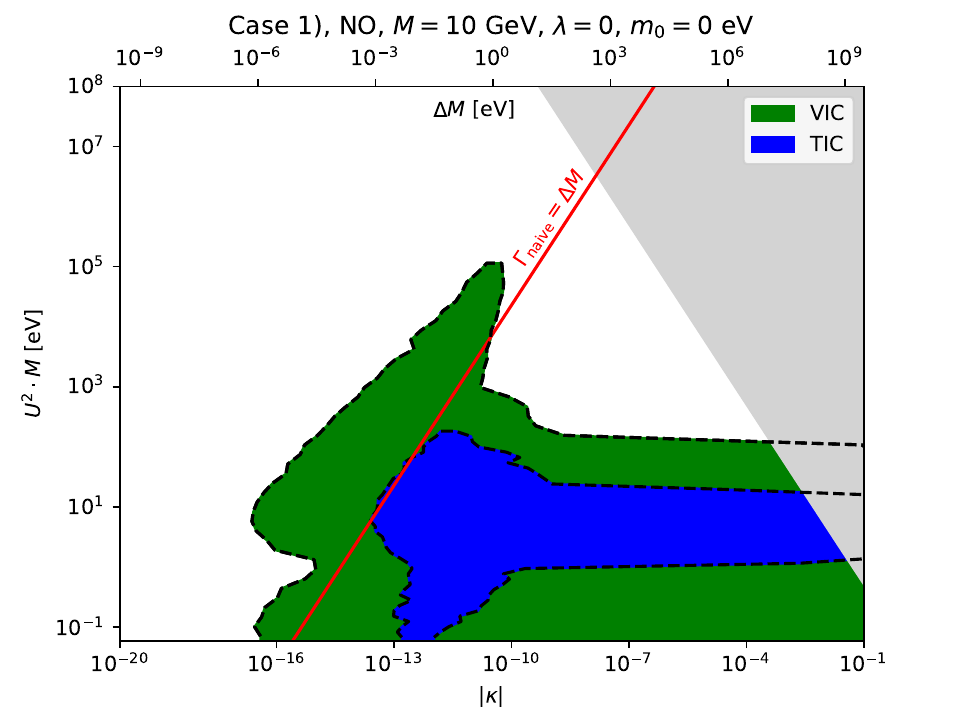}
    \includegraphics[width=.49\textwidth]{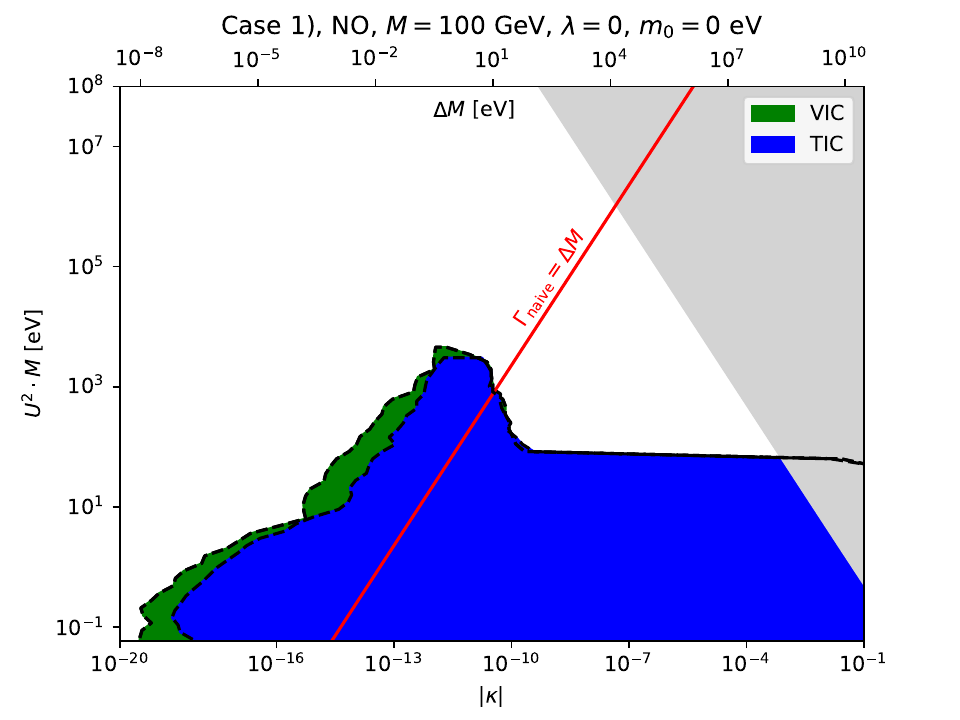}
    \centering\includegraphics[width=.49\textwidth]{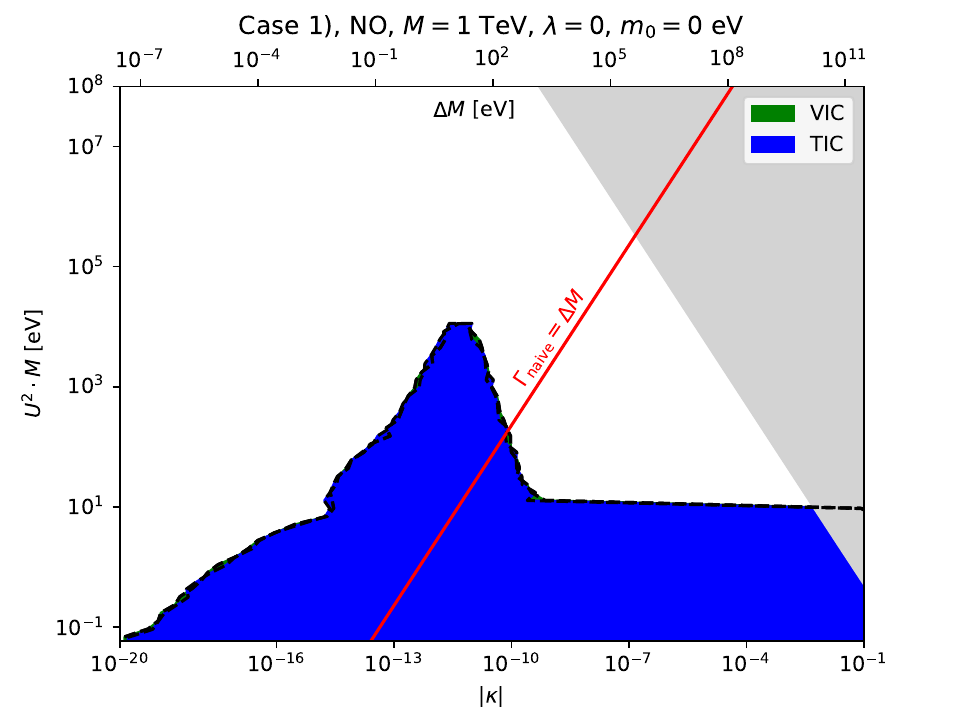}
    \caption{{\bf Case 1)} Parameter space leading to the successful generation of the BAU in the $|\kappa| \, (\Delta M)-U^2 \cdot M$-plane for three different values of the Majorana mass $M$, $M=10$ GeV (upper left plot), $M=100$ GeV (upper right plot) and $M=1$ TeV (bottom plot). Both types of initial conditions, VIC and TIC, are considered.
    We focus on light neutrino masses with strong NO. The red line indicates the resonance condition $\Gamma_\mathrm{naive} = \Delta M = 3\, \kappa \, M$.}
    \label{fig:kappavsU2}
\end{figure}
 
In Fig.~\ref{fig:CaseIMU2alpha} we display the results in the $M-U_\alpha^2$-planes for the different flavours, $\alpha=e, \mu, \tau$, for VIC and TIC as well as for light neutrinos with strong NO and strong IO in order to scrutinise the ability of different types of (future) experiments to probe the parameter space.\footnote{Note that the plot showing the parameter space for light neutrino masses with strong NO in the $M-U_\mu^2$-plane can also be found in the proceedings~\cite{Georis:2024qgk}.} In general, most of the parameter space can be tested for heavy neutrino masses below $M \lesssim 5$ GeV. Furthermore, we note that the exclusion potential is the largest in the $M-U_\mu^2$-plane for light neutrino masses with strong NO, while it is slightly reduced in the case of light neutrino masses with strong IO. For the latter, the exclusion potential in the $M-U_e^2$-plane is at the same time increased. This behaviour can be understood with the help of the ternary plots in Fig.~\ref{fig:Case1ternary}  that clearly show that $\frac{U_e^2}{U^2}$ is rather strongly suppressed for light neutrino masses with strong NO (we have $0.022$, $0.592$, $0.386$ for $\frac{U_\alpha^2}{U^2}$ with $\alpha=e, \mu, \tau$, respectively), whereas for strong IO $\frac{U_\mu^2}{U^2}$ turns out to be smaller and $\frac{U_e^2}{U^2}$ is enhanced; concretely, we find $\frac{U_e^2}{U^2} \approx 0.644$ and  $\frac{U_\mu^2}{U^2} \approx 0.074$.

In order to explore the maximal size of the combination $U^2 \cdot M$, compare also Tab.~\ref{tab:maxU2forallplots}, we choose three benchmark values of the Majorana mass $M$, $M=10 \, \mathrm{GeV}$, $M=100 \, \mathrm{GeV}$ and $M=1 \, \mathrm{TeV}$, see Fig. \ref{fig:kappavsU2}. Clearly, the largest
attainable value, $U^2 \cdot M \sim 10^5$ eV, requires small $M$, $|\kappa|$ of the order of $10^{-11}$ and VIC. For larger values of $M$ this maximum value is of the order of $10^4$ eV and can be obtained for both types
of initial conditions. At the same time, we confirm that for larger values of $M$ a smaller splitting $\kappa$ allows for successful leptogenesis, while large values of $\kappa$ are always compatible with the generation of the correct amount of BAU. This has also been emphasised in~\cite{Drewes:2022kap}. We observe a clear enhancement of the viable parameter space around the region in which the heavy neutrino decay width $\Gamma_\mathrm{naive}$ is comparable to the mass splitting $\Delta M=M_{R_{1}}-M_{R_{2,3}}=3 \, \kappa \, M$, compare red line in Fig.~\ref{fig:kappavsU2}; for more details see text at the end of section~\ref{prereqlepto}.

\paragraph{Case 2)}

\begin{figure}[!t]
    \centering
    \includegraphics[width=0.49\textwidth]{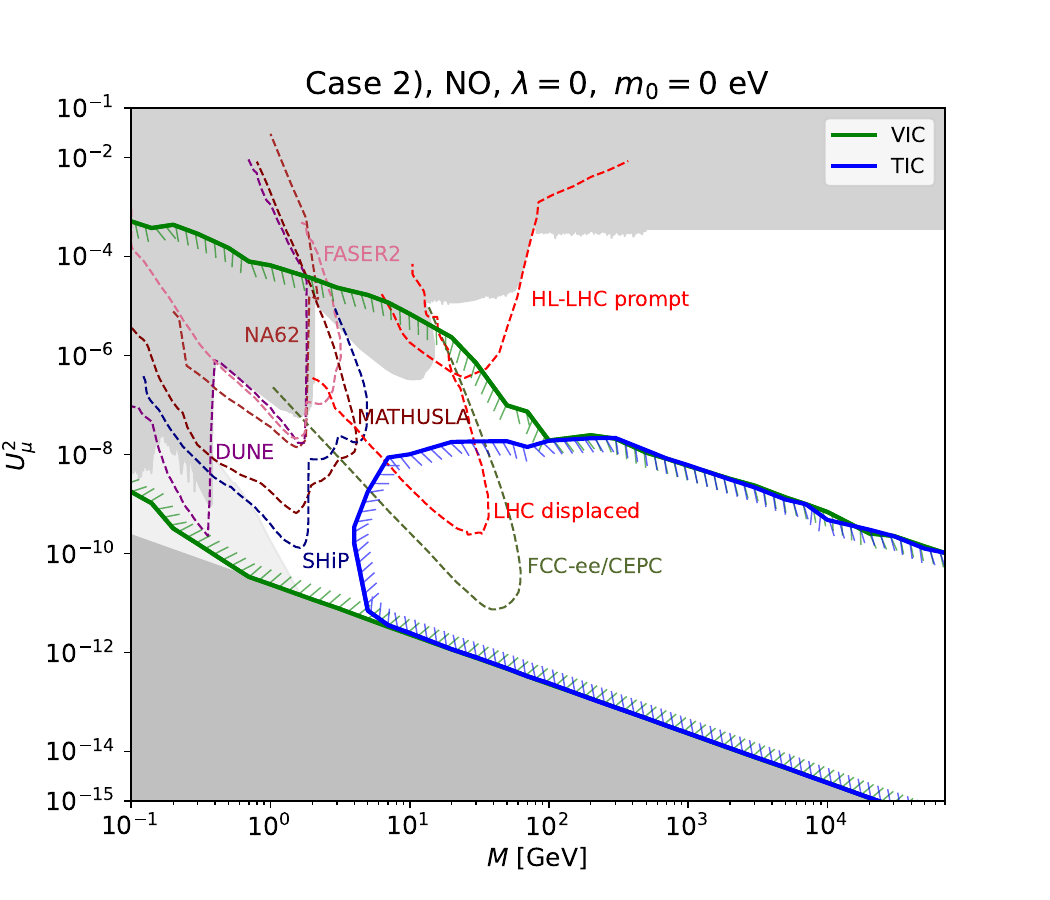}
    \includegraphics[width=0.49\textwidth]{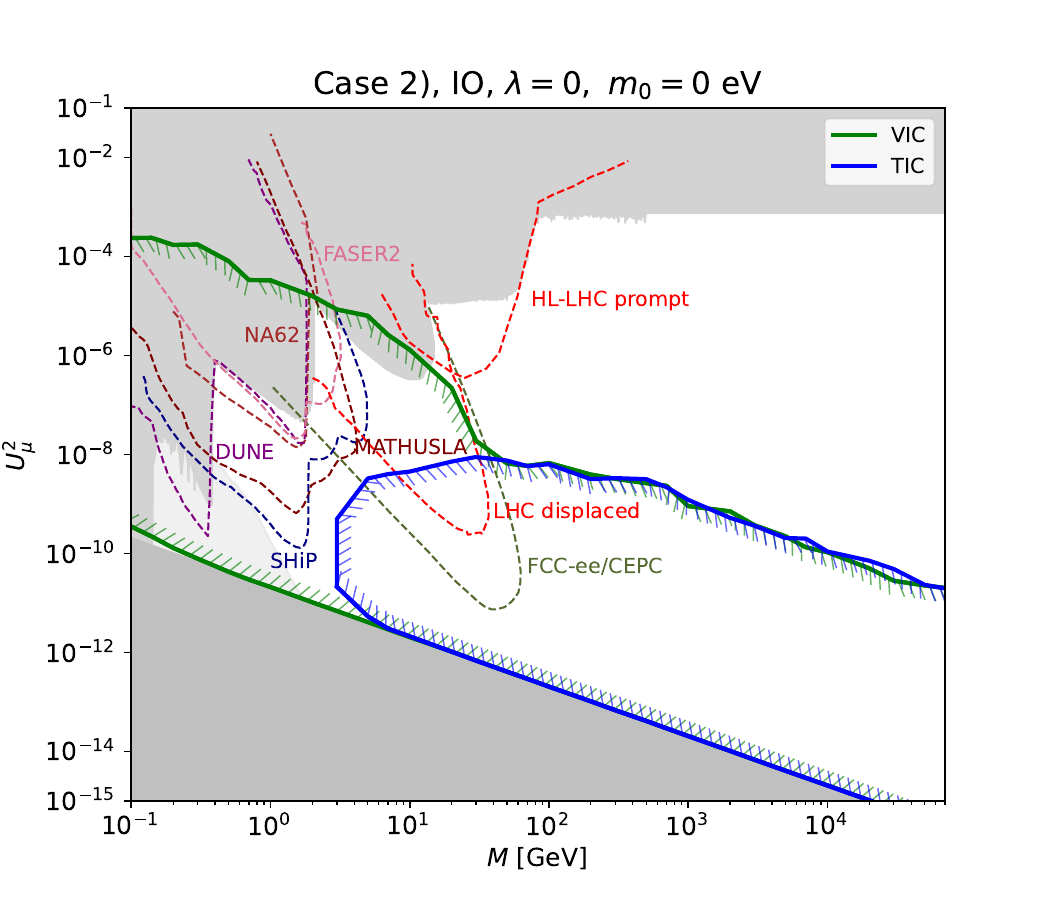}
    \caption{\label{CaseIIMU2m0light}
    {\bf Case 2)} Viable parameter space for leptogenesis displayed in the $M-U_\mu^2$-plane in the case of fully marginalising over the splitting $\kappa$ and the ratios $\frac un$ and $\frac vn$ in their allowed ranges, see Tab.~\ref{range of values parameters}. The left (right) plot refers to light neutrino masses with strong NO (IO). Both initial conditions, VIC and TIC, are considered. The same experimental sensitivities are shown as in Fig.~\ref{fig:CaseIMU2alpha}.}
\end{figure}

We display in Fig.~\ref{CaseIIMU2m0light} the parameter space leading to the successful generation of the BAU. In doing so, we marginalise over the splitting $\kappa$ and the ratios $\frac un$ and $\frac vn$ in their allowed ranges, see Tab.~\ref{range of values parameters}. We have checked that fixing
these ratios  to particular values, e.g.~$\frac un=0$, hardly impacts the size of the viable parameter space. This observation is consistent with the findings that can be deduced with
the help of the CP-violating combinations, presented in~\cite{Drewes:2022kap}.

As can be inferred from Figs.~\ref{fig:CaseIMU2alpha} and~\ref{CaseIIMU2m0light} as well as Tab.~\ref{tab:maxU2forallplots}, the parameter space consistent with leptogenesis is in Case 2) very similar to the one for Case 1).

\begin{figure}[!t]
    \centering
    \includegraphics[width=0.49\textwidth]{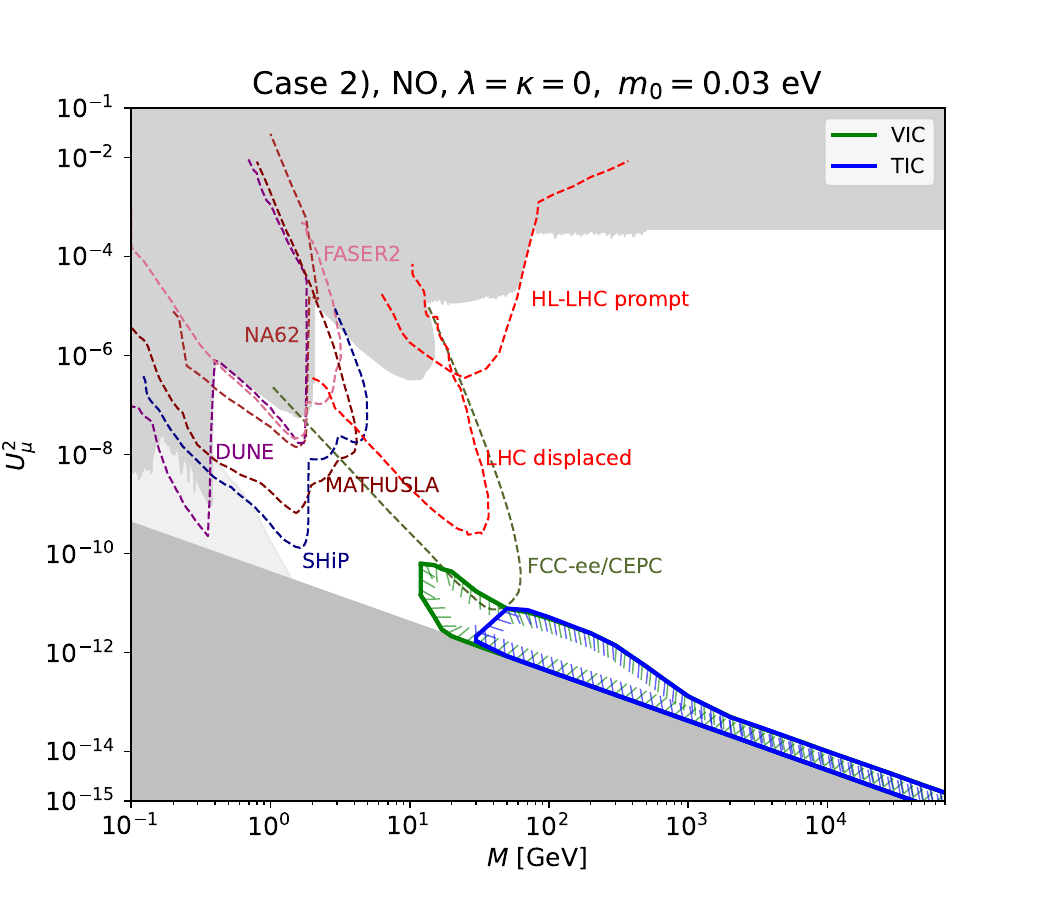}
    \includegraphics[width=0.49\textwidth]{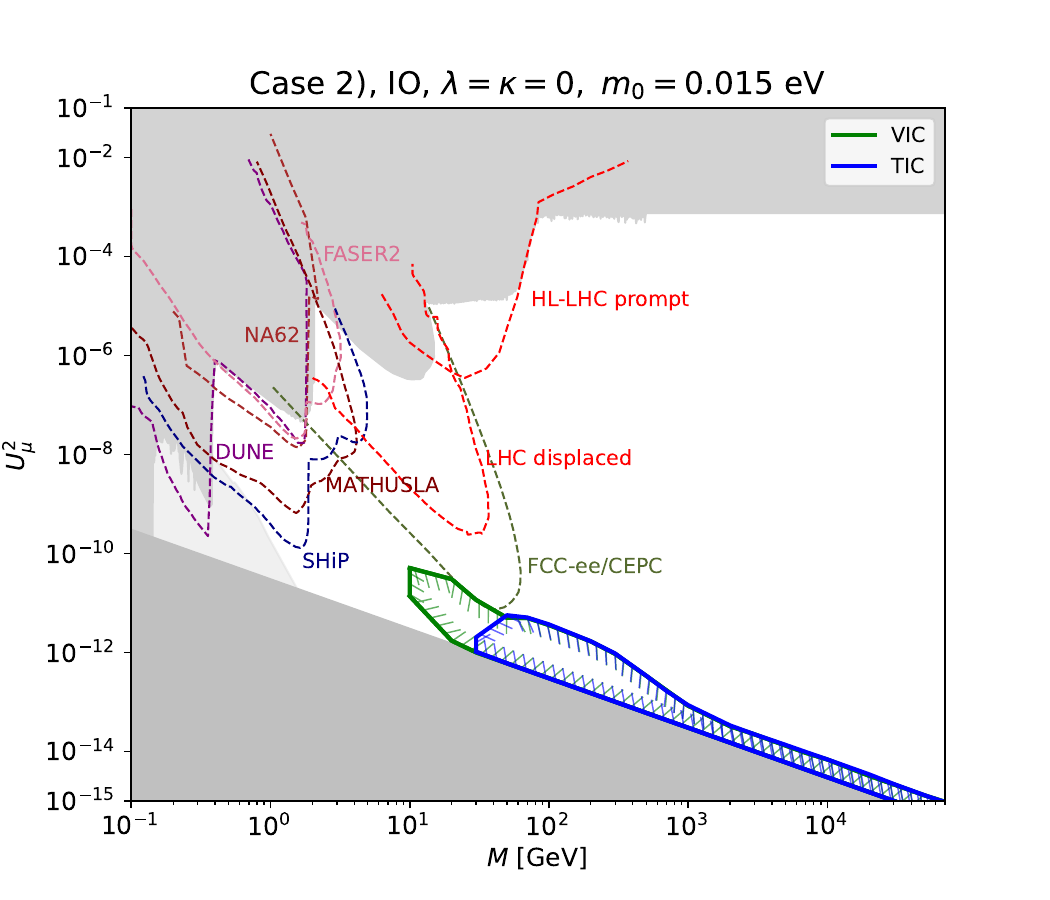}
    \caption{\label{CaseIIMU2m0heavy}
    {\bf Case 2)} Parameter space consistent with leptogenesis shown in the $M-U_\mu^2$-plane for vanishing splittings, $\kappa=0$ and $\lambda=0$, and light neutrino masses with NO (IO) and $m_0 = 0.03\, (0.015)$ eV in the left (right) plot. Results for VIC and TIC are displayed. For details about the experimental sensitivities, see text and Fig.~\ref{fig:CaseIMU2alpha}.}
\end{figure}

As is known~\cite{Antusch:2017pkq,Sandner:2023tcg}, leptogenesis remains viable in the case in which the RH neutrino masses are exactly degenerate, since non-zero mass splittings among the heavy neutrinos are generated by Higgs and thermal contributions. In~\cite{Drewes:2022kap} it has been, in particular, noticed that for the lightest neutrino mass $m_0$ being non-zero, a sufficient amount of the BAU can be generated in Case 2) even for vanishing splittings $\kappa$ and $\lambda$.
 We have, thus, explored the size of the viable parameter space for leptogenesis in this situation, choosing the lightest neutrino mass $m_0$ to be the largest value compatible with NO and IO, respectively. The results are displayed in Fig.~\ref{CaseIIMU2m0heavy}. As one can see, the findings for light neutrino masses with NO and IO are very similar -- an observation that is  confirmed by the maximally reached values of $U^2 \cdot M$ given in Tab.~\ref{tab:maxU2forallplots}. While this parameter space is, as expected, considerably reduced with respect to the case with non-zero splitting, it is still possible to probe a small part of it with FCC-ee/CEPC; however, more likely for VIC than for TIC.  At the same time, we emphasise that a rather large value of the Majorana mass $M$ is needed, i.e.~$M \gtrsim 10 \, \mathrm{GeV}$ for VIC and even larger for TIC, $M \gtrsim 30 \, \mathrm{GeV}$, for generating the correct amount of BAU. Indeed, for $\kappa = \lambda = 0$, the only relevant 
 CP-violating combination is $C_{\mathrm{DEG},\alpha}$ and, thus, the production of the BAU depends on a flavoured washout, which can convert a lepton flavour asymmetry into a lepton number one, see Eqs.~(\ref{eq:defLFVcombination}-\ref{eq:flvw}) and~\cite{Drewes:2022kap}. This explains why the amount of generated BAU is generally smaller. Furthermore, this type of asymmetry is only possible at intermediate temperatures $M/T \sim 1$, resulting in larger lower bounds for $M$.
 
\paragraph{Case 3 a)}

This case has not been considered numerically  in~\cite{Drewes:2022kap}, since it  requires a sizeable value of the index $n$ of the group $\Delta (3 \, n^2)$ or $\Delta (6 \, n^2)$, at least $n=16$, in order to appropriately accommodate the observed values of the lepton mixing angles, compare~\cite{Hagedorn:2014wha,Drewes:2022kap} and section~\ref{sec31}. Nevertheless, it is interesting to study the size of the parameter space consistent with leptogenesis and to confront it with the findings for the other cases.

Before coming to the numerical results, let us comment on the characteristics that can be derived from the CP-violating combinations, given in~\cite{Drewes:2022kap}. The form of these combinations is identical for Case 3 a) and Case 3 b.1), only the assignment of the couplings $y_f$, $f=1,2,3$, to the light neutrino masses $m_i$, $i=1,2,3$, differs, see~\cite{Hagedorn:2014wha,Drewes:2022kap} and section~\ref{sec31}. In particular, for $m$ and $s$ both even or both odd\footnote{We remind the reader that for these combinations of $m$ and $s$ a (large) enhancement of the values of the total mixing $U^2$ cannot be achieved, compare discussion in~\cite{Drewes:2022kap}.} a non-vanishing BAU requires that the different flavours are washed out at different rates as well as at least one of the splittings, $\kappa$ and $\lambda$, to be non-zero. Furthermore, we note that for light neutrino masses with strong IO the BAU should reveal a dependence on $\frac sn$ ($\phi_s$) of the form $\sin 3 \, \phi_s$. For the combinations of $m$ and $s$, where one is even and the other one odd, the CP-violating combination encoding LNV does not vanish and is proportional to $\sin 4\, \theta_R$, i.e.~for $\theta_R = 0, \frac \pi4, \ldots$ the generated BAU is  expected to be very suppressed. In the case of zero splittings, $\kappa=0$ and $\lambda=0$, the same behaviour is foreseen. Furthermore, the generated BAU should be proportional to $\sin 3 \, \phi_s$. Eventually, we note that for light neutrino masses  with strong NO a non-zero splitting $\kappa$ and/or $\lambda$ is necessary in order to produce a non-vanishing BAU.

\begin{figure}[!t]
    \centering
    \includegraphics[width=0.49\textwidth]{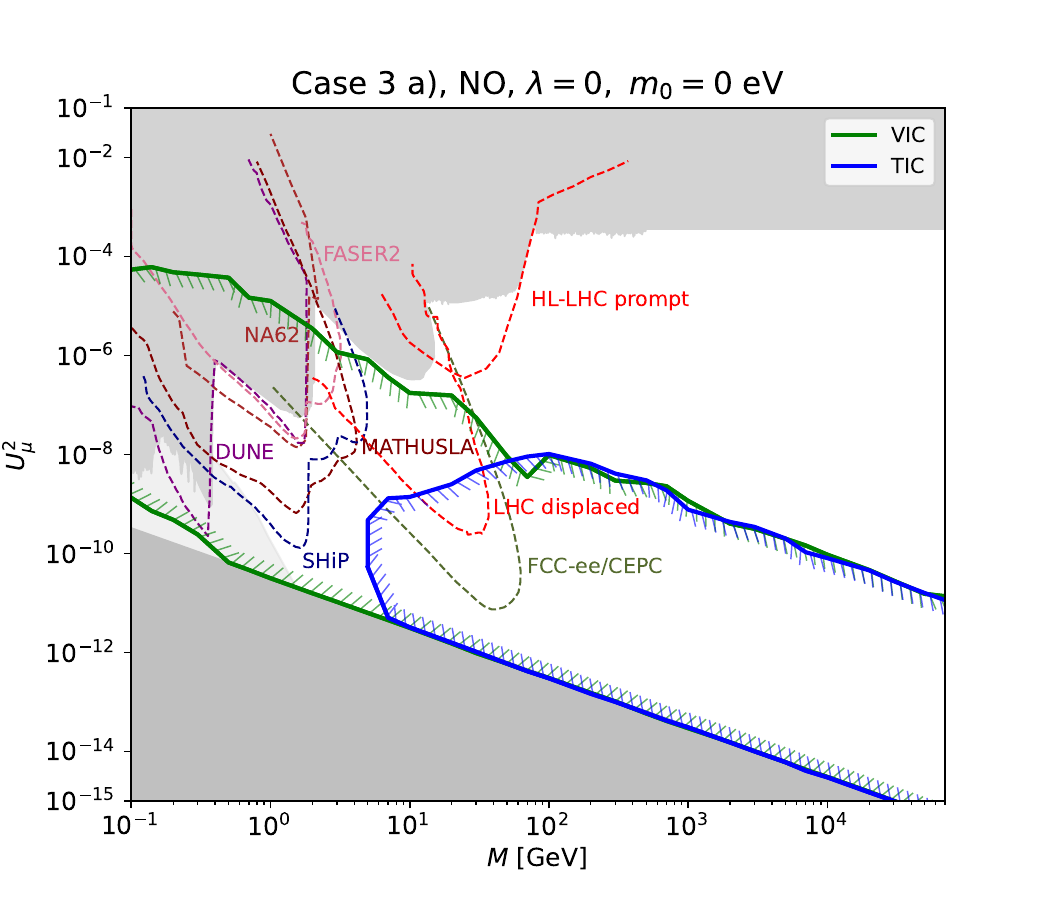}
    \includegraphics[width=0.49\textwidth]{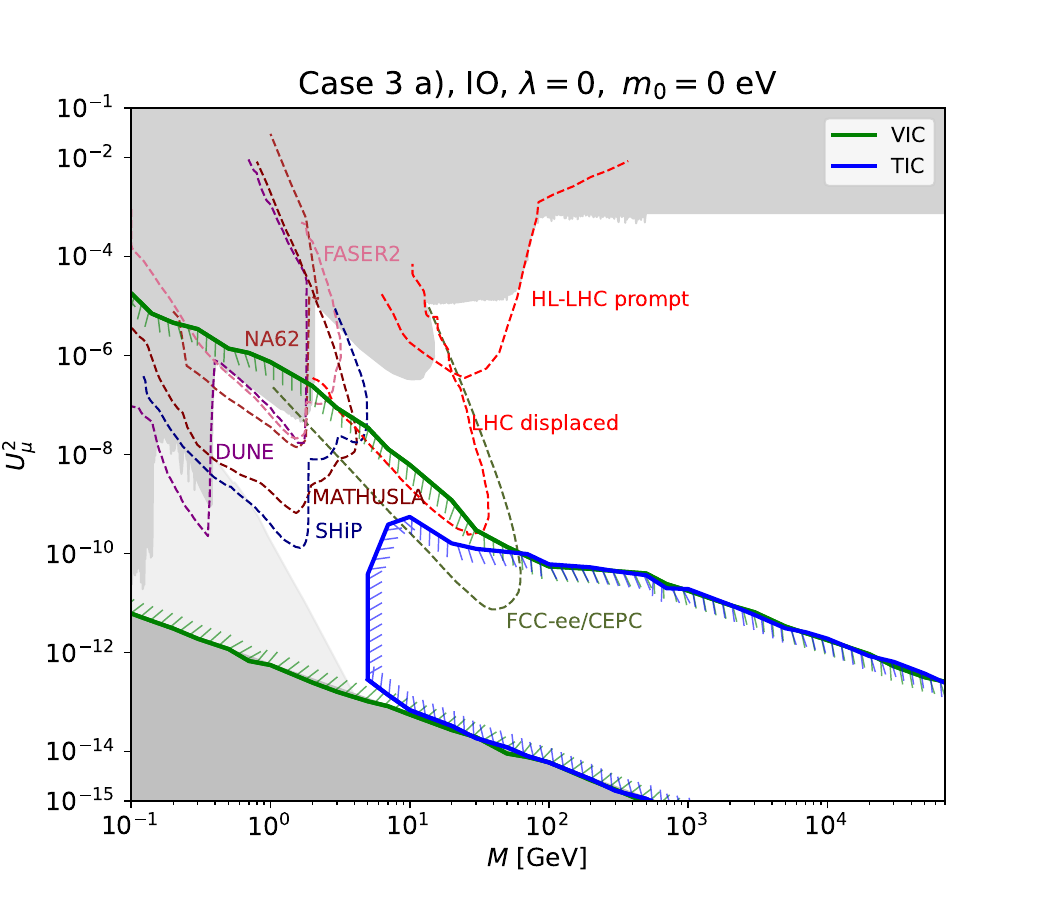}
    \caption{\label{CaseIIIaMU2m0light}
    {\bf Case 3 a)} Viable parameter space for leptogenesis in the $M-U_\mu^2$-plane in the case of fully marginalising over the splitting $\kappa$ and the ratios $\frac mn$ and $\frac sn$ in their allowed ranges, see Tab.~\ref{range of values parameters}.
    The left (right) plot shows the results for light neutrino masses with strong NO (IO). Both initial conditions, VIC and TIC, are considered. Different coloured dashed lines indicate expected sensitivities of various experiments, see text and Fig.~\ref{fig:CaseIMU2alpha} for details.}
\end{figure}

As shown in Fig.~\ref{CaseIIIaMU2m0light}, we obtain a (slightly) smaller parameter 
space compatible with successful leptogenesis for Case 3 a) than for Case 1) and Case 2). This is also reflected in Tab.~\ref{tab:maxU2forallplots} by the maximal value of $U^2\cdot M$ that can be reached. Nevertheless, the viable parameter space remains experimentally testable, in particular with displaced vertex searches, at SHiP, (HL-)LHC or FCC-ee/CEPC. One striking feature of Case 3 a) is the reduced parameter space which is excluded by the requirement to correctly reproduce the light neutrino masses, if these follow strong IO, compare right plot in Fig.~\ref{CaseIIIaMU2m0light}. Indeed, the ratio $\frac{U_\mu^2}{U^2}$ can be as small as $10^{-5}$, once the total mixing $U^2$ is about two orders of magnitude larger than the naive seesaw limit, found in Eq.~(\ref{eq:naiveseesawformula}). This is also confirmed by the corresponding ternary plot, see right plot in Fig.~\ref{fig:Case3aternary}.

\paragraph{Case 3 b.1)}

\begin{figure}[!t]
    \centering
    \includegraphics[width=0.49\textwidth]{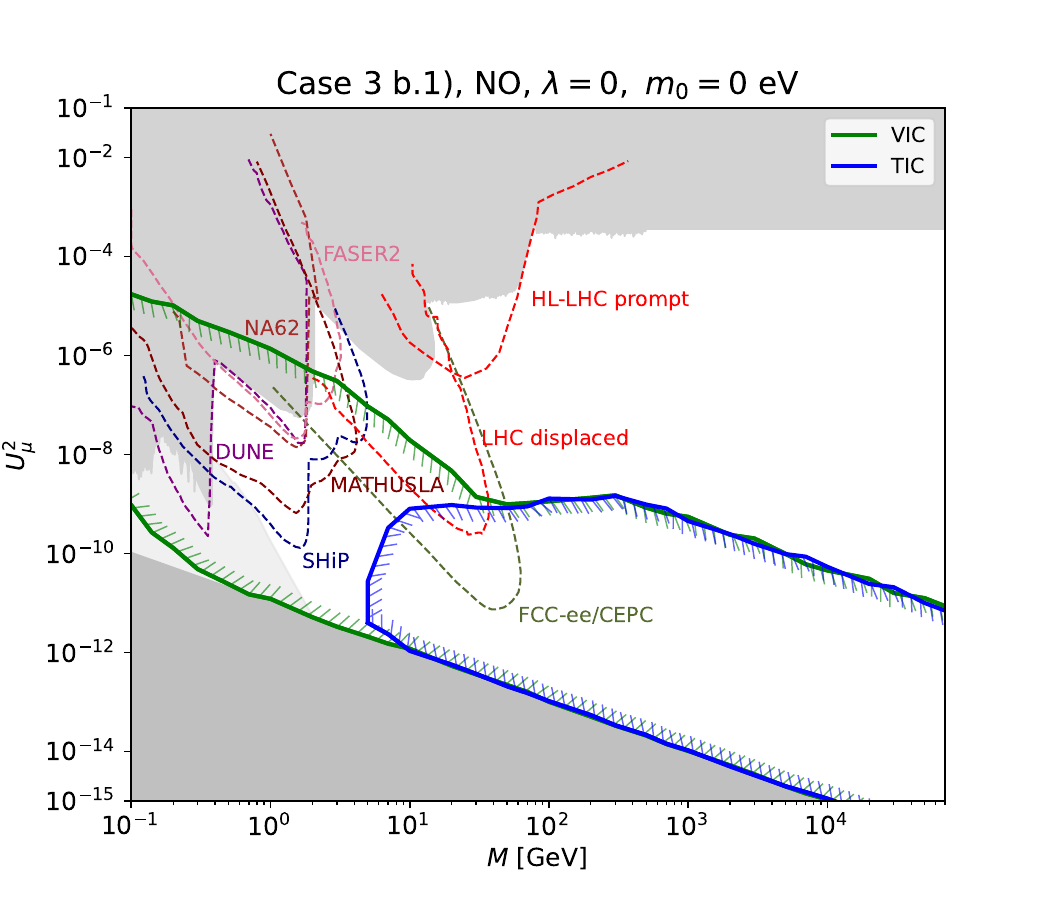}
    \includegraphics[width=0.49\textwidth]{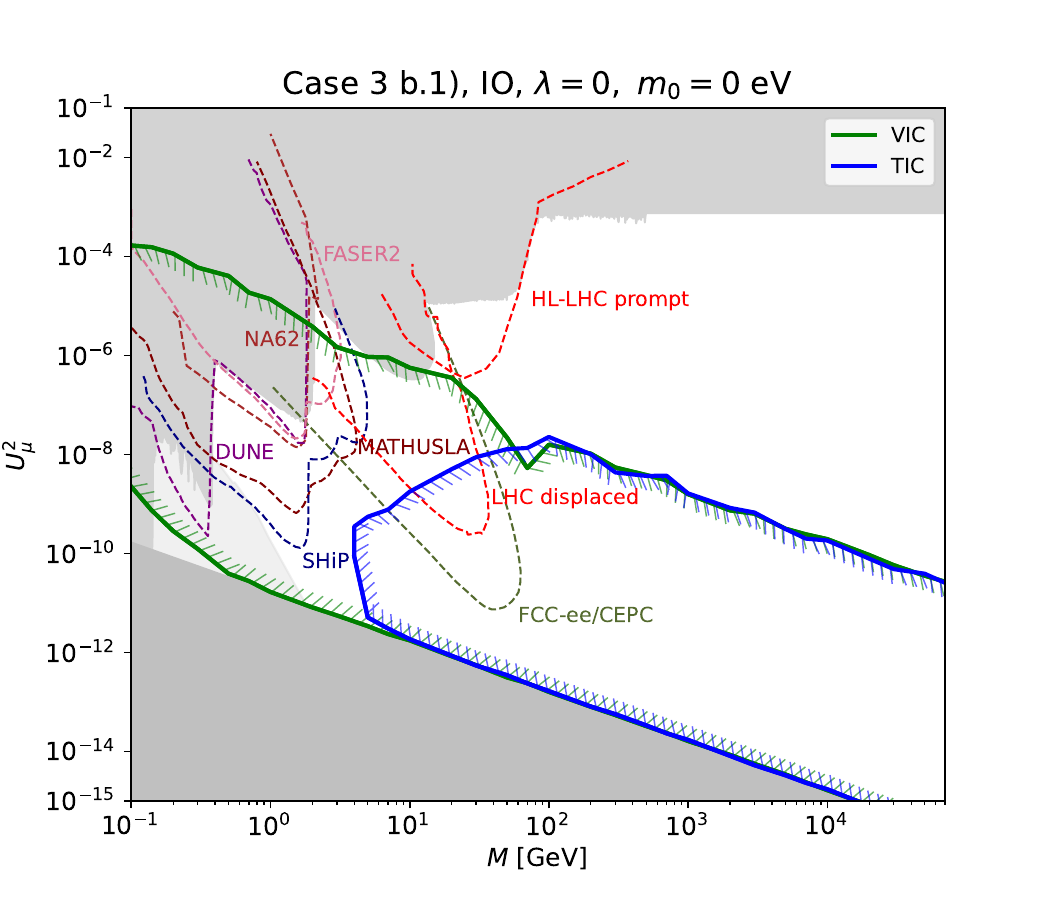}
    \caption{\label{CaseIIIb1MU2m0light}
    {\bf Case 3 b.1)} Parameter space consistent with leptogenesis shown in the $M-U_\mu^2$-plane, resulting from fully marginalising over the splitting $\kappa$ and the ratios $\frac mn$ and $\frac sn$ according to the ranges in Tab.~\ref{range of values parameters}. Light neutrino masses with strong NO (left plot) and strong IO (right plot) are considered. Furthermore, both types of initial conditions, VIC and TIC, are studied. The quoted experimental sensitivities are the same as in Fig.~\ref{fig:CaseIMU2alpha} for Case 1).}
\end{figure}

We begin the discussion of the results for Case 3 b.1) with Fig.~\ref{CaseIIIb1MU2m0light} which shows the parameter space consistent with leptogenesis in the $M-U_\mu^2$-plane. It is  achieved, when fully marginalising over the splitting $\kappa$ and the ratios $\frac mn$ and $\frac sn$ in the ranges quoted in Tab.~\ref{range of values parameters}. We note that these results are obtained for light neutrino masses with strong NO and strong IO, respectively. Comparing this figure with the corresponding plots for Case 1), see Fig.~\ref{fig:CaseIMU2alpha}, we find that for light neutrino masses with NO the available parameter space is sizeably reduced for Case 3 b.1), in particular in the case of VIC. This is in agreement with the maximally achieved value of the combination $U^2 \cdot M$ which is about a factor of 40 (20) smaller for Case 3 b.1) for VIC (TIC), compare Tab.~\ref{tab:maxU2forallplots}. For light neutrino masses with IO instead we observe similar results for Case 1) and Case 3 b.1), up to a slight reduction of the viable parameter space for VIC. Again, this is reflected in the maximal size of $U^2 \cdot M$, see Tab.~\ref{tab:maxU2forallplots}. 

\begin{figure}[!t]
    \centering
    \includegraphics[width=0.49\textwidth]{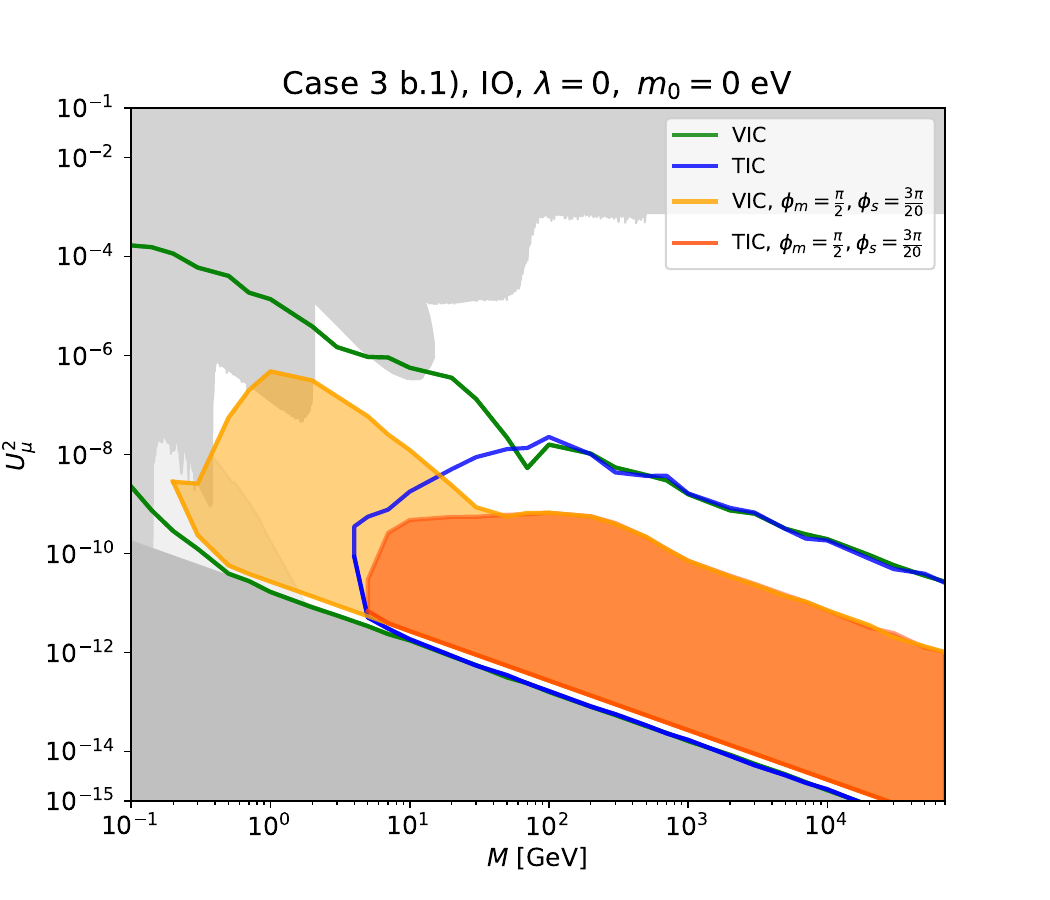}
    \includegraphics[width=0.49\textwidth]{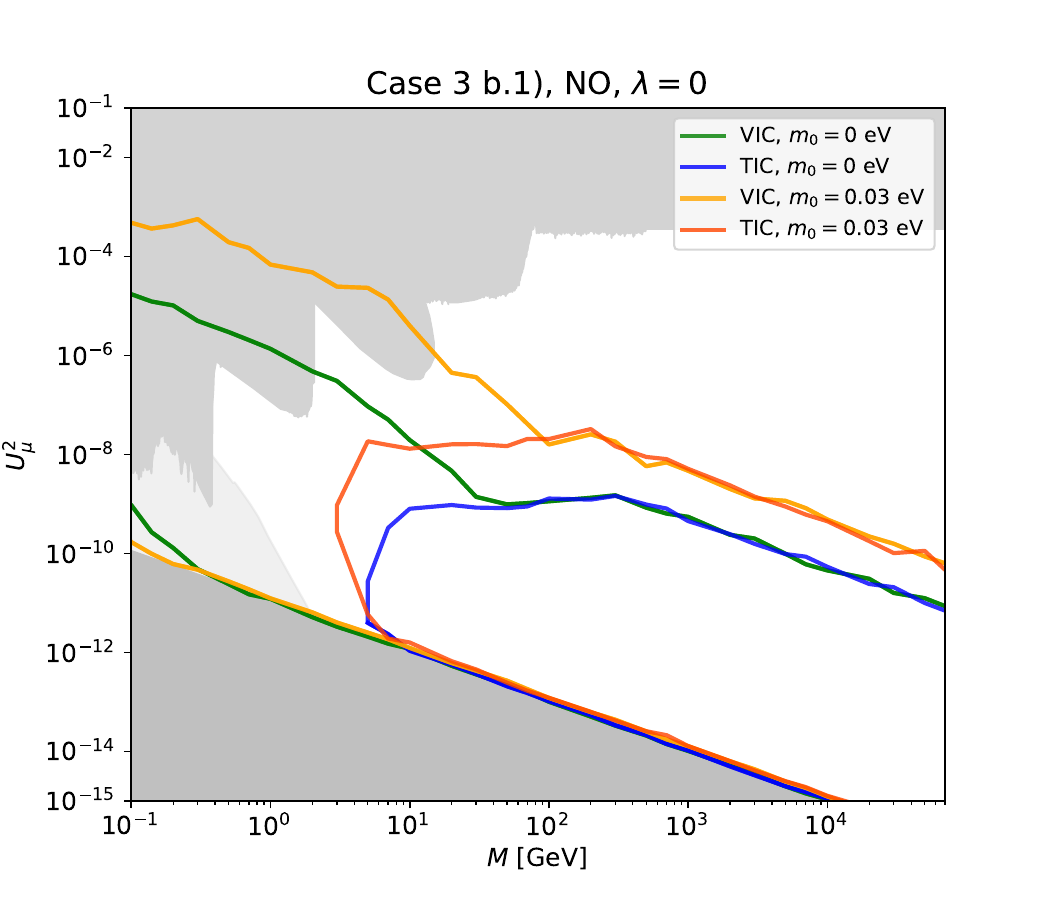}
    \caption{{\bf Case 3 b.1)} Left plot: Comparison of the viable parameter space in the $M-U_\mu^2$-plane resulting from fully marginalising over the splitting $\kappa$ and the ratios $\frac mn$ and $\frac sn$ in the ranges found in Tab.~\ref{range of values parameters} and
for fixed values of the ratios $\frac mn$ and $\frac sn$ (corresponding to $m=10$, $s=3$ and $n=20$, as used in~\cite{Drewes:2022kap}), while still marginalising over $\kappa$. Blue- and green-bordered areas are obtained for full marginalisation, while yellow- and orange-shaded areas are due to the specific choice of $\frac mn$ and $\frac sn$. The different colours refer to different initial conditions, VIC and TIC, as indicated. Light neutrino masses are constrained to have strong IO. Right plot: Comparison of the viable parameter space in the $M-U_\mu^2$-plane for the lightest neutrino mass $m_0$ being zero (blue- and green-bordered areas) and being $m_0=0.03$ eV, as maximally allowed by cosmology for light neutrino masses with NO (yellow- and orange-bordered areas). Again, different colours stand for VIC and TIC.}
    \label{CaseIIIb1smfixedcompare}
\end{figure}

For Case 3 b.1), it is, furthermore, interesting to analyse the impact of having fixed values of the parameters $m$ and $s$ in contrast to marginalising over these in their allowed ranges, see section~\ref{sec3} and Tab.~\ref{range of values parameters}, since in~\cite{Drewes:2022kap} Case 3 b.1) has led to a (much) smaller viable parameter space than Case 1) and Case 2). As one can see in Fig.~\ref{CaseIIIb1smfixedcompare} (left plot), for specific values of $m$ and $s$ for fixed $n$, namely $m=10$, $s=3$ and $n=20$ which is equivalent to $\phi_m=\frac \pi2$ and $\phi_s=\frac{3 \, \pi}{20}$, as employed in~\cite{Drewes:2022kap}, the parameter space consistent with successful leptogenesis, indeed, turns out to be considerably smaller -- regarding both the smallest viable heavy neutrino mass and the maximal value of the active-sterile mixing angle $U_\mu^2$ that can be achieved.\footnote{We note that smaller values of the ratio $\frac{U_\mu^2}{U^2}$ become viable, when marginalising over the allowed values of $\frac mn$ and $\frac sn$. This explains the difference in the parameter region excluded by the requirement to correctly reproduce the light neutrino masses.} We have also checked that there exists a fixed value of the ratio $\frac sn$ that leads to the same viable parameter space, as marginalising over this ratio.

\begin{figure}[!t]
    \centering
    \includegraphics[width=0.49\textwidth]{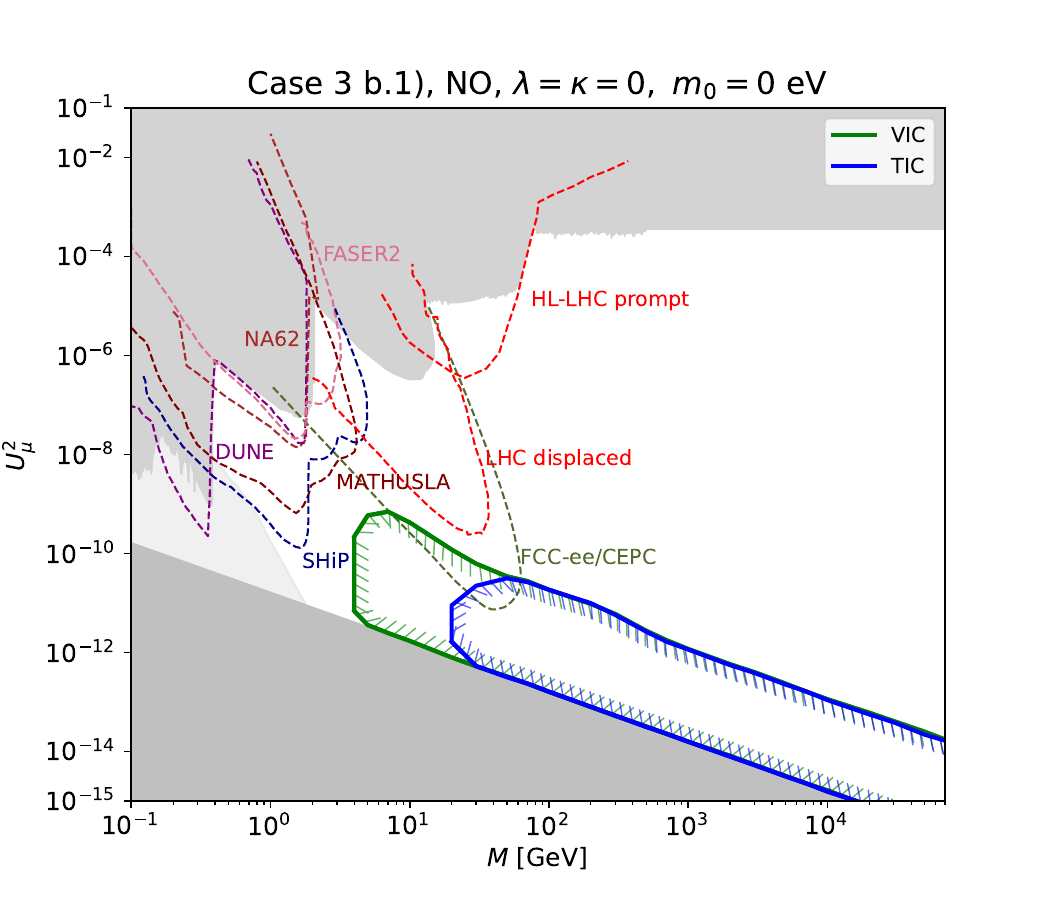}
    \includegraphics[width=.49\textwidth,height=.27\textheight]{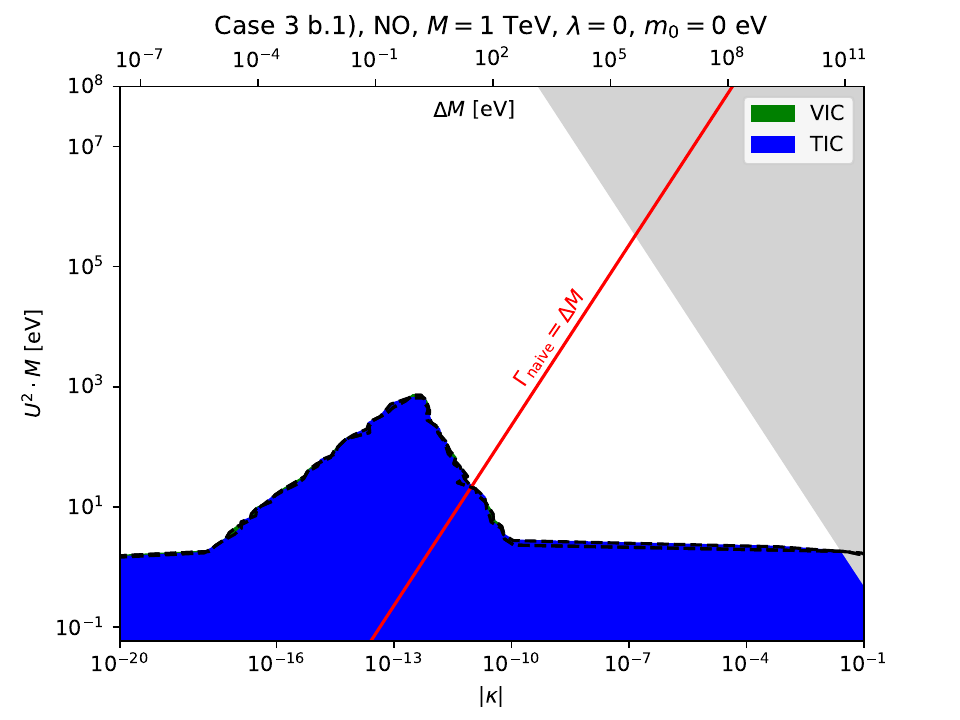}
    \caption{\label{CaseIIIb1MU2m0lightkappazero}
    {\bf Case 3 b.1)} Left plot: Parameter space consistent with leptogenesis in the $M-U_\mu^2$-plane for vanishing splittings, $\kappa=0$ and $\lambda=0$, and light neutrino masses with strong NO. Both VIC and TIC are considered. The different coloured dashed lines refer to various experiments, see text and Fig.~\ref{fig:CaseIMU2alpha} for details. Right plot: Viable parameter space for leptogenesis in the $|\kappa| (\Delta M)-U^2 \cdot M$-plane for the Majorana mass $M$ being fixed to $M=1$ TeV and light neutrino masses with strong NO. Also here results for both initial conditions, VIC and TIC, are shown. 
    The red line indicates the resonance condition $\Gamma_\mathrm{naive}= \Delta M= 3 \, \kappa \, M$.}
\end{figure}

It is also of relevance to study the effect of a non-vanishing lightest neutrino mass $m_0$ on the size of the allowed parameter space. As one can clearly see in Fig.~\ref{CaseIIIb1smfixedcompare} (right plot), for Case 3 b.1) this leads to an enlargement of the parameter space, for light neutrino masses with NO. Indeed, the maximal value of the active-sterile mixing angle $U_\mu^2$ can be increased by a factor of more than ten. This observation is confirmed by the corresponding increase of the maximal achievable size of the combination $U^2 \cdot M$, see Tab.~\ref{tab:maxU2forallplots}, that can amount to up to a factor of 70 (for VIC). This is in contrast to the findings for Case 1), compare Fig.~\ref{fig:bothm0Case1}, where non-zero $m_0$ only has a mild impact on the size of the viable parameter space for light neutrino masses with NO.

We complete the information available in Fig.~\ref{CaseIIIb1smfixedcompare} with Fig.~\ref{fig:comparefig21}, found in appendix~\ref{appB2}. The left plot of Fig.~\ref{fig:comparefig21} displays the same information as the left plot of Fig.~\ref{CaseIIIb1smfixedcompare}, however, this time for light neutrino masses with NO. As one clearly sees, in this case the viable parameter space for leptogenesis is very similar (for VIC and TIC), if one marginalises over both ratios $\frac mn$ and $\frac sn$, according to Tab.~\ref{range of values parameters}, or if one uses as fixed values $\frac mn = \frac 12$ and $\frac sn = \frac{3}{20}$ (corresponding to $\phi_m=\frac{\pi}{2}$ and $\phi_s=\frac{3 \, \pi}{20}$, respectively). Similarly, we display in the right plot of Fig.~\ref{fig:comparefig21} the same as in the right plot of Fig.~\ref{CaseIIIb1smfixedcompare}, but for light neutrino masses with IO and observe only a mild enlargement of the allowed parameter space for leptogenesis for the lightest neutrino mass being $m_0=0.015$ eV instead of vanishing. This is in contrast to the sizeable increase of the parameter space found for light neutrino masses with NO.

Similar to Case 2), it is possible for Case 3 b.1) to obtain a non-vanishing value of the BAU for zero splittings, $\kappa$ and $\lambda$. In contrast to Case 2), this can also happen for strong NO, i.e.~vanishing lightest neutrino mass $m_0$.
In Fig.~\ref{CaseIIIb1MU2m0lightkappazero} (left plot), we display the resulting viable parameter space. 
Comparing to the corresponding figure for Case 2), see Fig.~\ref{CaseIIMU2m0heavy}, we note that the parameter space for Case 3 b.1) is slightly larger. This is also reflected in the slightly larger value of the maximum of the combination $U^2 \cdot M$, as found in Tab.~\ref{tab:maxU2forallplots}. Nevertheless, we can only expect this parameter space to be tested by FCC-ee/CEPC. Furthermore, we note that a minimum value of the Majorana mass $M$ around $4 \, \mathrm{GeV}$ is required for VIC and $M \gtrsim 20 \, \mathrm{GeV}$ for TIC. These values turn out to be slightly smaller than those obtained for Case 2).

In Fig.~\ref{CaseIIIb1MU2m0lightkappazero} (right plot) we display the combination $U^2 \cdot M$ with respect to the absolute value of the splitting $\kappa$ (the mass splitting $\Delta M= M_{R_1}-M_{R_{2,3}}$) for the Majorana mass $M$ being fixed to $M=1$ TeV and light neutrino masses with strong NO. The features of this plot are similar to those of the corresponding plot for Case 1), compare Fig.~\ref{fig:kappavsU2} (bottom plot), apart from two crucial differences. The first one regards the possibility to have non-zero BAU for arbitrarily small values of the splitting $\kappa$ which is in accordance with the left plot of Fig.~\ref{CaseIIIb1MU2m0lightkappazero}. The second one is instead related to the maximal size of $U^2 \cdot M$ that is about a factor of ten smaller for Case 3 b.1) and also reached for smaller values of $|\kappa|$ compared to Case 1), since the heavy neutrino decay width is naively proportional to $\kappa$ and to the active-sterile  mixing $U^2$ as well. Eventually, we remind that only one combination of the parameters $m$ and $s$, namely $m$ odd and $s$ even, allows for non-zero BAU for large splitting $\kappa$, compare~\cite{Drewes:2022kap}.

\begin{figure}[!t]
    \centering
    \includegraphics[width=.49\textwidth]{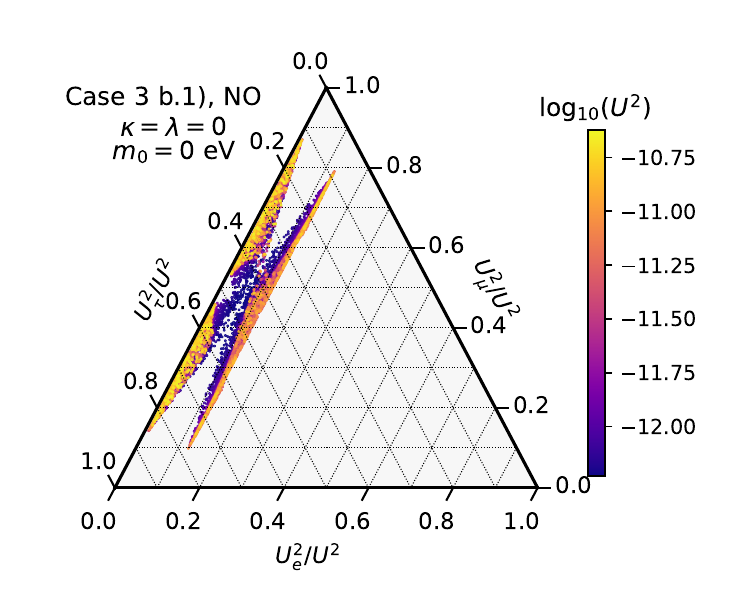}
    \caption{{\bf Case 3 b.1)} Parameter space consistent with leptogenesis highlighted in the ternary plot for vanishing splittings, $\kappa=0$ and $\lambda=0$, and light neutrino masses with strong NO. The different colours indicate the size of the active-sterile mixing $U^2$.}
    \label{fig:flavrationodM_leptogenesisimpact}
\end{figure}

We conclude the discussion of Case 3 b.1) with Fig.~\ref{fig:flavrationodM_leptogenesisimpact}. In this ternary plot we demonstrate the correlation among successful leptogenesis, the size of the active-sterile mixing $U^2$ and the branching ratios of the heavy neutrinos for light neutrino masses with strong NO and vanishing splittings $\kappa$ and $\lambda$. As one can see, the largest values of $U^2$, $U^2 \sim 10^{-11}$, can be reached for the smallest possible values of the flavour ratio $\frac{U_e^2}{U^2}$, consistent with the findings of \cite{Antusch:2017pkq}. The visible correlation is due to the CP phases relevant for leptogenesis and determining the ratios $\frac{U_\alpha^2}{U^2}$. The shown parameter space should be compared with the darker red region in the ternary plot in the left of Fig.~\ref{fig:Case3b1ternary}.

\paragraph{Impact of experiments searching for charged lepton flavour violation in \mathversion{bold}$\mu-e$\mathversion{normal} transitions} 

\begin{figure}[!t]
    \centering
    \includegraphics[width=0.49\textwidth]{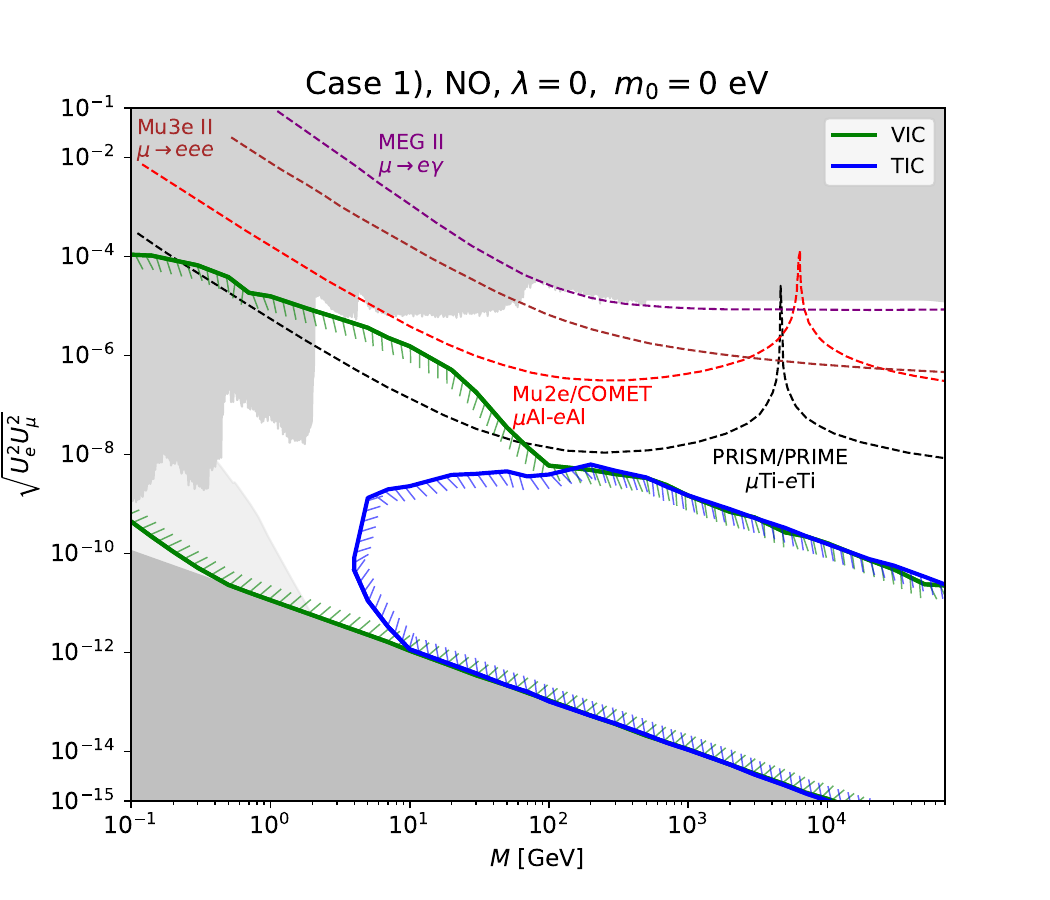}
    \includegraphics[width=0.49\textwidth]{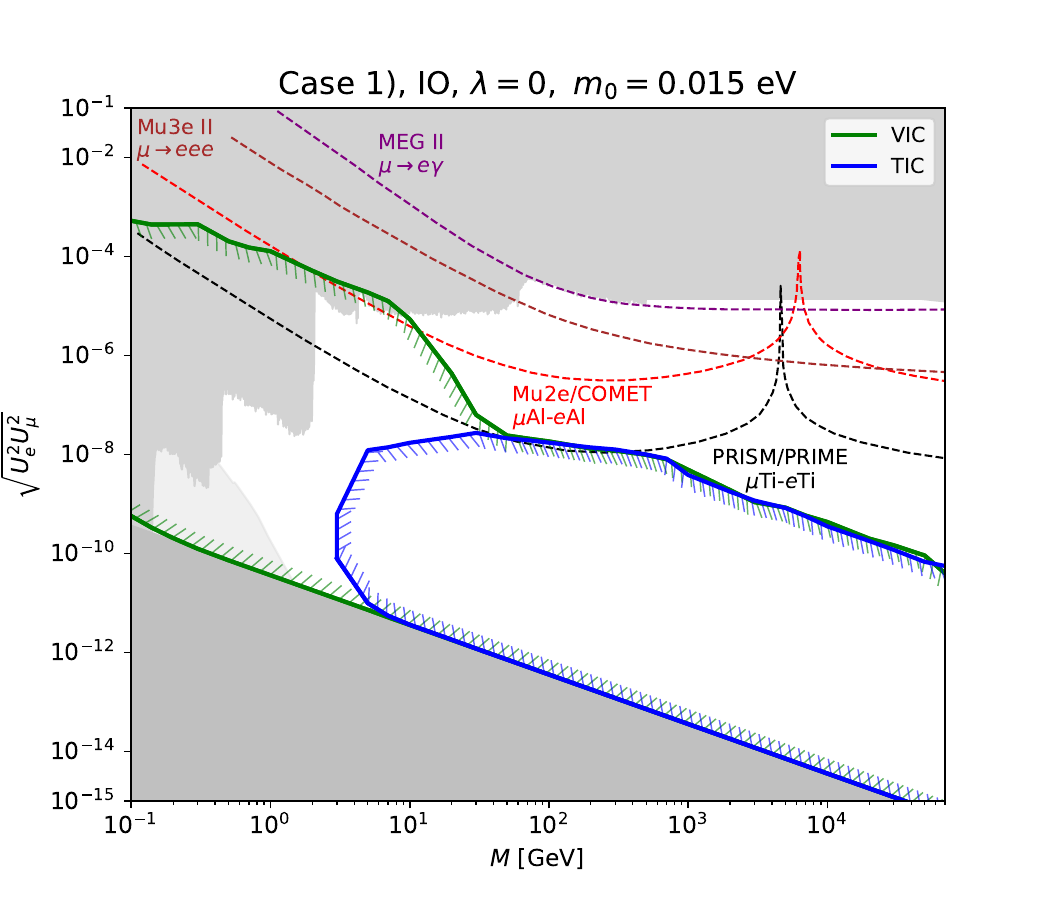}
    \caption{\label{CaseIMUeUmu}
     {\bf Constraints from searches for charged lepton flavour violation in \mathversion{bold}$\mu-e$\mathversion{normal} transitions} on the parameter space consistent with leptogenesis, shown in the $M-\sqrt{U_e^2 \, U_\mu^2}$-plane, for Case 1) (fully marginalised over the splitting $\kappa$ and the ratio $\frac sn$) and light neutrino masses with strong NO (left plot) and with IO and $m_0=0.015$ eV (right plot). Both initial conditions, VIC and TIC, are considered. For details regarding the different experimental sensitivities  see text.}
\end{figure}

Beyond direct searches mentioned previously, it is also interesting to consider the constraints arising from precision flavour experiments, searching for charged lepton flavour violation in $\mu-e$ transitions, in particular the processes $\mu \to e \gamma$, $\mu \to e e e$ and $\mu-e$ conversion in different nuclei, since these can strongly limit the combination $\sqrt{U_e^2 \, U_\mu^2}$. As regards the sensitivities of different (current and future) experiments, we use the following upper limits
\begin{subequations}
    \begin{align}
        \mathrm{BR}(\mu\rightarrow e\gamma) &< 6\cdot 10^{-14}~~~~~~~~~~~~~ (\mbox{MEG II}~\text{\cite{MEGII:2021fah}}),\\
        \mathrm{BR}(\mu\rightarrow eee) &<  10^{-16} ~~~~~~~~~~~~~~~~~(\mbox{Mu3e Phase II}~\text{\cite{Mu3e:2020gyw}}),\\
        \mathrm{CR}(\mu \mathrm{Al} - e \mathrm{Al}) &< 4.6\cdot 10^{-17} ~~~~~~~~~~~(\mbox{Mu2e/COMET} ~\text{\cite{Mu2e:2014fns,COMET:2018auw,KUNO:17072023}}),\\
        \mathrm{CR}(\mu \mathrm{Ti} - e \mathrm{Ti}) &< 10^{-19} ~~~~~~~~~~~~~~~~~(\mbox{PRISM/PRIME}~ \text{\cite{Barlow:2011zza,KUNO:17072023}}).
    \end{align}
\end{subequations}
The results for Case 1), fully marginalised over the splitting $\kappa$ and the ratio $\frac sn$, see Tab.~\ref{range of values parameters}, and for light neutrino masses with strong NO (left plot) as well as with IO and $m_0=0.015 \, \mathrm{eV}$
(right plot) can be found in Fig.~\ref{CaseIMUeUmu}, where the parameter space leading to the successful generation of the BAU is plotted in the $M-\sqrt{U_e^2 \, U_\mu^2}$-plane. The coloured dashed lines representing the different expected experimental sensitivities have been extracted from~\cite{Granelli:2022eru}.\footnote{See also~\cite{Urquia-Calderon:2022ufc} for a study of the impact of such experimental probes on the parameter space relevant for leptogenesis.} As one can see, only a future $\mu-e$ conversion experiment such as PRISM/PRIME~\cite{Barlow:2011zza} could  probe a small part of the available parameter space in the case of VIC, while the upcoming experiments
Mu2e~\cite{Mu2e:2014fns} and COMET~\cite{COMET:2018auw} can marginally constrain the allowed region in the $M-\sqrt{U_e^2 \, U_\mu^2}$-plane for light neutrino masses with IO and $m_0=0.015 \, \mathrm{eV}$. 
As expected, searches for charged lepton flavour violation involving the tau lepton cannot put any constraints on the discussed scenario. 
These experiments are, thus, of lesser importance
for testing the presented scenario, while being complementary to the direct searches and their reach, shown in the other figures in this section. 

We have checked that very similar conclusions hold 
 for a different light neutrino mass spectrum (with NO or IO and a distinct lightest neutrino mass $m_0$) as
well as for the other cases, Case 2) through Case 3 b.1). 

\vspace{0.2in}
\noindent We close this discussion with Tab.~\ref{tab:maxU2forallplots} which shows the maximum values of the combination $U^2 \cdot M$ for both types of initial conditions, VIC and TIC. We mention the values for each case, Case 1) through Case 3 b.1), for a vanishing lightest neutrino mass $m_0$
as well as, for comparison, for $m_0$ assuming its value maximally allowed by cosmology for Case 1), Case 2) and Case 3 b.1). We see that the largest maximal values of $U^2 \cdot M$, $U^2 \cdot M \lesssim 5 \cdot 10^5$ eV, are achieved for VIC and non-vanishing splitting $\kappa$ and, in particular, for Case 3 b.1) for non-zero $m_0$. 
Where possible, we also consider the situation in which both splittings, $\kappa$ and $\lambda$, are set to zero. Generally we observe that in these instances the maximum value of $U^2 \cdot M$ is several orders of magnitude smaller than for the other cases. This considerably reduces the prospects for testing them at current and future experiments, compare Figs.~\ref{CaseIIMU2m0heavy} and~\ref{CaseIIIb1MU2m0lightkappazero} (left plot).

\renewcommand{\arraystretch}{1.5}
\begin{table}[!t]
    \centering
    \begin{tabular}{|l|c|c|c|c|}
    \hline
         ~~~~~~~~~~~Case & \multicolumn{4}{c|}{ $\mbox{max}(U^2 \cdot M)$ [eV] } \\
          & \multicolumn{2}{c|}{VIC} & \multicolumn{2}{c|}{TIC}\\
          \hline
          ~~~~~~~~\underline{\mbox{$m_0 = 0$ eV}}~~~~~& NO & IO & NO & IO \\
         Case 1) & $1.7 \cdot 10^{5}$ & $2.6 \cdot 10^{5}$ & $1.5 \cdot 10^{4}$ & $8.3\cdot 10^{3}$\\
         Case 2) & $1.7 \cdot 10^{5}$ & $2.1 \cdot 10^{5}$ & $1.4 \cdot 10^{4}$ & $6.8\cdot 10^{3}$ \\
         Case 3 a) & $7.6 \cdot 10^{4}$ & $1.2\cdot 10^{4}$  & $4.4 \cdot 10^{3}$ & $5.6 \cdot 10^{1}$ \\
         Case 3 b.1) & $3.9 \cdot 10^{3}$ & $5.2 \cdot 10^{4}$ & $8.1\cdot 10^{2}$ & $6.8 \cdot 10^{3}$\\
         Case 3 b.1), $\kappa = 0$  & $6.1$& / & $2.5$ &/ \\
         \hhline{|=|=|=|=|=|}
         \underline{\mbox{$m_0 = 0.03 \, (0.015)$ eV}} & & & &\\
         Case 1) & $2.1 \cdot 10^{5}$ & $4.1 \cdot 10^{5}$ & $9.5 \cdot 10^{3}$ & $1.4\cdot 10^{4}$\\
          Case 2), $\kappa=0$ & $2.3$ & $2.2$ & $1.1$ & $1.4$ \\
          Case 3 b.1) & $2.8\cdot 10^{5}$& $3.5 \cdot 10^{5}$ & $8.8 \cdot 10^{3}$& $1.6 \cdot 10^{4}$\\
          \hline
          \end{tabular}
    \caption{\mathversion{bold}{\bf Summary table displaying the maximal value of the total mixing $U^2$ multiplied by the Majorana mass $M$}\mathversion{normal} for the different cases, Case 1) through Case 3 b.1), both initial conditions, VIC and TIC, and both light neutrino mass orderings, NO and IO. The lightest neutrino mass $m_0$ is always specified, with $m_0=0.03 \, (0.015)$ eV referring to light neutrino masses with NO (IO). In all considered instances, we have marginalised over the splitting $\kappa$, if not stated otherwise, and over all relevant parameters characterising the (residual) symmetries, e.g.~the ratio $\frac sn$ for Case 1), as detailed in Tab.~\ref{range of values parameters}. The splitting $\lambda$ is always set to zero. Situations in which the produced BAU always vanishes  are marked with the symbol /, see~\cite{Drewes:2022kap}.}
    \label{tab:maxU2forallplots}
\end{table}
\renewcommand{\arraystretch}{1}

\paragraph{Impact of the splitting \mathversion{bold}$\lambda$\mathversion{normal}}

\begin{figure}[!t]
    \centering
    \includegraphics[width=0.49\textwidth]{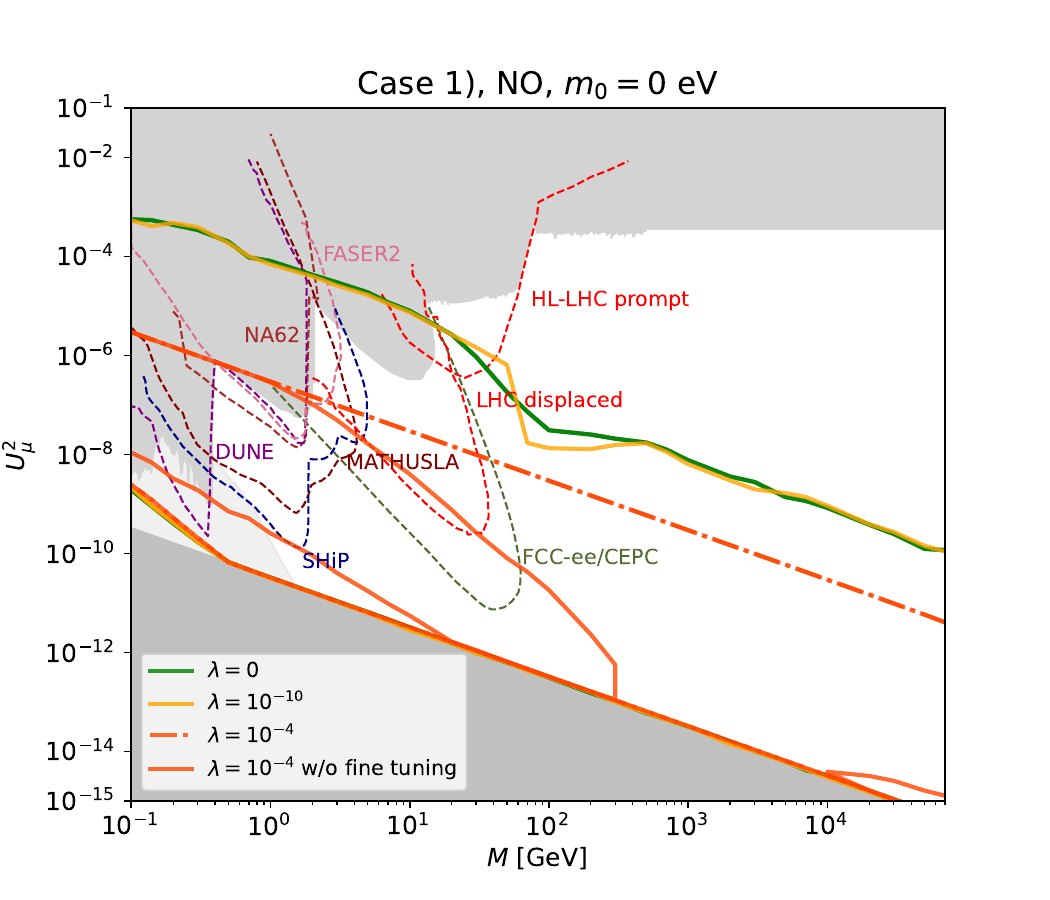}
    \caption{{\bf Impact of splitting \mathversion{bold}$\lambda$\mathversion{normal} on the parameter space available for leptogenesis}, shown in the $M-U_\mu^2$-plane, for Case 1) (fully marginalising over the splitting $\kappa$ and the ratio $\frac sn$), light neutrino masses with strong NO and VIC. We compare the parameter space 
    for $\lambda=0$ (green-bordered area), $\lambda=10^{-10}$ (yellow-bordered area) and $\lambda=10^{-4}$ (orange-bordered area). For the largest splitting $\lambda$, we distinguish whether $\lambda$ is fine-tuned, i.e.~$\lambda \approx 3 \, \kappa$,  
    (dot-dashed orange-bordered area) or not (orange-bordered area).
   For further details, see text.}
\label{fig:Case1lambdanonzero}
\end{figure}

As argued in section~\ref{sec2}, the splitting $\kappa$ can have its origin in the symmetry breaking of the flavour and CP symmetry to the residual group, present among charged leptons, while the splitting $\lambda$ is introduced ad hoc in order to generate a splitting among the masses of all three RH neutrinos. Nevertheless, it is interesting to study the impact of $\lambda$ on the size of the parameter space leading to viable leptogenesis. In the following, we exemplify its effect with the help of Case 1). As already remarked in~\cite{Drewes:2022kap}, larger values for the splitting $\lambda$ suppress the
generation of a sufficient amount of the BAU in a large part of the parameter space. This is clearly confirmed in Fig.~\ref{fig:Case1lambdanonzero}, where the viable parameter space for different benchmark values of $\lambda$ is plotted in the $M-U_\mu^2$-plane. Here, we fully marginalise over the splitting $\kappa$ and the ratio $\frac sn$, as quoted in Tab.~\ref{range of values parameters}, as well as assume light neutrino masses with strong NO and VIC.
 We observe that for $\lambda$ small enough, $\lambda=10^{-10}$ (yellow-bordered area), the reduction of the available parameter space is marginal compared to setting $\lambda$ to zero (green-bordered area), while for larger values, $\lambda=10^{-4}$ (orange-bordered area), the allowed parameter space considerably shrinks. In particular, for values of the RH neutrino mass larger than $300$ GeV we find that the successful generation of the BAU becomes unfeasible, unless the two splittings, $\kappa$ and $\lambda$, are fine-tuned, as indicated by the dash-dotted orange-bordered area. Such a fine-tuning requires the fulfilment of the condition $\lambda \approx 3 \, \kappa$, since in this case the RH neutrino masses $M_{R_1}$ and $M_{R_2}$ become degenerate, compare Eq.~(\ref{eq:Mlambda}). Concretely, the dash-dotted orange-bordered area is obtained, if the ratio $\frac{3 \, \kappa}{\lambda}$ is unconstrained, whereas the orange-bordered area requires that the absolute value of this ratio differs by more than $10\%$ in magnitude from 1, $|1-|\frac{3 \, \kappa}{\lambda}||>0.1$. Indeed, the parameter space resulting for $\lambda=10^{-4}$, while allowing for a possible fine-tuning of the splittings, only appears to be reduced compared to the one obtained for $\lambda=0$, since the condition in Eqs.~(\ref{eq:m3kappalambda}-\ref{eq:U2kappa}) is no longer fulfilled. As mentioned, this restriction, however, does not exclude the possibility of viable leptogenesis. In order to evidence better the  considerable impact of the fine-tuning, we show in Fig.~\ref{fig:Case1lambdanon-zeromoredetailed} in appendix~\ref{appB2} separate plots with the viable parameter space for $\lambda=10^{-4}$ with (left plot) and without fine-tuning (right plot).  

\begin{figure}[!t]
    \centering\includegraphics[width=.49\textwidth]{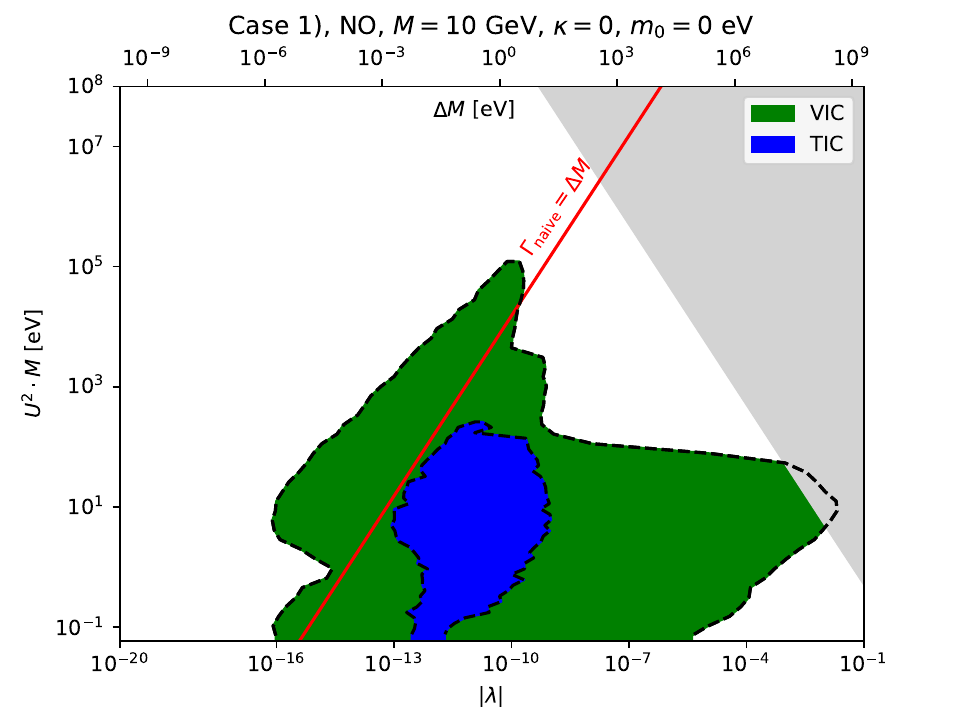}
    \includegraphics[width=.49\textwidth]{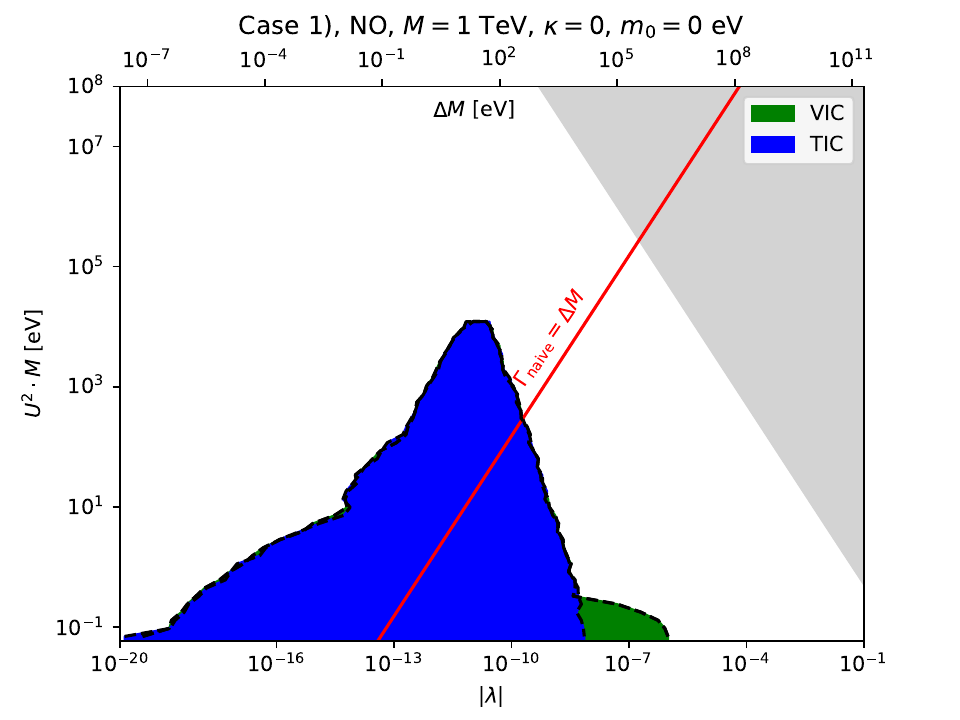}
    \caption{{\bf Impact of the splitting \mathversion{bold}$\lambda$\mathversion{normal} on the parameter space consistent with leptogenesis}, shown in the $|\lambda|(\Delta M)-U^2 \cdot M$-plane, for two different benchmark values of the Majorana mass $M$, $M=10$ GeV (left plot) and $M=1$ TeV (right plot). We take Case 1), assume light neutrino masses with strong NO, and set $\kappa=0$. Both initial conditions, VIC and TIC, are considered. The red line displays the resonance condition  that reads  $\Gamma_\mathrm{naive}= \Delta M= 2 \, \lambda \, M$.}
    \label{fig:LambdavsU21TeV}
\end{figure}

 The reduction of the viable parameter space for larger values of the splitting $\lambda$ is also underlined, when plotting the combination $U^2 \cdot M$ with respect to the absolute value of $\lambda$ (using as mass splitting $\Delta M=M_{R_2}-M_{R_3}= 2 \, \lambda \, M$) for two different benchmark values of the Majorana mass $M$, $M=10 \, \mathrm{GeV}$ (left plot) and $M=1 \, \mathrm{TeV}$ (right plot), in Fig.~\ref{fig:LambdavsU21TeV}. We take Case 1) as example, assume for light neutrino masses strong NO and set, for simplicity, $\kappa=0$. Both initial conditions, VIC and TIC, are considered. For small $M$, we see that a minimum value of $\lambda$, $|\lambda| \gtrsim 10^{-16}$,
is required for the successful generation of the BAU, while large values of $\lambda$ are only slightly compatible with achieving the correct amount of the BAU.\footnote{See also~\cite{Drewes:2012ma} for the observation that in the case of small RH neutrino masses even sizeable mass splittings can allow for the generation of a sufficient amount of the BAU.} In the case of TIC the available parameter space is considerably reduced and only values of $\lambda$ in the interval $10^{-13} \lesssim |\lambda| \lesssim 10^{-9}$ permit the production of a sufficient amount of BAU. Comparing this figure with the corresponding one in which the splitting $\kappa$ is varied and $\lambda=0$, see Fig.~\ref{fig:kappavsU2} (upper left plot), we clearly see the reduction of the available parameter space occurring for larger values of the splitting. Regarding the maximal values of $U^2 \cdot M$ that can be achieved, these are very similar for both splittings, $\lambda$ and $\kappa$, compare also Tab.~\ref{tab:maxU2forallplots}, and are obtained for a splitting with an absolute value of the order of $10^{-11}$ to $10^{-10}$. For larger $M$ instead the differences between the effect of the splitting $\lambda$ and $\kappa$ become more obvious, since in this case values of $\lambda$ larger than $|\lambda| \gtrsim 10^{-6}$ ($|\lambda| \gtrsim 10^{-8}$) for VIC (TIC) cannot lead to successful leptogenesis. Regarding small values of the splitting the results for both splittings $\lambda$, see Fig.~\ref{fig:LambdavsU21TeV} (right plot), and $\kappa$, see Fig.~\ref{fig:kappavsU2} (bottom plot), are very similar, including the possibility to have viable leptogenesis for very small values of the splitting. Also the maximal values that can be achieved for the combination $U^2 \cdot M$ are comparable. Similar to Fig.~\ref{fig:kappavsU2}, the red line highlights the (naive) resonance condition  that is in this case given by $\Gamma_\mathrm{naive} = 2 \, \lambda \, M$.

\section{Summary}
\label{summ}

We have scrutinised a scenario of low-scale type-I seesaw with three RH neutrinos, endowed with a flavour and a CP symmetry.
 These and their residuals determine the flavour structure of this scenario. In particular, light neutrino masses 
are fixed via three couplings, lepton mixing only depends on one (real) angle, the (quasi-)degenerate RH neutrino masses 
are given by the Majorana mass $M$ and potentially by splittings that originate from possible further symmetry breaking, while the active-sterile mixing and the generation of the 
BAU are also driven by the free angle $\theta_R$. This reduced number of
parameters allows to thoroughly study the phenomenology of this scenario for the different cases, Case 1) through
Case 3 b.1), which lead to distinct patterns of lepton mixing. 

In certain regions of parameter space, corresponding to special choices of the angle $\theta_R$ (that can be related to
an enhanced residual symmetry of the neutrino Yukawa coupling matrix $Y_D$ in some of the cases), the total mixing $U^2$
can be orders of magnitude larger than the naive seesaw limit such that this scenario becomes testable in laboratory experiments. For these regions, we have analysed both the heavy neutrino lifetimes and the branching ratios of their decays into the
different flavours $\alpha = e, \mu, \tau$. 

We have identified  three limiting situations, two large mass splittings ($|\lambda| \gg U^2$), one large mass splitting ($|\kappa| \gg U^2 \gg |\lambda|$) and small mass splittings ($U^2 \gg |\kappa|, |\lambda|$), that give rise to different ratios of the heavy neutrino mixing, $U_1^2:U_2^2:U_3^2$,
 and, consequently, to different lifetimes of the three heavy neutrinos. The resulting distributions of observed heavy neutrino events can 
potentially be distinguished from each other. At the same time,
they are distinct from the distribution arising from the situation in which the heavy neutrino mixing is identical for the three states, $U_1^2:U_2^2:U_3^2=1:1:1$.
 Furthermore, we have encountered occasions in which one of the heavy neutrinos becomes (nearly) decoupled, while the other two have sizeable mixing so that
they can be searched for at colliders. 

Given the potentially large event numbers at future lepton colliders, up to $10^5$ at FCC-ee/CEPC for $10^{12}$ produced $Z$ bosons, which
can allow for a relative accuracy of $1\%$ in the determination of the flavour ratios $\frac{U_\alpha^2}{U^2}$,
it might also be possible to differentiate among the cases, Case 1) through Case 3 b.1), with the help of the branching ratios of the 
heavy neutrinos to the different flavours $\alpha$. In particular, for Case 1) and to certain extent also for Case 2) the results for the flavour ratios $\frac{U_\alpha^2}{U^2}$
are (very) constrained (corresponding to a rather small region that is accessible in the ternary plot). Consequently, information on the light neutrino mass
spectrum, its ordering as well as the lightest neutrino mass $m_0$, could be obtained by measuring the branching ratios of the heavy neutrinos with
certain precision. Also for Case 3 a) and Case 3 b.1) peculiarities have been observed, e.g.~for the former $\frac{U_\mu^2}{U^2}$ is always
smaller than $0.45$, while for the latter $\frac{U_e^2}{U^2}$ turns out to be bounded from above by $0.35$.

Going beyond the study performed in~\cite{Drewes:2022kap}, we have explored as much as possible the parameter space in which the BAU can be 
successfully generated via leptogenesis, by marginalising over the splitting $\kappa$
(while setting $\lambda=0$) as well as over the parameters that characterise the flavour and CP symmetry and their residuals in the
different cases and that lead to a correct description of the measured lepton mixing angles. In doing so, we have considered both vanishing and thermal initial conditions. We emphasise that we have also presented results
 for Case 3 a), which has not been discussed in~\cite{Drewes:2022kap}. Indeed, these turn out to be qualitatively similar to those obtained for the other cases.
 We have demonstrated that for all cases, Case 1) through Case 3 b.1), a sizeable portion of the viable parameter space
 can be tested with various current and future accelerator-based experiments such as SHiP, MATHUSLA, (HL-)LHC and FCC-ee/CEPC, 
especially if vanishing initial conditions are assumed and as long as at least one of the splittings, $\kappa$ and $\lambda$, is non-zero. 
 Even if both these splittings vanish (for Case 2) and Case 3 b.1)), the viable parameter space might be explored to a certain degree at 
FCC-ee/CEPC. Interestingly enough, also precision flavour experiments searching for $\mu-e$ conversion in nuclei could test small parts of 
the allowed parameter space. Eventually, we have exemplified the different effects of the splittings $\kappa$ and $\lambda$ and have shown their
range compatible with successful leptogenesis, with and without fine-tuning among $\kappa$ and $\lambda$. 

We note that for light neutrinos with strong inverted ordering for Case 3 a) or strong normal ordering for Case 3 b.1), the considered scenario can effectively be reduced to the framework with only two heavy neutrinos -- both from the viewpoint of collider searches and of leptogenesis. 

Tab.~\ref{tab:summary} contains a summary of the parameters potentially accessible in measurements of the heavy neutrino lifetimes and their branching ratios as well as the constraints coming from the requirement of successful leptogenesis. For example, the value of the lightest neutrino mass $m_0$ could be extracted from the ratios $\frac{U_\alpha^2}{U^2}$ for Case 1) and Case 2), $t$ odd, while this is in general more challenging for the other cases. Nevertheless, it is, in principle, also possible for Case 3 a) and Case 3 b.1) to at least constrain a combination of parameters, e.g.~for Case 3 a) the lightest neutrino mass $m_0$ and the ratio $\frac sn$ (the allowed range for the ratio $\frac mn$ is already strongly limited by the measured value of the reactor mixing angle $\theta_{13}$).

Given the rich phenomenology ranging from collider searches and signals in precision flavour experiments to the generation 
of the BAU, it is worth to analyse variants of the presented scenario, in which e.g.~the RH neutrinos are
not (nearly) degenerate in mass and/or their number is varied. 

\renewcommand{\arraystretch}{1.5}
\begin{table}
\begin{center}
\begin{tabular}{|l|l|l|l|}
\hline 
Case & $\frac{U_i^2}{U^2}$ & $\frac{U_\alpha^2}{U^2}$ & Constraints due to leptogenesis\\
\hline
Case 1) & $3 \, \kappa - \lambda$ & $m_0$, NO/IO &
Figs.~\ref{fig:CaseIMU2}, \ref{fig:bothm0Case1}, \ref{fig:CaseIMU2alpha}, \ref{CaseIMUeUmu}, \ref{fig:Case1lambdanonzero} (\ref{fig:Case1lambdanon-zeromoredetailed})
\\ 
\hline
Case 2) & $3 \, \kappa - \lambda$ & $m_0$, NO/IO, $\frac un $ & 
Figs.~\ref{CaseIIMU2m0light}, \ref{CaseIIMU2m0heavy}
\\
\hline
Case 3 a) & $\lambda$, ($\kappa$) & $m_0$, $(\frac mn)$, $\frac sn$ &
Fig.~\ref{CaseIIIaMU2m0light}
\\
\hline
Case 3 b.1) & $\lambda$, ($\kappa$) & $m_0$, NO/IO, $\frac mn$, $\frac sn$ &
Figs.~\ref{CaseIIIb1MU2m0light}, \ref{CaseIIIb1smfixedcompare}, \ref{CaseIIIb1MU2m0lightkappazero}, \ref{fig:comparefig21}
\\
\hline
\end{tabular}
\end{center}
\caption{{\bf Summary} of parameters potentially accessible in measurements of heavy
neutrino lifetimes and branching ratios as well as references to plots showing constraints from successful leptogenesis for large active-sterile mixing $U^2$. The latter can be achieved for the angle $\theta_R$ being close to a special value, see section~\ref{sec32}, for Case 1), Case 2), $t$ odd as well as for Case 3 a) and Case 3 b.1) with $m$ even and $s$ odd or vice versa. Information on the closeness of $\theta_R$ to such a special value can be extracted from a measurement of $U^2$, compare~\cite{Drewes:2022kap}.  We mention the splitting $\kappa$ in parentheses for Case 3 a) and Case 3 b.1), since only for the combination $m$ odd and $s$ even it can also be accessed. With `NO/IO' we indicate the possibility to distinguish between the two light neutrino mass orderings. Furthermore, the ratio $\frac mn$ is already strongly constrained for Case 3 a), see section~\ref{sec31}, and thus also appears in parentheses.}
\label{tab:summary}
\end{table}

\section*{Acknowledgements}

Y.G. acknowledges the support of the French Community of Belgium through the FRIA grant No.~1.E.063.22F. C.H. is supported by the Spanish MINECO through the Ram\'o{}n y Cajal programme RYC2018-024529-I, by the national grant PID2023-148162NB-C21, by the Generalitat Valenciana through PROMETEO/2021/083, by the MCIU/AEI Severo-Ochoa-project CEX2023-001292-S as well as by the European Union's Horizon 2020 research and innovation programme under the Marie Sk\l{}odowska-Curie grant agreement No.~860881 (HIDDe$\nu$ network) and under the Marie Sk\l{}odowska-Curie Staff Exchange grant agreement No. 101086085 (ASYMMETRY).
Y.G. and C.H. thank the School of Physics of the University of New South Wales (UNSW) for its hospitality during part of this project. J.K. acknowledges the support of the Fonds de la Recherche Scientifique - FNRS under Grant No.~4.4512.10. Computational resources have been provided by the supercomputing facilities of the Université catholique de Louvain (CISM/UCL) and the Consortium des Équipements de Calcul Intensif en Fédération Wallonie Bruxelles (CÉCI) funded by the Fonds de la Recherche Scientifique de Belgique (F.R.S.-FNRS) under convention 2.5020.11 and by the Walloon Region.

\appendix

\section{\mathversion{bold}Matrices \texorpdfstring{$\Omega (3)$}{Omega3} and  \texorpdfstring{$\Omega ( 3^\prime)$}{Omega3p} and rotation matrices \texorpdfstring{$R_{ij} (\theta)$}{Rij}\mathversion{normal}}
\label{appA}

In this appendix, we collect the different forms of the matrices $\Omega ({\bf 3})$ and  $\Omega ({\bf 3^\prime})$ for the cases, Case 1) through Case 3 b.1).
Furthermore, we mention the form of the rotation matrices $R_{ij} (\theta_L)$ and $R_{kl} (\theta_R)$ for each combination.
More details can be found in~\cite{Drewes:2022kap}.

\noindent For Case 1) we can use as $\Omega ({\bf 3})$ and $\Omega ({\bf 3^\prime})$
\begin{eqnarray}
&&\Omega(s) ({\bf 3}) = e^{i \, \phi_s} \, U_{\mbox{\scriptsize{TB}}} \,
\left( \begin{array}{ccc}
1 & 0 & 0 \\
0 & e^{-3 \, i \, \phi_s} & 0\\
0 & 0 & -1
\end{array}
\right) \; , 
\\ 
&&
\Omega(s \, \mbox{even}) ({\bf 3^\prime}) =  U_{\mbox{\scriptsize{TB}}} 
\;\; \mbox{and} \;\;
\Omega(s \, \mbox{odd}) ({\bf 3^\prime}) =  U_{\mbox{\scriptsize{TB}}} \, \left(
\begin{array}{ccc}
i & 0 & 0\\
0 & 1 & 0\\
0 & 0 & i
\end{array}
\right)
\end{eqnarray}
with $U_{\mbox{\scriptsize{TB}}}$ describing tri-bimaximal (TB) mixing 
\begin{equation}
U_{\mbox{\scriptsize{TB}}} =
\left( \begin{array}{ccc}
\sqrt{2/3} & \sqrt{1/3} & 0\\
-\sqrt{1/6} & \sqrt{1/3} & \sqrt{1/2} \\
-\sqrt{1/6} & \sqrt{1/3} & -\sqrt{1/2}
\end{array}
\right)
\end{equation}
and $\phi_s$ being defined as $\phi_s=\frac{\pi \, s}{n}$. The rotation matrices $R_{ij} (\theta_L)$ and $R_{kl} (\theta_R)$ are
\begin{equation}
R_{13} (\theta_L) = \left(
\begin{array}{ccc}
\cos\theta_L & 0 & \sin\theta_L\\
0 & 1 & 0\\
-\sin\theta_L & 0 & \cos\theta_L
\end{array}
\right)
\;\; \mbox{and} \;\;
R_{13} (\theta_R) = \left(
\begin{array}{ccc}
\cos\theta_R & 0 & \sin\theta_R\\
0 & 1 & 0\\
-\sin\theta_R & 0 & \cos\theta_R
\end{array}
\right)\; .
\end{equation}
The matrices $\Omega ({\bf 3})$ and $\Omega ({\bf 3^\prime})$ in Case 2) read 
\begin{eqnarray}
&&\Omega (s,t) ({\bf 3}) = \Omega (u,v) ({\bf 3}) =  e^{i \phi_v/6} \, U_{\mbox{\scriptsize{TB}}} \, R_{13} \left( -\frac{\phi_u}{2} \right) \, \left( \begin{array}{ccc}
1 & 0 & 0\\
0 & e^{- i \phi_v/2} & 0\\
0 & 0 & -i
\end{array}
\right) \; ,
\\
&&\Omega (s \, \mbox{even},t \, \mbox{even}) ({\bf 3^\prime}) = U_{\mbox{\scriptsize{TB}}} \, \left( \begin{array}{ccc}
1 & 0 & 0\\
0 & 1 & 0\\
0 & 0 & i
\end{array}
\right) \; ,
\\
&&\Omega (s \, \mbox{even},t \, \mbox{odd}) ({\bf 3^\prime}) = e^{-i \pi/4} \, U_{\mbox{\scriptsize{TB}}} \, R_{13} \left( \frac{\pi}{4} \right) \, \left(
\begin{array}{ccc}
-i & 0 & 0\\
0 & e^{- i \pi/4} & 0\\
0 & 0 & 1
\end{array}
\right) \; ,
\\
&&\Omega (s \, \mbox{odd},t \, \mbox{even}) ({\bf 3^\prime}) = U_{\mbox{\scriptsize{TB}}} \, \left( \begin{array}{ccc}
i & 0 & 0\\
0 & 1 & 0\\
0 & 0 & 1
\end{array}
\right) \; ,
\\
&&\Omega (s \, \mbox{odd},t \, \mbox{odd}) ({\bf 3^\prime}) = e^{- 3 \, i \, \pi/4} \, U_{\mbox{\scriptsize{TB}}} \, R_{13} \left( \frac{\pi}{4} \right) \, \left(
\begin{array}{ccc}
-i & 0 & 0\\
0 & e^{i \, \pi/4} & 0\\
0 & 0 & 1
\end{array}
\right)
\end{eqnarray}
with $\phi_u$ and $\phi_v$ being $\phi_u=\frac{\pi \, u}{n}$ and $\phi_v=\frac{\pi \, v}{n}$, respectively. The rotation matrices $R_{ij} (\theta_L)$ and $R_{kl} (\theta_R)$
act both in the (13)-plane, i.e.~$R_{ij} (\theta_L)=R_{13} (\theta_L)$ and $R_{kl} (\theta_R)=R_{13} (\theta_R)$.

\noindent For Case 3 a) and Case 3 b.1), we find for the matrices $\Omega ({\bf 3})$ and $\Omega ({\bf 3^\prime})$
\begin{eqnarray}
&&\Omega (s, m) ({\bf 3}) =e^{i \, \phi_s} \,  \left( \begin{array}{ccc}
1 & 0 & 0\\
0 & \omega & 0 \\
0 & 0 & \omega^2
\end{array}
\right) \, U_{\mbox{\scriptsize{TB}}} \,
\left( \begin{array}{ccc}
1 & 0 & 0\\
0 & e^{-3 \, i \, \phi_s} & 0 \\
0 & 0 & -1
\end{array}
\right) \, R_{13} \left( \phi_m \right) \; ,
\\
&&\Omega (s \, \mbox{even}) ({\bf 3^\prime}) = \left(
\begin{array}{ccc}
1 & 0 & 0\\
0 & \omega & 0\\
0 & 0 & \omega^2
\end{array}
\right) \, U_{\mbox{\scriptsize{TB}}} \, \left(
\begin{array}{ccc}
1 & 0 & 0\\
0 & 1 & 0\\
0 & 0 & -1
\end{array}
\right) \; ,
\\
&&\Omega (s \, \mbox{odd}) ({\bf 3^\prime}) = \left(
\begin{array}{ccc}
1 & 0 & 0\\
0 & \omega & 0\\
0 & 0 & \omega^2
\end{array}
\right) \, U_{\mbox{\scriptsize{TB}}} \, \left(
\begin{array}{ccc}
i & 0 & 0\\
0  & -1 & 0\\
0 & 0 & -i
\end{array}
\right) 
\end{eqnarray}
with $\phi_s$ and $\phi_m$ being $\phi_s=\frac{\pi \, s}{n}$ and $\phi_m=\frac{\pi \, m}{n}$ as well as $\omega=e^{\frac{2 \pi i}{3}}$.
The rotation matrices $R_{ij} (\theta_L)$ and $R_{kl} (\theta_R)$ turn out to be
\begin{equation}
R_{12} (\theta_L) =  \left(
\begin{array}{ccc}
\cos\theta_L & \sin\theta_L & 0\\
-\sin\theta_L & \cos\theta_L & 0\\
0 & 0 & 1
\end{array}
\right) \; ,
\end{equation}
while the form of $R_{kl} (\theta_R)$ depends on whether $m$ is even or odd. For $m$ even it reads
\begin{equation}
R_{kl} (\theta_R)=R_{12} (\theta_R) \;\; \mbox{and} \;\; R_{kl} (\theta_R)=R_{23} (\theta_R)=\left(
\begin{array}{ccc}
1 & 0 & 0\\
0 &  \cos\theta_R & \sin\theta_R\\
0 & -\sin\theta_R & \cos\theta_R
\end{array}
\right)
\end{equation}
for $m$ odd. In the latter instance also the permutation matrix $P^{ij}_{kl}$ is needed and its form is
\begin{equation}
P^{12}_{23}=P_{13}= \left(
\begin{array}{ccc}
0 & 0 & 1\\
0 & 1 & 0\\
1 & 0 & 0
\end{array}
\right) \, .
\end{equation}

\section{Supplementary plots for leptogenesis and analytic formula}
\label{appB}

In this appendix, we display additional plots that complete the discussion of leptogenesis found in section~\ref{lepto}.
Furthermore, we mention some analytic formula for the CP-violating combinations that has  not been discussed in~\cite{Drewes:2022kap}.

\subsection{Supplementary plots for leptogenesis}
\label{appB2}

\begin{figure}[!t]
    \centering
      \includegraphics[width=.49\textwidth]{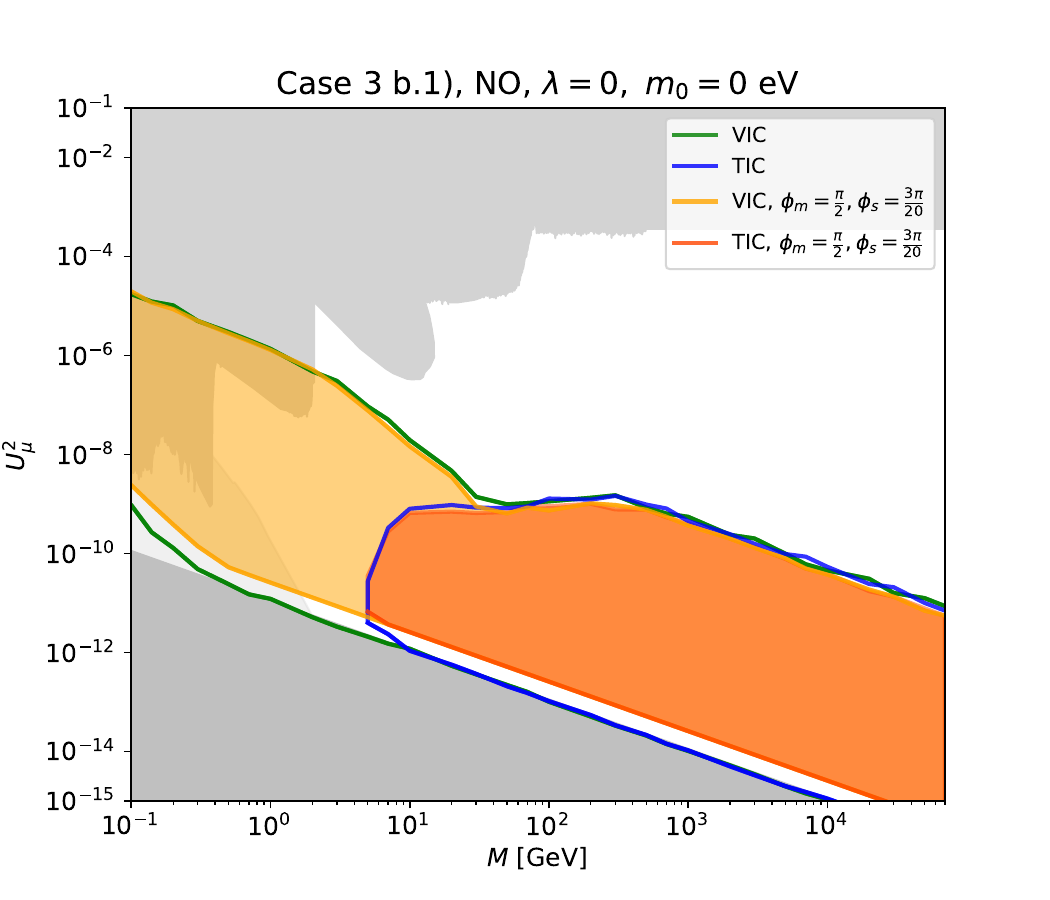}
    \includegraphics[width=.49\textwidth]{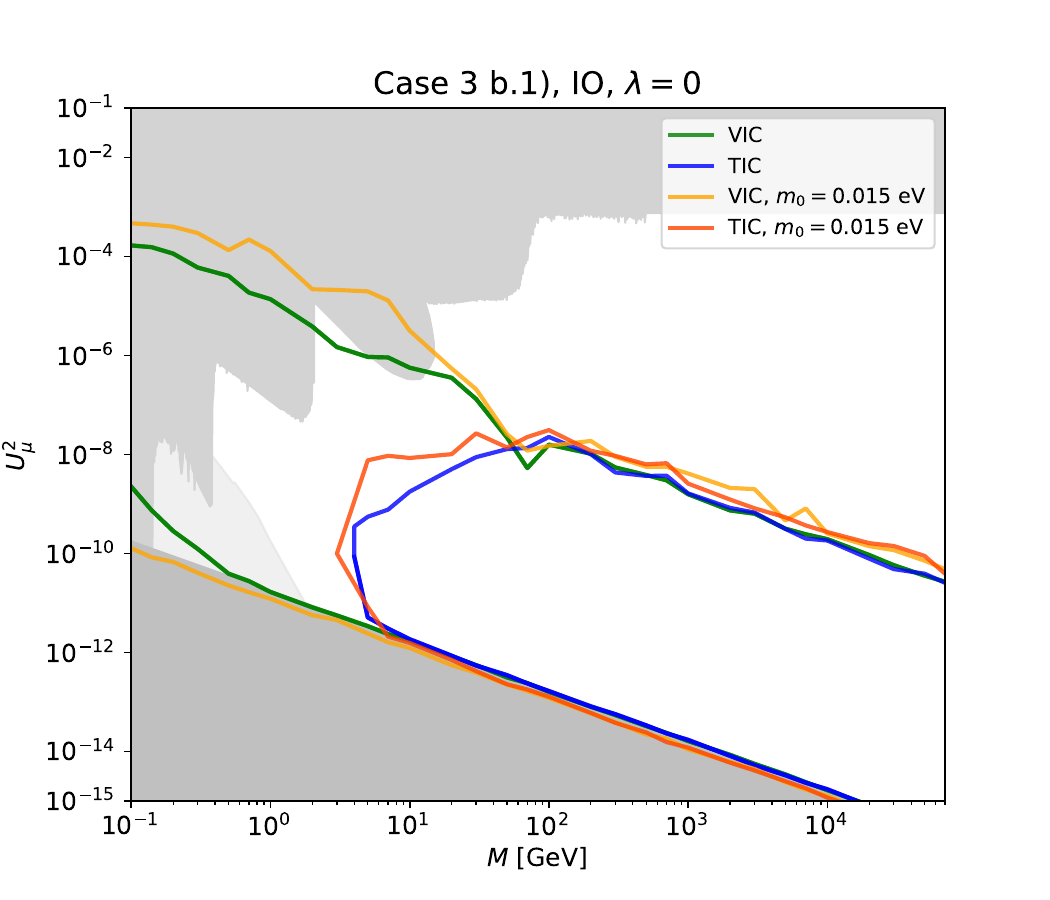}
    \caption{{\bf Case 3 b.1)} Plots similar to those given in Fig.~\ref{CaseIIIb1smfixedcompare} in the main text. Left plot: Comparison of the viable parameter space in the $M-U_\mu^2$-plane resulting from fully marginalising over the splitting $\kappa$ and the ratios $\frac mn$ and $\frac sn$ in the ranges found in Tab.~\ref{range of values parameters} and
for fixed values of the ratios $\frac mn$ and $\frac sn$ ($m=10$, $s=3$ and $n=20$), while still marginalising over $\kappa$. Here, light neutrino masses are constrained to follow strong NO. Right plot: Comparison of the viable parameter space in the $M-U_\mu^2$-plane for the lightest neutrino mass $m_0$ being zero and being $m_0=0.015$ eV, as maximally allowed by cosmology for light neutrino masses with IO. Note that the angular line shapes in the right plot are due to a reduced convergence, since the scan has not been optimised, unlike for the other results shown in this work.}
    \label{fig:comparefig21}
\end{figure}

\begin{figure}[!t]
    \centering
    \includegraphics[width=.49\textwidth]{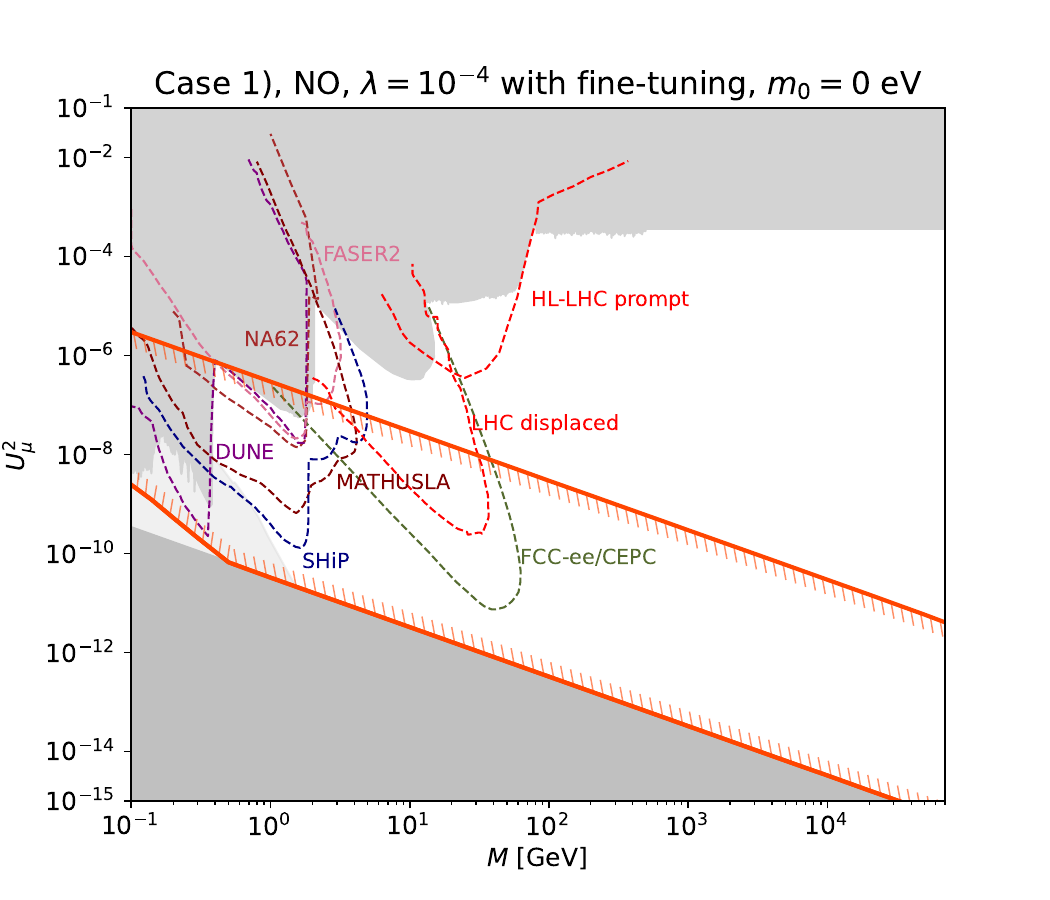}
    \includegraphics[width=.49\textwidth]{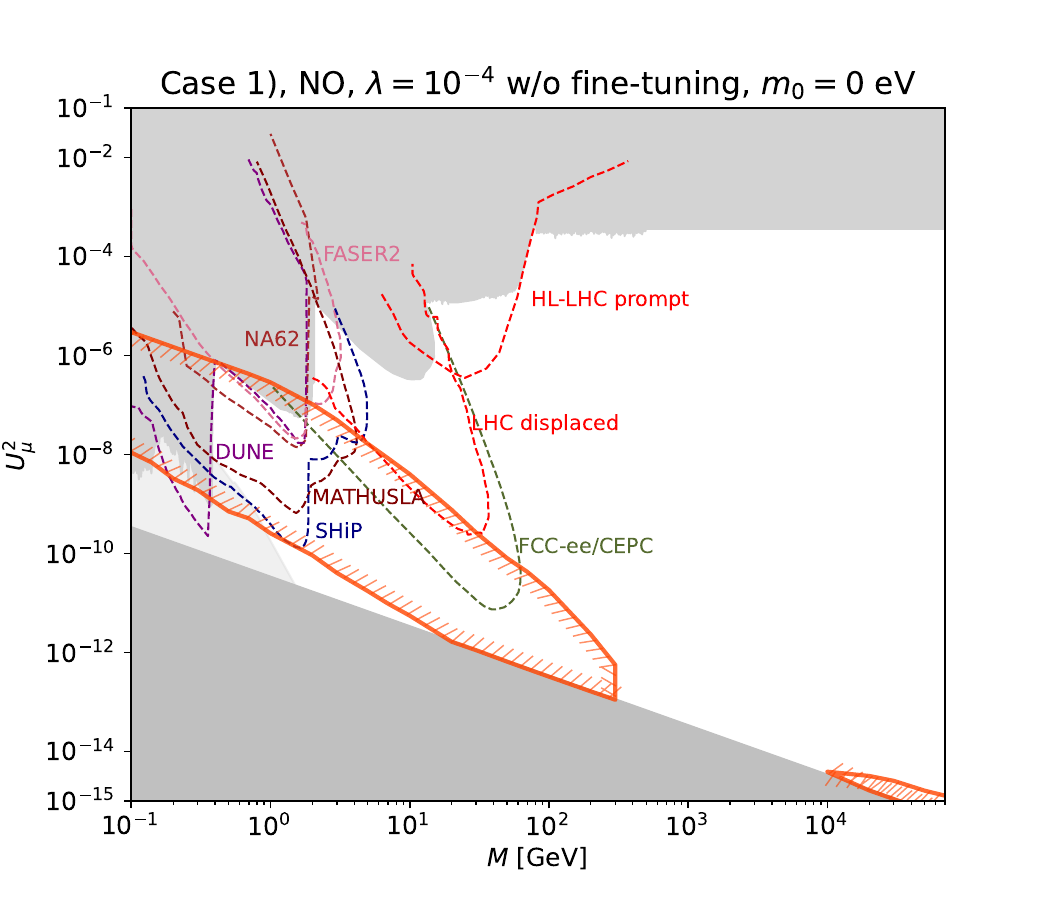}
        \caption{{\bf Impact of splitting \mathversion{bold}$\lambda$\mathversion{normal} on the parameter space available for leptogenesis}, shown in the $M-U_\mu^2$-plane, for Case 1) (fully marginalising over the splitting $\kappa$ and the ratio $\frac sn$), light neutrino masses with strong NO and VIC. Here, we display the results for $\lambda=10^{-4}$ with (left plot) and without fine-tuning (right plot), i.e.~$\lambda \approx 3 \, \kappa$ or not, separately, compare Fig.~\ref{fig:Case1lambdanonzero}.}
    \label{fig:Case1lambdanon-zeromoredetailed}
\end{figure}

We start by showing plots that complement the information given in section~\ref{lepto}. In Fig.~\ref{fig:comparefig21} we display the plots corresponding to the other choice of light neutrino mass ordering as chosen in the plots in Fig.~\ref{CaseIIIb1smfixedcompare}. Furthermore, we show in Fig.~\ref{fig:Case1lambdanon-zeromoredetailed} the allowed parameter space for leptogenesis for $\lambda=10^{-4}$ with (left plot) and without fine-tuning (right plot) separately in order to improve readability compared to Fig.~\ref{fig:Case1lambdanonzero}, given in the main text.

\subsection{Analytic formula}
\label{appB1}

For Case 2), $t$ odd, the CP-violating combination $C_{\mathrm{DEG},\alpha}$ weighted with the flavoured washout parameter $f_\alpha$ and summed over the lepton flavour $\alpha$
gives zero. A non-vanishing result is obtained at higher order, using $f_\alpha^2$ instead of $f_\alpha$ only,
\begin{equation}
\label{eq:sumalphaCDEGalphafalpha2Case1}
\sum_\alpha C_{\mathrm{DEG},\alpha} \, f_\alpha^2 =  \frac{1}{36} \, y_1 \, y_3 \, \Big( \frac{(\Delta y_{13}^2)^2}{ \Sigma y^2} \Big)^2 \, \cos^2 (2 \, \theta_L) \, \sin 4 \, \theta_R \, \sin 3 \, \phi_u \; ,
\end{equation}
with
\begin{equation}
\Delta y_{ij}^2 = y_i^2-y_j^2 \;\; \mbox{for} \;\; i < j \;\; \mbox{and} \;\;  \Sigma y^2=y_1^2+y_2^2+y_3^2 \; .
\end{equation}
The expressions for $C_{\mathrm{DEG},\alpha}$ and $f_\alpha$ can be found in~\cite{Drewes:2022kap}, see equations (139) and (146). We note that this result does not depend on whether $s$ is even or odd. We have numerically confirmed the dependence on $\theta_R$ and $\phi_u$, see e.g.~figure 17 (right plot) in~\cite{Drewes:2022kap} for an illustration of the dependence on $\theta_R$.

\bibliographystyle{JHEP}
\bibliography{main}

\providecommand{\href}[2]{#2}\begingroup\raggedright\begin{thebibliography}{100}

\bibitem{Minkowski:1977sc}
P.~Minkowski, \emph{{$\mu \to e\gamma$ at a Rate of One Out of $10^{9}$ Muon Decays?}}, \href{http://dx.doi.org/10.1016/0370-2693(77)90435-X}{\emph{Phys. Lett. B} {\bf 67} (1977) 421--428}.

\bibitem{Yanagida:1979as}
T.~Yanagida, \emph{{Horizontal gauge symmetry and masses of neutrinos}}, {\emph{Conf. Proc. C} {\bf 7902131} (1979) 95--99}.

\bibitem{Glashow:1979nm}
S.~L. Glashow, \emph{{The Future of Elementary Particle Physics}}, \href{http://dx.doi.org/10.1007/978-1-4684-7197-7_15}{\emph{NATO Sci. Ser. B} {\bf 61} (1980) 687}.

\bibitem{Gell-Mann:1979vob}
M.~Gell-Mann, P.~Ramond and R.~Slansky, \emph{{Complex Spinors and Unified Theories}}, {\emph{Conf. Proc. C} {\bf 790927} (1979) 315--321}, [\href{https://arxiv.org/abs/1306.4669}{{\tt 1306.4669}}].

\bibitem{Mohapatra:1979ia}
R.~N. Mohapatra and G.~Senjanovic, \emph{{Neutrino Mass and Spontaneous Parity Nonconservation}}, \href{http://dx.doi.org/10.1103/PhysRevLett.44.912}{\emph{Phys. Rev. Lett.} {\bf 44} (1980) 912}.

\bibitem{Shaposhnikov:2006nn}
M.~Shaposhnikov, \emph{{A Possible symmetry of the nuMSM}}, \href{http://dx.doi.org/10.1016/j.nuclphysb.2006.11.003}{\emph{Nucl. Phys. B} {\bf 763} (2007) 49--59}, [\href{https://arxiv.org/abs/hep-ph/0605047}{{\tt hep-ph/0605047}}].

\bibitem{Kersten:2007vk}
J.~Kersten and A.~Y. Smirnov, \emph{{Right-Handed Neutrinos at CERN LHC and the Mechanism of Neutrino Mass Generation}}, \href{http://dx.doi.org/10.1103/PhysRevD.76.073005}{\emph{Phys. Rev. D} {\bf 76} (2007) 073005}, [\href{https://arxiv.org/abs/0705.3221}{{\tt 0705.3221}}].

\bibitem{Moffat:2017feq}
K.~Moffat, S.~Pascoli and C.~Weiland, \emph{{Equivalence between massless neutrinos and lepton number conservation in fermionic singlet extensions of the Standard Model}},  \href{https://arxiv.org/abs/1712.07611}{{\tt 1712.07611}}.

\bibitem{Asaka:2005pn}
T.~Asaka and M.~Shaposhnikov, \emph{{The $\nu$MSM, dark matter and baryon asymmetry of the universe}}, \href{http://dx.doi.org/10.1016/j.physletb.2005.06.020}{\emph{Phys. Lett. B} {\bf 620} (2005) 17--26}, [\href{https://arxiv.org/abs/hep-ph/0505013}{{\tt hep-ph/0505013}}].

\bibitem{Asaka:2005an}
T.~Asaka, S.~Blanchet and M.~Shaposhnikov, \emph{{The nuMSM, dark matter and neutrino masses}}, \href{http://dx.doi.org/10.1016/j.physletb.2005.09.070}{\emph{Phys. Lett. B} {\bf 631} (2005) 151--156}, [\href{https://arxiv.org/abs/hep-ph/0503065}{{\tt hep-ph/0503065}}].

\bibitem{Canetti:2012kh}
L.~Canetti, M.~Drewes, T.~Frossard and M.~Shaposhnikov, \emph{{Dark Matter, Baryogenesis and Neutrino Oscillations from Right Handed Neutrinos}}, \href{http://dx.doi.org/10.1103/PhysRevD.87.093006}{\emph{Phys. Rev. D} {\bf 87} (2013) 093006}, [\href{https://arxiv.org/abs/1208.4607}{{\tt 1208.4607}}].

\bibitem{Ghiglieri:2020ulj}
J.~Ghiglieri and M.~Laine, \emph{{Sterile neutrino dark matter via coinciding resonances}}, \href{http://dx.doi.org/10.1088/1475-7516/2020/07/012}{\emph{JCAP} {\bf 07} (2020) 012}, [\href{https://arxiv.org/abs/2004.10766}{{\tt 2004.10766}}].

\bibitem{Abada:2018oly}
A.~Abada, G.~Arcadi, V.~Domcke, M.~Drewes, J.~Klaric and M.~Lucente, \emph{{Low-scale leptogenesis with three heavy neutrinos}}, \href{http://dx.doi.org/10.1007/JHEP01(2019)164}{\emph{JHEP} {\bf 01} (2019) 164}, [\href{https://arxiv.org/abs/1810.12463}{{\tt 1810.12463}}].

\bibitem{Drewes:2021nqr}
M.~Drewes, Y.~Georis and J.~Klari\'c, \emph{{Mapping the Viable Parameter Space for Testable Leptogenesis}}, \href{http://dx.doi.org/10.1103/PhysRevLett.128.051801}{\emph{Phys. Rev. Lett.} {\bf 128} (2022) 051801}, [\href{https://arxiv.org/abs/2106.16226}{{\tt 2106.16226}}].

\bibitem{daSilva:2022mrx}
P.~C. da~Silva, D.~Karamitros, T.~McKelvey and A.~Pilaftsis, \emph{{Tri-resonant leptogenesis in a seesaw extension of the Standard Model}}, \href{http://dx.doi.org/10.1007/JHEP11(2022)065}{\emph{JHEP} {\bf 11} (2022) 065}, [\href{https://arxiv.org/abs/2206.08352}{{\tt 2206.08352}}].

\bibitem{Kang:2022psa}
D.~W. Kang, J.~Kim, T.~Nomura and H.~Okada, \emph{{Natural mass hierarchy among three heavy Majorana neutrinos for resonant leptogenesis under modular A$_{4}$ symmetry}}, \href{http://dx.doi.org/10.1007/JHEP07(2022)050}{\emph{JHEP} {\bf 07} (2022) 050}, [\href{https://arxiv.org/abs/2205.08269}{{\tt 2205.08269}}].

\bibitem{Zhao:2024asa}
Z.-h. Zhao and Z.-C. Liu, \emph{{Tri-resonant leptogenesis from modular symmetry neutrino models}},  \href{https://arxiv.org/abs/2405.09363}{{\tt 2405.09363}}.

\bibitem{Drewes:2024bla}
M.~Drewes, Y.~Georis, J.~Klari\'c and A.~Wendels, \emph{{On the collider-testability of the type-I seesaw model with 3 right-handed neutrinos}},  \href{https://arxiv.org/abs/2407.13620}{{\tt 2407.13620}}.

\bibitem{Ishimori:2010au}
H.~Ishimori, T.~Kobayashi, H.~Ohki, Y.~Shimizu, H.~Okada and M.~Tanimoto, \emph{{Non-Abelian Discrete Symmetries in Particle Physics}}, \href{http://dx.doi.org/10.1143/PTPS.183.1}{\emph{Prog. Theor. Phys. Suppl.} {\bf 183} (2010) 1--163}, [\href{https://arxiv.org/abs/1003.3552}{{\tt 1003.3552}}].

\bibitem{King:2013eh}
S.~F. King and C.~Luhn, \emph{{Neutrino Mass and Mixing with Discrete Symmetry}}, \href{http://dx.doi.org/10.1088/0034-4885/76/5/056201}{\emph{Rept. Prog. Phys.} {\bf 76} (2013) 056201}, [\href{https://arxiv.org/abs/1301.1340}{{\tt 1301.1340}}].

\bibitem{Feruglio:2019ybq}
F.~Feruglio and A.~Romanino, \emph{{Lepton flavor symmetries}}, \href{http://dx.doi.org/10.1103/RevModPhys.93.015007}{\emph{Rev. Mod. Phys.} {\bf 93} (2021) 015007}, [\href{https://arxiv.org/abs/1912.06028}{{\tt 1912.06028}}].

\bibitem{Grimus:2011fk}
W.~Grimus and P.~O. Ludl, \emph{{Finite flavour groups of fermions}}, \href{http://dx.doi.org/10.1088/1751-8113/45/23/233001}{\emph{J. Phys. A} {\bf 45} (2012) 233001}, [\href{https://arxiv.org/abs/1110.6376}{{\tt 1110.6376}}].

\bibitem{Feruglio:2012cw}
F.~Feruglio, C.~Hagedorn and R.~Ziegler, \emph{{Lepton Mixing Parameters from Discrete and CP Symmetries}}, \href{http://dx.doi.org/10.1007/JHEP07(2013)027}{\emph{JHEP} {\bf 07} (2013) 027}, [\href{https://arxiv.org/abs/1211.5560}{{\tt 1211.5560}}].

\bibitem{Holthausen:2012dk}
M.~Holthausen, M.~Lindner and M.~A. Schmidt, \emph{{CP and Discrete Flavour Symmetries}}, \href{http://dx.doi.org/10.1007/JHEP04(2013)122}{\emph{JHEP} {\bf 04} (2013) 122}, [\href{https://arxiv.org/abs/1211.6953}{{\tt 1211.6953}}].

\bibitem{Chen:2014tpa}
M.-C. Chen, M.~Fallbacher, K.~T. Mahanthappa, M.~Ratz and A.~Trautner, \emph{{CP Violation from Finite Groups}}, \href{http://dx.doi.org/10.1016/j.nuclphysb.2014.03.023}{\emph{Nucl. Phys. B} {\bf 883} (2014) 267--305}, [\href{https://arxiv.org/abs/1402.0507}{{\tt 1402.0507}}].

\bibitem{Grimus:1995zi}
W.~Grimus and M.~N. Rebelo, \emph{{Automorphisms in gauge theories and the definition of CP and P}}, \href{http://dx.doi.org/10.1016/S0370-1573(96)00030-0}{\emph{Phys. Rept.} {\bf 281} (1997) 239--308}, [\href{https://arxiv.org/abs/hep-ph/9506272}{{\tt hep-ph/9506272}}].

\bibitem{Ecker:1983hz}
G.~Ecker, W.~Grimus and H.~Neufeld, \emph{{Spontaneous {CP} Violation in Left-right Symmetric Gauge Theories}}, \href{http://dx.doi.org/10.1016/0550-3213(84)90373-0}{\emph{Nucl. Phys. B} {\bf 247} (1984) 70--82}.

\bibitem{Ecker:1987qp}
G.~Ecker, W.~Grimus and H.~Neufeld, \emph{{A Standard Form for Generalized {CP} Transformations}}, \href{http://dx.doi.org/10.1088/0305-4470/20/12/010}{\emph{J. Phys. A} {\bf 20} (1987) L807}.

\bibitem{Neufeld:1987wa}
H.~Neufeld, W.~Grimus and G.~Ecker, \emph{{Generalized {CP} Invariance, Neutral Flavor Conservation and the Structure of the Mixing Matrix}}, \href{http://dx.doi.org/10.1142/S0217751X88000254}{\emph{Int. J. Mod. Phys. A} {\bf 3} (1988) 603--616}.

\bibitem{Harrison:2002kp}
P.~F. Harrison and W.~G. Scott, \emph{{Symmetries and generalizations of tri-bimaximal neutrino mixing}}, \href{http://dx.doi.org/10.1016/S0370-2693(02)01753-7}{\emph{Phys. Lett. B} {\bf 535} (2002) 163--169}, [\href{https://arxiv.org/abs/hep-ph/0203209}{{\tt hep-ph/0203209}}].

\bibitem{Grimus:2003yn}
W.~Grimus and L.~Lavoura, \emph{{A Nonstandard CP transformation leading to maximal atmospheric neutrino mixing}}, \href{http://dx.doi.org/10.1016/j.physletb.2003.10.075}{\emph{Phys. Lett. B} {\bf 579} (2004) 113--122}, [\href{https://arxiv.org/abs/hep-ph/0305309}{{\tt hep-ph/0305309}}].

\bibitem{Luhn:2007uq}
C.~Luhn, S.~Nasri and P.~Ramond, \emph{{The Flavor group $\Delta(3\, n^2)$}}, \href{http://dx.doi.org/10.1063/1.2734865}{\emph{J. Math. Phys.} {\bf 48} (2007) 073501}, [\href{https://arxiv.org/abs/hep-th/0701188}{{\tt hep-th/0701188}}].

\bibitem{Escobar:2008vc}
J.~A. Escobar and C.~Luhn, \emph{{The Flavor Group $\Delta(6 \, n^2)$}}, \href{http://dx.doi.org/10.1063/1.3046563}{\emph{J. Math. Phys.} {\bf 50} (2009) 013524}, [\href{https://arxiv.org/abs/0809.0639}{{\tt 0809.0639}}].

\bibitem{Ding:2013hpa}
G.-J. Ding, S.~F. King, C.~Luhn and A.~J. Stuart, \emph{{Spontaneous CP violation from vacuum alignment in $S_4$ models of leptons}}, \href{http://dx.doi.org/10.1007/JHEP05(2013)084}{\emph{JHEP} {\bf 05} (2013) 084}, [\href{https://arxiv.org/abs/1303.6180}{{\tt 1303.6180}}].

\bibitem{Feruglio:2013hia}
F.~Feruglio, C.~Hagedorn and R.~Ziegler, \emph{{A realistic pattern of lepton mixing and masses from $S_4$ and CP}}, \href{http://dx.doi.org/10.1140/epjc/s10052-014-2753-2}{\emph{Eur. Phys. J. C} {\bf 74} (2014) 2753}, [\href{https://arxiv.org/abs/1303.7178}{{\tt 1303.7178}}].

\bibitem{King:2014rwa}
S.~F. King and T.~Neder, \emph{{Lepton mixing predictions including Majorana phases from $\Delta (6\, n^2)$ flavour symmetry and generalised CP}}, \href{http://dx.doi.org/10.1016/j.physletb.2014.07.043}{\emph{Phys. Lett. B} {\bf 736} (2014) 308--316}, [\href{https://arxiv.org/abs/1403.1758}{{\tt 1403.1758}}].

\bibitem{Ding:2014ora}
G.-J. Ding, S.~F. King and T.~Neder, \emph{{Generalised CP and $\Delta(6\, n^2)$ family symmetry in semi-direct models of leptons}}, \href{http://dx.doi.org/10.1007/JHEP12(2014)007}{\emph{JHEP} {\bf 12} (2014) 007}, [\href{https://arxiv.org/abs/1409.8005}{{\tt 1409.8005}}].

\bibitem{Ding:2015rwa}
G.-J. Ding and S.~F. King, \emph{{Generalized CP and $\Delta (3n^2)$ Family Symmetry for Semi-Direct Predictions of the PMNS Matrix}}, \href{http://dx.doi.org/10.1103/PhysRevD.93.025013}{\emph{Phys. Rev. D} {\bf 93} (2016) 025013}, [\href{https://arxiv.org/abs/1510.03188}{{\tt 1510.03188}}].

\bibitem{Drewes:2022kap}
M.~Drewes, Y.~Georis, C.~Hagedorn and J.~Klari\'c, \emph{{Low-scale leptogenesis with flavour and CP symmetries}}, \href{http://dx.doi.org/10.1007/JHEP12(2022)044}{\emph{JHEP} {\bf 12} (2022) 044}, [\href{https://arxiv.org/abs/2203.08538}{{\tt 2203.08538}}].

\bibitem{Curtin:2018mvb}
D.~Curtin et~al., \emph{{Long-Lived Particles at the Energy Frontier: The MATHUSLA Physics Case}}, \href{http://dx.doi.org/10.1088/1361-6633/ab28d6}{\emph{Rept. Prog. Phys.} {\bf 82} (2019) 116201}, [\href{https://arxiv.org/abs/1806.07396}{{\tt 1806.07396}}].

\bibitem{Chauhan:2021xus}
G.~Chauhan and P.~S.~B. Dev, \emph{{Interplay between resonant leptogenesis, neutrinoless double beta decay and collider signals in a model with flavor and CP symmetries}}, \href{http://dx.doi.org/10.1016/j.nuclphysb.2022.116058}{\emph{Nucl. Phys. B} {\bf 986} (2023) 116058}, [\href{https://arxiv.org/abs/2112.09710}{{\tt 2112.09710}}].

\bibitem{Shaposhnikov:2008pf}
M.~Shaposhnikov, \emph{{The nuMSM, leptonic asymmetries, and properties of singlet fermions}}, \href{http://dx.doi.org/10.1088/1126-6708/2008/08/008}{\emph{JHEP} {\bf 08} (2008) 008}, [\href{https://arxiv.org/abs/0804.4542}{{\tt 0804.4542}}].

\bibitem{Drewes:2019byd}
M.~Drewes, J.~Klari\'c and P.~Klose, \emph{{On lepton number violation in heavy neutrino decays at colliders}}, \href{http://dx.doi.org/10.1007/JHEP11(2019)032}{\emph{JHEP} {\bf 11} (2019) 032}, [\href{https://arxiv.org/abs/1907.13034}{{\tt 1907.13034}}].

\bibitem{Hagedorn:2016lva}
C.~Hagedorn and E.~Molinaro, \emph{{Flavor and CP symmetries for leptogenesis and $0\nu\beta\beta$ decay}}, \href{http://dx.doi.org/10.1016/j.nuclphysb.2017.03.015}{\emph{Nucl. Phys. B} {\bf 919} (2017) 404--469}, [\href{https://arxiv.org/abs/1602.04206}{{\tt 1602.04206}}].

\bibitem{Hagedorn:2014wha}
C.~Hagedorn, A.~Meroni and E.~Molinaro, \emph{{Lepton mixing from $\Delta (3 \, n^2)$ and $\Delta (6 \, n^2)$ and CP}}, \href{http://dx.doi.org/10.1016/j.nuclphysb.2014.12.013}{\emph{Nucl. Phys. B} {\bf 891} (2015) 499--557}, [\href{https://arxiv.org/abs/1408.7118}{{\tt 1408.7118}}].

\bibitem{Esteban:2020cvm}
I.~Esteban, M.~C. Gonzalez-Garcia, M.~Maltoni, T.~Schwetz and A.~Zhou, \emph{{The fate of hints: updated global analysis of three-flavor neutrino oscillations}}, \href{http://dx.doi.org/10.1007/JHEP09(2020)178}{\emph{JHEP} {\bf 09} (2020) 178}, [\href{https://arxiv.org/abs/2007.14792}{{\tt 2007.14792}}].

\bibitem{Esteban:2024eli}
I.~Esteban, M.~C. Gonzalez-Garcia, M.~Maltoni, I.~Martinez-Soler, J.~a.~P. Pinheiro and T.~Schwetz, \emph{{NuFit-6.0: Updated global analysis of three-flavor neutrino oscillations}},  \href{https://arxiv.org/abs/2410.05380}{{\tt 2410.05380}}.

\bibitem{Atre:2009rg}
A.~Atre, T.~Han, S.~Pascoli and B.~Zhang, \emph{{The Search for Heavy Majorana Neutrinos}}, \href{http://dx.doi.org/10.1088/1126-6708/2009/05/030}{\emph{JHEP} {\bf 05} (2009) 030}, [\href{https://arxiv.org/abs/0901.3589}{{\tt 0901.3589}}].

\bibitem{Boyarsky:2009ix}
A.~Boyarsky, O.~Ruchayskiy and M.~Shaposhnikov, \emph{{The Role of sterile neutrinos in cosmology and astrophysics}}, \href{http://dx.doi.org/10.1146/annurev.nucl.010909.083654}{\emph{Ann. Rev. Nucl. Part. Sci.} {\bf 59} (2009) 191--214}, [\href{https://arxiv.org/abs/0901.0011}{{\tt 0901.0011}}].

\bibitem{Drewes:2013gca}
M.~Drewes, \emph{{The Phenomenology of Right Handed Neutrinos}}, \href{http://dx.doi.org/10.1142/S0218301313300191}{\emph{Int. J. Mod. Phys. E} {\bf 22} (2013) 1330019}, [\href{https://arxiv.org/abs/1303.6912}{{\tt 1303.6912}}].

\bibitem{Deppisch:2015qwa}
F.~F. Deppisch, P.~S. Bhupal~Dev and A.~Pilaftsis, \emph{{Neutrinos and Collider Physics}}, \href{http://dx.doi.org/10.1088/1367-2630/17/7/075019}{\emph{New J. Phys.} {\bf 17} (2015) 075019}, [\href{https://arxiv.org/abs/1502.06541}{{\tt 1502.06541}}].

\bibitem{Antusch:2016ejd}
S.~Antusch, E.~Cazzato and O.~Fischer, \emph{{Sterile neutrino searches at future $e^-e^+$, $pp$, and $e^-p$ colliders}}, \href{http://dx.doi.org/10.1142/S0217751X17500786}{\emph{Int. J. Mod. Phys. A} {\bf 32} (2017) 1750078}, [\href{https://arxiv.org/abs/1612.02728}{{\tt 1612.02728}}].

\bibitem{Cai:2017mow}
Y.~Cai, T.~Han, T.~Li and R.~Ruiz, \emph{{Lepton Number Violation: Seesaw Models and Their Collider Tests}}, \href{http://dx.doi.org/10.3389/fphy.2018.00040}{\emph{Front. in Phys.} {\bf 6} (2018) 40}, [\href{https://arxiv.org/abs/1711.02180}{{\tt 1711.02180}}].

\bibitem{Agrawal:2021dbo}
P.~Agrawal et~al., \emph{{Feebly-interacting particles: FIPs 2020 workshop report}}, \href{http://dx.doi.org/10.1140/epjc/s10052-021-09703-7}{\emph{Eur. Phys. J. C} {\bf 81} (2021) 1015}, [\href{https://arxiv.org/abs/2102.12143}{{\tt 2102.12143}}].

\bibitem{Abdullahi:2022jlv}
A.~M. Abdullahi et~al., \emph{{The present and future status of heavy neutral leptons}}, \href{http://dx.doi.org/10.1088/1361-6471/ac98f9}{\emph{J. Phys. G} {\bf 50} (2023) 020501}, [\href{https://arxiv.org/abs/2203.08039}{{\tt 2203.08039}}].

\bibitem{delAguila:2008ir}
F.~del Aguila, S.~Bar-Shalom, A.~Soni and J.~Wudka, \emph{{Heavy Majorana Neutrinos in the Effective Lagrangian Description: Application to Hadron Colliders}}, \href{http://dx.doi.org/10.1016/j.physletb.2008.11.031}{\emph{Phys. Lett. B} {\bf 670} (2009) 399--402}, [\href{https://arxiv.org/abs/0806.0876}{{\tt 0806.0876}}].

\bibitem{Bezrukov:2012sa}
F.~Bezrukov, M.~Y. Kalmykov, B.~A. Kniehl and M.~Shaposhnikov, \emph{{Higgs Boson Mass and New Physics}}, \href{http://dx.doi.org/10.1007/JHEP10(2012)140}{\emph{JHEP} {\bf 10} (2012) 140}, [\href{https://arxiv.org/abs/1205.2893}{{\tt 1205.2893}}].

\bibitem{Shaposhnikov:2007nj}
M.~Shaposhnikov, \emph{{Is there a new physics between electroweak and Planck scales?}},  in \emph{{Astroparticle Physics: Current Issues, 2007 (APCI07)}}, 8, 2007.
\newblock \href{https://arxiv.org/abs/0708.3550}{{\tt 0708.3550}}.

\bibitem{ATLAS:2022atq}
{\scshape ATLAS} collaboration, G.~Aad et~al., \emph{{Search for Heavy Neutral Leptons in Decays of W Bosons Using a Dilepton Displaced Vertex in $\sqrt{s}=13$ TeV pp Collisions with the ATLAS Detector}}, \href{http://dx.doi.org/10.1103/PhysRevLett.131.061803}{\emph{Phys. Rev. Lett.} {\bf 131} (2023) 061803}, [\href{https://arxiv.org/abs/2204.11988}{{\tt 2204.11988}}].

\bibitem{CMS:2023jqi}
{\scshape CMS} collaboration, A.~Hayrapetyan et~al., \emph{{Search for long-lived heavy neutral leptons with lepton flavour conserving or violating decays to a jet and a charged lepton}}, \href{http://dx.doi.org/10.1007/JHEP03(2024)105}{\emph{JHEP} {\bf 03} (2024) 105}, [\href{https://arxiv.org/abs/2312.07484}{{\tt 2312.07484}}].

\bibitem{CMS:2024ita}
{\scshape CMS} collaboration, A.~Hayrapetyan et~al., \emph{{Search for long-lived heavy neutrinos in the decays of B mesons produced in proton-proton collisions at $ \sqrt{s} $ = 13 TeV}}, \href{http://dx.doi.org/10.1007/JHEP06(2024)183}{\emph{JHEP} {\bf 06} (2024) 183}, [\href{https://arxiv.org/abs/2403.04584}{{\tt 2403.04584}}].

\bibitem{CMS:2024xdq}
{\scshape CMS} collaboration, A.~Hayrapetyan et~al., \emph{{Search for heavy neutral leptons in final states with electrons, muons, and hadronically decaying tau leptons in proton-proton collisions at $ \sqrt{s} $ = 13 TeV}}, \href{http://dx.doi.org/10.1007/JHEP06(2024)123}{\emph{JHEP} {\bf 06} (2024) 123}, [\href{https://arxiv.org/abs/2403.00100}{{\tt 2403.00100}}].

\bibitem{CMS:2024ake}
{\scshape CMS} collaboration, A.~Hayrapetyan et~al., \emph{{Search for long-lived heavy neutral leptons decaying in the CMS muon detectors in proton-proton collisions at $\sqrt{s}=13$ TeV}}, \href{http://dx.doi.org/10.1103/PhysRevD.110.012004}{\emph{Phys. Rev. D} {\bf 110} (2024) 012004}, [\href{https://arxiv.org/abs/2402.18658}{{\tt 2402.18658}}].

\bibitem{CMS:2024hik}
{\scshape CMS} collaboration, A.~Hayrapetyan et~al., \emph{{Search for long-lived heavy neutral leptons in proton-proton collision events with a lepton-jet pair associated with a secondary vertex at $\sqrt{s}$ = 13 TeV}},  \href{https://arxiv.org/abs/2407.10717}{{\tt 2407.10717}}.

\bibitem{T2K:2019jwa}
{\scshape T2K} collaboration, K.~Abe et~al., \emph{{Search for heavy neutrinos with the T2K near detector ND280}}, \href{http://dx.doi.org/10.1103/PhysRevD.100.052006}{\emph{Phys. Rev. D} {\bf 100} (2019) 052006}, [\href{https://arxiv.org/abs/1902.07598}{{\tt 1902.07598}}].

\bibitem{Belle:2024wyk}
{\scshape Belle} collaboration, M.~Nayak et~al., \emph{{Search for a heavy neutral lepton that mixes predominantly with the tau neutrino}}, \href{http://dx.doi.org/10.1103/PhysRevD.109.L111102}{\emph{Phys. Rev. D} {\bf 109} (2024) L111102}, [\href{https://arxiv.org/abs/2402.02580}{{\tt 2402.02580}}].

\bibitem{NA62:2020mcv}
{\scshape NA62} collaboration, E.~Cortina~Gil et~al., \emph{{Search for heavy neutral lepton production in $K^+$ decays to positrons}}, \href{http://dx.doi.org/10.1016/j.physletb.2020.135599}{\emph{Phys. Lett. B} {\bf 807} (2020) 135599}, [\href{https://arxiv.org/abs/2005.09575}{{\tt 2005.09575}}].

\bibitem{Pascoli:2018heg}
S.~Pascoli, R.~Ruiz and C.~Weiland, \emph{{Heavy neutrinos with dynamic jet vetoes: multilepton searches at $ \sqrt{s}=14 $ , 27, and 100 TeV}}, \href{http://dx.doi.org/10.1007/JHEP06(2019)049}{\emph{JHEP} {\bf 06} (2019) 049}, [\href{https://arxiv.org/abs/1812.08750}{{\tt 1812.08750}}].

\bibitem{Drewes:2019fou}
M.~Drewes and J.~Hajer, \emph{{Heavy Neutrinos in displaced vertex searches at the LHC and HL-LHC}}, \href{http://dx.doi.org/10.1007/JHEP02(2020)070}{\emph{JHEP} {\bf 02} (2020) 070}, [\href{https://arxiv.org/abs/1903.06100}{{\tt 1903.06100}}].

\bibitem{Beacham:2019nyx}
J.~Beacham et~al., \emph{{Physics Beyond Colliders at CERN: Beyond the Standard Model Working Group Report}}, \href{http://dx.doi.org/10.1088/1361-6471/ab4cd2}{\emph{J. Phys. G} {\bf 47} (2020) 010501}, [\href{https://arxiv.org/abs/1901.09966}{{\tt 1901.09966}}].

\bibitem{Antel:2023hkf}
C.~Antel et~al., \emph{{Feebly-interacting particles: FIPs 2022 Workshop Report}}, \href{http://dx.doi.org/10.1140/epjc/s10052-023-12168-5}{\emph{Eur. Phys. J. C} {\bf 83} (2023) 1122}, [\href{https://arxiv.org/abs/2305.01715}{{\tt 2305.01715}}].

\bibitem{MATHUSLA:2020uve}
{\scshape MATHUSLA} collaboration, C.~Alpigiani et~al., \emph{{An Update to the Letter of Intent for MATHUSLA: Search for Long-Lived Particles at the HL-LHC}},  \href{https://arxiv.org/abs/2009.01693}{{\tt 2009.01693}}.

\bibitem{Ariga:2018uku}
{\scshape FASER} collaboration, A.~Ariga et~al., \emph{{FASER\textquoteright{}s physics reach for long-lived particles}}, \href{http://dx.doi.org/10.1103/PhysRevD.99.095011}{\emph{Phys. Rev. D} {\bf 99} (2019) 095011}, [\href{https://arxiv.org/abs/1811.12522}{{\tt 1811.12522}}].

\bibitem{Feng:2024zfe}
J.~L. Feng, A.~Hewitt, F.~Kling and D.~La~Rocco, \emph{{Simulating heavy neutral leptons with general couplings at collider and fixed target experiments}}, \href{http://dx.doi.org/10.1103/PhysRevD.110.035029}{\emph{Phys. Rev. D} {\bf 110} (2024) 035029}, [\href{https://arxiv.org/abs/2405.07330}{{\tt 2405.07330}}].

\bibitem{Drewes:2018gkc}
M.~Drewes, J.~Hajer, J.~Klaric and G.~Lanfranchi, \emph{{NA62 sensitivity to heavy neutral leptons in the low scale seesaw model}}, \href{http://dx.doi.org/10.1007/JHEP07(2018)105}{\emph{JHEP} {\bf 07} (2018) 105}, [\href{https://arxiv.org/abs/1801.04207}{{\tt 1801.04207}}].

\bibitem{Krasnov:2019kdc}
I.~Krasnov, \emph{{DUNE prospects in the search for sterile neutrinos}}, \href{http://dx.doi.org/10.1103/PhysRevD.100.075023}{\emph{Phys. Rev. D} {\bf 100} (2019) 075023}, [\href{https://arxiv.org/abs/1902.06099}{{\tt 1902.06099}}].

\bibitem{Ballett:2019bgd}
P.~Ballett, T.~Boschi and S.~Pascoli, \emph{{Heavy Neutral Leptons from low-scale seesaws at the DUNE Near Detector}}, \href{http://dx.doi.org/10.1007/JHEP03(2020)111}{\emph{JHEP} {\bf 03} (2020) 111}, [\href{https://arxiv.org/abs/1905.00284}{{\tt 1905.00284}}].

\bibitem{Alekhin:2015byh}
S.~Alekhin et~al., \emph{{A facility to Search for Hidden Particles at the CERN SPS: the SHiP physics case}}, \href{http://dx.doi.org/10.1088/0034-4885/79/12/124201}{\emph{Rept. Prog. Phys.} {\bf 79} (2016) 124201}, [\href{https://arxiv.org/abs/1504.04855}{{\tt 1504.04855}}].

\bibitem{SHiP:2018xqw}
{\scshape SHiP} collaboration, C.~Ahdida et~al., \emph{{Sensitivity of the SHiP experiment to Heavy Neutral Leptons}}, \href{http://dx.doi.org/10.1007/JHEP04(2019)077}{\emph{JHEP} {\bf 04} (2019) 077}, [\href{https://arxiv.org/abs/1811.00930}{{\tt 1811.00930}}].

\bibitem{Gorbunov:2020rjx}
D.~Gorbunov, I.~Krasnov, Y.~Kudenko and S.~Suvorov, \emph{{Heavy Neutral Leptons from kaon decays in the SHiP experiment}}, \href{http://dx.doi.org/10.1016/j.physletb.2020.135817}{\emph{Phys. Lett. B} {\bf 810} (2020) 135817}, [\href{https://arxiv.org/abs/2004.07974}{{\tt 2004.07974}}].

\bibitem{Antusch:2023nqd}
S.~Antusch, J.~Hajer and J.~Rosskopp, \emph{{Decoherence effects on lepton number violation from heavy neutrino-antineutrino oscillations}}, \href{http://dx.doi.org/10.1007/JHEP11(2023)235}{\emph{JHEP} {\bf 11} (2023) 235}, [\href{https://arxiv.org/abs/2307.06208}{{\tt 2307.06208}}].

\bibitem{Abada:2022wvh}
A.~Abada, P.~Escribano, X.~Marcano and G.~Piazza, \emph{{Collider searches for heavy neutral leptons: beyond simplified scenarios}}, \href{http://dx.doi.org/10.1140/epjc/s10052-022-11011-7}{\emph{Eur. Phys. J. C} {\bf 82} (2022) 1030}, [\href{https://arxiv.org/abs/2208.13882}{{\tt 2208.13882}}].

\bibitem{Tastet:2021vwp}
J.-L. Tastet, O.~Ruchayskiy and I.~Timiryasov, \emph{{Reinterpreting the ATLAS bounds on heavy neutral leptons in a realistic neutrino oscillation model}}, \href{http://dx.doi.org/10.1007/JHEP12(2021)182}{\emph{JHEP} {\bf 12} (2021) 182}, [\href{https://arxiv.org/abs/2107.12980}{{\tt 2107.12980}}].

\bibitem{Hernandez:2016kel}
P.~Hern\'andez, M.~Kekic, J.~L\'opez-Pav\'on, J.~Racker and J.~Salvado, \emph{{Testable Baryogenesis in Seesaw Models}}, \href{http://dx.doi.org/10.1007/JHEP08(2016)157}{\emph{JHEP} {\bf 08} (2016) 157}, [\href{https://arxiv.org/abs/1606.06719}{{\tt 1606.06719}}].

\bibitem{Drewes:2016jae}
M.~Drewes, B.~Garbrecht, D.~Gueter and J.~Klaric, \emph{{Testing the low scale seesaw and leptogenesis}}, \href{http://dx.doi.org/10.1007/JHEP08(2017)018}{\emph{JHEP} {\bf 08} (2017) 018}, [\href{https://arxiv.org/abs/1609.09069}{{\tt 1609.09069}}].

\bibitem{Chrzaszcz:2019inj}
M.~Chrzaszcz, M.~Drewes, T.~E. Gonzalo, J.~Harz, S.~Krishnamurthy and C.~Weniger, \emph{{A frequentist analysis of three right-handed neutrinos with GAMBIT}}, \href{http://dx.doi.org/10.1140/epjc/s10052-020-8073-9}{\emph{Eur. Phys. J. C} {\bf 80} (2020) 569}, [\href{https://arxiv.org/abs/1908.02302}{{\tt 1908.02302}}].

\bibitem{Krasnov:2023jlt}
I.~Krasnov, \emph{{HNL see-saw: lower mixing limit and pseudodegenerate state}},  \href{https://arxiv.org/abs/2307.01190}{{\tt 2307.01190}}.

\bibitem{Gorbunov:2007ak}
D.~Gorbunov and M.~Shaposhnikov, \emph{{How to find neutral leptons of the $\nu$MSM?}}, \href{http://dx.doi.org/10.1088/1126-6708/2007/10/015}{\emph{JHEP} {\bf 10} (2007) 015}, [\href{https://arxiv.org/abs/0705.1729}{{\tt 0705.1729}}].

\bibitem{Blondel:2022qqo}
A.~Blondel et~al., \emph{{Searches for long-lived particles at the future FCC-ee}}, \href{http://dx.doi.org/10.3389/fphy.2022.967881}{\emph{Front. in Phys.} {\bf 10} (2022) 967881}, [\href{https://arxiv.org/abs/2203.05502}{{\tt 2203.05502}}].

\bibitem{Drewes:2022rsk}
M.~Drewes, \emph{{Distinguishing Dirac and Majorana Heavy Neutrinos at Lepton Colliders}}, \href{http://dx.doi.org/10.22323/1.414.0608}{\emph{PoS} {\bf ICHEP2022} (2022) 608}, [\href{https://arxiv.org/abs/2210.17110}{{\tt 2210.17110}}].

\bibitem{FCC:2018evy}
{\scshape FCC} collaboration, A.~Abada et~al., \emph{{FCC-ee: The Lepton Collider}: {Future Circular Collider Conceptual Design Report Volume 2}}, \href{http://dx.doi.org/10.1140/epjst/e2019-900045-4}{\emph{Eur. Phys. J. ST} {\bf 228} (2019) 261--623}.

\bibitem{CEPCStudyGroup:2018ghi}
{\scshape CEPC Study Group} collaboration, M.~Dong et~al., \emph{{CEPC Conceptual Design Report: Volume 2 - Physics \& Detector}},  \href{https://arxiv.org/abs/1811.10545}{{\tt 1811.10545}}.

\bibitem{Blondel:2014bra}
{\scshape FCC-ee study Team} collaboration, A.~Blondel, E.~Graverini, N.~Serra and M.~Shaposhnikov, \emph{{Search for Heavy Right Handed Neutrinos at the FCC-ee}}, \href{http://dx.doi.org/10.1016/j.nuclphysbps.2015.09.304}{\emph{Nucl. Part. Phys. Proc.} {\bf 273-275} (2016) 1883--1890}, [\href{https://arxiv.org/abs/1411.5230}{{\tt 1411.5230}}].

\bibitem{Bondarenko:2019yob}
K.~Bondarenko, A.~Boyarsky, M.~Ovchynnikov and O.~Ruchayskiy, \emph{{Sensitivity of the intensity frontier experiments for neutrino and scalar portals: analytic estimates}}, \href{http://dx.doi.org/10.1007/JHEP08(2019)061}{\emph{JHEP} {\bf 08} (2019) 061}, [\href{https://arxiv.org/abs/1902.06240}{{\tt 1902.06240}}].

\bibitem{Drewes:2019vjy}
M.~Drewes, A.~Giammanco, J.~Hajer and M.~Lucente, \emph{{New long-lived particle searches in heavy-ion collisions at the LHC}}, \href{http://dx.doi.org/10.1103/PhysRevD.101.055002}{\emph{Phys. Rev. D} {\bf 101} (2020) 055002}, [\href{https://arxiv.org/abs/1905.09828}{{\tt 1905.09828}}].

\bibitem{Antusch:2017pkq}
S.~Antusch, E.~Cazzato, M.~Drewes, O.~Fischer, B.~Garbrecht, D.~Gueter et~al., \emph{{Probing Leptogenesis at Future Colliders}}, \href{http://dx.doi.org/10.1007/JHEP09(2018)124}{\emph{JHEP} {\bf 09} (2018) 124}, [\href{https://arxiv.org/abs/1710.03744}{{\tt 1710.03744}}].

\bibitem{Ajmal:2024kwi}
S.~Ajmal, P.~Azzi, S.~Giappichini, M.~Klute, O.~Panella, M.~Presilla et~al., \emph{{Searching for type I seesaw mechanism in a two Heavy Neutral Leptons scenario at FCC-ee}},  \href{https://arxiv.org/abs/2410.03615}{{\tt 2410.03615}}.

\bibitem{Antusch:2017ebe}
S.~Antusch, E.~Cazzato and O.~Fischer, \emph{{Resolvable heavy neutrino\textendash{}antineutrino oscillations at colliders}}, \href{http://dx.doi.org/10.1142/S0217732319500615}{\emph{Mod. Phys. Lett. A} {\bf 34} (2019) 1950061}, [\href{https://arxiv.org/abs/1709.03797}{{\tt 1709.03797}}].

\bibitem{Cvetic:2018elt}
G.~Cveti\v{c}, A.~Das and J.~Zamora-Sa\'a, \emph{{Probing heavy neutrino oscillations in rare $W$ boson decays}}, \href{http://dx.doi.org/10.1088/1361-6471/ab1212}{\emph{J. Phys. G} {\bf 46} (2019) 075002}, [\href{https://arxiv.org/abs/1805.00070}{{\tt 1805.00070}}].

\bibitem{Tastet:2019nqj}
J.-L. Tastet and I.~Timiryasov, \emph{{Dirac vs. Majorana HNLs (and their oscillations) at SHiP}}, \href{http://dx.doi.org/10.1007/JHEP04(2020)005}{\emph{JHEP} {\bf 04} (2020) 005}, [\href{https://arxiv.org/abs/1912.05520}{{\tt 1912.05520}}].

\bibitem{Cottin:2021lzz}
G.~Cottin, J.~C. Helo, M.~Hirsch, A.~Titov and Z.~S. Wang, \emph{{Heavy neutral leptons in effective field theory and the high-luminosity LHC}}, \href{http://dx.doi.org/10.1007/JHEP09(2021)039}{\emph{JHEP} {\bf 09} (2021) 039}, [\href{https://arxiv.org/abs/2105.13851}{{\tt 2105.13851}}].

\bibitem{Boyarsky:2020dzc}
A.~Boyarsky, M.~Ovchynnikov, O.~Ruchayskiy and V.~Syvolap, \emph{{Improved big bang nucleosynthesis constraints on heavy neutral leptons}}, \href{http://dx.doi.org/10.1103/PhysRevD.104.023517}{\emph{Phys. Rev. D} {\bf 104} (2021) 023517}, [\href{https://arxiv.org/abs/2008.00749}{{\tt 2008.00749}}].

\bibitem{Sabti:2020yrt}
N.~Sabti, A.~Magalich and A.~Filimonova, \emph{{An Extended Analysis of Heavy Neutral Leptons during Big Bang Nucleosynthesis}}, \href{http://dx.doi.org/10.1088/1475-7516/2020/11/056}{\emph{JCAP} {\bf 11} (2020) 056}, [\href{https://arxiv.org/abs/2006.07387}{{\tt 2006.07387}}].

\bibitem{Anamiati:2016uxp}
G.~Anamiati, M.~Hirsch and E.~Nardi, \emph{{Quasi-Dirac neutrinos at the LHC}}, \href{http://dx.doi.org/10.1007/JHEP10(2016)010}{\emph{JHEP} {\bf 10} (2016) 010}, [\href{https://arxiv.org/abs/1607.05641}{{\tt 1607.05641}}].

\bibitem{Anamiati:2017rxw}
G.~Anamiati, R.~M. Fonseca and M.~Hirsch, \emph{{Quasi Dirac neutrino oscillations}}, \href{http://dx.doi.org/10.1103/PhysRevD.97.095008}{\emph{Phys. Rev. D} {\bf 97} (2018) 095008}, [\href{https://arxiv.org/abs/1710.06249}{{\tt 1710.06249}}].

\bibitem{Das:2017hmg}
A.~Das, P.~S.~B. Dev and R.~N. Mohapatra, \emph{{Same Sign versus Opposite Sign Dileptons as a Probe of Low Scale Seesaw Mechanisms}}, \href{http://dx.doi.org/10.1103/PhysRevD.97.015018}{\emph{Phys. Rev. D} {\bf 97} (2018) 015018}, [\href{https://arxiv.org/abs/1709.06553}{{\tt 1709.06553}}].

\bibitem{Dib:2017iva}
C.~O. Dib, C.~S. Kim and K.~Wang, \emph{{Signatures of Dirac and Majorana sterile neutrinos in trilepton events at the LHC}}, \href{http://dx.doi.org/10.1103/PhysRevD.95.115020}{\emph{Phys. Rev. D} {\bf 95} (2017) 115020}, [\href{https://arxiv.org/abs/1703.01934}{{\tt 1703.01934}}].

\bibitem{Abada:2019bac}
A.~Abada, C.~Hati, X.~Marcano and A.~M. Teixeira, \emph{{Interference effects in LNV and LFV semileptonic decays: the Majorana hypothesis}}, \href{http://dx.doi.org/10.1007/JHEP09(2019)017}{\emph{JHEP} {\bf 09} (2019) 017}, [\href{https://arxiv.org/abs/1904.05367}{{\tt 1904.05367}}].

\bibitem{Blondel:2021mss}
A.~Blondel, A.~de~Gouv\^ea and B.~Kayser, \emph{{Z-boson decays into Majorana or Dirac heavy neutrinos}}, \href{http://dx.doi.org/10.1103/PhysRevD.104.055027}{\emph{Phys. Rev. D} {\bf 104} (2021) 055027}, [\href{https://arxiv.org/abs/2105.06576}{{\tt 2105.06576}}].

\bibitem{Antusch:2023jsa}
S.~Antusch, J.~Hajer and B.~M.~S. Oliveira, \emph{{Heavy neutrino-antineutrino oscillations at the FCC-ee}}, \href{http://dx.doi.org/10.1007/JHEP10(2023)129}{\emph{JHEP} {\bf 10} (2023) 129}, [\href{https://arxiv.org/abs/2308.07297}{{\tt 2308.07297}}].

\bibitem{Antusch:2024otj}
S.~Antusch, J.~Hajer and B.~M.~S. Oliveira, \emph{{Discovering heavy neutrino-antineutrino oscillations at the Z-pole}}, \href{http://dx.doi.org/10.1007/JHEP11(2024)102}{\emph{JHEP} {\bf 11} (2024) 102}, [\href{https://arxiv.org/abs/2408.01389}{{\tt 2408.01389}}].

\bibitem{Arbelaez:2017zqq}
C.~Arbela\'ez, C.~Dib, I.~Schmidt and J.~C. Vasquez, \emph{{Probing the Dirac or Majorana nature of the Heavy Neutrinos in pure leptonic decays at the LHC}}, \href{http://dx.doi.org/10.1103/PhysRevD.97.055011}{\emph{Phys. Rev. D} {\bf 97} (2018) 055011}, [\href{https://arxiv.org/abs/1712.08704}{{\tt 1712.08704}}].

\bibitem{Balantekin:2018ukw}
A.~B. Balantekin, A.~de~Gouv\^ea and B.~Kayser, \emph{{Addressing the Majorana vs. Dirac Question with Neutrino Decays}}, \href{http://dx.doi.org/10.1016/j.physletb.2018.11.068}{\emph{Phys. Lett. B} {\bf 789} (2019) 488--495}, [\href{https://arxiv.org/abs/1808.10518}{{\tt 1808.10518}}].

\bibitem{Antusch:2022ceb}
S.~Antusch, J.~Hajer and J.~Rosskopp, \emph{{Simulating lepton number violation induced by heavy neutrino-antineutrino oscillations at colliders}}, \href{http://dx.doi.org/10.1007/JHEP03(2023)110}{\emph{JHEP} {\bf 03} (2023) 110}, [\href{https://arxiv.org/abs/2210.10738}{{\tt 2210.10738}}].

\bibitem{Hernandez:2018cgc}
P.~Hern\'andez, J.~Jones-P\'erez and O.~Suarez-Navarro, \emph{{Majorana vs Pseudo-Dirac Neutrinos at the ILC}}, \href{http://dx.doi.org/10.1140/epjc/s10052-019-6728-1}{\emph{Eur. Phys. J. C} {\bf 79} (2019) 220}, [\href{https://arxiv.org/abs/1810.07210}{{\tt 1810.07210}}].

\bibitem{Drewes:2022akb}
M.~Drewes, J.~Klari\'c and J.~L\'opez-Pav\'on, \emph{{New benchmark models for heavy neutral lepton searches}}, \href{http://dx.doi.org/10.1140/epjc/s10052-022-11100-7}{\emph{Eur. Phys. J. C} {\bf 82} (2022) 1176}, [\href{https://arxiv.org/abs/2207.02742}{{\tt 2207.02742}}].

\bibitem{Planck:2018vyg}
{\scshape Planck} collaboration, N.~Aghanim et~al., \emph{{Planck 2018 results. VI. Cosmological parameters}}, \href{http://dx.doi.org/10.1051/0004-6361/201833910}{\emph{Astron. Astrophys.} {\bf 641} (2020) A6}, [\href{https://arxiv.org/abs/1807.06209}{{\tt 1807.06209}}].

\bibitem{Garbrecht:2018mrp}
B.~Garbrecht, \emph{{Why is there more matter than antimatter? Calculational methods for leptogenesis and electroweak baryogenesis}}, \href{http://dx.doi.org/10.1016/j.ppnp.2019.103727}{\emph{Prog. Part. Nucl. Phys.} {\bf 110} (2020) 103727}, [\href{https://arxiv.org/abs/1812.02651}{{\tt 1812.02651}}].

\bibitem{Beneke:2010wd}
M.~Beneke, B.~Garbrecht, M.~Herranen and P.~Schwaller, \emph{{Finite Number Density Corrections to Leptogenesis}}, \href{http://dx.doi.org/10.1016/j.nuclphysb.2010.05.003}{\emph{Nucl. Phys. B} {\bf 838} (2010) 1--27}, [\href{https://arxiv.org/abs/1002.1326}{{\tt 1002.1326}}].

\bibitem{Buchmuller:2005eh}
W.~Buchmuller, R.~D. Peccei and T.~Yanagida, \emph{{Leptogenesis as the origin of matter}}, \href{http://dx.doi.org/10.1146/annurev.nucl.55.090704.151558}{\emph{Ann. Rev. Nucl. Part. Sci.} {\bf 55} (2005) 311--355}, [\href{https://arxiv.org/abs/hep-ph/0502169}{{\tt hep-ph/0502169}}].

\bibitem{Garbrecht:2019zaa}
B.~Garbrecht, P.~Klose and C.~Tamarit, \emph{{Relativistic and spectator effects in leptogenesis with heavy sterile neutrinos}}, \href{http://dx.doi.org/10.1007/JHEP02(2020)117}{\emph{JHEP} {\bf 02} (2020) 117}, [\href{https://arxiv.org/abs/1904.09956}{{\tt 1904.09956}}].

\bibitem{Ghiglieri:2017gjz}
J.~Ghiglieri and M.~Laine, \emph{{GeV-scale hot sterile neutrino oscillations: a derivation of evolution equations}}, \href{http://dx.doi.org/10.1007/JHEP05(2017)132}{\emph{JHEP} {\bf 05} (2017) 132}, [\href{https://arxiv.org/abs/1703.06087}{{\tt 1703.06087}}].

\bibitem{Klaric:2021cpi}
J.~Klari\'c, M.~Shaposhnikov and I.~Timiryasov, \emph{{Reconciling resonant leptogenesis and baryogenesis via neutrino oscillations}}, \href{http://dx.doi.org/10.1103/PhysRevD.104.055010}{\emph{Phys. Rev. D} {\bf 104} (2021) 055010}, [\href{https://arxiv.org/abs/2103.16545}{{\tt 2103.16545}}].

\bibitem{Biondini:2017rpb}
S.~Biondini et~al., \emph{{Status of rates and rate equations for thermal leptogenesis}}, \href{http://dx.doi.org/10.1142/S0217751X18420046}{\emph{Int. J. Mod. Phys. A} {\bf 33} (2018) 1842004}, [\href{https://arxiv.org/abs/1711.02864}{{\tt 1711.02864}}].

\bibitem{Laine:2022pgk}
M.~Laine, \emph{{Sterile neutrino rates for general M, T, \ensuremath{\mu}, k: Review of a theoretical framework}}, \href{http://dx.doi.org/10.1016/j.aop.2022.169022}{\emph{Annals Phys.} {\bf 444} (2022) 169022}, [\href{https://arxiv.org/abs/2203.05772}{{\tt 2203.05772}}].

\bibitem{Hernandez:2015wna}
P.~Hern\'andez, M.~Kekic, J.~L\'opez-Pav\'on, J.~Racker and N.~Rius, \emph{{Leptogenesis in GeV scale seesaw models}}, \href{http://dx.doi.org/10.1007/JHEP10(2015)067}{\emph{JHEP} {\bf 10} (2015) 067}, [\href{https://arxiv.org/abs/1508.03676}{{\tt 1508.03676}}].

\bibitem{Hernandez:2022ivz}
P.~Hernandez, J.~Lopez-Pavon, N.~Rius and S.~Sandner, \emph{{Bounds on right-handed neutrino parameters from observable leptogenesis}}, \href{http://dx.doi.org/10.1007/JHEP12(2022)012}{\emph{JHEP} {\bf 12} (2022) 012}, [\href{https://arxiv.org/abs/2207.01651}{{\tt 2207.01651}}].

\bibitem{Hernandez:2014fha}
P.~Hernandez, M.~Kekic and J.~Lopez-Pavon, \emph{{$N_{\rm eff}$ in low-scale seesaw models versus the lightest neutrino mass}}, \href{http://dx.doi.org/10.1103/PhysRevD.90.065033}{\emph{Phys. Rev. D} {\bf 90} (2014) 065033}, [\href{https://arxiv.org/abs/1406.2961}{{\tt 1406.2961}}].

\bibitem{Vincent:2014rja}
A.~C. Vincent, E.~F. Martinez, P.~Hern\'andez, M.~Lattanzi and O.~Mena, \emph{{Revisiting cosmological bounds on sterile neutrinos}}, \href{http://dx.doi.org/10.1088/1475-7516/2015/04/006}{\emph{JCAP} {\bf 04} (2015) 006}, [\href{https://arxiv.org/abs/1408.1956}{{\tt 1408.1956}}].

\bibitem{Domcke:2020ety}
V.~Domcke, M.~Drewes, M.~Hufnagel and M.~Lucente, \emph{{MeV-scale Seesaw and Leptogenesis}}, \href{http://dx.doi.org/10.1007/JHEP01(2021)200}{\emph{JHEP} {\bf 01} (2021) 200}, [\href{https://arxiv.org/abs/2009.11678}{{\tt 2009.11678}}].

\bibitem{Mastrototaro:2021wzl}
L.~Mastrototaro, P.~D. Serpico, A.~Mirizzi and N.~Saviano, \emph{{Massive sterile neutrinos in the early Universe: From thermal decoupling to cosmological constraints}}, \href{http://dx.doi.org/10.1103/PhysRevD.104.016026}{\emph{Phys. Rev. D} {\bf 104} (2021) 016026}, [\href{https://arxiv.org/abs/2104.11752}{{\tt 2104.11752}}].

\bibitem{Mastrototaro:2019vug}
L.~Mastrototaro, A.~Mirizzi, P.~D. Serpico and A.~Esmaili, \emph{{Heavy sterile neutrino emission in core-collapse supernovae: Constraints and signatures}}, \href{http://dx.doi.org/10.1088/1475-7516/2020/01/010}{\emph{JCAP} {\bf 01} (2020) 010}, [\href{https://arxiv.org/abs/1910.10249}{{\tt 1910.10249}}].

\bibitem{Arguelles:2021dqn}
C.~A. Arg\"uelles, N.~Foppiani and M.~Hostert, \emph{{Heavy neutral leptons below the kaon mass at hodoscopic neutrino detectors}}, \href{http://dx.doi.org/10.1103/PhysRevD.105.095006}{\emph{Phys. Rev. D} {\bf 105} (2022) 095006}, [\href{https://arxiv.org/abs/2109.03831}{{\tt 2109.03831}}].

\bibitem{Kelly:2021xbv}
K.~J. Kelly and P.~A.~N. Machado, \emph{{MicroBooNE experiment, NuMI absorber, and heavy neutral leptons}}, \href{http://dx.doi.org/10.1103/PhysRevD.104.055015}{\emph{Phys. Rev. D} {\bf 104} (2021) 055015}, [\href{https://arxiv.org/abs/2106.06548}{{\tt 2106.06548}}].

\bibitem{Bondarenko:2021cpc}
K.~Bondarenko, A.~Boyarsky, J.~Klaric, O.~Mikulenko, O.~Ruchayskiy, V.~Syvolap et~al., \emph{{An allowed window for heavy neutral leptons below the kaon mass}}, \href{http://dx.doi.org/10.1007/JHEP07(2021)193}{\emph{JHEP} {\bf 07} (2021) 193}, [\href{https://arxiv.org/abs/2101.09255}{{\tt 2101.09255}}].

\bibitem{Izaguirre:2015pga}
E.~Izaguirre and B.~Shuve, \emph{{Multilepton and Lepton Jet Probes of Sub-Weak-Scale Right-Handed Neutrinos}}, \href{http://dx.doi.org/10.1103/PhysRevD.91.093010}{\emph{Phys. Rev. D} {\bf 91} (2015) 093010}, [\href{https://arxiv.org/abs/1504.02470}{{\tt 1504.02470}}].

\bibitem{Das:2017gke}
A.~Das, P.~Konar and A.~Thalapillil, \emph{{Jet substructure shedding light on heavy Majorana neutrinos at the LHC}}, \href{http://dx.doi.org/10.1007/JHEP02(2018)083}{\emph{JHEP} {\bf 02} (2018) 083}, [\href{https://arxiv.org/abs/1709.09712}{{\tt 1709.09712}}].

\bibitem{MEGII:2021fah}
{\scshape MEG II} collaboration, A.~M. Baldini et~al., \emph{{The Search for $\mu^+ \to e^+\gamma$ with $10^{-14}$ Sensitivity: The Upgrade of the MEG Experiment}}, \href{http://dx.doi.org/10.3390/sym13091591}{\emph{Symmetry} {\bf 13} (2021) 1591}, [\href{https://arxiv.org/abs/2107.10767}{{\tt 2107.10767}}].

\bibitem{Mu3e:2020gyw}
{\scshape Mu3e} collaboration, K.~Arndt et~al., \emph{{Technical design of the phase I Mu3e experiment}}, \href{http://dx.doi.org/10.1016/j.nima.2021.165679}{\emph{Nucl. Instrum. Meth. A} {\bf 1014} (2021) 165679}, [\href{https://arxiv.org/abs/2009.11690}{{\tt 2009.11690}}].

\bibitem{Mu2e:2014fns}
{\scshape Mu2e} collaboration, L.~Bartoszek et~al., \emph{{Mu2e Technical Design Report}},  \href{https://arxiv.org/abs/1501.05241}{{\tt 1501.05241}}.

\bibitem{COMET:2018auw}
{\scshape COMET} collaboration, R.~Abramishvili et~al., \emph{{COMET Phase-I Technical Design Report}}, \href{http://dx.doi.org/10.1093/ptep/ptz125}{\emph{PTEP} {\bf 2020} (2020) 033C01}, [\href{https://arxiv.org/abs/1812.09018}{{\tt 1812.09018}}].

\bibitem{Barlow:2011zza}
R.~J. Barlow, \emph{{The PRISM/PRIME project}}, \href{http://dx.doi.org/10.1016/j.nuclphysbps.2011.06.009}{\emph{Nucl. Phys. B Proc. Suppl.} {\bf 218} (2011) 44--49}.

\bibitem{KUNO:17072023}
Y.~Kuno, ``\textit{Charged lepton flavour violation}.'' Plenary talk at the 31st International Symposium on Lepton Photon Interactions at High Energies, Melbourne, Australia, July, 2023.

\bibitem{Calibbi:2017uvl}
L.~Calibbi and G.~Signorelli, \emph{{Charged Lepton Flavour Violation: An Experimental and Theoretical Introduction}}, \href{http://dx.doi.org/10.1393/ncr/i2018-10144-0}{\emph{Riv. Nuovo Cim.} {\bf 41} (2018) 71--174}, [\href{https://arxiv.org/abs/1709.00294}{{\tt 1709.00294}}].

\bibitem{Davidson:2022jai}
S.~Davidson, B.~Echenard, R.~H. Bernstein, J.~Heeck and D.~G. Hitlin, \emph{{Charged Lepton Flavor Violation}},  \href{https://arxiv.org/abs/2209.00142}{{\tt 2209.00142}}.

\bibitem{Georis:2024qgk}
Y.~Georis, \emph{{Recent Developments in Testable Leptogenesis}}, \href{http://dx.doi.org/10.5506/APhysPolBSupp.17.2-A22}{\emph{Acta Phys. Polon. Supp.} {\bf 17} (2024) 2--A22}, [\href{https://arxiv.org/abs/2401.04840}{{\tt 2401.04840}}].

\bibitem{Sandner:2023tcg}
S.~Sandner, P.~Hernandez, J.~Lopez-Pavon and N.~Rius, \emph{{Predicting the baryon asymmetry with degenerate right-handed neutrinos}}, \href{http://dx.doi.org/10.1007/JHEP11(2023)153}{\emph{JHEP} {\bf 11} (2023) 153}, [\href{https://arxiv.org/abs/2305.14427}{{\tt 2305.14427}}].

\bibitem{Granelli:2022eru}
A.~Granelli, J.~Klari\'c and S.~T. Petcov, \emph{{Tests of low-scale leptogenesis in charged lepton flavour violation experiments}}, \href{http://dx.doi.org/10.1016/j.physletb.2022.137643}{\emph{Phys. Lett. B} {\bf 837} (2023) 137643}, [\href{https://arxiv.org/abs/2206.04342}{{\tt 2206.04342}}].

\bibitem{Urquia-Calderon:2022ufc}
K.~A. Urqu\'\i{}a-Calder\'on, I.~Timiryasov and O.~Ruchayskiy, \emph{{Heavy neutral leptons \textemdash{} Advancing into the PeV domain}}, \href{http://dx.doi.org/10.1007/JHEP08(2023)167}{\emph{JHEP} {\bf 08} (2023) 167}, [\href{https://arxiv.org/abs/2206.04540}{{\tt 2206.04540}}].

\bibitem{Drewes:2012ma}
M.~Drewes and B.~Garbrecht, \emph{{Leptogenesis from a GeV Seesaw without Mass Degeneracy}}, \href{http://dx.doi.org/10.1007/JHEP03(2013)096}{\emph{JHEP} {\bf 03} (2013) 096}, [\href{https://arxiv.org/abs/1206.5537}{{\tt 1206.5537}}].

\end{thebibliography}\endgroup

\end{document}